\documentstyle[12pt,epsf]{report} 
\begin{document}   
\pagestyle{empty}

 
\def\mathrm{\rm} 
\newcommand{\ra}{\mbox{$\rightarrow$}}
\newcommand{\pr}{\mbox{\small $\prime$}}
\newcommand{\half}{\mbox{$\frac{1}{2}$}}
\newcommand{\third}{\mbox{$\frac{1}{3}$}}
\newcommand{\thovsxt}{\mbox{$\frac{3}{16}$}}
\newcommand{\thoveit}{\mbox{$\frac{3}{8}$}}
\newcommand{\thovfr}{\mbox{$\frac{3}{4}$}}
\newcommand{\eitovth}{\mbox{$\frac{8}{3}$}}
\newcommand{\sxtovth}{\mbox{$\frac{16}{3}$}}
\newcommand{\onovsij}{\mbox{$\frac{1}{\sij}$}}
\newcommand{\onovsi}{\mbox{$\frac{1}{\sigma_i}$}}
                                                   
\newcommand{\tn}{\mbox{$\tau$}}
\def\t{\tn}
\newcommand{\tm}{\mbox{$\tau^-$}}
\newcommand{\tp}{\mbox{$\tau^+$}}
\newcommand{\nut}{\mbox{$\nu_{\tau}$}}
\newcommand{\nutb}{\mbox{$\bar{\nu}_{\tau}$}}
 
\newcommand{\eln}{\mbox{\rm e}}
\newcommand{\elm}{\mbox{\rm e$^-$}}
\newcommand{\elp}{\mbox{\rm e$^+$}}
\newcommand{\nue}{\mbox{$\nu_e$}}
\newcommand{\nueb}{\mbox{$\bar{\nu}_e$}}
 
\newcommand{\mun}{\mbox{$\mu$}}
\newcommand{\mum}{\mbox{$\mu^-$}}
\newcommand{\mup}{\mbox{$\mu^+$}}
\newcommand{\num}{\mbox{$\nu_{\mu}$}}
\newcommand{\numb}{\mbox{$\bar{\nu}_{\mu}$}}
 
\newcommand{\elln}{\mbox{$\ell$}}
\newcommand{\ellm}{\mbox{$\ell^-$}}
\newcommand{\ellp}{\mbox{$\ell^+$}}
\newcommand{\nul}{\mbox{$\nu_{\ell}$}}
\newcommand{\nulb}{\mbox{$\bar{\nu}_{\ell}$}}      
 
\newcommand{\pin}{\mbox{$\pi$}}
\newcommand{\pim}{\mbox{$\pi^-$}}
\newcommand{\pip}{\mbox{$\pi^+$}}
\newcommand{\piz}{\mbox{$\pi^0$}}
\newcommand{\piK}{\mbox{$\pi$(K)}}
 
\newcommand{\rhon}{\mbox{$\rho$}}
\newcommand{\rhom}{\mbox{$\rho^-$}}
\newcommand{\rhop}{\mbox{$\rho^+$}}
 
\newcommand{\Kstn}{\mbox{\rm K$^{\ast}$}}
\newcommand{\Kstm}{\mbox{\rm K$^{\ast +}$}}
\newcommand{\Kstp}{\mbox{\rm K$^{\ast -}$}}

\newcommand{\ksh}{\mbox{\rm K$^0_s$}}
\newcommand{\Xp}{\mbox{\rm X$^+$}}
\newcommand{\Xm}{\mbox{\rm X$^-$}}

\newcommand{\aonep}{\mbox{\rm a$_1^-$}}
\newcommand{\aonem}{\mbox{\rm a$_1^+$}}
 
\newcommand{\Z}{\mbox{\rm Z$^0$ }}
\def\tt{\mbox{\tp \tm}}
\newcommand{\tautau}{\mbox{\tp \tm}}
\newcommand{\ee}{\mbox{\elp \elm} }
\newcommand{\mumu}{\mbox{\mup \mum} }
\newcommand{\ggmm}{\mbox{$\gamma\gamma$\ra\mumu}}
\newcommand{\led}    {\elln\nulb\nut}
\newcommand{\pid}    {\piK \nut}
 
\newcommand{\tmu}{\mbox{\tn \ra \mun \numb \nut}}
\newcommand{\tel}{\mbox{\tn \ra \eln \nueb \nut}}
\newcommand{\tle}{\mbox{\tn \ra \elln \nulb \nut}}
\newcommand{\tpi}{\mbox{\tn \ra $\pi$ \nut}}
\newcommand{\tpiK}{\mbox{\tn \ra \piK \nut}}
\newcommand{\tro}{\mbox{\tn \ra \rhon\nut}} 
\def\thp{\t \ra h\pi^0\nu}
\newcommand{\tkstr}{\mbox{\tn \ra K$^{\ast}$ \nut}}
\newcommand{\taone}{\mbox{\tn \ra a$_1$\nut}}
\newcommand{\thnpiz}{\mbox{\tn \ra h $>$2\piz \nut}}
\newcommand{\eemm}{\mbox{\ee \ra \mumu} }
\newcommand{\eett}{\mbox{\ee \ra \tautau} }
\newcommand{\pzgg}{\mbox{\piz \ra $\gamma \gamma$}}
\def\p0gg{\pi ^0 \ra \gamma \gamma}   
 
\newcommand{\efswsq}
{\mbox{$\sin^2\theta_{W}^{\mbox{\small \it eff}}$}}
\def\swsq{\sin^2\theta_W} 
\def\cost{\cos\theta}
\newcommand{\cosst}  {\cos^2\theta}
\newcommand{\cst}{\mbox{$\cos\theta$}}
\newcommand{\cstp}{\mbox{$\cos\theta^{\pr}$}}
\newcommand{\csts}{\mbox{$\cos\theta^{\ast}$}}
\newcommand{\cpsi}{\mbox{$\cos\psi$}}
\newcommand{\sspsi}{\mbox{$\sin^2\psi$}}
\newcommand{\cspsi}{\mbox{$\cos^2\psi$}}
\newcommand{\seta}{\mbox{$\sin\eta$}}
\newcommand{\ceta}{\mbox{$\cos\eta$}}
\newcommand{\cseta}{\mbox{$\cos^2\eta$}}
\newcommand{\ctsh}{\mbox{$\cos\frac{\theta^{\ast}}{2}$}}
\newcommand{\stsh}{\mbox{$\sin\frac{\theta^{\ast}}{2}$}}
\newcommand{\cstsh}{\mbox{$\cos^2\frac{\theta^{\ast}}{2}$}}
\newcommand{\sstsh}{\mbox{$\sin^2\frac{\theta^{\ast}}{2}$}}
\newcommand{\act}{\mbox{$|\cos\theta |$}}
\newcommand{\cstsq}{\mbox{$\cos^{2}\theta$}}
\newcommand{\eb}{\mbox{$E_{beam}$}}
\newcommand{\elep}{\mbox{$E_{\ell}$}}
\newcommand{\ehad}{\mbox{$E_h$}}
\newcommand{\ecm}{\mbox{$E_{cm}$}}  
\def\ecma{\langle\ecm\rangle}
\newcommand{\cof}{\mbox{$\chi^2/D.O.F.$}}
\newcommand{\chs}{\mbox{$\chi^2$}}
\newcommand{\xih}{\mbox{$\xi_{had}$}}
\newcommand{\xxi}{\mbox{$x_i$}}
\newcommand{\xxj}{\mbox{$x_j$}}
\newcommand{\xip}{\mbox{$x_i$}}
\newcommand{\xjp}{\mbox{$x_j$}}
\newcommand{\xp}{\mbox{$x$}}
\newcommand{\ptr}{\mbox{$p_{\scriptscriptstyle T}$}}
\newcommand{\xtr}{\mbox{$x_{\scriptscriptstyle T}$}}
\newcommand{\xtp}{\mbox{$x^{\pr}_{\scriptscriptstyle T}$}}
\newcommand{\tpr}{\mbox{$\theta^{\pr}$}}

\newcommand{\fix}{\mbox{$f_i$}}
\newcommand{\fjx}{\mbox{$f_j$}}
\newcommand{\gix}{\mbox{$g_i$}}
\newcommand{\gjx}{\mbox{$g_j$}}
\newcommand{\rxx}{\mbox{${\cal R}_i(\xxi^0,\xip)$}}
\newcommand{\mhs}{\mbox{$m_h^2$}}
\newcommand{\mts}{\mbox{$m_{\tau}^2$}}
\newcommand{\mrs}{\mbox{$m_{\rho}^2$}}
\newcommand{\mpc}{\mbox{$m_{\pi^{\pm}}$}}
\newcommand{\mpz}{\mbox{$m_{\pi^0}$}}
\newcommand{\mt}{\mbox{$m_{\tau}$}}
\newcommand{\mr}{\mbox{$m_{\rho}$}}
\newcommand{\mzsq}{\mbox{$m_{\mbox \rm Z}^2$}}
\newcommand{\mz}{\mbox{$m_Z$}}
\def\mzs{s=m_Z^2}   
\newcommand{\lamt}{\mbox{${\cal A}_{\tau}$}}
\newcommand{\lame}{\mbox{${\cal A}_e$}}
\newcommand{\laml}{\mbox{${\cal A}_{\ell}$}}
\newcommand{\vovat}{\mbox{\rm g$_v^{\tau}$/g$_a^{\tau}$}}
\newcommand{\vovae}{\mbox{\rm g$_v^e$/g$_a^e$}}
\newcommand{\voval}{\mbox{ \rm g$_v^{\ell}$/g$_a^{\ell}$} }
\newcommand{\epstij}
{\mbox{${\cal E}^{\pm}_{ij}(\xip,\xjp,\cst)$}}
\newcommand{\epsti}
{\mbox{${\cal E}^{\pm}_{i}(\xip,\cst)$}}
\newcommand{\epsdi}{\mbox{$\epsilon_{i}(\xip,\cst)$}}
\newcommand{\epsdj}{\mbox{$\epsilon_{j}(\xjp,\cst)$}}
\def\epstpi{{\cal E}^{\pr\pm}_{i}(\xip,\cst)}
\newcommand{\hi}{\mbox{$h^{\pm}_i(\xip)$}}
\newcommand{\hj}{\mbox{$h^{\pm}_j(\xjp)$}}
\newcommand{\beti}{\mbox{$\beta^{\pm}_{i} (\xip,\cst)$}}
\newcommand{\betj}{\mbox{$\beta^{\pm}_{j} (\xjp,\cst)$}}
\def\bbetij{\bar{\beta}^\pm_{ij}}
\def\bbeti{\bar{\beta}^\pm_{i}}
\newcommand{\bntij}{\mbox{$\beta^{non-\tau}_{ij}(\xip,\xjp,\cst)$}}
\newcommand{\ri}{\mbox{$r^{\pm}_i(\xxi)$}}
 
\newcommand{\ptm}{\mbox{$\langle P_{\tau^-}\rangle$}}
\newcommand{\ptp}{\mbox{$\langle P_{\tau^+}\rangle$}}
\newcommand{\pta}{\mbox{$\langle P_{\tau}\rangle$}}
\newcommand{\pfa}{\mbox{$\langle P_{f}\rangle$}}
\newcommand{\ptat}{\mbox{$\langle P_{\tau}\rangle_{\theta}$}}
\newcommand{\dpta}{\mbox{$\Delta \langle P_{\tau}\rangle$}}
\newcommand{\Pt}{\mbox{\rm P$_{\tau}$}}
\newcommand{\ptau}   {P_{\tau}}
\def\pt{P_{\t}}   
\newcommand{\afb}{\mbox{\rm A$_{FB}$}}
\newcommand{\aplfb}{\mbox{\rm A$_{pol}^{FB}$}}
\newcommand{\daplfb}{\mbox{$\Delta$ \rm A$_{pol}^{FB}$}}
\newcommand{\apolf}  {\langle\ptau\rangle^F}
\newcommand{\apolb}  {\langle\ptau\rangle^B}
\newcommand{\sigp}{\mbox{$\sigma_+$}}
\newcommand{\sigm}{\mbox{$\sigma_-$}}
\newcommand{\sigtot}{\mbox{$\sigma_{tot}$}}
\newcommand{\sigtotp}{\mbox{$\sigma_{tot}^{\pr}$}}
\newcommand{\spr}{\mbox{$\sigma^{\pr}$}}
\newcommand{\sij}{\mbox{$\sigma_{ij}$}}
\newcommand{\sijp}{\mbox{$\spr_{ij}$}}
\newcommand{\sip}{\mbox{$\spr_i$}}
\def\sigpp{\sigma(++)}
\def\sigpm{\sigma(+-)}
\def\sigmp{\sigma(-+)}
\def\sigmm{\sigma(--)}      
\newcommand{\sigf}   {\sigma^F}
\newcommand{\sigb}   {\sigma^B}  
\newcommand{\ptrans}{\mbox{\bf p$_{\scriptscriptstyle T}$}}
\def\dsigtw{\frac{d^2\sigma_i}{d\cst\:d\xxi}}
\def\dsigtwp{\frac{d^2\sip}{d\cst\:d\xip}}
\def\dsigth{\frac{d^3\sij}{d\cos\theta\:d\xxi\:d\xxj}}
\def\dsigthp{\frac{d^3\sijp}{d\cst\:d\xip\:d\xjp}}

\def\fz{F_0(s)}
\def\fo{F_1(s)}
\def\ft{F_2(s)}
\def\fth{F_3(s)}
\newcommand{\fos}    {F_0(s)}
\newcommand{\fis}    {F_1(s)}
\newcommand{\fts}    {F_2(s)}
\newcommand{\frs}    {F_3(s)}

\def\qe{q_{e}}
\def\qt{q_{\tau}}
\def\ae{a_{e}}
\def\at{a_{\tau}}
\def\ve{v_{e}}
\def\vt{v_{\tau}}
 
\newcommand{\dzero}{\mbox{$| d_0|$}}
\newcommand{\zzero}{\mbox{$| z_0|$}}
\def\d0{|d_0|}
\def\z0{|z_0|}  
\newcommand{\Excess}{\mbox{\rm E$_{excess}$}}
\newcommand{\dphimax}{\mbox{$\delta \phi_{max}$}}
\newcommand{\Nhl}{\mbox{\rm N$^{HC}_{hits/layer}$}}
\newcommand{\Nhc}{\mbox{\rm N$^{HC}_{layers}$}}
\newcommand{\Nhm}{\mbox{\rm N$^{HC/MU}_{layers}$}}
 
\newcommand{\degree} {$^\circ$}
\newcommand{\cth}    {$|\cos\theta|$}
\newcommand{\id}{identification}
\newcommand{\goro}[1]{\multicolumn{2}{c|}{#1}}
\newcommand{\bb}{\mathrm{b}\overline{\mathrm{b}}}
\def\dedx{\mbox{$\mathrm{d}E/\mathrm{d}x$}}
 
\newcommand{\etal}{\mbox{\it et al.,}~}
\newcommand {\beq} {\begin{equation}}
\newcommand {\eeq} {\end{equation}}
\newcommand {\bea} {\begin{eqnarray}}
\newcommand {\eea} {\end{eqnarray}}
\newcommand{\ind}[1]{#1\index{#1}}

\newcommand{\Ntpair}{30663}
 
\newcommand{\PTEMP}{-13.0}
\newcommand{\PTEMPF}{-13.5}
\newcommand{\PTEMPST}{2.9}
\newcommand{\PTEMPSY}{2.2}
\newcommand{\PTRHO}{-15.7}
\newcommand{\PTRHOST}{2.4}
\newcommand{\PTRHOSY}{1.5}
\newcommand{\PTALL}{-14.9}
\newcommand{\PTALLST}{1.9}
\newcommand{\PTALLSY}{1.3}
\newcommand{\VOVAT}{0.077}
\newcommand{\VOVATSI}{0.012}
\newcommand{\VOVAE}{0.062}
\newcommand{\VOVAESI}{0.016}
\newcommand{\APFBEMP}{-11.1}
\newcommand{\APFBEMPF}{-11.0}
\newcommand{\APFBEMPST}{3.5}
\newcommand{\APFBEMPSY}{0.5}
\newcommand{\APFBRHO}{-7.1}
\newcommand{\APFBRHOST}{2.8}
\newcommand{\APFBRHOSY}{1.5}
\newcommand{\APFBALL}{-8.9}
\newcommand{\APFBALLST}{2.2}
\newcommand{\APFBALLSY}{0.9}
\newcommand{\LAMBDALEP}{*.*}
\newcommand{\LAMBDALEPSIG}{*.*}
\newcommand{\SINWALL}{0.2331}
\newcommand{\SINWALLSI}{0.0023}
 
\newcommand{\MFPTAU}{11.8}  
\newcommand{\MFDPTAU}{3.2}  
 
\newcommand{\MFAPOLFB}{11.8}  
\newcommand{\MFDAPOLFB}{3.7}  
 
\newcommand{\MFXI}{0.83}  
\newcommand{\MFDXI}{0.18}  
 
\newcommand{\MFXIH}{1.13}  
\newcommand{\MFDXIH}{0.16}  
 
\newcommand{\MFRHO}{0.77}  
\newcommand{\MFDRHO}{0.01}  
 
\newcommand{\MFDELTA}{0.87}  
\newcommand{\MFDDELTA}{0.13}  
%
\newcommand{\MDPTAU}{13.1}  
\newcommand{\MDDPTAU}{3.2}  
 
\newcommand{\MDAPOLFB}{11.1}  
\newcommand{\MDDAPOLFB}{3.7}  
 
\newcommand{\MDXI}{0.95}  
\newcommand{\MDDXI}{0.17}  
 
\newcommand{\MDXIH}{1.06}  
\newcommand{\MDDXIH}{0.18}  
 
\newcommand{\MCPTAU}{12.3}  
\newcommand{\MCDPTAU}{2.8}  
 
\newcommand{\MCAPOLFB}{11.4}  
\newcommand{\MCDAPOLFB}{3.4}  
 
\newcommand{\MCXI}{-1.06}  
\newcommand{\MCDXI}{0.02}  
%

\bigskip
 
\begin{center} {\Large\bf
    {\Huge A}NALYSIS OF THE  {\Huge$\t$} {\Huge P}OLARIZATION AND ITS
    {\Huge F}ORWARD-{\Huge B}ACKWARD {\Huge A}SYMMETRY
    ON THE {\Huge\Z}    \\             }
\vspace{10.cm}
        { Thesis submitted to the Senate  
                of the Tel-Aviv University \\
                as part of the requirements for 
                the degree "Doctor of Philosophy"   \\

          by  }
\end{center}

\begin{center}{\Large \bf Erez Etzion}
\end{center}
\vfill
 
\begin{center}
        July 1994
\end{center}
 
\renewcommand{\Huge}{\huge}

\newpage
\vspace{24cm}
{\tiny .}
\newpage
{\tiny .}
\vspace{8cm}

\begin{center}
{\Large
      The research work for this thesis was carried out  \\
      in the Experimental High Energy  Group \\
      of the School of Physics and Astronomy \\
      under the supervision of {\bf Prof. Gideon Alexander} }\\
\end{center}
\newpage
\vspace{24cm}
{\tiny .}
\newpage
\pagestyle{headings}
\pagenumbering{roman} \vspace{8 cm}
\section*{Abstract}
\baselineskip 18pt
This thesis describes a new a measurement of
the tau lepton polarization and its forward-backward asymmetry at the
\Z ~resonance using the OPAL detector.  This measurement is
based on analyses of the $\tel$ , $\tmu$ and $\tpiK$
decays from a sample of $\Ntpair$ $\eett$ events collected
in the polar angle range of $\act <0.68$
during the 1990-1992 data taking period.
Taking then the Standard Model with the
V--A structure of the tau lepton decay,
we measure the average \tn ~polarization to be
 
\[ \pta=(\PTEMPF \pm \PTEMPST \pm \PTEMPSY)\% \]
and the \tn ~polarization forward-backward
asymmetry to be
\[ \aplfb=(\APFBEMPF \pm \APFBEMPST \pm \APFBEMPSY)\% \]
where the first error is statistical and the second systematic.
Combining these figures with the OPAL $\pta$ and $\aplfb$ measured with
the $\tro$ channel we get an average $\tn$ polarization,
\[ \pta= (\PTALL \pm \PTALLST \pm \PTALLSY)\% \]
 and for the asymmetry in the polarization,
 \[ \aplfb=(\APFBALL \pm \APFBALLST \pm \APFBALLSY)\% \]
These results are
consistent with  lepton universality.
Combining the two results we obtain for the electroweak mixing angle
the value
\[ \efswsq=\SINWALL \pm \SINWALLSI \]
within the context of the Standard Model, where the error includes both
statistical and systematic uncertainties.
\newpage

\tableofcontents
\listoffigures
\listoftables
\newpage 
\pagestyle{empty}
\vspace{24cm}
{\tiny .}
\newpage 
\pagestyle{headings}
\pagenumbering{arabic}
\chapter{Introduction}
One of the central phenomena characterizing weak interactions
is the non-conservation of parity.
This effect, originally established for
weak charged-current interactions, is also embedded in the Standard
Model~\cite{GWS1,GWS2,GWS3} to exist in neutral-current interactions,
resulting in
different $\Z$ couplings to left-handed and right-handed fermions.
Consequently, the $\Z$ boson produced by $\ee$ annihilation
is expected to be polarized because of its
different couplings to the incoming left-handed and right-handed electrons.
Similarly, fermions produced in $\Z$ decay are expected to
have a degree of polarization depending on their coupling constants.
Another consequent of this $\Z$ polarization is a
forward-backward asymmetry in the polarization of the outgoing fermions.
Hence, some of the best Standard Model tests in the
annihilation
of $e^+e^-$ at the $\Z$ pole provided  by  the
following asymmetries measurements:
\begin{itemize}
\item {\bf The Forward Backward asymmetry, $\afb$,} which can be measured in
all the $e^+e^-\to f\bar{f}$ channels.
\item {\bf The Left Right Asymmetry, $A_{LR}$,} which can be measured in the
annihilation of longitudinal polarized $e^+e^-$ beams (at the moment
available only at SLAC's  SLD experiment).
\item {\bf The Final Lepton Polarization Asymmetry, $\pfa$}
which can be measured only in the $\tn$ decay ($\pta$).
\item {\bf The Forward Backward Asymmetry in the Lepton Polarization,
$\aplfb$,} which
again can be measured only in the  $\tn$ decay .
\end{itemize}
The $\pta$ and the $\aplfb$ asymmetries can be
studied in the process
\eett ~using the energy distribution of the $\tn$ decay
products in the laboratory frame~\cite{CALASY}.
 
The mixing angle between the electromagnetic and the weak interaction, $\swsq$,
plays a central role in the theory of the Standard Model.
Being so it is measured by the LEP experiments in several  processes and
by various methods.
The Standard Model gives predictions for $\afb$, $\pta$ and $\aplfb$
as a functions of $s=\ecm^2$ in terms of the $\Z$
parameters (mass and width) and its vector ($v_l$) and
axial-vector ($a_l$) couplings to the electron and the tau.
Using the improved Born
approximation~\cite{bib-IBA} which accounts for most weak radiative
corrections,
our measurements of $\pta$ and $\aplfb$
provide a test of e-\tn ~universality in the neutral current which is
independent of lepton universality tests performed
by studying the line
shapes and the forward-backward asymmetries of the Bhabha,
\mun -pair  and
\tn -pair cross-sections~\cite{bib-z0par}.
 
Our results   supercede  OPAL's
first $\tn$ polarization measurements~\cite{OPALPL}
which was based solely on the 1990 data.
The other LEP collaborations have also reported their measurements of
the $\t$ polarization~\cite{ALE1PL}-\cite{L32}
forward-backward asymmetry~\cite{ALE2PL}-\cite{L32}, based on their 1990 and 1991
data.

The present thesis describes a measurement of the $\tn$ polarization, ~\pta,
and its forward-backward asymmetry, ~\aplfb ,
using the $\tn$-pair events collected with the OPAL
detector at LEP during the period 1990-1992. It is based on a sample of
\Ntpair ~\eett events which were detected within the polar angle range
of\footnote{The coordinate system is
defined with $z$ along the e$^-$ beam direction, $\theta$ and $\phi$ being
the polar and azimuthal angles, respectively.}
$|\cos\theta|<0.68$.
Most of these events (91.5\%) were measured on the $\Z$
peak and the remainder at center-of-mass energies (\ecm)
of 1, 2, and 3 ~$GeV$ below and
above the peak of $\Z$ resonance.
The decay channels $\tel$, $\tmu$ and $\tpiK$
are used.
 
These results combined with the analysis of the
$\tro$ decay channel~\cite{TN203,TN142} were presented as an OPAL
publication~\cite{PN126,TN202}.
Therefore for the sake of completeness
we present here also the combined results.
 
The analysis is
based on an event-by-event Maximum Likelihood  fit to
the theoretical energy distributions of the $\tn$ decay product.
The energy is corrected for
radiative effects and detector response, in which all correlations
between the
two tau decays are taken into account. We
apply our `global fit'  method to the
three $\tn$ decay channels from which the polarization is extracted using
distributions in simple observables.
This method has the advantage
of taking explicitly into account the correlations between polarization
observables of the two taus introduced by the selection and identification
criteria.
This is particularly important for the leptonic channels where
requirements are made on the whole event in order to suppress backgrounds
from Bhabha and  mu-pair events. In addition, the method correctly
extracts information using the tau-tau spin correlations in those
events where both taus have identified decays.

 
The thesis is organized as follows:
A description of the LEP accelerator and the OPAL detector
are presented in the following two Chapters.
A brief description of the Standard Model and a
discussion of the polarization formalism, are given in the
Chapter~\ref{chap-theory}. It includes the definition of the
various observables used in this analysis and the relations between them.
 
Chapter~\ref{chap-select}
presents the data and Monte Carlo samples used in the
analysis, the event selection and $\t$-decay identification.
Chapter~\ref{chap-anal} describes the fit procedure
and all the corrections used in our analysis,
starting with radiative effects (Sect.~\ref{sect-RCC}), detector
resolution (Sect.~\ref{sect-DR}), $\t$-pair selection efficiency
(Sect.~\ref{sect-TSE}), $\t$ decay identification efficiency
(Sect.~\ref{sect-TDI}), and completing this list with background from other $\t$
decay channels (Sect.~\ref{BOD}) and from non-$\t$ events
(Sect.~\ref{BNT}). Each correction is described along with the
associated systematic studies. The results, including a summary of the
systematic errors, and a presentation of the consistency checks
done on the analysis, are included in Chapter~\ref{chap-RCC}.
Finally the analysis is summarized in Chapter~\ref{chap-SUM}.
 

\chapter{The LEP Accelerator}
\label{chap-LEP}
\section{General Features}
\begin{figure}[htbp]
\epsfysize=16cm.
\epsffile[  0 70 800 770  ]{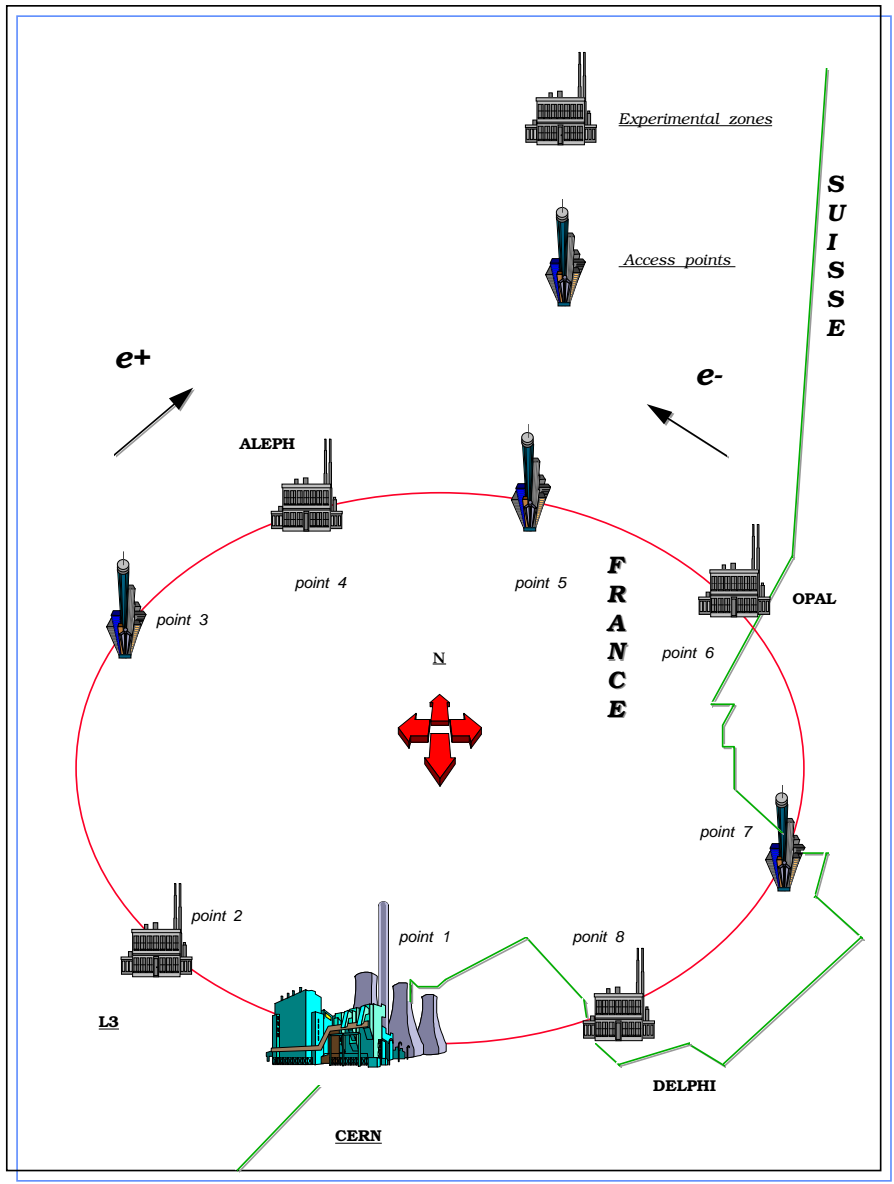}
\caption [ A schematic layout of the LEP storage ring.]
{ A schematic layout of the LEP storage ring. Point 1 is the
          injection point,
          while the even points are the four detectors:
          L3, ALEPH, OPAL, and DELPHI.
          The circumference of this ring is 26.6$~km$ and it
          lies between $43~m$ and $140~m$
          under the ground. }
\label{fig:LEP1}
\end{figure}

The electrons and positrons which annihilate in the experiment
are produced in the
Large Electron Positron ring LEP,  the current worlds largest $e^+e^-$
collider~\cite{LEP1}-\cite{LEP4}.
 At the present stage LEP-I can provide colliding beams of energies
upto $\sim47$~$GeV$. With the machine upgrade planed for the next years
LEP-II will accelerate the beam to $\sim 100$~$GeV$.
The circumference of LEP is 26.66~$km$, partly in France
and partly in Switzerland. The collider tunnel lies
between about 43 $m$ and
140 $m$ below the surface. As injectors, LEP uses existing
CERN machines.
The LEP ring, has four interaction regions and was designed to reach
a maximum luminosity of about L $\sim 1.7 \times 10^{31}~cm^{-2}s^{-1}$ at a
center of mass
energy around $\ecm=90$ $GeV$.
The storage ring has eight long straight sections, four of which
are used for the particle physics experiments: L3, ALEPH, OPAL, and
DELPHI.
Fig.~\ref{fig:LEP1}  shows a schematic layout of the  LEP installations.
The construction of LEP and its four detectors were completed in
July 1989, and the first $Z^0$ event was seen in OPAL on the $13^{th}$ of
August  1989.

\subsection{Structure}
The LEP storage ring is constructed out of
 four basic types of components, namely:
\begin{itemize}
\item {\em The arc.}  The arc bends the beam in a circle and focuses
      it at the same time.
      These regions consists of  regular arrays of
      quadrupoles and bending magnets, and most of the beam parameters
      are determined by its optics.
\item {\em Dispersion suppressor.} Particles with different energies have
      different orbits in the ring. This can increase the beam size in
      the interaction point which reduces the luminosity.
      In order to avoid the luminosity reduction the dispersion
      suppressor are introduced
      in the RF section and in the low beta insertion.
\item {\em RF cavities.} The function of it is to
      replace the energy lost by the
      particles through synchrotron radiation and other effects, and to
      provide a phase focussing.
\item {\em Interaction region.} In theses regions the two beams collide.
      In order to get a high rate of interactions the
      beams are strongly focussed at these collision points.
      This is obtained by the low beta quadrupoles and a set of
      additional electro-magnetic lenses which optically match the insertion
      to the rest of the
      machine.
\end{itemize}
The focussing and bending property of the beam optics leads to two
effects.  First, a particle with nominal energy but with an
uncertainty in position $\Delta x$ and/or
in angle $\Delta \varphi$, will get focussed toward the nominal
orbit and
oscillate around it. This oscillation is called the
{\em betatron oscillation} and it has the form
\beq
x(\ell)=\alpha \sqrt{\beta(\ell)} \cos(\phi(\ell)+\phi_0).
\eeq
where $\ell$ is the longitudinal position in the ring.
The square root of the periodic function $\beta (\ell)$ gives the envelope
of the betatron oscillation over many turns.
The $\beta$ function value is large in a
focussing quadrupole and small in a defocussing one.
 
The additional properties of the focussing and bending
is to restrict the particles to an energy region $\Delta$E around the
nominal energy . On one hand such particles
with an excess energy do bend less in the dipole magnets,
but on the other hand they go off center through the quadrupoles and
therefore receive an extra bending.
The result is, that the off-energy closed orbit particles have a
horizontal displacement $\Delta x_h$
\beq
\Delta x_h = D_x \frac{\Delta E}{E}
\eeq
where $D_x$ is called the periodic dispersion,
and is a function of the longitudinal
position in the ring, $D_x=D_x(\ell)$. The dispersion is typically
large at the horizontally focussing quadrupoles and
typically small at the defocussing ones. In LEP it is arranged so that
the dispersion vanishes in all the straight sections
by using proper transition to the arc, namely, the dispersion
suppressors.
In the arc section, where there is a finite dispersion, the
deviation of the circumference is proportional to the energy deviation.
The RF acceleration works in such a periodic function that a particle
with an excess energy, which has a longer path length than a nominal
energy one, will arrive later to the RF voltage area. In its time of
arrival the voltage is smaller, and therefore it will suffer an energy loss.
All this leads to oscillation around a Gaussian equilibrium  of the LEP
particles in their energy, longitudinal position,
and transverse angle and position.
\section{Luminosity}
The reaction rate is proportional to the luminosity, $L$,
and the reaction cross section, $\sigma$. The luminosity is a machine
parameter, given by:
\beq
        L=\frac{N^2 k f_0}{4\pi  \sigma_x \sigma_y}
\eeq
where $N$ is the number of particles per bunch,
$k$ the number of bunches,$f_0$ the revolution frequency
and $\sigma_x$ and $\sigma_y$ the
RMS values of the transverse radii at the collision points.
 
However there is an important limitation imposed by the beam-beam
effect, due to the electromagnetic forces between the electrons
and the
positrons in the two crossing beams. This force has a linear part
which leads to a
change of the betatron frequency called the beam-beam tune shift.
The particles oscillating with small amplitudes
compared to $\sigma_x$ and $\sigma_y$ at the interaction points,
suffer from the largest change in their tune. This shift
is given by:
\beq
    \xi_i=\frac{N r_e \beta_i}{2 \pi y \sigma_x(\sigma_x+\sigma_i)}
\eeq
where the index $i$ stands for $x$ or $y$, $r_e$ is the classical
electron radius, $\beta_x$ and $\beta_y$ are the transverse
beta function
at the crossing  points. It turns out that
for most storage rings experiments,
that the upper limit on $\xi_ y$ is between 0.03 and 0.06.
To optimize the luminosity estimation we can choose
a tune shift as large as possible and this leaves us with
\beq
      L=\frac{Nf_0k_y\xi_y}{2r_e\beta_y}.
\eeq
Therefore, the experimental collision points are arranged in
low $\beta$ insertions where $\beta_x$ and $\beta_y$ are much smaller
than in the rest of LEP, namely that
$\sigma_x$ and $\sigma_y$ become a fraction of a millimeter.
 
The cross section $\sigma$ is a property of the reaction
itself, and one often compares it to the QED process
 $e^+e^-\stackrel{\gamma}{\rightarrow} \mu^+\mu^-$.
The cross section to this process decreases with  $1/s$.
It value is $\sim 10^{-35}~cm^{2}$ when $ E_{beam}=M_Z/2$,
which means that with  a typical LEP
luminosity of $10^{31}~cm^{-2} s^{-1}$ one will obtain a
$\Z$ counting rate of $\sim 0.36~hour^{-1}$.
On the $Z^0$ mass the main annihilation channel is
via the formation of Z boson, $e^+e^-\stackrel{Z^0}{\rightarrow} X$,
with a cross section  higher by a factor of $5 \times 10^3$ than
the propagation via a $\gamma$.
Thus we  expect on the peak about 1800 events per hour.
 
A typical luminosity for LEP during 1990 was about
$0.4 \times 10^{31}~cm^{-2}s^{-1}$ at the beginning of a fill.
This leads to an average of one multihadronic event\footnote{
\Z decays to quark-antiquark pairs
produce final states with many hadrons, refers in the following as MH}
every $8~sec$, forward Bhabha event every $5~sec$, and $Z^0$ decay into
charge leptons every $50~sec$.
The \Z~production rate has further increased in the following years 
as  can be seen
in Table~\ref{lumi} which gives  
the OPAL integrated luminosity in the years 1990-1992. 
In 1993 the four LEP experiments together saw 3 million $\Z$s,
making a total of some 8 million since LEP began operating.
 
\begin {table} [htb]
\begin{center}
\begin{tabular}{||c|c|c||}  \hline  \hline
 Year         & Int. luminosity  &No. of MH    \\  \hline
 1990         & 5.637 $pb^{-1}$ & 147425   \\
 1991         & 12.236 $pb^{-1}$ & 353324   \\
 1992         & 21.6319 $pb^{-1}$ & 767156   \\
  \hline \hline
\end{tabular}
\caption[Integrated luminosity in OPAL.]
{The integrated luminosity  and the number of
MH events measured  with the OPAL detector in the years 1990-1992.}
\label{lumi}
\end{center}
\end{table}

\subsection{Energy Loss Due to Synchrotron Radiation}
The energy and the number of particles in a bunch are both limited
by the synchrotron radiation. Electrons and positrons circulating in an
orbit with radius $\rho$ lose energy due to synchrotron radiation.
This loss per turn is given by:
\beq
       U_s=\frac{4 \pi}{3}\frac{ r_e m_0 c^2 E^4}{\rho}
\eeq
where $\rho$ is the bending radius of the arc which is 3096 $m$ in LEP.
From this follows that the energy loss is 140 $MeV$ and 2330 $MeV$ for
LEP operated at energies of 47 $GeV$ and 95 $GeV$ respectively.
This loss has to be replaced by the RF system. The RF requirements
are dominated by the losses in the cavities, which in
storage rings increase with the beam energy.
Therefore, synchrotron radiation causes sharp upper limits on both
current and energy.
 
\section{The Injection Chain}
\begin{figure}[htbp]
\epsfysize=12cm.
\epsffile[0 50 622 711]{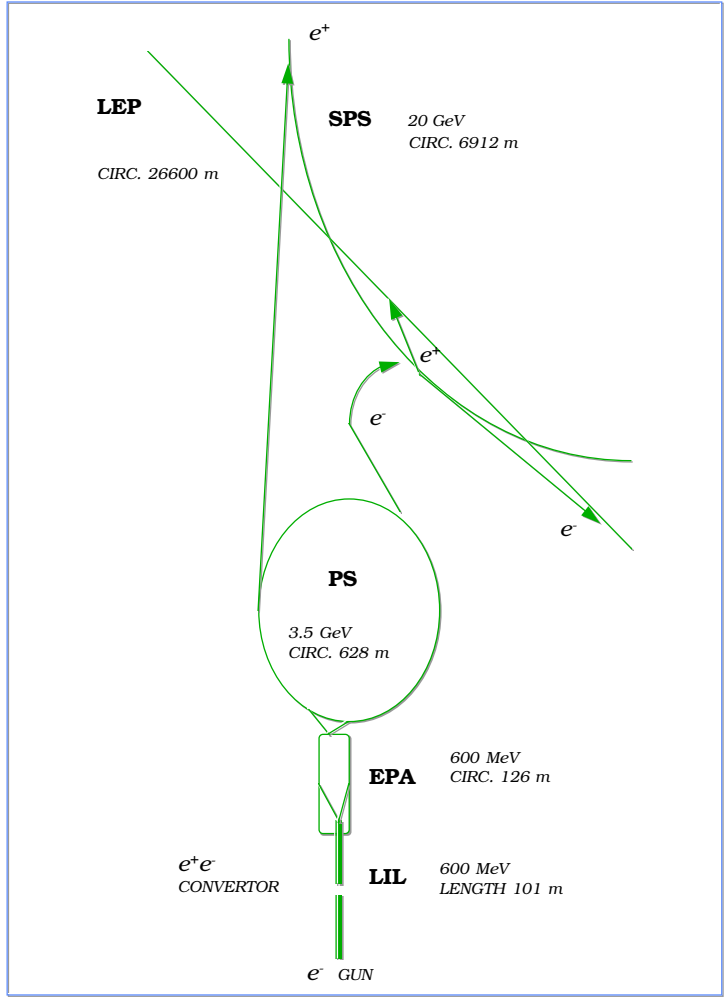}
\caption [LEP injection chain]
{Injection into LEP. The LIL output are $600~MeV$ electrons and
         positrons. The EPA transfer the $e^-$ and $e^+$ as appropriate
         into the $3.5~GeV$ PS ring. From the PS the $e^-$ and $e^+$ are
         injected to the SPS ring where they are accelerated to $20~GeV$,
         and finally injected into LEP in eight equally
         spaced bunches of electrons and positrons.}
\label{fig:LEP2}
\end{figure}
The scheme for injection into LEP is shown in Fig.~\ref{fig:LEP2}.
An electron linac {\it (LINear ACcelerator)}
which is a high current, 200 $MeV$ machine producing an output current of
2.5 $A$ for electron-to-positron conversion in a tungsten target.
From there a positron current of 12 $mA$ is produced in a form suitable
for subsequent acceleration in another linac which is used to accelerate
these positrons up to an energy of 600 $MeV$. The first linac is also
used to provide electrons for LEP by detuning its electron gun to
produce 110 $mA$ at the linac output for acceleration to 600 $MeV$ in
the second linac. The next stage in this chain is transferring the
electrons or positrons, as appropriate, into an electron-positron
accumulator ring (EPA) where they are stored and accumulated before
injection into the Proton Synchrotron (PS).
Acceleration to 3.5 $GeV$ takes place in the PS, followed by transfer to
the Super Proton Synchrotron (SPS) for acceleration to 20 $GeV$ and
finally injection into LEP. The 20 $GeV$ electrons and positrons are
stacked in LEP in eight (four until the middle of 1992) equally
spaced bunches for each sign of particles. This whole process is repeated
successively until the required currents are reached.

\subsection{Magnet System}
For bending, magnets at LEP have low fields
between $0.05~T$ and $0.1~T$.
The accelerated $e^+$ and $e^-$ trajectories are bent by 3304
dipole cores installed over the whole $20~km$ of the arcs.
At each end of an arc, there is a special weak-field dipole where the
magnetic field has about $\frac{1}{10}$ of its normal value,
to shield the experiments from synchrotron
photons emitted in the main bending dipoles.
At regular intervals  488 quadrupoles magnets are installed,
to focus the beam around the arc. These quadrupoles, $1.6~m$
magnetic length, have a maximum gradient of 9.7 $T m^{-1}$.
In the long straight sections,
there are 288 stronger quadrupoles 2 $m$ long
with a maximum gradient of 10.9 $T m^{-1}$.
On each side of the of the four experiments a stronger
2 $m$ superconducting quadrupole is installed providing a field gradient
of 36 $T m^{-1}$ to squeeze the transverse beam dimension as much as
possible in order to increase the luminosity.
 
LEP, as other colliders, suffers from chromaticity, which is
a phenomenon of particles with off momentum values which are focused
differently in the quadrupoles.
In order to correct the machine's chromaticity, 504 sextupoles magnets
of two different lengths 0.4 $m$ and 0.76 $m$,  with a maximum strength
$B''= 180~T m^{-2}$, were installed near the quadrupoles in the arc.
This produce a quadratic field variation across the apparatus.
The closest orbit distortions resulting from the remaining field, and
alignment errors
of the dipoles and quadrupoles, are corrected by means of $616\times0.5\: m$
long orbit correctors, which have a maximum field between 0.04 $T$ and
0.08 $T$ and are also placed near the quadrupoles.
Other magnets exist in smaller numbers, mainly near the interaction and
the injection points.
Most of the magnet coils are water-cooled to remove the electrical power
dissipated.
\subsection{RF System}
There are eight
$e^-$ bunches turning counter-clockwise,
and eight $e^+$ bunches
turning clockwise symmetrically spaced in the LEP ring (four$\times$four
at until 1992).
The beam intensity is built up by multiple injections from the SPS
into the same eight bunches. Once the required intensity is reached, the
two beams are simultaneously accelerated to their final energy by the RF
system and kept in their respective trajectories throughout the run.
The acceleration structure consists of 128 coupled cavity units, each
containing a five-cell accelerating cavity of 2.12 $m$ active length,
which can give a peak gain of 3.1 $MeV$ to an electron or positron, i.e.
the whole structure gives a maximum accelerating gradient of 1.47
$MV m^{-1}$. The nominal value of peak RF voltage per revolution is 400
$MV$, and to drive the cavities at their full rating, a total of 16
$MW$ RF power is required.
\subsection{Vacuum System}
In order to limit the particle loss rate due to beam-gas interactions,
LEP requires a vacuum of $10^{-9}~Torr$ when the beam is circulating,
which is equivalent to the requirement of $10^{-11}~Torr$ without beam.
The main difficulty of the vacuum system, is the synchrotron radiation
which causes an outgassing of gas molecules, and can in this way raise
the pressure by orders of magnitude during the beginning of a run.
The outgassing is highly reduced
with the increasing of the beam.
This effect, called the beam cleaning, allows the  installation
of a smaller pumping system.
The huge 26.66 $km$ vacuum chamber of LEP is divided into
sectors of up to 474 $m$ length. Each sector consists of two pumping
systems,
one for obtaining the starting vacuum of $\sim 10^{-5}$ $Torr$, and
the second for reaching and maintaining the needed ultra-high vacuum.
The synchrotron radiation also causes the
heating of the vacuum chambers.
Therefore the chambers
are made of aluminum which has a
good thermal conductivity and these chambers are also surrounded
by water-cooling channels.
The heat which escapes the aluminum can cause
severe radiation problems and therefore the aluminum is covered with
lead of up to 8 $mm$ in thickness.
\subsection{Electrostatic Separators}
During the injection and acceleration, it is necessary to separate the
$e^+$ and the $e^-$ beams.
To achieve this, each of the eight
collision points is equipped with four electrostatic separators. These
separators can force a total
distance of 0.49 $mm$ between the $e^+$ and the  $e^-$ bunches at
55 $GeV$ in the low beta insertions, and a 3.25 $mm$ at the high beta
insertions. The system  is designed so that at higher LEP energies,
sufficient separation can be obtained
by adding two more separators at each experimental collision region.
After the needed energy is reached, the beams are brought into collision
in the four experimental points, while being kept separated elsewhere
in LEP.
 
\subsection{LEP Parameters}
\nopagebreak
Table~\ref{tablep} gives some general details on the LEP accelerator,
and its magnetic and RF system. In one
column are the LEP-I parameters as they are at present,
and for a comparison, the LEP-II designed
parameters are also given.
 
\begin{table}[h]
\begin{center}
\begin{tabular}{|l|c|c|c|}  \hline 
    &units &phase 1  &phase 2    \\
\hline
Circumference of the machine   & $km$    &26.658   &26.658  \\
Inner diameter of the tunnel   & $m$     &3.80     &3.80    \\
Radius of curvature in one dipole  &$m$  &3096     &3096    \\
Injection energy               &$GeV$    &20       &20      \\
Number of $e^+$ and $e^-$ bunches &    &8(4)        &4       \\
Bunch length                   &$m$      &0.013 to 0.04 &0.012 to 0.04 \\
Maximum luminosity
                               &$cm^{-2}s^{-1}$  &$1.7\times10^{31}$
                                                &$7\times10^{31}$ \\
Circulating current per beam  &$mA$     &3.0      &3.0      \\
{\bf Magnetic System}         &        &        &         \\
Dipoles magnetic field at injections  &$T$  &0.0215  &0.0215 \\
Maximum dipoles magnetic field &$T$     &0.0590  &0.1075     \\
Number of iron dipoles       &        &3304    &3304       \\
Number of dipoles with weak fields   &  &64    &64         \\
Number of dipoles in the injection region & &24 &24       \\
Number of quadrupoles        &    &816   &800 to 816     \\
Number of focussing sextupoles  &  &248  &248   \\
Number of defocussing sextupoles &  &256  &256  \\  \hline
{\bf Radio-Frequency System}    &  &    &       \\
Number of equipped straight sections &  &2  &4   \\
Number of klystrons            &    &16   &20    \\
RF frequency                   &$MHz$   &352.21    &352.21  \\
Necessary circumferential voltage  &$MV/turn$ &360 &3224    \\
Synchrotron energy loss per particle &$MeV/turn$   &260   &2830 \\
Total synchrotron radiation power &$MW$  &1.6 &17.1  \\ \hline 
\end{tabular}
\caption[The design LEP parameters.]
{ Design LEP parameters. The third column
shows LEP-I parameters, the LEP-II designed
parameters are given in the fourth column.}
\end{center}
\label{tablep}
\end{table}

\chapter{The {OPAL Detector}}
 
OPAL ({\bf O}mni-{\bf P}urpose {\bf A}pparatus for {\bf L}EP) is
a multipurpose apparatus designed to reconstruct efficiently
and identify all types of e$^+$e$^-$ events.
Furthermore, OPAL was designed so that it will  has a
good acceptance for $Z^0$ decays over a solid angle of nearly $4\pi$.
A general layout of the detector is shown
in Fig.~\ref{fig:FOPAL1}, indicating
the location and relative size of the
various components.
 
\begin{figure}[p]
\epsfysize=15cm.
\epsffile[130 0  1390 680]{fig3_1.eps}
 
\caption [ A schematic view of the OPAL detector.]
{ A schematic view of the entire OPAL detector.
          The incident beams
          travel along the central beam pipe and collide at the center
          of the detector. In each direction particles traverse the
          central detector, followed by the electromagnetic and
          hadronic calorimeters and finally the muon chambers.
          Particles produced at small angles are detected by the
          forward detectors. }
 
\label{fig:FOPAL1}
\end{figure}
 
The main features of the detector are:
\begin{itemize}
\item Tracking of charged particles, performed by the central detector,
and providing
 measurements of the particles direction and momentum,
 their identification by $dE/dx$ and reconstruction of primary
 and secondary vertices at and near the interaction region.

\item Identification of photons and electrons by the Electromagnetic
Calorimeter and measurement of their energy.
 
\item Measurement of hadronic energy by total absorption
using the magnet yoke instrumented as a  Hadron Calorimeter.
 
\item Identification of muons by measurement of their position
and direction  within and behind the hadron absorber.
 
\item Measurement of absolute machine luminosity
  using  Bhabha scattering events in the
  very forward direction with respect to the beam line.
\end{itemize}
 
The OPAL has a 3D cartesian coordinate system
whose origin is at the nominal interaction point. It has
the $z$-axis along the nominal electron beam direction (this is anticlockwise
around LEP when viewed from above),  the $x$-axis horizontal and directed
towards the center of LEP, and  the $y$-axis normal to the $z$-$x$ plane.
 
 
Full
details of the OPAL detector can be found in Ref.~\cite{ROPAL1,ROPAL2}.
Only a very brief introduction is given in
the successive sections  describing the parts of the detector
most important to this analysis. In
particular we concentrate on
the methods  the sub-detectors information are utilized
in the event selection
and the physics analysis.
 
 
\section{The Central Detector}
 
The Central Detector\index{CD} (CD) consists of a
Silicon Microvertex detector and
three drift chamber devices,
the vertex detector, jet chamber and surrounding Z-chambers situated
inside a pressure vessel of 4 bar.
The central detector is inside a solenoid supplying
a uniform axial magnetic field of 0.435~$T$\index{Magnetic field}.
Originally and until 1991 there was no Silicon detector and the inner wall of
the pressure vessel at 7.8~$cm$ radius formed the beam pipe.
This beam pipe\index{Beam pipe} consisted of 0.13~$cm$ thick carbon
fibre with a $100~\mu m$ aluminium inner lining.
In 1991 a second beam pipe was added at a radius of 5.35~$cm$, consisting of
0.11~$cm$ thick Beryllium,  and
the Silicon detector inserted between the pipes.
 

\subsection{\ind{Jet Chamber}}
 
The jet chamber (\ind{CJ}) is a cylindrical drift chamber of
length 400~$cm$  with an
outer radius of 185~$cm$ and inner of 25~$cm$. The chamber consists of
24 identical sectors each containing a sense wire plane of 159 wires
strung parallel to the beam direction. The end planes are conical
and can be described by $|z|=147+0.268\times R$.
 
\noindent
The coordinates of wire hits in the $r-\phi$ plane are
determined from a
measurement of drift time. The $z$ coordinate is measured using a
charge division technique and by summing the charges received at each
end of a wire allows the energy loss, ${\rm d}E/{\rm d}x$, to be
calculated.
 
 
\subsection{\ind{Z-Chambers}}
 
The Z-chambers (\ind{CZ}) provide a precise measurement of the
$z$ coordinate of
tracks as they leave the jet chamber. They consist of a layer of 24 drift
chambers 400~$cm$ long, 50~$cm$ wide and 5.9~$cm$
thick covering 94\% of the
azimuthal angle at the polar angle range of \cth$<0.72$. Each
chamber is divided in  the $z$ axis into 8 cells
of $50 \times 50~cm^2$,
 each of which has 6 sense wires spaced at 0.4~$cm$.
 
\subsection{\ind{Vertex Detector}}
 
The vertex detector (\ind{CV}) is a high precision cylindrical jet
drift chamber.
It is 100~$cm$ long with a radius of 23.5~$cm$
and consists of two layers of  36
sectors each. The inner layer contains the axial sectors, each containing
a plane of 12 sense wires strung parallel to the beam direction.
The wires range radially from 10.3 to 16.2~$cm$ with a spacing of
0.583~$cm$.
The outer layer contains the stereo sectors each containing a plane
of 6 sense wires inclined at a stereo angle\index{Stereo angle}
 of $\sim$4\degree. The
stereo wires lie between the radii 18.8 and 21.3~$cm$ with
a spacing of 0.5~$cm$.
 
\noindent
A precise measurement of the drift time on to the axial sector
sense wires allows the $r-\phi$ position to be calculated.
Measuring the time difference
between signals at either end of the sense wires allows a fast but
relatively coarse $z$ coordinate measurement
that is used by the OPAL track trigger
and in pattern recognition. A more precise $z$ measurement is then
obtained offline
by combining axial and stereo drift time information offline.
Multiple hits on a wire can be recorded.

\subsection{\ind{Silicon Microvertex Detector}}
 
The Silicon Microvertex Detector (\ind{SI})
consists of two barrels of single sided Silicon Microstrip Detectors
at radii of 6 and 7.5~$cm$. The inner layer consists of 11 ladders
and the outer of 14. Each ladder\index{Ladder}
 is 18~$cm$ long and consists of 3
silicon wafers daisy chained together. There are 629 strips per detector
at 25~$\mu m$ pitch and every other strip is read out at 50~$\mu m$ pitch.
The detector was originally installed in OPAL in 1991 and had
$r-\phi$ readout only. In 1993 an upgraded detector was installed that
has $r-\phi$ and $r-z$ wafers glued back to back.

 
 
\section{\ind{The Electromagnetic Calorimeter}}
 
The function of the electromagnetic calorimeter (ECAL) is to detect and
identify electrons and photons. It consists of a lead glass total
absorption calorimeter split into a barrel
and two end cap arrays. This arrangement together with two forward lead
scintillator calorimeters of the forward detector makes
the OPAL acceptance for electron and photon detection almost
99\% of the solid angle.
 
\noindent
The presence of $\sim$2 radiation lengths of
material in front of the calorimeter
(mostly due to the solenoid and pressure vessel), results  most of the
electromagnetic showers to start before reaching the lead glass.
Presampling devices are therefore installed in front of the lead glass
in the barrel and endcap regions to measure the position and energy
of showers to improve overall spatial and energy resolution
and give additional $\gamma/\pi^{0}$
and electron/hadron discrimination. In front of the Barrel Presampler
is the Time Of Flight detector.
 
 
\subsection{Barrel \ind{Lead Glass} Calorimeter}
 
The barrel lead glass calorimeter (\ind{EB}) consists of a cylindrical
array of 9440
lead glass blocks at a radius of 246~$cm$ covering the polar angle
range \cth$<0.81$.
Each block is 24.6 radiation lengths,
37~$cm$ in depth and $\sim10\times 10~{\rm cm}^{2}$.
In order to
maximize detection efficiency the longitudinal axis of each block
is angled to point at the interaction region. The focus of this
pointing geometry is slightly offset from the e$^+$e$^-$
collision point
in order to reduce particle losses in the gaps between blocks.
 
\noindent
\v{C}erenkov light from the passage of relativistic charged particles
through the lead glass is detected by 3~$inch$ diameter phototubes at
the base of each block.
 
\subsection{Endcap Electromagnetic Calorimeter}
 
The endcap electromagnetic calorimeter (\ind{EE})
consists of two dome-shaped arrays of 1132 lead glass blocks
located in the region between the pressure bell and the pole tip hadron
calorimeter. It has an acceptance coverage of the full azimuthal angle
and $0.81<$\cth$<0.98$.
 
\noindent
As opposed to the barrel calorimeter, the endcap lead glass blocks
follow a non-pointing geometry being mounted
coaxial with the beam line.
The lead glass blocks provide typically 22~radiation lengths
of material and come in three lengths (38, 42 and 52~$cm$) to form the
domed structure following the external contours of the pressure bell.
 
\noindent
The blocks are read out by special
Vacuum Photo Triodes (\ind{VPT}s) operating in the full OPAL
magnetic field.
 
\subsection{Barrel Electromagnetic \ind{Presampler}}
 
The Barrel Electromagnetic Presampler (\ind{PB}) consists of 16 chambers
forming a cylinder of radius 239~$cm$ and length 662~$cm$ covering the
polar angle range \cth$<0.81$.
Each chamber
consists of two layers of drift tubes operated in the limited streamer
mode with the anode wires running parallel to the beam direction. Each
layer of tubes contains 1~$cm$ wide
cathode strips on both sides at $\pm45$\degree~ to the
wire direction. Spatial positions can then be determined
by reading out the strips in conjunction with a measurement of the
charge collected at each end of the wires to give a $z$ coordinate
by charge division. The hit multiplicity is approximately proportional
to the energy deposited in the material in front of the presampler
allowing the calorimeter shower energy to be corrected with a
corresponding improvement in resolution.
 
 
\subsection{Endcap Electromagnetic \ind{Presampler}}
 
The endcap electromagnetic presampler (\ind{PE}) is 
an umbrella shaped arrangement of 32 chambers in 16
wedges (sectors). It is located between the pressure bell of the central 
tracking system and the endcap lead glass
calorimeter, covering the full azimuthal angle 
in the polar angle range $0.83<$\cth$<0.95$.


\subsection{\ind{Time-Of-Flight Counters}}
 
The time-of-flight (\ind{TOF}) system provides charged particle
identification
in the range 0.6 to 2.5~$GeV$, fast triggering information
and an effective rejection of cosmic rays.
 
\noindent
The TOF system consists of 160 scintillation counters forming a barrel
layer 684~$cm$ long at a  mean radius of 236~$cm$ surrounding
the OPAL coil covering the polar angle range \cth$<0.82$.
 


 
\section{The \ind{Hadron Calorimeter}}
 
The hadron calorimeter (HCAL) is built in three sections - the barrel, the
endcaps and the pole-tips. By positioning detectors between the layers
of the magnet return
yoke a sampling calorimeter is formed covering a solid angle
of $0.97\times 4\pi$ and offering at least 4 interaction lengths of
iron absorber to particles emerging from the electromagnetic
calorimeter. Essentially all hadrons are absorbed at this stage
leaving only muons to pass on into the surrounding muon chambers.
 
\noindent
To correctly measure the hadronic energy, the hadron calorimeter
information must be used in combination with that from the preceding
electromagnetic calorimeter. This is necessary due to the
likelihood of
hadronic interactions occurring in
the 2.2 interaction lengths of material that exists in front of the
iron yolk.
 
 
\subsection{Hadron Endcap and Barrel Calorimeter}
 
The barrel region (\ind{HB}) contains 9 layers of
chambers sandwiched between
8 layers of 10~$cm$ thick iron. The barrel ends are then
closed off by
toroidal endcap regions (\ind{HE})
which consist of 8 layers of chambers
sandwiched between 7 slabs of iron.
 
\noindent
The chambers themselves are limited streamer tube devices strung with
anode wires 1~$cm$ apart in a gas mixture of isobutane (75\%) and
argon (25\%) that is continually flushed through the system. The signals
from the wires themselves are used only for monitoring purposes. The
chamber signals result from induced charge collected on pads and strips
located on the outer and inner surfaces of the chambers respectively.
 
\noindent
The layers of pads are grouped together to form towers that divide up
the detector volume into 48 bins in $\phi$ and 21 bins in $\theta$. The
analogue signals from the 8 or so pads in each chamber are then summed
to produce an estimate of the energy in hadronic showers.
 
\noindent
The strips consist of 0.4~$cm$ wide aluminium  that run the full length of
the chamber, centered above the anode wire positions. They hence run
parallel to the beam line in the barrel region and in a
plane perpendicular
to this in the endcaps. Strip hits thus provide muon tracking information
with positional accuracy limited by the 1~$cm$ wire spacing. Typically,
the hadronic shower initialized by a normally incident 10~$GeV$ pion
produces 25 strip hits and generates a charge of 600~$pc$.
 
 
\subsection{Hadron \ind{Pole-Tip Calorimeter}}

The pole-tip hadron calorimeter 
(\ind{HP}) complements the barrel and endcap ones by 
extending the solid angle
coverage 0.91 $<$ \cth $<$ 0.99. Here the gap between the iron plates,
available for detectors,
was reduced to 10~$mm$ to avoid perturbing the magnetic field. 
In order to improve the energy
resolution in the forward direction, where the momentum resolution of 
the central detector is falling
off, the distance between samplings was reduced to 8~$cm$ and the number 
of samplings increased
to 10.

%
 
\noindent
The detectors themselves are 0.7~$cm$ thick 
containing  a gas mixture of CO$_{2}$ (55\%) and n-pentane (45\%), strung
with anode wires at a spacing of 0.2~$cm$.
Again, the chambers have pads on
one side (of typical area 500~$cm^{2}$) and strips on the other.
Corresponding pads from the 10 layers then form towers analogous to the
treatment in the rest of the calorimeter.
 
 
\section{The \ind{Muon Detector}}
 
The muon detector aims to identify muons in an unambiguous way from
a potential hadron background. To make the background manageable,
particles incident  on the detector
have traversed the equivalent of 1.3~$m$ of iron so 
\nopagebreak
reducing
the probability of a pion not interacting to be less than 0.001.
 
 
\subsection{Barrel Muon Detector}
 
The barrel region (\ind{MB}) consists of 110 drift chambers that
cover the acceptance
\cth$\,<\,0.68$ for four layers and \cth$\,<\,0.72$ for one or
more layers.
The chambers range in length between 10.4~$m$ and 6~$m$ in order to
fit between the magnet support legs and all have the same cross sectional
area
of $120 \times 9~cm^2$.
 
\noindent
Each chamber is split into two adjoining cells each containing an anode
signal wire
running the full length of the cell, parallel to the beam line.
The inner
surfaces of the cells  have 0.75~$cm$ cathode strips etched in them to
define the drift field and in the regions directly opposite the anode
wires are diamond shaped cathode pads. In all, six signals are read out
from each cell namely, one from each end of the anode wire and four from
the cathode pads and these are digitized via an 8-bit FADC.
 
\noindent
Spatial position in the $\phi$ plane is derived using the drift time
onto the
anode and can be reconstructed to an accuracy of better than 0.15~$cm$. A
rough estimate of the $z$ coordinate is also achieved by using the
difference
in time and pulse height of the signals arriving at both ends of
the anode
wire. A much better measure of the $z$ coordinate is given by using
induced signals on two sets of cathode pads whose diamond shape repeats
every 17.1~$cm$ and 171~$cm$ respectively. This results in a $z$
coordinate accurate of 0.2~$cm$, modulo 17.1~$cm$ or accurate to 3~$cm$
modulo 171~$cm$.
 
 
\subsection{Endcap Muon Detector}
 
Each endcap muon detector (\ind{ME}) consists of two layers of four
quadrant chambers ($6 \times 6\;\;m^2$)
and two layers of two
patch chambers (3~$\times 2.5~m^2$), for an angular
coverage of $0.670<$\cth$<0.985$.
Each chamber is an arrangement of two layers
of limited streamer tubes in the plane perpendicular to the beam line,
where one layer has its wires horizontal and the other vertical.
 
\noindent
The basic streamer tube used has a cross section of
$0.9 \times 0.9~cm^2$
with the inner walls coated with a carbon-suspension cathode.
Each plane of
tubes is open on one side and closed on the other to rows of aluminium
strips 0.8~$cm$ wide.
The strips on the open side, run perpendicular to the tube anode
wires and typically have charge induced over five or so strips. By
finding a weighted average using the recorded pulse heights, the streamer
is located to better than 0.1~$cm$. The strips on the closed side run
parallel to the tube wires and so can only give that coordinate to the
nearest wire or 0.9/$\sqrt{12}~cm$.
 
\noindent
Within each chamber therefore, with two layers of tubes each with two
layers of strips, the $x$ and $y$ coordinates of a track can be
measured once accurately and once relatively coarsely. As with the barrel
region, the actual position of the strips is known to about 0.1~$mm$ via
survey information.
 
 
\section{The \ind{Forward Detector}}
 
The forward detector (\ind{FD}) consists
of an array of devices, listed below,
whose primary objective is to detect low angle Bhabha scattering events
as a way of determining the LEP luminosity
for the normalization of
measured reaction rates from Z$^0$ decays.
 
\noindent
To achieve this, the forward detector enjoys a relatively clean
acceptance
for particles between 47 and 120~$mrad$
from the interaction point, with
the only obstructions being the beam pipe and 2~$mm$ of aluminium from
the central detector pressure vessel.
\begin{itemize}
\item        {\em Calorimeter}. The forward calorimeter consists of 35
sampling layers\index{Forward calorimeter}
of lead-scintillator sandwich divided into a  presampler of
4 radiation lengths and the main
calorimeter of 20 radiation lengths.
 
\item         {\em \ind{Tube Chambers}}.
There are three layers of proportional tube
chambers positioned between the presampler and main sections
of the calorimeter. The positioning is known to
$\pm$0.05~$cm$ and they can give the position of a shower
centroid to $\pm 0.3~cm$.
 
\item         {\em \ind{Gamma Catcher}}.  The gamma catcher is a ring of
lead scintillator sandwich sections of 7 radiation lengths
thickness. They plug the hole in acceptance between the
inner edge of EE and the start of the forward calorimeter.
 
\item  {\em  \ind{Far Forward Monitor}}. The far forward monitor counters
are small lead-scintillator calorimeter modules, 20 radiation lengths
thick, mounted either side of the beam pipe 7.85~$m$ from the
intersection region. They detect electrons scattered in the
              range 5 to 10~$mrad$ that are deflected outwards by the
              action of LEP quadrupoles.
\end{itemize}
 
 
\section{The \ind{Silicon Tungsten Detector}}
 
The silicon tungsten detector (\ind{SW}) is a sampling calorimeter designed to
detect
low angle Bhabha scattering events in order to measure the luminosity.
There are 2 calorimeters at $\pm238.94~cm$ in $z$ from the interaction
point
with an angular acceptance of 25~ to 59~$mrad$.
Each calorimeter consists of 19 layers of silicon detectors and 18
layers of tungsten. At the front of each calorimeter is a bare layer
of silicon to detect preshowering.  The next 14 silicon layers are
each behind 1 radiation length (3.8~$mm$) of tungsten and the final 4
layers are behind 2 radiation lengths (7.6~$mm$) of tungsten.
 
Each silicon layer consists of 16 wedge shaped silicon detectors. The
wedges cover $22.5^{o}$ in $\phi$
with an inner radius at $6.2~cm$ and an outer one at $14.2~cm$.
The wedges are subdivided into 64 pads (32 in $r$ and 2 in $\phi$)
giving  a total of 38912 channels which are read out individually.
Adjacent wedges in a  layer are offset by 800~$\mu m$ in $z$ and
positioned in such a  way that there is no gap in the active area of
the silicon.
Consecutive layers in the detector are offset in $\phi$ by half a
wedge ($11.25$\degree) so that any cracks between the tungsten
half--rings do not line up.
 
 
\section{The \ind{Trigger}}
 
Events are only recorded by the data acquisition system if they satisfy
certain trigger conditions (and since 1992 pretrigger
conditions).
From a bunch crossing rate of 45~$kHz$, in the original  4+4 bunch mode,
the trigger system reduces the
event rate to 2--3~$Hz$
at a typical luminosity of
0.5$\times$10$^{31}$~$cm^{-2}s^{-1}$.
The data are subsequently processed by a software ``filter'' 
which uses a partial event reconstruction and some preliminary
event classification to reduce the event rate by
a further $\sim$30\%.
 
\noindent
Subdetector trigger signals divide into two categories, ``stand-alone''
signals such as multiplicity counts or energy sums, and lower threshold
signals from a $6\times 24$ bins in $\theta$ and $\phi$ respectively.
The trigger processor makes its decision by forming correlations in
space between subdetectors in $\theta/\phi$ together with the
stand-alone signals.
 
\noindent
The original trigger system was designed as a single stage trigger,
because the time between a bunch crossing and the decision being made
plus the time required by detector components to clear
their electronics (``reset time'') amounts to
$\sim$20$\,\mu$s, which
is less than the interbunch time of
22$\,\mu$s in 4+4 bunch mode.
 
From the start of data-taking in 1992,
the detector has run with a two-stage trigger system suitable
for operation with more than 4+4 bunches.
With 8+8 bunches in LEP (and potentially even more
bunches),
the pretrigger performs a deadtime-free
first level decision
without compromising trigger efficiency or acceptance,
reducing the 90~kHz bunch crossing rate down
to a rate of positive pretriggers of 1--2 kHz.
Similarly to the trigger system,
subdetector pretrigger signals divide into ``stand-alone''
signals from energy sums, and lower threshold signals from 12 bins
in $\theta/\phi$.  The pretrigger processor makes its decision by
multiplicity counting and possibly forming correlations in $\theta/\phi$
between subdetectors, together with the stand--alone signals.

\noindent
 
 
\section{Online Dataflow}\index{Online dataflow}\index{Dataflow}
 
When a beam crossing is selected by the trigger as containing a
potentially
interesting event, the subdetectors are read out.
Each one of the sixteen subdetectors is read out separately by its own
special front-end readout electronics into its local system crate(s)
(``\ind{LSC}'')\index{Local system crate}~\cite{FILTER}.
The subevent structures from the different LSCs (eighteen of them,
including the trigger and track trigger) are assembled by the
event builder (``\ind{EVB}'')\index{Event builder}.
 
\noindent
When the complete events have been assembled by the EVB, they are passed
in sequence to the filter\index{Filter} program.
In the filter, the events are checked, analyzed, monitored and compressed,
where some obvious background, typically 15-35\,\% of all triggers, are
rejected.
At this stage, a MH event
typically occupies 210~ $kbytes$
of data storage and other events 60~ $kbytes$. This is reduced by the filter
by an average factor of five, and than events are written to a buffer disk.
The buffer
with a capacity of several hours of data taking is used to decouple the data
acquisition from subsequent event reconstruction and data recording.
As a backup the events are also copied from the filter disk buffer to
IBM cartridges tapes.
 
The event  reconstruction program, ``ROPE'' based on a network of HP Apollo
workstations, fetches the event files from the filter buffer disk, and records
them on optical disks for long term storage. The full
reconstruction is performed as soon as up-to-date calibration data from the
LSC's are available, usually within an hour of the events being taken.
 
The event data are in ZEBRA~\cite{ZEBRA} format data structure. Each ZEBRA
structure of a complete event includes a ``header''
 with 64 words of basic event
information, such as trigger pattern and filter event classification. it can
be updated during the event processing and is used for fast selection of
events of a particular type.
 
Monitoring is performed at various levels in the data acquisition chain.
The LSC's perform detailed monitoring of subdetectors. At filter level,
complete event are given, so correlation between subdetectors can be
made and event classification can be used. The results are immediately
available. The ROPE monitors the combined detector performance after
full processing has been performed. The events are available for analysis
when moved to a permanent storage area. Data are often ready for analysis
within a few hours from the end of a run.
 

\chapter{\label{chap-theory} Theoretical Background}
 
\section{ A Brief Review of the Standard Model}
A common aim of all science is to explain as many facts as possible with
a few simple principles.
This leads to the efforts to relate the known phenomena and
the attempts to reach unification in our theories
(see for example reference~\cite{unity}).
 
Our experimental
evidences nowadays suggest that all matter is composed of structureless,
point like, quarks and leptons, both obey Fermi statistic rules,
and the interactions between them
are mediated by gauge massive and massless bosons.
The theories describing these interactions are all gauge theories which
can be described as a change from  global symmetry to  local symmetry.
The nature of transmission of forces is the interaction
between a gauge field, and a conserved matter current.
 
The fermions are grouped in
three families ~\cite{generations} (sometimes refers as generations),
each consist of two quarks, a massive charged lepton,
and its light (or even massless) uncharged partner, the neutrino.
\pagebreak
In rising weight order the families are:
 
$\left( \begin{array}{c} u \\ d   \end{array}
\right ) $ \hspace{2.2cm}
$\left( \begin{array}{c} c \\ s   \end{array}
\right ) $  \hspace{2.2cm}
$\left( \begin{array}{c} t \\ b  \end{array}
\right ) {\bf Quarks} $
 
$\left( \begin{array}{c}  e \\ \nu_e  \end{array} \right ) $
\hspace{2cm}
$\left( \begin {array}{c} \mu \\ \nu_{\mu}  \end{array}
\right ) $  \hspace{2cm}
$\left( \begin{array}{c} \tau \\ \nu_{\tau} \end{array}
\right ) {\bf Leptons}$

 
\hspace{.2cm} (I) \hspace{2.8cm} (II) \hspace{2.6cm} (III)
 
The forces of nature are traditionally  reduced to  4 fundamental
forces:
\begin{itemize}
\item {\em Electromagnetic}, carried by the uncharged massless particle,
       the photon, $\gamma$.
\item {\em Weak nuclear force}, mediated by the massive bosons $Z^0$,
       $W^+$, and $W^-$.
\item {\em Strong nuclear force}, carried by the massless boson, the
       \mbox{gluon, g}.
\item {\em Gravitation},
      carried by the massless particle with spin 2 the \mbox
      {graviton, $g_r$}.
\end{itemize}
 
QED, quantum electrodynamics, is the exact theory of electromagnetism.
The agreement between the theory and experiment is excellent, in fact
QED is the most precise theory in the field of physics.
 
The weak nuclear force,
is the force which is responsible, among other things, for the
$\beta$ decay:
\mbox{$n \to p\: e\: \bar{\nu_e}$} and
the muon decay \mbox{$\mu \to e \nu_{\mu}\bar{\nu_e}$}.
The weak force is known to be of a short range ($\sim 10^{-17}~m$) with
a maximal violation of parity.
 
Quantum-Chromo-Dynamics, QCD, the strong field theory,
is the gauge field theory which describes the
interaction of colored quarks and gluons. The principle of "asymptotic
freedom"
determines that the renormalized QCD coupling to be small only at high
energies,
and therefore only at that region high precision tests can be
performed using perturbation theory.
 
QCD, the weak force, and QED are the three components of the
Standard Model (SM) \mbox{SU(3) $\times $ SU(2) $\times $ U(1)} theory.
\pagebreak
 
{\bf The Glashow-Weinberg-Salam Theory}
 
 The $V-A$ weak interaction is only a phenomenological theory, because it
is not renormalizable. Attempts to construct renormalizable weak
interaction theory have failed, until Glashow (1961)~\cite{GWS1},
Weinberg and Salam (1967)~\cite{GWS2,GWS3}
constructed a spontaneously broken gauge theory, which unified
the weak and the electromagnetic interactions.
After a large number of different versions of this
theory were tested during the 1970's,         it turned out that it is
possible to describe
all the experimental data using
Glashow-Weinberg-Salam theory.
One of the
shortcomings of this model is its many parameters
which their values are not predicted by the  theory.
 
The electroweak interactions
are mediated by 4 massless vector  bosons
$W_1$, $W_2$, $W_3$ and $B$. The $W$ triplet couples to the weak isospin
of the fermions, and the singlet $B$ couples to the "hypercharge"
$(\;Y\;)$ a
combination of weak isospin
$(\;I\;)$, and the electric charge $(\;q_f\;)$: $Y=2q_f-I_3$.
 
The theory formulated in this way describes interactions between massless
fermions, a postulate which clearly disagrees with the experimental
observations. It is also involves with massless gauge bosons and therefore
long range forces, which again is not realistic because it is experimentally
known that the weak interaction are short-ranged.
For these reasons a scalar complex "Higgs" field,
which spontaneously breaks the $SU(2) \times U(1)$ symmetry, was introduced
to the theory.
The spontaneous symmetry breaking leads to four physical particles:
\begin {itemize}
\item Two massive charged particles $W^+$ and $W^-$, which are responsible
for the charged weak currents.
\item Massive $Z^0$ which is responsible for weak neutral interactions.
\item The massless particle $\gamma$, which is responsible for electromagnetic
interactions.
\end{itemize}
The photon,$\gamma$, and the $\Z$
are correlated through:
\beq
\left( \begin{array}{c} \gamma \\
                           \Z  \end{array}\right)=
\left( \begin{array}{cc}\sin\theta_W &\cost_W \\
                        \cost_W &-\sin\theta_W \end{array}
                                           \right)
\left( \begin{array}{c} W_3 \\B \end{array} \right )
%
\eeq
where the mixing angle $\theta_W$ (Weinberg angle) is a free parameter.
The masses of the four real bosons are related through the relations:
\beq
m_W=\frac{37.3 ~[GeV]}{\sin\theta_W},\:\:\:
m_{Z^0}=\frac{37.3~ [GeV]}{\sin\theta_W\cost_W},\:\:\:
m_{\gamma}=0.
\eeq
 
The insertion of the Higgs field generates masses to the fermions.
However, the fermions mass values are not predicted by the theory, and
are left as free parameters. Another value which is almost unlimited is
the mass of the Higgs boson which generates this spontaneously symmetry
breaking.
 
There is a slight difference between the quarks listed at the
beginning of this section
which are  essentially the mass eigenstates and the eigenstates of
the elctroweak model. The $d^{\prime}$-type quarks
($d^{\prime},s^{\prime},b^{\prime}$) which are the eigenstates of EW are
mixed with the mass eigenstates via the Cabbibo Kobayashi Maskawa (CKM)
matrix:
 
\beq
\left( \begin{array}{c} d^{\prime} \\
                        s^{\prime} \\
                        b^{\prime}
      \end{array}\right) =
\left( \begin{array}{ccc} V_{ud} & V_{us} & V_{ub} \\
                          V_{cd} & V_{cs} & V_{cb} \\
                          V_{td} & V_{ts} & V_{tb}
      \end{array} \right)
\left( \begin{array}{c} d \\ s \\ b \end{array} \right )
\eeq
The CKM matrix elements are not predicted by the SM hence they must be
input from the experimental measurements.
 
In addition to the fermion masses and the CKM elements there  are still
three more degrees of freedom left to be experimentally obtained,
and there are several parameter combinations one can choose to fix
the theory with. A common set of three parameters is for example:
\begin{enumerate}
\item $\theta_W$ - the Weinberg mixing angle defined with
$\tan \theta_W=g^\prime/g$ the ratio between the U(1) and the SU(2)
coupling constants.
\item $e$ - the  electric charge of the electron which before radiative
correction $e^2=g^2\swsq$.
\item $M_W$ or $M_Z$ - the mass of the $W^{\pm}$  or $\Z$ bosons where
without radiative correction hold the relation $\swsq=1-M_W^2/M_Z^2$.
\end{enumerate}
 
The weak charged currents couple to fermions by a $V-A$ interaction.
The electromagnetic interaction has only  a vector coupling.
The neutral currents are a mixture of $V$ and $A$, given by the couplings
\beq
    g^{L,R}=\frac{I_3^{L,R}-q_f\sin^2\theta_W}{\sin\theta_W\cost_{W}}
\eeq
where $I^{L,R}$ is the weak isospin of the left or right handed fermions.
$I_3=0$ for all right handed fermions, and for the left handed doublets:
 
$\left( \begin {array}{c}  \nu_e \\ e \end{array} \right )_L\; $
$\left( \begin {array}{c}  \nu_{\mu} \\ \mu \end{array} \right )_L\; $
$\left( \begin {array}{c}  \nu_{\tau} \\ \tau \end{array} \right )_L\;\;\;\;\;\;$
$\left( \begin {array}{c}   u \\ d \end{array} \right )_L\; $
$\left( \begin {array}{c}   c \\ s \end{array} \right )_L\; $
$\left( \begin {array}{c}   t  \\ b \end{array} \right )_L $
 
the upper components have
$I_3=+\frac{1}{2}$ and the lower components
have
\newline
$I_3=-\frac{1}{2}$.
 
The vector, $v$, and the axial-vector, $a$, couplings are
combinations of the left and right couplings:
\beq
v=\frac{1}{2}(g^R+g^L),\;\;\;a=\frac{1}{2}(g^L-g^R)
\eeq
The difference between the left and right coupling causes parity
violation  and helicity effects.

Table~\ref{tab-const} presents the SM weak
coupling constants for the lepton and antilepton doublets.

\begin {table} [htb]
\begin{center}
\begin{tabular} {| l | l | l || l | l | } \hline
 
         &$g^R$ &$g^L$ &$v$  &$a$    \\ \hline
\elm,\tm,\mum
&$\;\:\:\swsq$  &$-\frac{1}{2}+\swsq$ &$-\frac{1}{4}+\swsq$ &$-\frac{1}{4}$ \\
\elp,\tp,\mup
&$-\swsq$  &$\;\:\:\frac{1}{2}-\swsq$ &$\;\:\:\frac{1}{4}-\swsq$ &$\;\:\:\frac{1}{4}$ \\
\nue,\num,\nut
&$\;\:\:0$  &$\;\:\:\frac{1}{2}$   &$\;\:\:\frac{1}{4}$       &$-\frac{1}{4}$ \\
\nueb,\numb,\nutb
&$\;\:\:0$  &$-\frac{1}{2}$  &$-\frac{1}{4}$      &$\;\:\:\frac{1}{4}$ \\
\hline
\end{tabular}
\caption[The leptons weak coupling constants.]
{The weak coupling constants of the leptons doublets in units of
$1/\sin\theta_W \cos\theta_W$
}
\label{tab-const}
\end{center}
\end{table}

 
The electroweak model is currently consistent with all experimental
findings and is
generally accepted as giving the correct description of these
interactions, even though some ingredients are still not
well tested.
The LEP experiments in general, and the $\tau$ decays tests in
particular, represent a potential field for making rigorous
tests of the SM predictions, measurements of its free
parameters, and searching for possible deviations from the model.

\section{\label{sec-tauformalism} The $\tau$-Asymmetry Formalism}
 
The annihilation of $e^+e^-$ into tau pair  supplies information on the
relative strength of the neutral vector and axial-vector couplings to the
charged heavy lepton.
This tallies with the fact that one of the main physics goals of the
LEP experiments, is a precision measurement of the electroweak
mixing angle, $\swsq$, in many different ways.
 
The $\tau$ pairs are produced with a
forward backward asymmetry, a longitudinal polarization and a forward
backward asymmetry of its polarization,
due to the interference of electromagnetic,
and neutral currents
~\cite{Tsai}-
\cite{TN031}.
The presence of the $\tau$ polarzation is a purely
parity violating effect (arising from the neutral current),
while the forward backward asymmetry can arise  also
from higher order QED effects (interference of two photons).

The $\tt$ pair, as a system of two fermions, can be produced in four
polarization (helicity) combinations,
\beq
(P_{\tm},P_{\tp})=(+,+),\;\;(+,-),\;\;(-,+),\;\;(-,-),
\eeq
with the corresponding cross sections,
\beq
\sigpp,\;\; \sigpm,\;\; \sigmp,\;\; \sigmm.
\eeq
 
This leads to the following definitions of the average $\tm$ and $\tp$
polarizations,
\bea
\label{ptmptp}
\ptm=\frac{[\sigpp+\sigpm]-[\sigmp+\sigmm]}{\sigtot},   \\ \nonumber
\ptp=\frac{[\sigpp+\sigmp]-[\sigpm+-\sigmm]}{\sigtot},
\eea
where,
\beq
\sigtot=\sigpp+\sigpm+\sigmp+\sigmm.
\eeq
 
However, in $\ee$ annihilations, the $\tt$ pair are known to be produced
through an intermediate state of a spin-one boson which can be a photon
or a $\Z$.
Looking first at QED scattering of an electron,
it is known that the helicity of the electron is conserved at high
energies, whereas at low energies the direction of the spin
with respect to a fixed coordinate system in space is preserved.
The  analogous behavior is present
in $e^+e^- \to \tau^+ \tau^-$, namely:
 
\begin{itemize}
\item {\bf In the  $\frac{m_{\tau}}{E} \to 0$ region}:
the helicities of the $\tau^+$ and the
      $\tau^-$ tend to be opposite to each
other $(\uparrow \downarrow)$.
\item {\bf While in the $\frac{m_{\tau}}{E} \to 1$ region }:
the helicities of
      the $\tau^+$ and the $\tau^-$ tend to be parallel
      $(\uparrow \uparrow)$.
\end{itemize}
 
In this case, the helicity conservation rule at high energies
restricts the number of helicity combinations to two, namely $(+,-)$ and
$(-,+)$ or \newline
 {\mbox{$(P_{\t^-}=-P_{\t^+})$}},
whereas the cross sections for the other two combinations,
$\sigpp$ and $\sigmm$ are negligibly small. The $\t^-$ and $\t^+$
helicity relation stands also for their average polarization values.
For these reasons we will denote in the following :
 
\bea
\label{defpta}
\pt &\equiv P_{\tm}&=-P_{\tp} , \\   \nonumber
\pta&\equiv\ptm&=-\ptp.
\eea

The general expression for  a differential cross section of the reaction
$e^+e^-\rightarrow\tau^+\tau^-$ intermediates by a vector boson
is given by,
\begin{eqnarray}
\label{sigcos}
\frac{d\sigma}{d\cst}(s,\cst,\pt)=
(1+\cstsq)\fz+2\cst\fo \nonumber \\
-\pt[(1+\cstsq)\ft+2\cst\fth].
\end{eqnarray}
Where the  $F_i(s)$ are form factors.
 
\begin{figure}[htb]
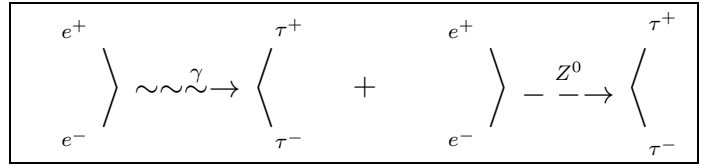

\begin{center}
\fbox{
$\left. \begin{array}{c} \\ \\ \\ \end{array} \right.^{e^+}_{e^-}$
$ \left \rangle \begin{array}{c}
       \\ \sim \sim \stackrel{\gamma}{\sim} \rightarrow \\  \\  \end{array}
 \right \langle^{\tau^+}_{\tau^-}  \:\:\:\:\:\:+\:\:\:\: $
$\left. \begin{array}{c} \\ \\ \\ \end{array} \right.^{e^+}_{e^-}$
$ \left \rangle \begin{array}{c}
        \\ - \stackrel{Z^0}{-} \rightarrow \\  \\  \end{array}
 \right \langle^{\tau^+}_{\tau^-}    $
     }
\end{center}
 
\caption [Feynman diagrams for the $\tau$ pair production]
{ Feynman diagrams for the $\tau$ pair production in $e^+e^-$
annihilation through the single photon and $Z^0$ exchange.
}
\label{fig_feynman}
\end{figure}

Fig.~\ref{fig_feynman} presents the two
SM $e^+e^-$ annihilation diagrams for the production of a $\tau$ pair:
one via $\gamma$  and the other through a $Z^0$ exchange (while the
 Higgs exchange channel can be neglected~\cite{Higgs})
 
The tree level equation  for the process
$e^{+} e^{-} \rightarrow \tau^{+} \tau^{-}$
is a linear combination of the three terms~\cite{CALASY,Ellis}:
\begin{itemize}
\item Pure QED contribution :
 
$\sigma_{\gamma}=1+\cos^2 \theta $.
 
\item Electro-weak interference contribution:
 
      $\sigma_{\gamma Z^0}=-2q_f[ \ve \vt
      (1+\cos^2 \theta )+2\ae \at \cos \theta ]Re \chi $.
 
\item Pure weak interaction contribution:
 
      $\sigma_{Z^0}=[(\ve^2+\ae^2)(\vt^2+\at^2)(1+\cos^2\theta)
      +8\ve\vt\ae\at\cos\theta]|\chi^2|$.
\end{itemize}
where the $\Z$ propagator,
\begin{equation}
\label{chis}
\chi(s)=\frac{s}{s-M_Z^2+is\Gamma_Z/M_Z}.
\end{equation}
describes the $\Z$ resonant shape without radiative corrections
and is symmetric arounnd $\sqrt{s}=M_Z$.
With radiative corrections
$\sigma$ has a smaller value at the peak and an unsymmetric shape.
$M_Z$ and $\Gamma_Z$ are the $\Z$ mass and width.
 
The $q_f,v_f$ and $a_f$ ($f=e$ or $\t$) are the electric charge,
the vector and axial-vector coupling constants of the fermions
to the $\Z$  respectively. The strength of the vertices
following the SM structure~\cite{halzen}
are:
\begin{itemize}
\item $\underline{ e\;\gamma\;e}$ : \hspace{1.3cm}
$V^{\mu}_{e\;\gamma\;e}   = -i\qe\gamma^{\mu}$
\item $\underline{ \t\;\gamma\;\t}$ :\hspace{1.3cm}
$V^{\mu}_{\t\;\gamma\;\t} = -i\qt\gamma^{\mu}$
\item $\underline{ e\;\Z\;e}$ :\hspace{1.1cm}
$V^{\mu}_{e\;\Z\;e}       = -i\frac{g_{weak}}{\cst_W}\gamma^{\mu}(\ve-\ae\gamma^5)$
\item $\underline{ \t\;\Z\;\t}$ :\hspace{1.1cm}
$V^{\mu}_{\t\;\Z\;\t}     = -i\frac{g_{weak,\t}}{\cst_W}\gamma^{\mu}(\vt-\at\gamma^5)$
\item $\underline{ \t\;W\;\t}$ :\hspace{1.2cm}
$V^{\mu}_{\t\;W\;\t}      = -i\frac{g_{weak,\t}}{\cst_W}\gamma^{\mu}(1-\gamma^5)$
\end{itemize}
 
The corresponding four possible differential cross-sections are:
\bea
\frac{d\sigma }{d\cst } (e^-_Le^+_R\to \t^-_L\t^+_R) =
\frac{2\pi\alpha^2}{s}\frac{(1+\cst)^2}{4}|\qe\qt+T^{-+}_e
T^{-+}_{\t}\chi(s)|^2
\nonumber \\
\frac{d\sigma }{d\cst } (e^-_Le^+_R\to \t^-_R\t^+_L)  =
\frac{2\pi\alpha^2}{s}\frac{(1-\cst)^2}{4}|\qe\qt+T^{-+}_e
T^{+-}_{\t}\chi(s)|^2
\nonumber \\
\frac{d\sigma }{d\cst } (e^-_Re^+_L\to \t^-_L\t^+_R) =
\frac{2\pi\alpha^2}{s}\frac{(1-\cst)^2}{4}|\qe\qt+T^{+-}_e
T^{-+}_{\t}\chi(s)|^2
 \\
\frac{d\sigma }{d\cst } (e^-_Re^+_L\to \t^-_R\t^+_L)   =
\frac{2\pi\alpha^2}{s}\frac{(1+\cst)^2}{4}|\qe\qt+T^{+-}_e
T^{+-}_{\t}\chi(s)|^2 \nonumber
\eea
 
Here the $\qe\qt$ are QED contribution while the coupling of the weak neutral
current
are
 
\beq
T^{\pm\mp}_{f}=v_f\pm\beta a_f.
\eeq
where $\beta$ is the velocity of $f$
in speed of light units.
 
Therefore according to the SM
the form-factors $F_i(s)$ of Eq.~\ref{sigcos} are given by,
 
\begin{eqnarray}
\label{forms}
\fz & = & \frac{\pi\alpha^2}{4s}[\qe^2\qt^2+
2Re\chi(s)\qe\qt \ve\vt+|\chi(s)|^2(\ve^2+\ae^2)(\vt^2
+\at^2)] \nonumber \\
\fo & = & \frac{\pi\alpha^2}{4s}[2Re\chi(s)\qe\qt \ae\at
+|\chi(s)|^2 2\ve\ae 2\vt\at] \nonumber \\
\ft & = & \frac{\pi\alpha^2}{4s}[2Re\chi(s)\qe\qt \ve\at
+|\chi(s)|^2 (\ve^2+\ae^2) 2\vt\at] \nonumber \\
\fth & = & \frac{\pi\alpha^2}{4s}[2Re\chi(s)\qe\qt \ae\vt
+|\chi(s)|^2 2\ve\ae (\vt^2+\at^2)].
\end{eqnarray}

From Eq.~\ref{sigcos}, four independent
cross sections can be constructed:
 
\bea
\sigf (s,\ptau\!=\!+1)=\frac{4}{3}[\fos-\fts]+[\fis-\frs],\nonumber \\
\sigf (s,\ptau\!=\!-1)=\frac{4}{3}[\fos+\fts]+[\fis+\frs], \\
\sigb (s,\ptau\!=\!+1)=\frac{4}{3}[\fos-\fts]-[\fis-\frs],\nonumber \\
\sigb (s,\ptau\!=\!-1)=\frac{4}{3}[\fos+\fts]-[\fis+\frs],\nonumber
\eea
where the indices F (B) refer to events with $\tau^-$
scattering into the forward (backward) hemisphere,
namely, $0\leq\cost\leq 1$ ($-1\leq\cost\leq 0$).
 
These four independent cross sections can be combined
into four other cross sections which are simpler to
measure, but are no longer independent,
\bea
\sigf (s)\equiv\sigf(s,\ptau\!=\!+1)+\sigf(s,\ptau\!=\!-1)=\frac{8}{3}\fos+2\fis,
\nonumber \\
\sigb (s)\equiv\sigb(s,\ptau\!=\!+1)+\sigb(s,\ptau\!=\!-1)=\frac{8}{3}\fos-2\fis,
\\
\sigma(s,\ptau\!=\!+1)\equiv\sigf(s,\ptau\!=\!+1)+\sigb(s,\ptau\!=\!+1)=\frac{8}{3}
[\fos-\fts], \nonumber \\
\sigma(s,\ptau\!=\!-1)\equiv\sigf(s,\ptau\!=\!-1)+\sigb(s,\ptau\!=\!-1)=\frac{8}{3}
[\fos+\fts], \nonumber
\eea
and the total cross section is given by,
\begin{equation}
\label{sigtot}
\sigma(s)=\sigf(s)+\sigb(s)=
\sigma(s,\ptau\!=\!+1)+\sigma(s,\ptau\!=\!-1)=\frac{16}{3}\fos.
\end{equation}
 
Summing over both helicity states of Eq.~\ref{sigcos}
one obtains the following
$\cst$ distribution,
\beq
\label{dsigdcst}
\frac{1}{\sigtot}\frac{d\sigma}{d\cst}=\thoveit(1+\cstsq+
\eitovth\afb\cst),
\eeq
 
utilizing the forward-backward asymmetry, $\afb$,
defined as,
\begin{equation}
\label{afbs}
\afb(s)\equiv\frac{\sigf(s)-\sigb(s)}{\sigma(s)}=
\frac{3}{4}\frac{\fis}{\fos}.
\end{equation}

In the same way  the polarization asymmetry, $\pta$,
($\equiv \pta(s)$)
can be written as,
\begin{equation}
\label{apols}
 \pta \equiv
\frac{\sigma(s,\ptau\!=\!+1)-\sigma(s,\ptau\!=\!-1)}
{\sigma(s)}=\frac{\fts}{\fos},
\end{equation}
consequently, the cross-section can be divided into
\bea
\sigp & = & \half(1+\pta)\sigtot  \nonumber  \\
\sigm & = & \half(1-\pta)\sigtot.
\eea

The polarization asymmetry can be also defined 
separately
for each hemisphere,
\bea
\label{apolfs}
\langle\ptau\rangle^F \equiv
 \frac{\sigf(s,\ptau\!=\!+1)-\sigf(s,\ptau\!=\!-1)}
{\sigf(s)} \nonumber \\
\\
=-\frac{4\fts+3\frs}{4\fos+3\fis}, \nonumber \\  \nonumber
\eea
whereas for the backward hemisphere,
\bea
\label{apolbs}
\langle\ptau\rangle^B \equiv
\frac{\sigb(s,\ptau\!=\!+1)-\sigb(s,\ptau\!=\!-1)}
{\sigb(s)} \nonumber \\
\\
= -\frac{4\fts-3\frs}{4\fos-3\fis}. \nonumber \\  \nonumber
\eea
 
The forward-backward polarization asymmetry is then defined as,
\bea
\label{aplfbs}
\aplfb \equiv\frac{[\sigf(s,\ptau\!=\!+1)-\sigf(s,\ptau\!=\!-1)]
-[\sigb(s,\ptau\!=\!+1)-\sigb(s,\ptau\!=\!-1)]}{\sigma(s)} \nonumber \\
\\
=-\frac{3}{4}\frac{\frs}{\fos}. \nonumber \\ \nonumber
\eea
 
\begin{figure}[htb]
\epsfysize=9 cm.
\epsffile[80 400 750 700]{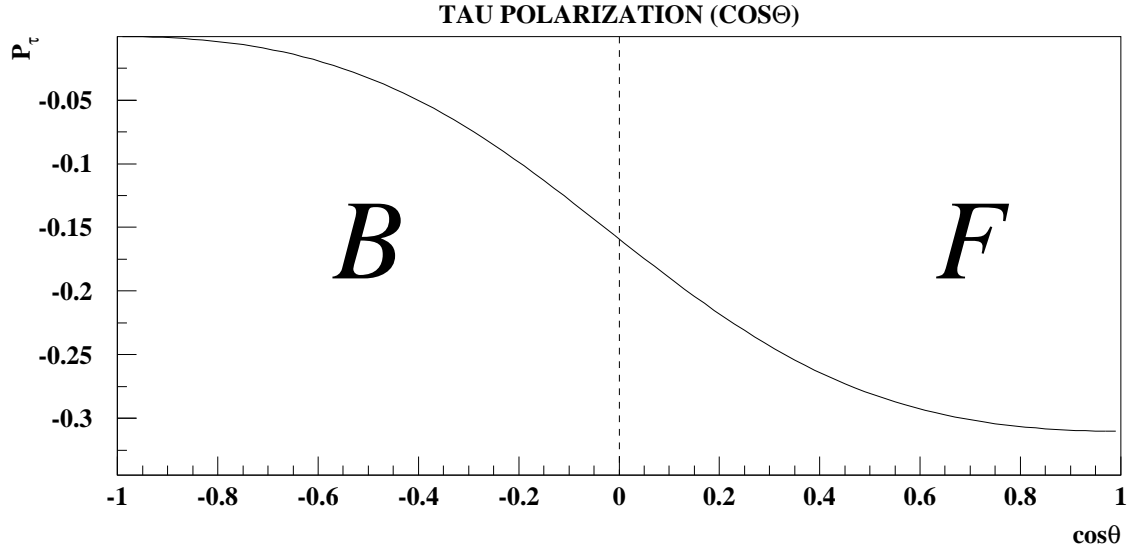}
\caption[The average tau polarization as a function of the polar
angle.]
{The average tau polarization as a function of the polar
angle. The {\bf B} and {\bf F} stand for the forward and backward regions.}
\label{fig_ptcost}
\end{figure}

Using these definitions one obtains,
 \bea
 \label{dspmdcst}
 \frac{1}{\sigtot}\frac{d\sigp}{d\cst} & = & \thovsxt
 [(1+\pta)(1+\cstsq)+\eitovth(\afb+\aplfb)\cst ] \nonumber \\
 \frac{1}{\sigtot}\frac{d\sigm}{d\cst} & = & \thovsxt
 [(1-\pta)(1+\cstsq)+\eitovth(\afb-\aplfb)\cst].
 \eea
 From the expressions above, the average polarization for a given polar
 angle $\theta$ is given by,
 \beq
 \label{ptt}
 \ptat=\frac{\pta(1+\cstsq)+\eitovth\aplfb\cst}{(1+\cstsq)+
             \eitovth\afb\cst}.
             \eeq
which is illustrated in Fig.~\ref{fig_ptcost}.

Utilizing the three asymmetries ($\pta,\;\afb$ and $\aplfb$)
as were determind here, the expression in  Eq.~\ref{sigcos} of
the differential cross section for an explicit $\tm$ helicity state
($\pt=\pm 1$) can also be writen as,
\bea
\label{sigcosa}
\frac{d\sigma}{d\cost}(s,\cost,\ptau) &=&\fos
\{(1+\cosst)+\frac{8}{3}\afb(s)\cost \\ \nonumber
& &+\ptau[\pta(s)(1+\cosst)+\frac{8}{3}\aplfb(s)\cost ]\}.
\eea
 
The distributions described by Eqs.~\ref{dspmdcst} cannot be directly
measured, because it is not possible to determine the $\t$ helicity
on an event-by-event basis. Instead, since the $\tau^+$ and the $\tau^-$
decay via weak interaction,
where parity conservation is  maximally violated,
the angular distribution of the decay
products depends strongly upon the spin orientation of the $\tau$.
As the $\tau^-$ is expected to be produced with a negative
polarization and the $\tau^+$ with a positive one, we expect to be
able to measure in the lab.\ system
a deviation from an unpolarized distribution, while measuring the
momentum spectrum of the  $\t$ decay products.
Thus for the determination of the average polarization we are using
the kinematical
distributions of the $\t$ decay products, depending on the helicity
and the decay mode.

\begin{figure}[htb]
\begin{picture}(300,100)
  \put(50,50){\circle{15}}
  \put(50,42){\vector(0,1){16}}
  \put(55.3,56){\vector(1,1){20}}
  \put(44.7,56){\vector(-1,1){20}}
  \put(50,42){\vector(-1,-1){20}}
  \put(20,85){$\elln$}
  \put(30,65){$\Downarrow$}
  \put(75,85){$\nulb$}
  \put(65,65){$\Uparrow$}
  \put(20,15){$\nu_{\tau}$}
  \put(35,30){$\Uparrow$}
  \put(48,10){(a)}
  \put(62,50){$\tau$}
\put(150,50){\circle{15}}
\put(150,42){\vector(0,1){16}}
\put(155.3,56){\vector(1,1){20}}
\put(144.7,56){\vector(-1,1){20}}
\put(150,42){\vector(-1,-1){20}}
\put(125,85){$\ellp$}
\put(130,65){$\Downarrow$}
\put(175,85){$\nul$}
\put(165,65){$\Uparrow$}
\put(120,15){$\overline{\nu_{\tau}}$}
\put(135,30){$\Uparrow$}
\put(148,10){(b)}
\put(162,50){$\tau^+$}
\put(250,50){\circle{15}}
\put(250,58){\vector(0,-1){16}}
\put(255.3,56){\vector(1,1){20}}
\put(244.7,56){\vector(-1,1){20}}
\put(250,42){\vector(1,-1){20}}
\put(225,85){$\nul$}
\put(230,65){$\Downarrow$}
\put(275,85){$\ellp$}
\put(265,65){$\Uparrow$}
\put(275,15){$\overline{\nu_{\tau}}$}
\put(260,30){$\Uparrow$}
\put(248,10){(c)}
\put(262,50){$\tau^+$}
\end{picture}
 
\caption [ $\tm$  and $\tp$ decay into leptons .]
{
 $\tm$ and $\tp$ decay configurations including their spin
orientation.
a) Is an arbitrary configuration of $\tau^-$ decays to $\elln$.
b) Is charge conjugation of (a) which is not a physically state
      because the $\ellp,\;\;\nul,\:\:\overline{\nu_{\tau}}$,
      have forbidden helicities.
c) Is an allowed state, the mirror image of (b).
 
}
\label{fig_tsai1}
\end{figure}
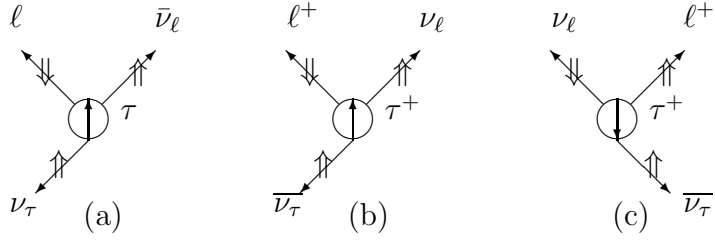

The drawings in Fig.~\ref{fig_tsai1}
illustrate the $\tm$ and $\tp$ decay configurations including their spin
orientation. In these  we are
looking at an arbitrary decay of $\t$ to lepton, $\tle$ ($\ell=e,\mu$).
Letting $m_{\ell} \to 0$,
leads  to negative $\ell,\nul,\nu_{\tau}$  helicities,
while the $\ellp,\nulb,\overline{\nu_{\tau}}$ get
positive helicities.

If CP invariance is valid in these processes then the rate of (a) is
equal to that of (c). Therefore, the distribution of the
decay $\tau^+ \to \ellp$ can be obtained by changing the sign of the
polarization vector in the distribution for $\tau^- \to \elln$.
Following this argumentation one can also relate $\tp \to \pi^+$ and
the $\tm \to \pi^-$ in the same way.

For $\tle$, 
the relevant kinematical variable for measring the $\t$ polarization is the
lepton energy, scaled by the beam energy, $x=\elep/\eb$.
The general form of energy and angular distribution of the lepton
decaying from
an arbitrarily polarized tau, can be written in the $\tau$ rest frame
as~\cite{Tsai}:
\beq
\Gamma (\tle)=
\frac{g_{\ell}g_{\tau}m_{\tau}^5}
{384 \pi^4}\int d\Omega_{\ell} \int^{1}_{0} dx\:x^2[3-2x-
({\em \vec{P}_{\tn} \cdot \hat{p_{\ell}}})(2x-1)]
\eeq
where ${\em \vec{P}_{\tn} }$ is the polarization vector of the $\tau$,
${\em \hat{p_{\ell}}}$ is the unit vector along the direction of the lepton.
The polarization dependent term
${\em \vec{P}_{\tn} \cdot \hat{p_{\ell}}}$ is due to the parity violation.
 
\underline{Near x=1}, the lepton is at its  highest energy value,
and the relative
magnitude of the parity violation term is maximal.
Here the $\elln$ tends to be emitted  opposite to the
direction of the $\tau^-$ spin, whereas the $\ellp$ tends to be emitted
parallel to the $\tau^+$ spin direction.
 
\underline{Near x=0}, exactly the opposite holds.
 
To illustrate this behavior let us look at the following diagrams.
 
At $\underline{x=1}$,
kinematics require that the $\nu$, and the $\overline
{\nu}$ are both emitted in the opposite direction of the lepton
(see Fig.~\ref{fig_tsai2}). Since the
component of the orbital angular momentum is zero along
the $\ell$ direction,
and the two neutrino spins add up to zero, the spin of the lepton
must be parallel to that of the $\tau$. Since the $\elln$ has a negative
helicity and the $\ellp$ has a positive helicity,
the $\elln$ tends to be
emitted opposite to the $\tau^-$ spin.
 
\begin{figure}[htb]
\begin{picture}(400,150)
\put(50,100){\circle{15}}
\put(50,92){\vector(0,1){16}}
\put(50,108){\vector(0,1){20}}
\put(46,93.07){\vector(0,-1){20}}
\put(54,93.07){\vector(0,-1){20}}
\put(52,135){$\ellp$}
\put(35,60){$\overline{\nu_{\tau}}$}
\put(58,60){$\nul$}
\put(44,35){x=1}
\put(30,5){{\em favored}}
\put(62,100){$\tau^+$}
\put(150,100){\circle{15}}
\put(150,108){\vector(0,-1){16}}
\put(150,108){\vector(0,1){20}}
\put(146,93.07){\vector(0,-1){20}}
\put(154,93.07){\vector(0,-1){20}}
\put(152,135){$\ellp$}
\put(135,60){$\overline{\nu_{\tau}}$}
\put(158,60){$\nul$}
\put(144,35){x=1}
\put(130,5){{\em forbidden}}
\put(162,100){$\tau^+$}
\put(250,100){\circle{15}}
\put(250,108){\vector(0,-1){16}}
\put(250,108){\vector(0,1){20}}
\put(246,93.07){\vector(0,-1){20}}
\put(254,93.07){\vector(0,-1){20}}
\put(252,135){$\elln$}
\put(235,60){$\nu_{\tau}$}
\put(258,60){$\nulb$}
\put(244,35){x=1}
\put(230,5){{\em favored}}
\put(262,100){$\tau^-$}
\put(350,100){\circle{15}}
\put(350,92){\vector(0,1){16}}
\put(350,108){\vector(0,1){20}}
\put(346,93.07){\vector(0,-1){20}}
\put(354,93.07){\vector(0,-1){20}}
\put(352,135){$\elln$}
\put(335,60){$\nu_{\tau}$}
\put(358,60){$\nulb$}
\put(344,35){x=1}
\put(330,5){{\em forbidden}}
\put(362,100){$\tau^-$}
\end{picture}
 
\caption[ $\tle$ decay in the $\tn$ rest frame ( $x=1$)]
{ $\tle$ decay in the $\tn$ rest frame for $x=1$.}
\label{fig_tsai2}
\end{figure}
 
When $\underline{x \simeq 0}$ (Fig.~\ref{fig_tsai3}),
the kinematics require that the $\nu$ and
$\overline{\nu}$ come out in opposite directions to each other,
hence their
 net spin is equal to unity and points toward  the direction of
the $\overline{\nu}$. In order to conserve angular momentum, the
$\elln$ has to move in the direction
of the $\overline{\nu}$ and the spin of the
$\tau^-$ points in the direction of $\overline{\nu}$. Hence, near $x=0$,
the $\elln$ tends to come out along the direction of spin of the
$\tau^-$, which is exactly the opposite to the $x=1$ case.
\begin{figure}[htb]
\begin{picture}(400,150)
\put(50,100){\circle{15}}
\put(50,92){\vector(0,1){16}}
\put(50,108){\vector(0,1){20}}
\put(46,93.07){\vector(0,-1){20}}
\put(54,93.07){\vector(0,-1){20}}
\put(52,135){$\overline{\nu_{\tau}}$}
\put(35,60){$\nul$}
\put(58,60){$\ellp$}
\put(44,35){$x\simeq 0$}
\put(30,5){{\em favored}}
\put(62,100){$\tau^+$}
\put(150,100){\circle{15}}
\put(150,108){\vector(0,-1){16}}
\put(150,108){\vector(0,1){20}}
\put(146,93.07){\vector(0,-1){20}}
\put(154,93.07){\vector(0,-1){20}}
\put(152,135){$\overline{\nu_{\tau}}$}
\put(135,60){$\nul$}
\put(158,60){$\ellp$}
\put(144,35){$x\simeq 0$}
\put(130,5){{\em forbidden}}
\put(162,100){$\tau^+$}
\put(250,100){\circle{15}}
\put(250,108){\vector(0,-1){16}}
\put(250,108){\vector(0,1){20}}
\put(246,93.07){\vector(0,-1){20}}
\put(254,93.07){\vector(0,-1){20}}
\put(252,135){$\nu_{\tau}$}
\put(235,60){$\nulb$}
\put(258,60){$\elln$}
\put(244,35){$x \simeq 0$}
\put(230,5){{\em favored}}
\put(262,100){$\tau^-$}
\put(350,100){\circle{15}}
\put(350,92){\vector(0,1){16}}
\put(350,108){\vector(0,1){20}}
\put(346,93.07){\vector(0,-1){20}}
\put(354,93.07){\vector(0,-1){20}}
\put(352,135){$\nu_{\tau}$}
\put(335,60){$\nulb$}
\put(358,60){$\elln$}
\put(344,35){$x \simeq 0$}
\put(330,5){{\em forbidden}}
\put(362,100){$\tau^-$}
\end{picture}
\caption[ $\tle$ decay in the $\tn$ rest frame (low $x$)]
{ $\tle$ decay in the $\tn$ rest frame for
very low $x$.}
\label{fig_tsai3}
\end{figure}

Boosting the angular distribution of the
$\tle$ from the $\t$ into the lab.\ system one obtains the $x$
distribution as given
by~\cite{Tsai},
 
\beq
\label{wl}
\frac{1}{\Gamma_\ell}\frac{d\Gamma_\ell}{dx}=\third (5-9x^2+4x^3)+
 \pt\third  (1-9x^2+8x^3). \;\;\;\;\;\;\;\;\;\;  (0\leq x\leq 1)
 \eeq

This distribution, as well as the following expressions for the
other decay channels, holds for $\tm$ as well as for $\tp$
provided that $\pt$ is always taken as the $\tm$ helicity.
Terms of order $m_\ell/m_{\t}$ have been neglected. This approximation
is fully justified for $\ell=e$, whereas for $\ell=\mu$ there is a
threshold effect around $x=0.005$ which has been
accounted for in our analysis.
 
The angular distribution of $\pi(K)^{\pm}$ from a polarized $\t^{\pm}$
is easier to understand. As it is a two-body decay the energy of
each particle in the $\t$ rest frame is fixed:
$E_{h}=\frac{\mts+m^{2}_{h^{\pm}}}{2\mt}$ and
$E_{\nu_{\t}}=\frac{\mts-m^2_{h^{\pm}}}{2\mt}$ ($h=\pi,K$).
Since the helicity of the $\nu_{\t}$ is negative it prefers to be
emitted opposite to the $\t$ spin direction, Hence the $\pi(K)^-$ prefers
to be emitted in the $\t^-$ spin direction (Fig.~\ref{fig_tsai4}).

\begin{figure}[htb]
\begin{picture}(400,150)
\put(50,100){\circle{15}}
\put(50,92){\vector(0,1){16}}
\put(50,108){\vector(0,1){20}}
\put(50,93.07){\vector(0,-1){20}}
\put(52,135){$\pi^-$}
\put(52,60){$\nu_{\tau}$}
\put(30,20){{\em favored}}
\put(62,100){$\tau^-$}
\put(150,100){\circle{15}}
\put(150,108){\vector(0,-1){16}}
\put(150,108){\vector(0,1){20}}
\put(150,93.07){\vector(0,-1){20}}
\put(152,135){$\pi^-$}
\put(152,60){$\nu_{\t}$}
\put(130,20){{\em forbidden}}
\put(162,100){$\tau^-$}
\put(250,100){\circle{15}}
\put(250,108){\vector(0,-1){16}}
\put(250,108){\vector(0,1){20}}
\put(250,93.07){\vector(0,-1){20}}
\put(252,135){$\pi^+$}
\put(252,60){$\overline{\nu_{\t}}$}
\put(230,20){{\em favored}}
\put(262,100){$\tau^+$}
\put(350,100){\circle{15}}
\put(350,92){\vector(0,1){16}}
\put(350,108){\vector(0,1){20}}
\put(350,93.07){\vector(0,-1){20}}
\put(352,135){$\pi^+$}
\put(352,60){$\overline{\nu_{\t}}$}
\put(330,20){{\em forbidden}}
\put(362,100){$\tau^+$}
\end{picture}
 
\caption[\tpiK~ decay and the  $\tn$ helicity orientation.]
{ \tpiK~ decay and  the $\tn$ helicity direction.}
\label{fig_tsai4}
\end{figure}

The same kinematical variable used in the leptonic case  can be used
also for $\tpiK$ decay.
Here it is related to $\csts$, where $\theta^*$ is the decay angle
of the hadron in the $\t$ rest frame,
\beq
\label{cstsd}
\csts=\frac{2x-1-\mhs/\mts}{\beta(1-\mhs/\mts)}. \;\;\;\;\;\;\;\;
\eeq
Henceforward, the $\t$-velocity term, $\beta$, in the denominator will be
approximated to 1 ($1-\beta=7.6\cdot10^{-4}$ for $\ecm=\mz$).
The $\csts$ distribution is given by~\cite{Tsai},
\beq
\label{wch}
\frac{1}{\Gamma_h}\frac{d\Gamma_h}{d\csts}=\half (1+  \pt\csts)
\;\;\;\;\;\;\;\;\;\;\; (-1\leq\csts\leq 1),
\eeq
corresponding to the following $x=p_h/\eb$ distributions,
\beq
\label{wxh}
\frac{1}{\Gamma_h}\frac{d\Gamma_h}{dx}=\frac{1}{1-\mhs/\mts}
\left(1+ \pt\frac{2x-1-\mhs/\mts}{1-\mhs/\mts}\right) \;\;\;\;\;\;\;\;\;\;
(\mhs/\mts\leq x\leq 1)
\eeq
 
\begin{figure}[htb]
\epsfysize=9.5 cm.
\epsffile[80 383 800 660]{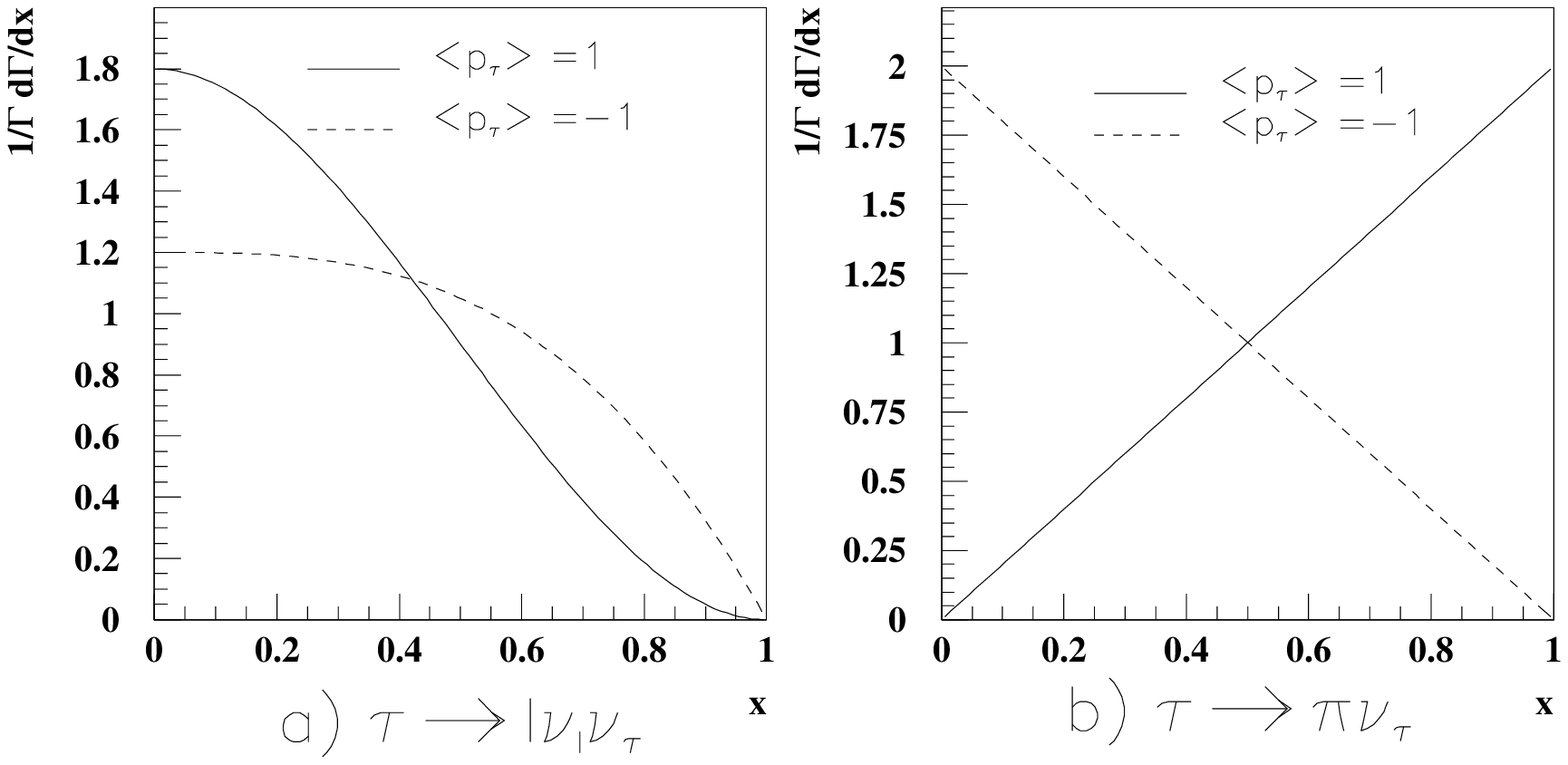}
\caption[The momentum distribution of $\tle$ and $\tpiK$.]
{The $x$ distribution $\frac{1}{\Gamma}\frac{d\Gamma}{dx}$ of
a) $\tle$ events, b) $\tpiK$ events. The solid (dashed) line is for
positive (negative) $\tn$ helicity.}
\label{fig_dsdx}
\end{figure}

Fig.~\ref{fig_dsdx} presents the $x$ distribution of $\tle$ and $\tpiK$
events for positive and negative $\tn$ helicity states, as parametrised in
Eq.~\ref{wl} for the leptonic decays
and in Eq.~\ref{wxh} for the $\pi(K)$ $\tn$ decays.
In this analysis we measure the hadron momentum, $p_h$, and do not
distinguish between $\pi$ and $K$. Therefore, $\csts$ cannot be
calculated event-by-event using Eq.~\ref{cstsd} and hence we choose
the variable $x$  taking into account the $\pi/K$
mixture by using the distribution,
\beq
\label{wxpik}
\frac{1}{\Gamma_h}\frac{d\Gamma_h}{dx}=
(1-b_K)\frac{1}{\Gamma_{\pi}}\frac{d\Gamma_{\pi}}{dx}+
 b_K\frac{1}{\Gamma_K}\frac{d\Gamma_K}{dx}
\eeq
where,
$b_K\equiv BR(\t\ra K\nu)/[BR(\t\ra\pi\nu)+BR(\t\ra K\nu)]=0.064\pm 0.007$
\cite{OPALPK}. The uncertainty in $b_k$ is
included in our systematic study (see Section~\ref{sect-RCC}).
The difference between the momentum and energy of the hadron is
also accounted for in our analysis.
 
In all  $\t$ decay channels considered above, the
 kinematical distribution is linear in $\pt$, and can be
written in the general form,
\beq
\label{wx}
\frac{1}{\Gamma}\frac{d\Gamma}{dx}=f(x)+ \pt g(x).
\eeq
Here, and in the remainder of this section, $x$ is a generic name
of the relevant kinematical variable. As shown above, $f(x)$ and
$g(x)$ may differ from one decay channel to another, but they
always satisfy the following normalization conditions,
\beq
\label{fgnor}
\int f(x) dx=1 \;\;\;\;\;\;\;\;  \int g(x) dx=0.
\eeq
 
Since in $\Z\to \tp\tm$ the helicities of the two tau leptons are expected
(assuming CP invariance) to be
fully anti-correlated~\cite{Tsai},
their measurement carry less information
compared with the case of $\tau$'s from two different events.
Until now previous analyses have neglected this
effect~\cite{OPALPL}-\cite{L32}.
In order to take these anti-correlations
into account, one has to analyze the double identified $\t$
events using the triple differential cross section with
respect to $\cst$, $\xxi$ and $\xxj$  of the
two $\tau$-decay products.

The joint distributions of the $\t$-pair production and decay is
obtained from Eqs.~\ref{dspmdcst} by combining them with the
corresponding decay distributions of the $\tm$ and the $\tp$
and summing up the two helicity configurations, resulting in,
\bea
\label{dsig3}
\dsigth=&\thovsxt \sij\sum_{hel=\pm\pm}\{[(1\pm\pta)(1+\cstsq)+
        \eitovth(\afb\pm\aplfb)\cst]  \nonumber \\
        &\times(\fix\pm\gix)(\fjx\pm\gjx)\}.
\eea
Here, $\sij$ is the cross section to produce $\t$'s decaying into
channels $i,j$.
(In the following we shall drop out the arguments $\xxi$ and $\xxj$
of the functions $\fix$ ($\gix$) and $\fjx$ ($\gjx$).
This expression accounts for the correlation between the decay
distributions of the two $\t$'s, as it should be utilized
when analyzing events where both $\t$ decay channels are identified.
 
When one of the $\t$-decay channels (e.g. the $\tp$) is not identified,
we have to integrate over its kinematical variable ($\xxj$) and we are
left with,
\bea
\label{dsig2}
\dsigtw=\thoveit
\sigma_i\sum_{hel=\pm}\{[(1+\cstsq)+\eitovth\afb\cst]\fix \nonumber \\
       \pm[\pta(1+\cstsq)+\eitovth\aplfb\cst]\gix\}.
\eea
When {\em both} $\t$-decay channels cannot be identified, we are left
with Eq.~\ref{dsigdcst} and those events contribute information only to
the forward backward asymmetry, $\afb$.
 
Defining $\lame$ and $\lamt$ as,
\begin{equation}
\label{lamet}
\lame\equiv \frac{2v_ea_e}{v_e^2+a_e^2} \;\;\;\;\;
\lamt\equiv \frac{2\vt\at}{\vt^2+\at^2},
\end{equation}
then on the $Z^0$-peak, the SM in the improved
Born approximation, neglecting the contribution of the intermediate
photon yields the following relations for $\pta$, $\afb$ and
$\aplfb$,
\bea
\label{asymsm}
\afb(\mzs)&=&\frac{F_1(\mzsq)}{F_0(\mzsq)}=
\frac{3}{4}\lame\lamt \nonumber \\
\nonumber  \\
\pta(\mzs)&=&-\frac{F_2(\mzsq)}{F_0(\mzsq)}=-\lamt \\
\nonumber  \\
\aplfb(\mzs)&=&-\frac{3}{4}\frac{F_3(\mzsq)}{F_0(\mzsq)}
=-\frac{3}{4}\lame.
    \nonumber \\ \nonumber
\eea
Assuming the SM with lepton universality, one
can write within the framework of the improved Born
approximation,
\begin{equation}
\label{lambda}
\lame=\lamt=\frac{2(1-4\swsq)}{1+(1-4\swsq)^2},
\end{equation}
or the ratio between the vector and the axial-vector coupling,
$v_l/a_l$, can  than be given as
\beq
\label{vovad}
v_l/a_l=1-4\swsq
\eeq

\begin{figure}[p]
\epsfysize=16cm.
\epsffile[60 112 1440 657]{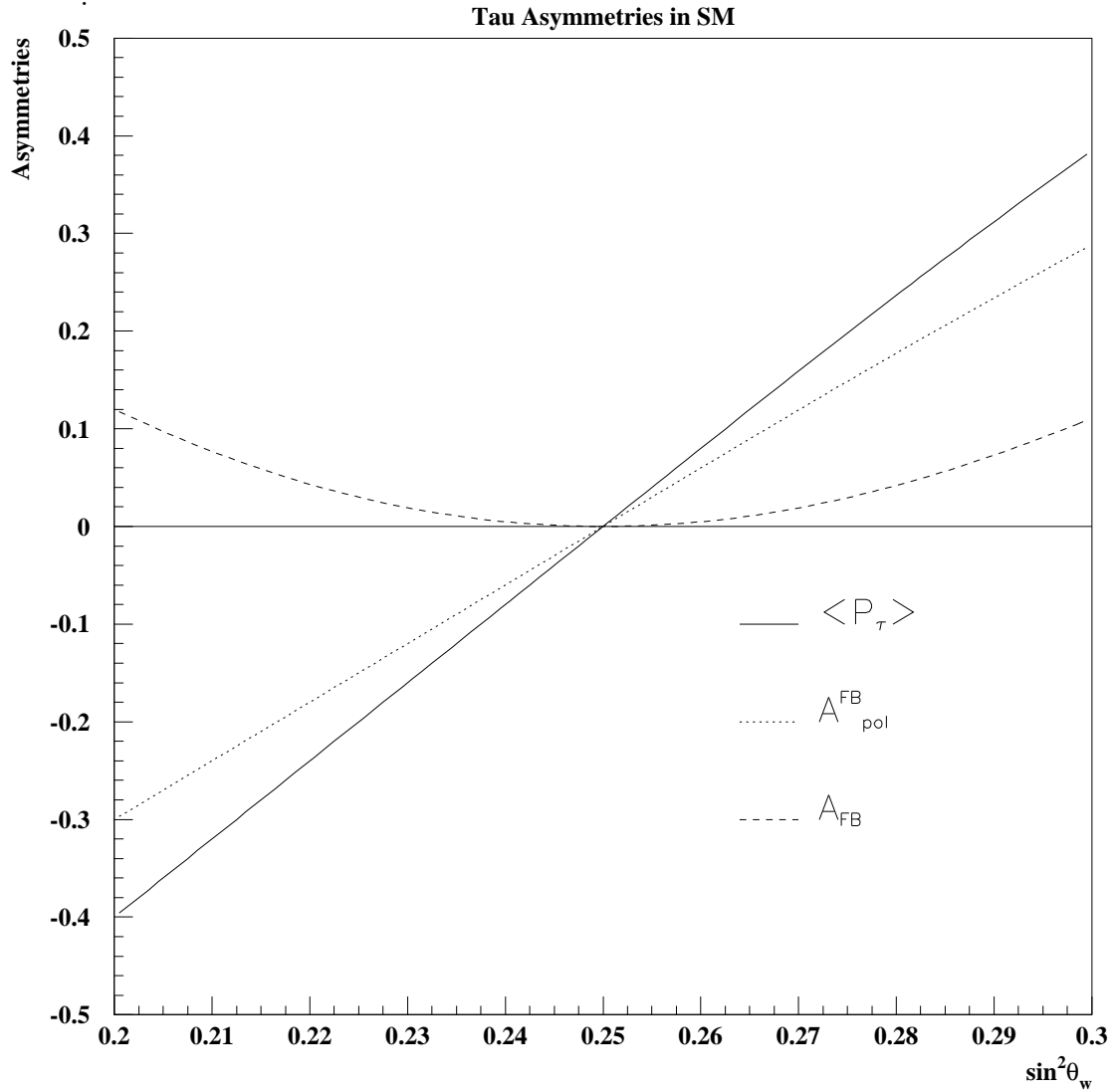}
\caption
[The $\t$ asymmetries as a function of the Weinberg
mixing angle.]
{The $\t$ asymmetries as a function of the Weinberg
mixing angle. The $\pta$, $\afb$ and $\aplfb$ are presented by
a solid line, dotted and dashed lines respectively.}
\label{fig_polsq}
\end{figure}

\begin{figure}[p]
\epsfysize=18cm.
\epsffile[40 87 1420 700]{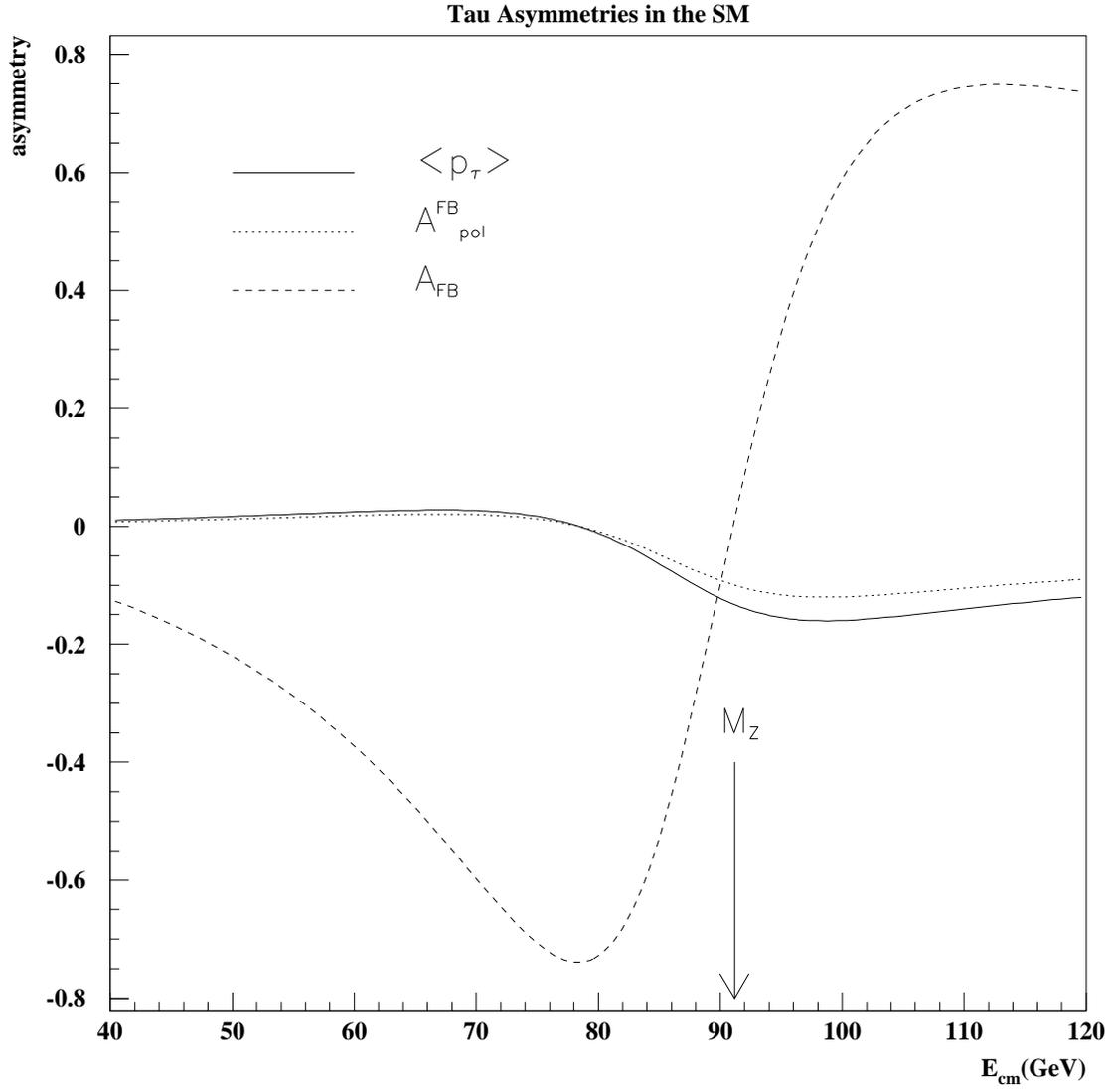}
\caption
[The $\t$ asymmetries as a function of $E_{cm}$]
{The $\t$ asymmetries as a function of $E_{cm}$.
The $\pta$, $\afb$ and $\aplfb$ are presented by
a solid line, dotted and dashed lines respectively.}
\label{fig_assecm}
\end{figure}

Fig.~\ref{fig_polsq}  shows the relations between the
three asymmetries,$\pta$, $\afb$ and $\aplfb$,
and the mixing angle, $\swsq$.
Fig.~\ref{fig_assecm} presents the asymmetries dependence on the
center of mass energy.
 
\section{\label{sec-wformalism}The Charged Weak Decay Structure}
 
In the framework of the SM the decay of the \t~ proceeds via
a charged '$V-A$'  type current.  A current of this kind has been
choosen to reproduce the observed maximal parity violation in the charged
weak interactions such as the $\beta$ decays. The $V-A$ characteristic of the
current was well tested in $\mu$ decays. As the \t~ assymetries are
extracted from the kinematical distribution of the ~\t~ decay products,
in the discussion of the polarization measurements it was implicity
assumed that the ~\t~ decay has also the same pure $V-A$
decay structure. However, any deviation from this behavior can directly
effect our measurement. Thus, in the following we shall present again the
Born term of the ~\t~ decay differential cross section in a more general form,
without restricting ourselves to the $V-A$ assumption.
The matrix element of a ~\t~ leptonic decay  can be written as
a  four fermion interaction in the following way~\cite{WMAT}
\beq
\label{MATLEP}
{\cal M}=\frac{4G_F}{\sqrt{2}}\sum_{\gamma,\epsilon,\mu,n,m}
         {g^{\gamma}_{\epsilon \mu}
         \langle\bar{\ell_{\epsilon}}|\Gamma^{\gamma}|\nul^n \rangle
         \langle \bar{\nu}^m_{\t}|\Gamma_{\gamma}|\t_{\mu} \rangle},
\eeq
where the index $\gamma$ labels the different types of interactions: scalar,
vector or tensor, $\epsilon$ and
$\mu$ stand for the chiral projection of the $\ell$ and $\t$ lepton
spinors ,  $n$ and $m$ represent the handness of the
$\nulb$ and the $\nu_{\t}$. There are 19 independent combinations
which can be measured by  experiment. The SM corresponds to
$g^V_{LL}=1$ (vector interaction of left handed $\ell$ and $\t$ leptons)
and all the other $g^{\gamma}_{\epsilon \mu}$  being zero.
 
In this parametrization the lepton momentum spectrum of the  \t~ decay
particles is modified to
\bea
\label{w2l}
\frac{1}{\Gamma_{\ell}}\frac{d\Gamma_{\ell}}{dx}&=&
f+ \xi \cdot \pt \cdot g \nonumber
\eea
where
\bea
f    &=& f_1+ \rho \cdot f_2 \nonumber \\
g    &=& g_1+ \delta \cdot g_2
\eea
and
\bea
f_1  &=& 2-6x^2+4x^3  \nonumber \\
f_2  &=& \frac{4}{9}(-1+9x^2-8x^3) \nonumber \\
g_1  &=&-\frac{2}{3}+4x-6x^2+\frac{8}{3}x^3 \nonumber \\
g_2  &=& \frac{4}{9}(1-12x+27x^2-16x^3)] \nonumber
 \eea
The $\rho,\eta,\delta$ and $\xi$ are the familiar
Michel parameters~\cite{Michel},
 and are function of the coupling constants as given in Ref.~\cite{Privitera}.
 Terms proportional to $\eta(m_{\ell}/m_{\t})$ have been neglected
 as only a vector interaction will be considered here {\mbox($\eta=0$)}.
 
 Table~\ref{Tab-Michel} gives the predicted values of the Michel parameters
 for the simple cases where only vector and axial-vector components are
 allowed at each of the $\t\nu_{\t}$ and the $\ell\nulb$ vertices.
 
\begin {table} [htb]
\begin {center}
\begin {tabular}{|c|c|c|c|c|c|} \hline
$\ell\nulb$  & $\t\nu_{\t}$ & $\rho$ & $\xi$ & $\delta$ & $\eta$ \\
\hline
$V-A$              & $V-A$        & $\frac{3}{4}$ & 1 & $\frac{3}{4}$ & 0 \\
$V$                & $V$          & $\frac{3}{8}$ & 0 & 0             & 0 \\
$A$                & $A$          & $\frac{3}{8}$ & 0 & 0             & 0  \\
$V+A$              & $V+A$        & $\frac{3}{4}$ & -1 & $\frac{3}{4}$ & 0 \\
$V-A$              & $V$          & $\frac{3}{8}$ & 2 & $\frac{3}{16}$ & 0 \\
$V-A$              & $A$          & $\frac{3}{8}$ & 2 & $\frac{3}{16}$ & 0 \\
$V-A$              & $V+A$        & 0             & 3 & 0              & 0 \\
\hline
\end{tabular}
\caption[The Michel parameters for some different $V$ and $A$ combinations]
{The Michel parameters for some different $V$ and $A$ component
combinations at the  \t~ and it lepton decay vertices.}
\label{Tab-Michel}
\end{center}
\end{table}

One can see that there are  combinations other than $(V-A)(V-A)$ which
lead to a value of 3/4 for the Michel parameter $\rho$.
For this reason one needs also to
measure the sign of the Michel polarization parameter $\xi$
in order to verify
 the $(V-A)(V-A)$ structure.
Note that when extracting the assymetries
looking separately at each \t~, one really measures the product
$\xi \cdot \pta$
rather than $\pta$.
 
The Matrix element of an hadronic \t~ decay is given by
\beq
\label{MATHAD}
{\cal M}\propto \bar{\nu_{\t}}\gamma_{\mu}(g^W_V+g^W_A\gamma^5)
\t^-\cdot J^{\mu}
\eeq
where $J^{\mu}$ stands for the hadronic current and $g^W_V,g^W_A$ are
the vector and axial-vector coupling constants.
The parameter $\xih$ is the chirality parameter defined as
\beq
\label{xih}
\xih=-\frac{2g^W_Vg^W_A}{(g^W_V)^2+(g^W_A)^2},
\eeq
and $-\xih$ can be interpeted as twice the  {\bf $\nu_{\t}$ helicity}.
 
Using this definition of the charged weak interaction, the $x$ distributions
of the $\tpiK$ decay (Eq.~\ref{wxh}) is modified to
\beq
\label{wxh2}
\frac{1}{\Gamma_h}\frac{d\Gamma_h}{dx}=\frac{1}{1-\mhs/\mts}
\left(1+\xih \cdot \pt\frac{2x-1-\mhs/\mts}{1-\mhs/\mts}\right)
\eeq
 
For the most general case one needs to have the following
modifications in Eq.~\ref{dsig3}
\begin{itemize}
\item Incorporate  $\xi_i$ and $\xi_j$ where both can stand for
       either $\xi$ or $\xih$ depending on the decay channel.
\item  Use the modified $f$ and $g$ terms in  a leptonic decay channel
      including explicity the $\rho$ and $\delta$ parameters as in
      Eq.~\ref{w2l}.
\end{itemize}
 
Rearranging  expression~\ref{dsig3}, one  gets  the
following form,
\bea
\label{dsig3p}
\dsigth&=&\thovsxt\sij \sum_{hel=\pm\pm}\{
[(1+\cstsq)+\eitovth\afb\cst][1+(\mp\xi_i)\cdot(\mp\xi_j)] \nonumber \\
     & & -[\pta(1+\cstsq)+\eitovth\aplfb\cst][(\pm\xi_i)+(\pm\xi_j)] \}
         \nonumber \\
    & &  \times (\fix(x,\rho)\pm\gix(x,\delta))\cdot(\fjx(x,\rho)\pm\gjx(x,\delta))
\eea
giving the possibility to extract the three
asymmetries $\afb$, $\pta$ and $\aplfb$, as well as the $\xi$, $\xih$,
$\rho$ and $\delta$
parameters by fitting it to the data.
 
The $\rho$ parameter was already measured precise enough to exlude interactions
different from $(V-A)$ at least in the $\ell\nulb$ vertex. Therefore,
one can take
 a $(V-A)$ at the $\ell\nulb$ vertex
and allow a combination of $(V-A)$ and $(V+A)$ at the \t~ one.
This assumption reduces the
number of parameters for the charged weak interaction and
implies~\cite{Privitera}:
\bea
\rho=\frac{3}{8}(1-h_{\nu_{\t}}) \nonumber \\
\delta=\frac{3}{16}\frac{(1-h_{\nu_{\t}})}{(1+\frac{h_{\nu_{\t}}}{2})} \\
\xi=2+h_{\nu_{\t}}  \nonumber \\
\xih=-h_{\nu_{\t}}. \nonumber
\label{helndef}
\eea
 
We see that the only way to extract the three $\t$ asymmetries and the
charged weak parameters all together, is to measure the triple differential
distribution of the variables $\xxi,\xxj,\cst$ using events where both
$\t$ decays are identified. However, one usually takes $\xi=\xih=1$,
$\rho=\delta=3/4$,
then the $\t$ asymmetries can be obtained also from the double
differential distributions of the variables $\xxi$ (or $\xxj$) and $\cst$,
using events where only one of the $\t$ decays was identified.
 
It is evident from Eq.~\ref{dsig3p} that $\xi$ can be obtained also at
low energies where the $\t$ asymmetries vanish. This was done
by the ARGUS collaboration for events where one $\t$ decays to electron
and the other $\t$ decays into muon, with the preliminary
result~\cite{Gulotv} $\xi=\sqrt{\xi_e\xi_{\mu}}$=0.90$\pm$0.13$(sys.)
\pm 0.13(stat.)$.

In the following unless it is  explicity stated
we shell assume  {\mbox{$(V-A)(V-A)$}} couplings,
and hence the  $\xi$ and $\xih$
as well as the other Michel parameters ($\rho,\;\delta$ and $\eta$)
we will substitute by their SM values.

\section{Correlations Between $\afb$, $\pta$ and $\aplfb$}
At first sight, the measurement of the
$\tau$-polarization carried via the
momentum distribution of the $\tau$-decay products
is independent of the measurement of the
$\tau$-forward-backward asymmetry  based on the $\tau$-
angular distribution.
This statement is however wrong, since the angular and
momentum distributions are not independent. It is a
direct consequence of the $\tau$-polarization dependence
on the $\tau$-scattering angle, which can be seen for
example by comparing the expressions for the polarization
asymmetries in the forward and backward hemispheres
(Eqs.~\ref{apolfs} and~\ref{apolbs}).
 
In order to obtain the correlations between $\afb$,
$\pta$ and $\aplfb$, they must be expressed in terms of
other variables which are mutually independent.
We choose $\apolf$, $\apolb$ and the fraction $\zeta$ of
the forward scattering events. $\apolf$ and $\apolb$ are
clearly mutually independent since they relate to
different event samples. They also do not depend on the
relative size of each sample which is determined by
$\zeta$. The expressions of $\afb$, $\pta$ and $\aplfb$
in terms of $\apolf$, $\apolb$ and $\zeta$ are
straightforward
\begin{eqnarray}
\label{trans}
\afb & = & 2\zeta-1, \nonumber \\
\pta & = & \zeta\apolf+(1-\zeta)\apolb, \\
\aplfb & = & \zeta\apolf-(1-\zeta)\apolb. \nonumber
\end{eqnarray}
 
Using the standard error propagation procedure, the
non-diagonal elements of the covariance matrix can be
calculated in terms of the variables $\zeta$, $\pta$,
$\aplfb$ and their statistical errors,
\begin{eqnarray}
\label{covar}
\langle\Delta\afb\Delta\pta\rangle &=&
2(\apolf-\apolb)\Delta\zeta^2, \nonumber  \\
\langle\Delta\afb\Delta\aplfb\rangle &=&
2(\apolf+\apolb)\Delta\zeta^2, \\
\langle\Delta\pta\Delta\aplfb\rangle &=&
[(\apolf)^2-(\apolb)^2]\Delta\zeta^2  \nonumber \\
& &+\zeta^2(\Delta\apolf)^2 -(1-\zeta)^2(\Delta\apolb)^2. \nonumber
\end{eqnarray}
 
Dividing those covariances by the corresponding statistical
error products yields the correlation coefficients,
\begin{eqnarray}
\label{correl}
\rho_{\afb\pta}&=&\langle\Delta\afb\Delta\pta\rangle/
                 (\Delta\afb\Delta\pta), \nonumber  \\
\rho_{\afb\aplfb}&=&\langle\Delta\afb\Delta\aplfb\rangle/
                  (\Delta\afb\Delta\aplfb), \\
\rho_{\pta\aplfb}&=&\langle\Delta\pta\Delta\aplfb\rangle/
                   (\Delta\pta\Delta\aplfb), \nonumber
\end{eqnarray}
and in order to obtain their numerical values, we can take
the SM values for the asymmetries, using the
$\swsq$ value of 0.2337 yielding,
\[\begin{array}{lcr}
\label{sma}
\apolf=0.225\; & \;\apolb=0.032\; &
\;\zeta=\frac{1}{2}(\afb+1)=0.506.
\end{array}\]
These numbers correspond to the full solid angle and they
must be modified if a cut on $|\cost|$ is performed on
the data.
 
It is straightforward to calculate the statistical
uncertainty in $\zeta$ using the binomial distribution
expression,
\begin{equation}
\label{detas}
\Delta\zeta^2=\frac{\zeta(1-\zeta)}{N},
\end{equation}
where N is the total number of events. Hence,
\begin{equation}
\label{dafb}
\Delta\afb=2\Delta\zeta=2\sqrt{\frac{\zeta(1-\zeta)}{N}}.
\end{equation}
On the other hand, the statistical errors of the polarization
asymmetries are not so easy to calculate and they depend on
the way that these asymmetries are measured. The calculations will
be performed in the next section, however a rough
estimate can be done already with former values obtained in the
experimental analyses.
For example  we made before our analysis (see Ref.~\cite{ABE}) an estimate
of the  correlation using the OPAL 1990+1991 values~\cite{OPALPL}
 of $\Delta\apolf$, $\Delta\apolb$, $\Delta\pta$ and
$\Delta\aplfb$ obtained within the angular region of
$|\cost|<.68$
The correlation results are given in  Table~\ref{correxp}
show that all the correlation coefficients turn
out to be at the level of few percents.
\begin {table} [htb]
\begin {center}
\begin {tabular}{|l|c|c|c|c|} \hline
$\tau$-decay mode&number of $\tau$'s&$\rho_{\afb\pta}$
 &$\rho_{\afb\aplfb}$  &$\rho_{\pta\aplfb}$ \\
 &  & (\%)  & (\%)  & (\%)  \\
\hline
$\tel$ & 1809 & 1.77 & 2.76 & 1.82 \\
$\tmu$ & 1729 & 1.64 & 2.39 & 1.82 \\
$\tpiK$ &  553 & 4.00 & 6.10 & 1.91 \\
\hline
\end{tabular}
\caption[The correlation coefficients between $\tau$-
asymmetries.]{The SM correlation coefficients between $\tau$-
asymmetries,  calculated using  the OPAL 1990-1991 results and their
uncertainties.}
\label{correxp}
\end{center}
\end{table}
 
\section{The Fit Method}
In order to choose and decide about the method of extraction
the $\t$ asymmetris
and to show the strength of our selected fit tecnique
we present in the following results of a comparison between
several approaches (for details see Ref.~\cite{ABE}).
\subsection{Events with only one Identified $\tau$-decays}
 
In order to obtain \pta~ and $\aplfb$ from the experimental
data, the total (normalized) momentum, $x$, of the $\tau$-decay products
(excluding the $\nu$'s) has to be measured.
However, as shown in Eq.~\ref{dsig2} the $\tau$-asymmetries $\afb$, $\pta$,
$\aplfb$ and their full
covariance matrix can be obtained by fitting that
expression to the data.
One can also reduce Eq.~\ref{dsig2} to a one parameter expression, namely
$\swsq$, which is related by Eq.~\ref{asymsm} and Eq.~\ref{lambda}
to \pta~ and $\aplfb$.
 
For an estimate of the theoretical expected errors prior to our
measurements, we have normalized both sides of Eq.~\ref{dsig2} to unity,
and wrote it in the shorthanded form,
\begin{equation}
\label{signor}
\frac{1}{\sigma}\frac{d^2\sigma}{d\cost dx}(s,\cost,x)=
\sum_{i,j=0}^{1}A_{ij}f_i(x)h_j(\cost),
\end{equation}
where,
\[\begin{array}{cccc}
\label{aij}
A_{00}=1\; & \;A_{01}=\afb\; & \;A_{10}=\pta\; & \;A_{11}=\aplfb,
\end{array}\]
and,
\[\begin{array}{lr}
\label{hij}
h_0(\cost)=\frac{3}{8}(1+\cosst)\; & \;h_1(\cost)=\cost.
\end{array}\]
 
In order to estimate the covariance matrix,
we divide the interval of $x$ into $n$ bins and define
$f_i^k$ as the integral of $f_i(x)$ over the bin $k$
($k=1\ldots n$).  In the same way, the interval of $\cost$ is
divided into $m$ bins and $g_j^l$ is defined as the integral
of $h_j(\cost)$ over the bin $l$ ($l=1\ldots m$). The quantity to
minimize is then,
\begin{equation}
\label{wmin}
W=\sum_{k=1}^{n}\sum_{l=1}^{m}[y^{kl}-
\sum_{i,j=0}^{1}A_{ij}f_i^kh_j^l]^2/(\Delta y^{kl})^2,
\end{equation}
where $y^{kl}$ is the fraction of events in the two-dimensional
bin (k,l) and its statistical error, $\Delta y^{kl}$, can be
approximated for the case of large number of bins (namely,
$y^{kl}\ll1$) by,
\begin{equation}
\label{dykl}
\Delta y^{kl}=\sqrt{y^{kl}/N},
\end{equation}
with N being the total number of $\tau$'s. The inverse of the
covariance matrix is a $3\times 3$ matrix given by,
\begin{equation}
\label{vmo}
V^{-1}_{\;\;\;ij,i'j'}=\frac{1}{2}\frac{\partial^2W}
{\partial A_{ij}\partial A_{i'j'}}=
N\sum_{k=1}^{n}\sum_{l=1}^{m}\frac{f_i^kh_j^lf_{i'}^kh_{j'}^l}
{y^{kl}}
\end{equation}
where $i\!=\!j\!=\!0$ and $i'\!=\!j'\!=\!0$ are excluded,
since $A_{00}\!=\!1$.
Apart from the constant $N$, the matrix $V^{-1}$ can be calculated even
without data, by using the SM estimates for $y^{kl}$.
 
The results listed in Table~\ref{covmat1} are obtained from the
elements of the covariance matrix $V$, using $100\times100$ bins.
The Table shows that the errors of $\afb$ are smaller than
those of $\pta$ and $\aplfb$.
By comparing between the two decay modes, one can see that
the errors on $\pta$ and $\aplfb$ are smaller for the
 $\tpiK$ decay mode.  The
factor between the two errors is 2.7. As expected, the errors
on $\afb$ do not depend on the $\tau$-decay mode.
\begin {table} [htb]
\begin {center}
\begin {tabular}{|l|c|c|c|} \hline
decay mode &
$\sqrt{N} \Delta\afb$  & $\sqrt{N} \Delta\pta$ &
$\sqrt{N} \Delta\aplfb$ \\   \hline
$\tle$ & 0.935 & 4.58 & 4.28 \\
$\tpiK$ & 0.934 & 1.72 & 1.61 \\
\hline
\end{tabular}
\caption[An estimate of the  \tn~ asymmetries statistical errors]
{The statistical errors of the $\tau$-asymmetries
 obtained from a two-dimensional fit
using the SM estimates.}
\label{covmat1}
\end{center}
\end{table}
 
All the errors listed in Table~\ref{covmat1} are for the ideal
case. They were calculated with the implicit assumption of no
background and a uniform efficiency over all $x$- and $\cost$-
values ($0\leq x \leq 1$, $|\cost| \leq 1$). In practice however,
this is not the case and therefore, our calculated errors should
be considered as lower limits.
 
Table~\ref{covmat2} lists  the correlation coefficients
obtained from the non-diagonal elements of the covariance
matrix. They are slightly higher than the values obtained in
the last section using a simpler method and the
experimentally measured statistical errors. This is mainly
because those correlation coefficients are inversely
proportional to the statistical errors, and the values
used here are for the ideal case.
 
\begin {table} [htb]
\begin {center}
\begin {tabular}{|l|c|c|c|} \hline
decay mode
& $\rho_{\afb\pta}$ & $\rho_{\afb\aplfb}$ &$\rho_{\pta\aplfb}$  \\
   & (\%) & (\%) & (\%) \\ \hline
$\tle$  &2.24 &2.77 & 4.38 \\
$\tpiK$  &5.96 &7.44 & 0.18 \\
\hline
\end{tabular}
\caption[An estimate of the \tn~ asymmetries correlations]
{The correlations in the $\tau$-asymmetries
obtained from a two-dimensional fit,
using the SM estimates.}
\label{covmat2}
\end{center}
\end{table}

An alternative method to the two-dimensional fit is
the use of
two independent one-dimensional fits for the  forward and backward
scattering events. This method is based on the integration of
Eq.~\ref{dsig2} over $\cost$,
separately in the forward ($0\leq\cost\leq1$) and in the
backward ($-1\leq\cost\leq0$) regions to obtain $\apolf$ and
$\apolb$.
From $\apolf$ and $\apolb$ one calculates then
$\pta$ and $\aplfb$. This approach has been adopted in the
previous published OPAL analysis~\cite{OPALPL}.
 
A comparison between the two methods done in Ref.~\cite{ABE}
has used 1241 $\tmu$ decays from the OPAL 1990 data~\cite{TN031}.
Performing a two-dimensional Maximum Likelihood (ML) fit,
the result reached was,
\begin{equation}
\label{eresult}
\efswsq=0.2306\pm0.0085.
\end{equation}
On the other hand, separate one-dimensional
fits on forward and backward scattering events gave
~\cite{TN031},
\begin{equation}
\label{erezcom}
\efswsq=0.2342^{+0.0110}_{-0.0095}.
\end{equation}
The statistical error of the last result is higher than
the uncertainty of the value from the two-dimensional fit
(Eq.~\ref{eresult}). As one could expect, some information in
the data was lost by changing from a
two-dimensional fit to two separate one-dimensional fits.
 
For a study of the dependence of the error on the grid size,
 we repeated the two-dimensional fit
by varying the number of bins. The results
are listed in Table~\ref{varbin}. The $\efswsq$ error values were
obtained by combining  the errors from $\afb$, $\pta$, and
$\aplfb$, taking into account their correlations.
\begin {table} [htb]
\begin {center}
\begin {tabular}{|l c|c|c|c|c|c|} \hline
\multicolumn{7}{|c|}{ $\tle$}   \\
\hline
       & $x$  & $\cst$ & \multicolumn{4}{|c|}{$\sqrt{N}\:\times$} \\
       & bins & bins  & $\Delta\afb$ & $\Delta\pta$
                              & $\Delta\aplfb$ & $\Delta\efswsq $ \\
\hline
$\chs$ & 10  &  2  & 1.000 & 4.61 & 4.61 & 0.390 \\
       & 10  & 10  & 0.937 & 4.61 & 4.32 & 0.376 \\
       & 100 & 100 & 0.935 & 4.58 & 4.28 & 0.373 \\
\hline
\multicolumn{3}{|l|}{ML} &
                   0.936 & 4.39 & 4.09 & 0.359 \\
\hline \hline
 \multicolumn{7}{|c|}{$\tpiK$}      \\
 \hline
       & $x$  & $\cst$ & \multicolumn{4}{|c|}{$\sqrt{N}\:\times$} \\
       & bins & bins  & $\Delta\afb$ & $\Delta\pta$
                              & $\Delta\aplfb$ & $\Delta\efswsq $ \\
\hline
$\chs$ & 10 &  2 & 1.000 & 1.73 & 1.73 & 0.173 \\
       & 10 & 10 & 0.937 & 1.73 & 1.62 & 0.168 \\
       & 100 &100 & 0.934 & 1.72 & 1.61 & 0.167 \\
\hline
\multicolumn{3}{|l|}{ML} &
                   0.932 & 1.73 & 1.62 & 0.169 \\ \hline
\end{tabular}
\caption[The errors in the $\tn$ asymmetries for different fit approachs]
{The statistical errors in the $\tau$-asymmetries
as obtained from a two-dimensional fit, using the least
square method for different numbers of bins and the
ML method.}
\label{varbin}
\end{center}
\end{table}
 
As expected, all errors decrease with increasing number of bins.
Using only two bins in $\cost$
leads to errors which are 6-7\% higher in $\afb$
and $\pta$, corresponding to a 3-4\% higher error in the
combined value of $\efswsq$. The error on $\pta$ is unaffected
by varying the number of $\cost$-bins, as expected. 10 bins in
$\cost$ and $x$ seem to be sufficient, since the improvement
obtained by using 100 bins in each variable is marginal. The
number of possible bins is of course limited by statistics.
However, replacing our least square method with the
(event by event) ML approach,
one should get effectively an accuracy
corresponding to infinite number of bins.
This was also tested by performing this ML
fit on 50000 Monte Carlo $\tau$-pair events.
The results are listed in Table~\ref{varbin}.
They turn out to be very close to those of the least
square method with $100\times100$ bins, except for the errors
in $\pta$ and $\aplfb$ from the $\tle$-decay channel
which are lower. Hence, using a ML fit for
this decay channel, may reduce the statistical uncertainty in
$\efswsq$ even further, to a value which is about 10\% below
the value obtained by the one-dimensional fits. This
 improvement was also obtained  at the beginning of this
section, using fits on real data.
 
\subsection{Events with Two Identified $\tau$-Decays}
With a probability of 30\%
to identify a $\tau$-decay (to e, $\mu$, $\pi(K)$), about 9\%
of the events have two  identified $\tau$'s.
Since both helicities are
fully anti-correlated, their measurement carries less information
compared with the case of two $\tau$'s of different events.
This effect was neglected in the previous LEP analyses.
In order to take the anticorrelation into account,
one has to analyze these
events using the triple differential cross section of Eq.~\ref{dsig3p}.
With this expression, the covariance matrix of the three
$\tau$-asymmetries can be estimated from a three-dimensional
fit, in an analogous way to the two-dimensional fit of the
last subsection. The results are listed in
Tables~\ref{twotau1} and~\ref{twotau2}.
Comparing these with the results of events with one identified
$\tau$ (Tables~\ref{covmat1} and~\ref{covmat2}), one can see that
the errors in $\pta$ and $\aplfb$ are smaller by a factor
of $\sqrt{2}$ from that expected for  the case
of two uncorrelated $\tau$'s. The errors in $\afb$ are
the same in both cases, but the correlations between $\afb$
and $\pta$ or $\aplfb$ become now more significant.
\begin {table} [htb]
\begin {center}
\begin {tabular}{|l|l|c|c|c|} \hline
\multicolumn{2}{|c|}{decay modes} & \multicolumn{3}{|c|}{$\sqrt{N}\; \times$} \\
\multicolumn{2}{|c|}{} &$\Delta\afb $ & $\Delta\pta$  &
$\Delta\aplfb$ \\
$\tau_1$ & $\tau_2$ &  &  & \\ \hline
$\led$ & $\led$ & 0.937 & 3.31 & 3.10  \\
$\led$ & $\pid$ & 0.937 & 1.66 & 1.56  \\
$\pid$ & $\pid$ & 0.937 & 1.37 & 1.28  \\
\hline
\end{tabular}
\caption[Estimated errors for events where both decays are identified]
{The statistical errors in the $\tau$-asymmetries
 for events where both $\tau$-decay
are identified.}
\label{twotau1}
\end{center}
\end{table}
 
\begin {table} [htb]
\begin {center}
\begin {tabular}{|l|l|c|c|c|} \hline
\multicolumn{2}{|c|}{decay modes} &
$\rho_{\afb\pta}$ &
$\rho_{\afb\aplfb}$ &$\rho_{\pta\aplfb}$  \\
$\tau_1$ & $\tau_2$  & (\%) & (\%) & (\%) \\ \hline
$\led$ & $\led$ & 3.09 &3.84 & 3.71 \\
$\led$ & $\pid$ & 6.15 &7.69 & 0.31 \\
$\pid$ & $\pid$ & 7.48 &9.36 & 0.05 \\
\hline
\end{tabular}
\caption[Estimated correlations for events where both
decays are identified]
{The statistical correlations in the $\tau$-asymmetries
for events where both $\tau$-decay
are identified.}
\label{twotau2}
\end{center}
\end{table}
 
All our results lead to the conclusion that the data should
be analyzed using a simultaneous fit to
all the relevant kinematical variables
which can be measured. The only way to realize it,
especially for data with limited statistics, is the event
by event ML method which has the following
advantages:
\begin{itemize}
\item All the information included in the data is used and
nothing gets lost through binning.
\item All the correlations between the three asymmetries are
obtained by this method. If one chooses to assume the
SM and lepton universality, taking $\efswsq$ as
the fit variable, all the correlations between the three
asymmetries are implicitly taken into account.
\item For events which are not exactly on the $Z^0$-peak, the
asymmetry values vary as a function of $s$
(see Fig.~\ref{fig_assecm}). In this case it is advantageous
to express the asymmetries as a function of
$\efswsq$   which can then be fitted over the whole $\Z$ line shape.
\item It is possible to use in the same fit events with
different information content. For example, events where no
$\tau$-decay mode was identified have only the $\cost$-
information. Events where only one $\tau$-mode was
identified will have the $\cost$ and $x$ variables.
Events
with both $\tau$-decay modes identified will have the
$\cost$, $\xxi$ and $\xxj$ variables.
\end{itemize}

\chapter{Data Selection}
\label{chap-select}
\section{Data}
\label{data}
This analysis uses \tn-pair events selected by the Tau Platform (TP)
program (TP103 version)\footnote{
The Tau Platform (TP) is an OPAL set of programs that create and manipulate
"mini DST" (Data Summary Tapes) files, designed for the \tn~ decay analysis.
The first version contributors are: K. Riles, M. Sasaki and E. Etzion.
The version used in our analysis is TP103. A primer which gives an
overview of the programs, its details, the variables and the event
selection is available for version 1.02~\cite{TP103}.}
using data taken by OPAL in
1990 (PASS 4) and 1991-1992 (PASS 5)\footnote{PASS4 and PASS5 are stages
of the events reconstruction  done with ROPE, a collection of modules
designed to reconstruct events from digits produced by the OPAL detector,
or by the OPAL simulation program, GOPAL.
Its aim is to produce the DST files, and it also provides
the event viewing facility through its graphic processor GROPE.}.
The detector and trigger status requirements are summarized in
Table~\ref{tab-status}.
They are identical to those required by TP103 except for the status cut
on FD which is not imposed here, since the FD is not used in any stage
of the event selection and the luminosity measured by this detector,
is not needed in this analysis.
 
\begin {table} [hbt]
\begin {center}
\begin {tabular}{|l|c|c|c|c|c|c|c|c|} \hline
         & CV & CJ & TB & PB & EB & EE & HS & MB \\ \hline
Detector & 3  & 3$^\ast$ & 3$^\dagger$ & 2 & 3 & 3 & 3 & 3 \\
Trigger  &    & 2  &    &    & 2  & 3  &    &     \\ \hline
\multicolumn{9}{l} { } \\
\multicolumn{9}{l} {$^\ast$ CJ$\geq$2 for runs 2001-2293.} \\
\multicolumn{9}{l} {$^\dagger$ only for 1991-1992 data.} \\
\end {tabular}
\caption[Detector and trigger status requirements.]
{The detector and trigger minimal status requirement.
The trigger and detector status of each subdetector is coded in
the following way: \\
{\bf 0} - Trigger / detector status unknown.\\
{\bf 1} - Trigger / detector status not on.\\
{\bf 2} - Detector partially on (low voltage or readout problems).\\
{\bf 3} - Detector in a good operation mode.
 }
\label{tab-status}
\end{center}
\end{table}
 
\begin{figure}[htbp]
\epsfysize=16cm.
\epsffile[40 87 622 692]{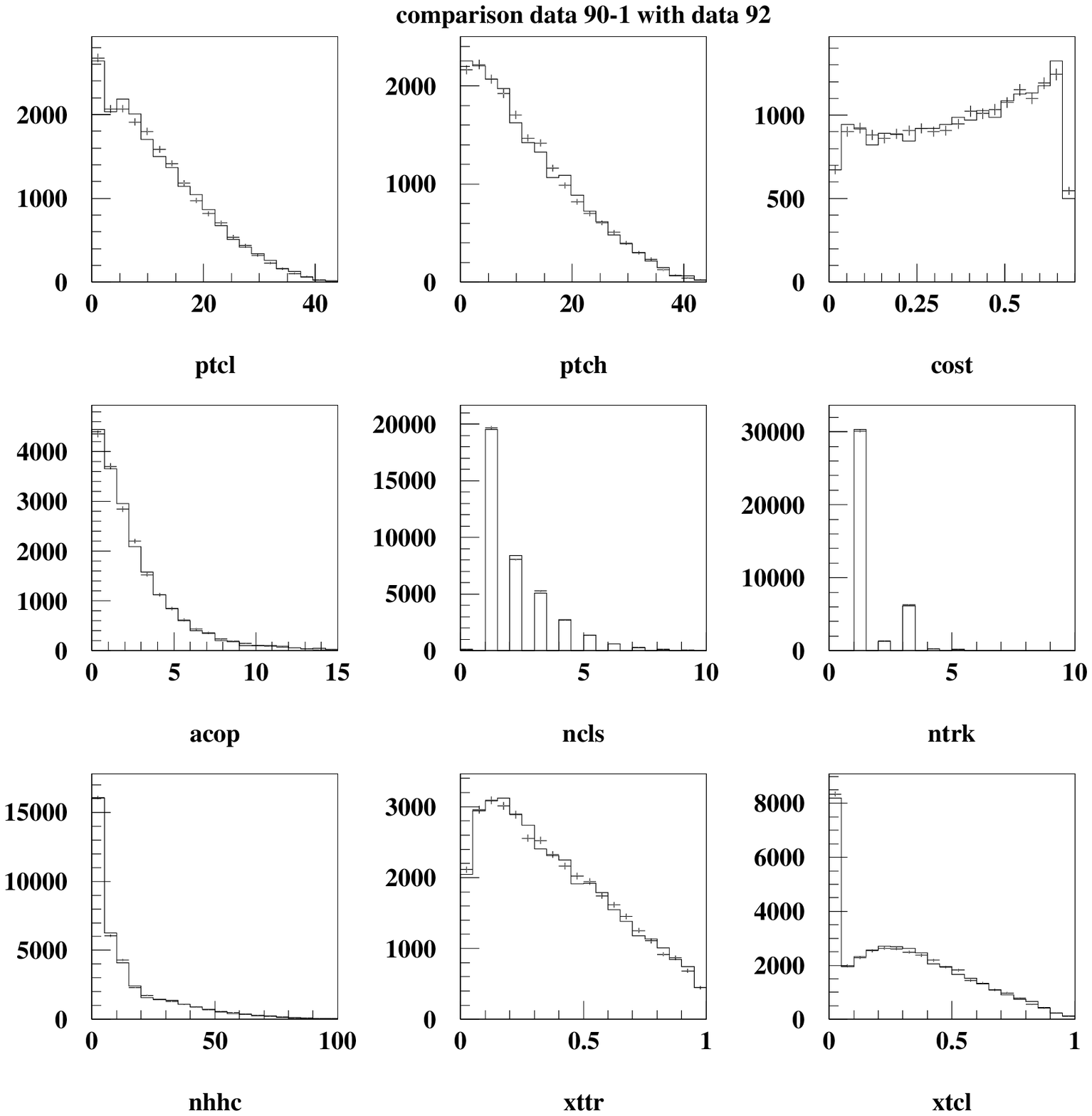}
\caption[Comparison of 1990+1991 with 1992 data samples.]
{(a) Comparison of 1990+1991 (histograms) with 1992 data
samples (error bars). The variables compared here are:
{\bf ptcl} - transverse momentum using ECAL clusters,
{\bf ptch} - transverse momentum using charged tracks,
{\bf cost} - $\cost$,
{\bf acop} - acoplanarity,
{\bf ncls} - no. of ECAL clusters,
{\bf ntrk} - no. of charged tracks,
{\bf nhhc} - no. of HCAL strip hits assigned to a cone,
{\bf xttr} - total track momentum/$\eb$,
{\bf xtcl} - total ECAL cluster momentum/$\eb$.
}
\label{fig-compd1}
\end{figure}

\begin{figure}[htbp]
\epsfysize=16cm.
\epsffile[40 87 622 692]{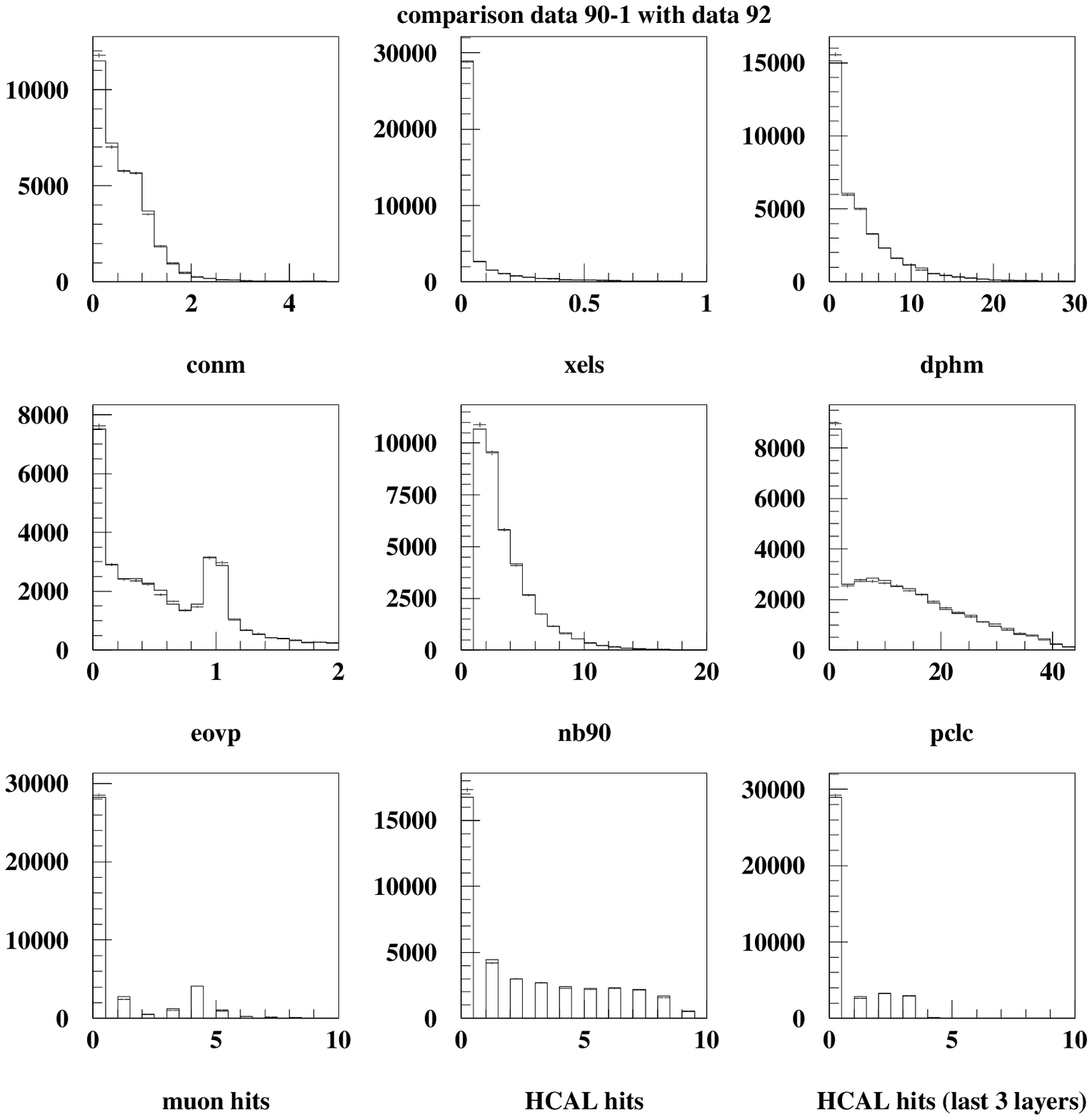}
\caption[Comparison of 1990+1991  with 1992 data samples.]
{(b) Comparison of 1990+1991 (histograms) with 1992 data samples
(error bars). The variables compared here are:
{\bf conm} - cone's invariant mass,
{\bf xels} - ECAL energy excluding the clusters associated to
track/$\eb$,
{\bf dphm} - $\dphimax$,
maximum difference between the track and the
presampler signals,
{\bf eovp} - E/P,
{\bf nb90} - no. of blocks containing 90\% of ECAL energy,
{\bf pclc} - ECAL energy,
{\bf MUON hits},
{\bf HCAL hits},
{\bf HCAL hits (last 3 layers)},
 
}
\label{fig-compd2}
\end{figure}

In Figures~\ref{fig-compd1} and~\ref{fig-compd2} we compare the data
samples of 1992 with the combined
1990+1991  by looking at the
distributions of those measured variables which are used to select
$\t$-pairs and identify their decays. No statistically significant
difference was found between the two samples. We therefore combine
them into one sample.
 
\section{Monte Carlo and Control Samples}
\label{MC}
In order to simulate the various processes which potentially
contribute to the selected tau-pair data sample,
several Monte Carlo (MC) data sets were used.
The response of the OPAL detector
to the generated particles in each case was modelled fully
using GOPAL, a simulation program~\cite{bib-gopal}
based on the CERN GEANT3~\cite{bib-geant} package.
We have further smeared the  particles momenta and their electromagnetic
energies of these MC events in order to account
for effects not well modeled in GOPAL.
In all cases, the MC
and real data were reconstructed and analyzed in an identical manner.
 
\begin{figure}[htbp]
\epsfysize=17cm.
\epsffile[40 93 622 716]{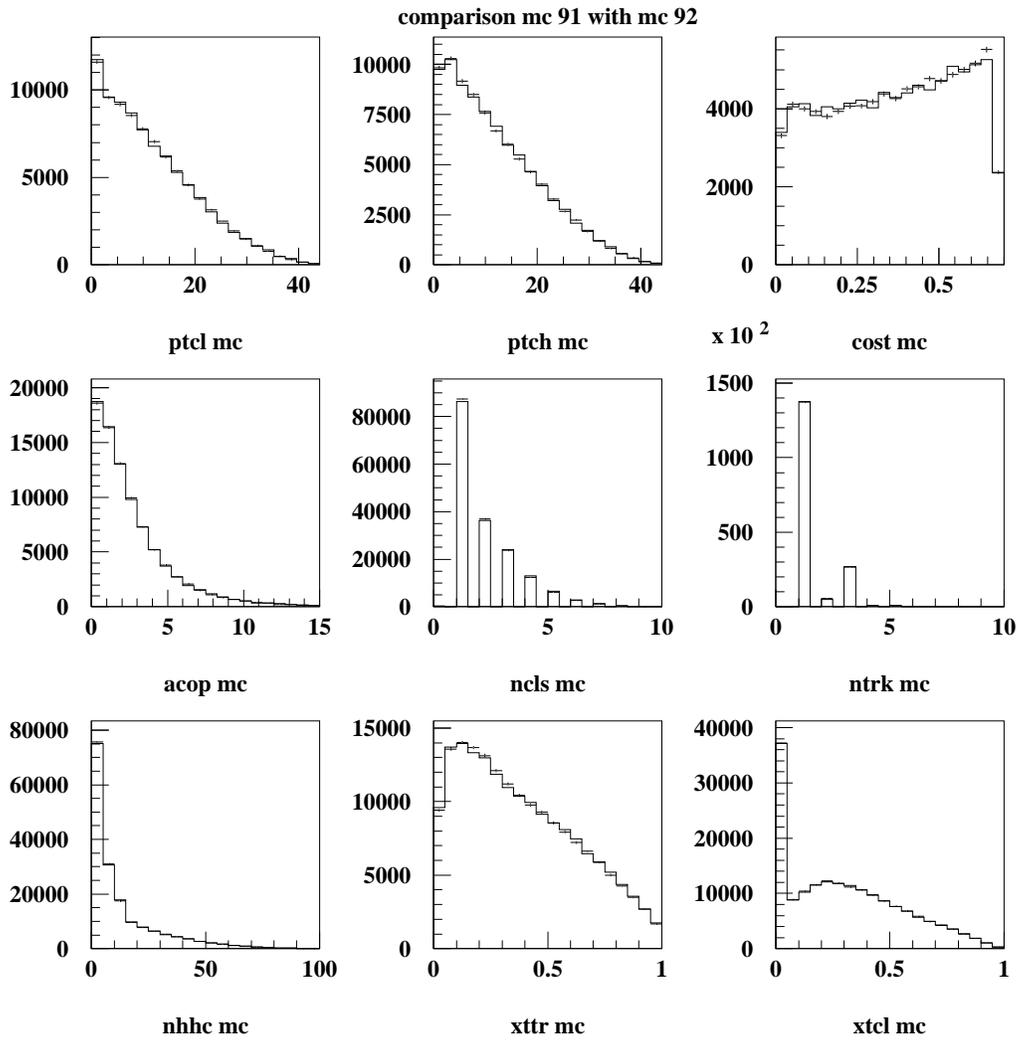}
\caption[Comparison of MC samples of 1991 with 1992 detector setups.]
{(a) Comparison of MC samples with 1991 (histograms) and 1992 (error bars)
detector setups.
The variables shown are identical to those given in Fig. 5.1}
\label{fig-compmc1}
\end{figure}

\begin{figure}[htbp]
\epsfysize=17cm.
\epsffile[40 93 622 716]{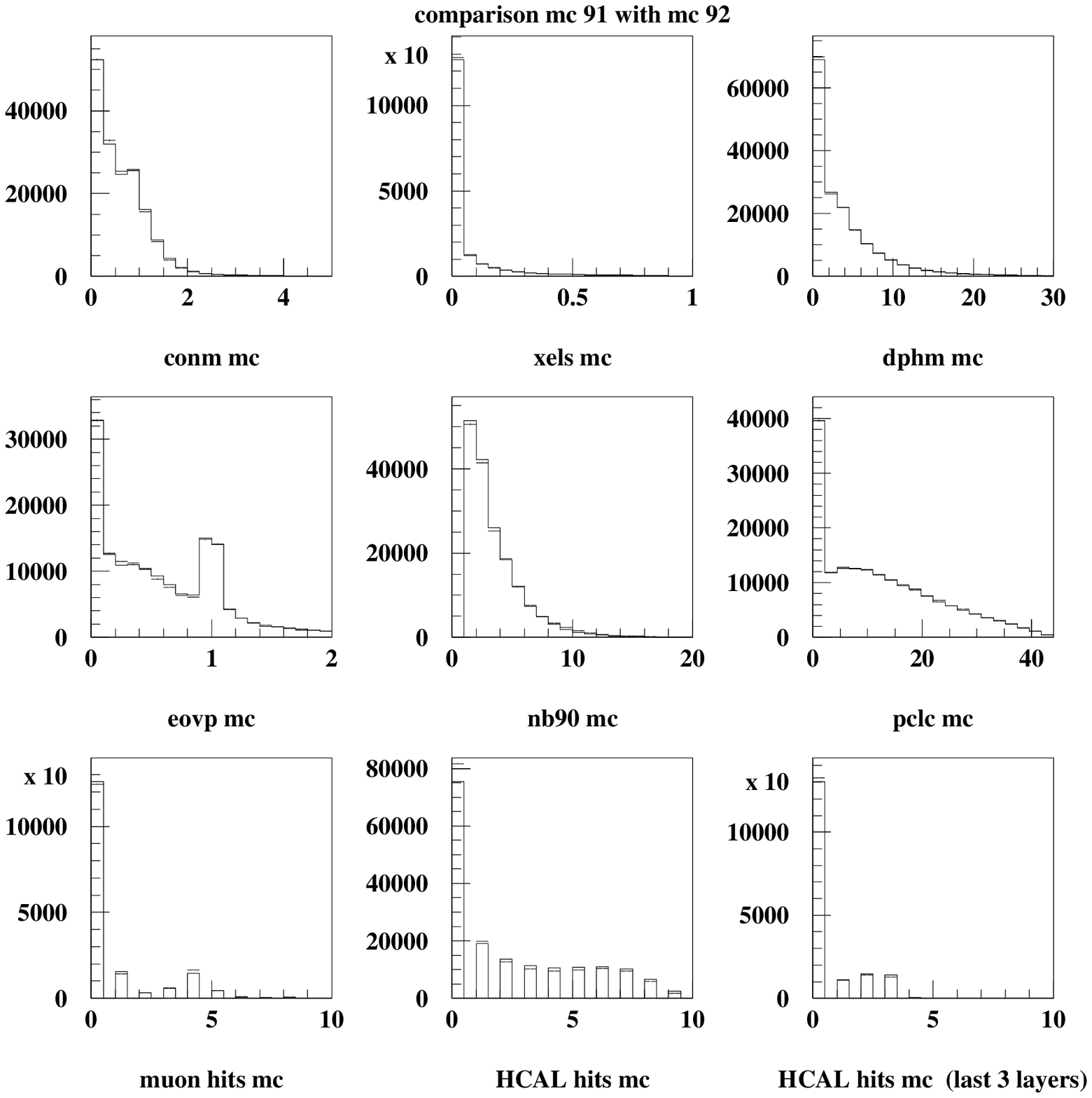}
\caption[Comparison of MC samples of 1991 with 1992 detector setups.]
{Comparison of MC samples with 1991 (histograms) and 1992 (error bars)
detector setups.
The variables shown are identical to those given in Fig. 5.2}
\label{fig-compmc2}
\end{figure}

Initial estimates of the efficiencies and purities of the selections for
the individual tau decay channels were obtained using
the KORALZ~3.8 MC generator~\cite{bib-koralz}.
This takes into
account initial state bremsstrahlung (with exclusive exponentiation)
up to ${\cal O}(\alpha^2)$, final state
bremsstrahlung and  electroweak corrections up to
${\cal O}(\alpha)$, It accounts also for
single bremsstrahlung (in the leading logarithmic approximation) in tau
decay for the decay modes used in this analysis.
Tau polarization and its effect on the decay spectra, as well as
the correlation between the two $\tn$'s are included (see Sect.~\ref{sect-RCC}).
372K MC tau pair events were generated for these studies, which
is a sample approximately six times larger than the data sample.
72K events  were generated with the
1991 detector setup (GOPAL127/ROPE401) and the rest, 300K,
were generated with the 1992 setup (GOPAL129/ROPE403).
The MC samples illustrated in Figs.~\ref{fig-compmc1}
and ~\ref{fig-compmc2} show no significant difference between these  two
samples.
Hence, also the MC samples are combined.
 
In order to estimate the  ``non-tau background''
coming from the multihadronic decays of the Z$^0$,
the JETSET MC~\cite{bib-Jetset} was used with the
parameters tuned to fit the global event shape
distributions of OPAL MH data~\cite{bib-OPALQCD}.
Backgrounds from radiative dimuon and Bhabha events
were estimated by using events generated with the
KORALZ~3.8 MC program~\cite{bib-koralz}
and the BABAMC MC program~\cite{bib-babamc}, respectively.
Non-resonant t-channel two-photon processes were simulated with
the generator described in reference~\cite{bib-vermas}.
 
Other event samples were also used for further corrections
in the analysis. Some of them were measured data samples
of other processes such as $\mu$-pairs, Bhabha events,
single electrons and photon-photon interactions. For this we used
events recorded in 1990-1992  corresponding to our $\t$-pair sample.
The events
were used as control samples to investigate some corrections
needed in our analysis, as will be later  described.
MC simulations of these processes were used to test the
quality of the detector simulation and to estimate the background to
our signal. More details about these event samples will be given
in the next chapters.
 
\section{Tau Pair Selection}
\label{sec-tausel}
In this analysis we initially select
$\tn$-pair candidates from which
samples of \tel , \tmu ~and \tpiK  ~candidates are subsequently
identified. Detailed description of most of the  \tn~ selection
and identification criteria can be found in the
OPAL Tau Platform manual~\cite{TP103}.
At \ecm =91~$GeV$, \eett ~events have sizeable
missing energy and unbalanced transverse momentum
appearing with a topology characterized by a pair of back-to-back,
narrow jets with low particle multiplicity.
In  the terminology of $\t$ recognition
we do not  restrict ourselves  to the  conventional  ``jet'' definition,
in place of that it is used as a general name which stands
for  the low  particle multiplicity
jets of the $\t$ hadronic decays, as well as  for the single tracks
characterizing the  $\t$ leptonic decay channels.
These characteristics are exploited by dividing
the event into jets as defined in
Ref.~\cite{bib-LINESHAPE0} where charged tracks and
ECAL clusters are assigned
to a cone of $35^\circ$ half-angle . Signals occurring in the presampler,
HCAL or MUON subdetectors within the solid angle defined by a jet cone
are also assigned to the jet. Only presampler signals associated to an
ECAL shower are included.
 
Charged tracks and electromagnetic clusters are defined
as being ``good'' in this analysis
if certain criteria on the quality of the reconstruction
are met. A good track must
have
\begin{itemize}
\item at least 20 hits registered in the jet chamber, the first one at a
radius smaller than 75~$cm$,
\item a distance of closest approach to the nominal
beam axis in the r-$\phi$ plane ($d_0$) of less than 2~$cm$,
\item the $z$  at this point ($z_0$) within
20~$cm$ of the nominal interaction point,
\end{itemize}
and
\begin{itemize}
\item a momentum
transverse to the beam axis of greater than 100~$MeV$.
\end{itemize}
 
A good barrel ECAL cluster, which is a contiguous group of lead-glass blocks
registering the deposition of some energy, must have  a minimum energy
of 100~$MeV$. A good endcap ECAL cluster has more than 200~$MeV$  deposited
energy contained within at least two lead-glass blocks, no one of which
may contribute more than 99\% to the cluster's energy.
The energy of clusters used in the \tel ~analysis
is corrected for the expected energy loss of electrons in the material
in front of the calorimeter.
The visible energy of the cone is defined as the
maximum between the (scalar) sum of the charged track momenta and the
sum of the ECAL energies associated to the cone.
 
A $|\cos\theta|$ is calculated for each tau jet and the average
of the two jets, $\overline{|\cos\theta|}$,
must be less than 0.68.
Background from two-photon processes is then suppressed
by requiring that the event has a minimum total energy and
significant missing transverse momentum when the total energy
in the event is low. After removing cosmic ray backgrounds,
the events which remain are almost entirely lepton-pairs.
The tau-pair events are further isolated by identifying and
removing the mu-pair and Bhabha events using their high
$E_{visible}/ \ecm$
characteristics.
 
An event is selected as a \eett ~candidate if each of the following
$\t$ ({\bf $T_{i}$}) requirements is met.
\begin{description}
\item T1.] In order to reject MH the event is required to
 contain between one and six good charged tracks and no more than 10
 ECAL clusters.
\item[T2.] The event must have exactly two tau jets and the
average $\act$ of the jets has to be below 0.68. This geometrical
requirement, although  introducing a 42.0\% loss in efficiency,
removes regions of non-uniform calorimeter response where the separation
of different tau decay modes is degraded.
\item[T3.] Cosmic rays events are removed by requiring that
\begin{description}
 \item[($i$)] at least one track in the event has $\dzero$
  and $\zzero$ below 0.5~$cm$
  and 20~$cm$ respectively,
 \item[($ii$)] the average value of $\zzero$ of all the
  good charged tracks at their point of closest approach to the
  beam is below 20~$cm$,
 \item[($iii$)] there is at least one TOF signal within 10~$ns$ of the
  expected value assuming particles in the event to originate
  at the \ee ~interaction point,
 \item[($iv$)] the event is discarded if all
  pairs of TOF signals which are separated by more than $165^\circ$ in
  the azimuthal plane have a time difference of more than 10~$ns$.
\end{description}
\item[T4.] Events from two-photon processes and events with
extreme radiation are suppressed by requiring that
the acollinearity between the two tau jets is below $15^\circ$.
The two-photon background is further reduced by rejecting
events if:
\begin{description}
\item[($i$)] the sum of the visible energies of the jets
 (defined as the maximum of the scalar sum of the charged track momenta and the
sum of the ECAL energies in the jet) is below $3\%$ of \ecm,
\end{description}
or,
\begin{description}
\item[($ii$)]
the sum of the visible energies of the tau jets
is less than $20\%$ of \ecm ~and
the missing transverse momentum of the event is below 2~$GeV$.
 This requirement on transverse momentum must be satisfied by both
the transverse momentum  calculated
from the charged tracks and that calculated
from the ECAL clusters.
\end{description}
\end{description}
 
The events passing these T1-T4 criteria are almost entirely leptonic
decays of the $\Z$.
The last stage in the tau-pair selection is the identification
and removal of \ee\ra\ee ~and \eemm ~events through the following cuts.
\begin{description}
\item[T5.]
An event is identified as \ee\ra\ee
~if
\begin{description}
\item[($i$)] the total ECAL energy in the event exceeds $80\%$
of \ecm ,
\end{description}
and,
\begin{description}
\item[($ii$)] the total of ECAL energy plus 30\% of the
scalar sum of the charged track momenta in the event is greater than $\ecm$.
\end{description}
\item[T6.] An event is classified as \eemm ~if each of the two
jets is a muon candidate  and if
the scalar sum of the track momenta in the event plus the sum of the
energies of the most energetic ECAL cluster in each jet is greater
than $60\%$ of $\ecm$.
A muon candidate jet must have at least one charged track and
satisfy one or more of the following three requirements:
\begin{description}
 \item[($i$)] there are at least two signals in the MUON chambers
  which are associated to the highest momentum track in the jet,
 \item[($ii$)] there are  more than three signals in different HCAL layers
   in the jet, at least one of which is in the last three layers, and the
   total number of HCAL signals divided by the
   number of HCAL layers hit in the jet, $\Nhl$, is less than two,
 \item[($iii$)] the ECAL energy of the jet is less than 2~$GeV$.
\end{description}
\end{description}
 
\begin{figure}[htbp]
\epsfxsize=16.8cm.
\epsfysize=18cm.
\epsffile[90 150 782 840]{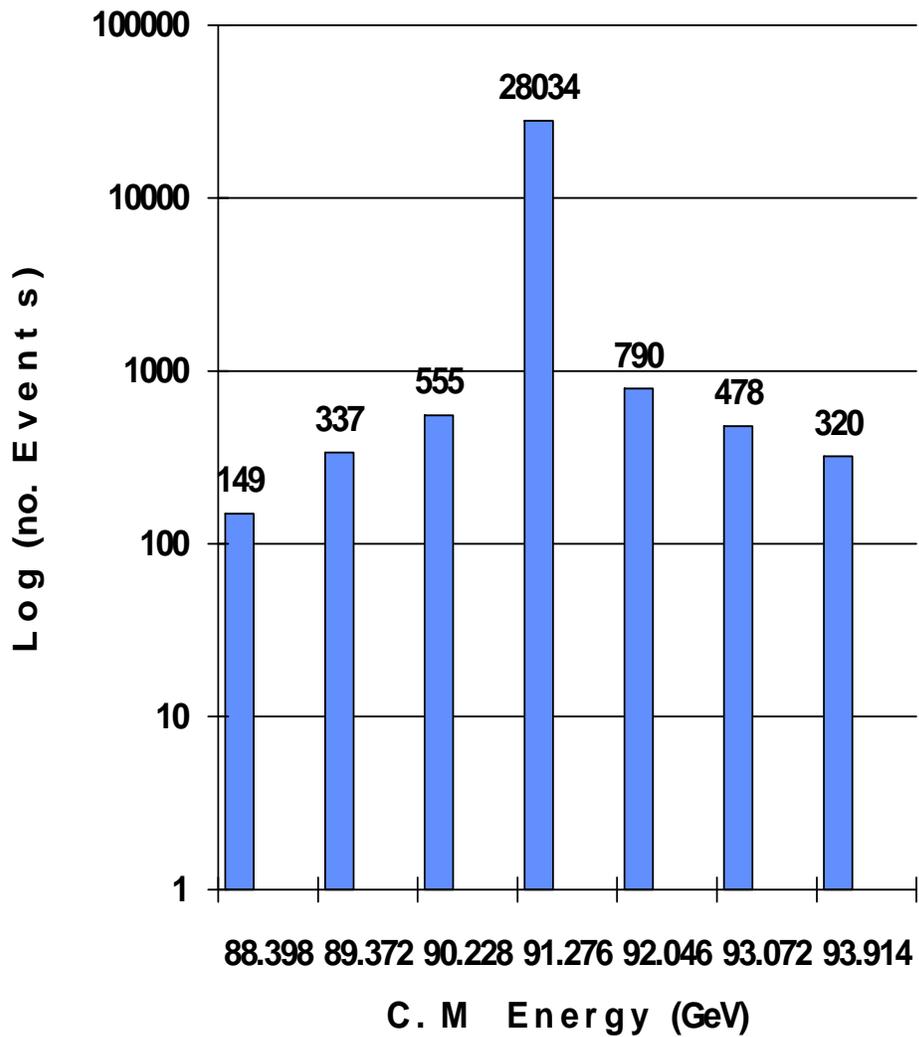}
\caption[$\ecm$  distribution of the data.]
{A logarithmic plot of the $\ecm$ distribution
of the $\tn$-pairs selected in 1990-92. The numbers on top of
each bin present the exact number of events. Most of the events
(91.5\%) are at the central energy bin (the $\Z$ peak).
}
\label{figecm}
\end{figure}

We select \Ntpair
~tau-pair events by these criteria.
The \eett ~selection is estimated from MC
studies \cite{bib-koralz,bib-gopal,bib-geant} to have
93\%  efficiency within the geometrical acceptance used in our analysis
(which is about 54.1\% of all the $\Z \to \tt$ events)
and a background level of 1.7\%.
These selected events are spread over 7 center-of mass energy bins on
the $\Z$ peak and 1,2 or 3 $GeV$ above or below the $\Z$ resonance.
The $\ecm$ distribution of the selected events is plotted in
Fig.~\ref{figecm} which shows that most of the events are measured
on the peak bin itself.
 
Details of the background in each $\t$-decay
channel  are discussed in the following chapters.
Note that the criteria designed to remove two-photon
 (T4), Bhabha (T5) and mu-pair (T6)
events introduce kinematic biases which must be studied
for the measurement of the polarization.
The treatment of the above selection criteria
and the systematic uncertainties associated with them
are considered in Chapter~\ref{chap-anal}.

\section{Tau Decay Identification}
\label{sec-tauid}
After selecting a $\tn$-pair event, an attempt is made to
classify the $\tn$ decay channels of its two jets.
Our \tel ~identification algorithm selects jets containing
a track which deposits almost all of its energy within a highly localized
region of the ECAL.
The \tmu ~selection requires an isolated charged track with
energy deposition in the ECAL and HCAL
consistent with the passage of a
minimum-ionizing particle and signals in the MUON chambers
associated with the track.
 
Unfortunately, the signature for a \tpiK ~decay is not as distinct as
that for  a \tel ~or a \tmu ~decay because  hadronic
interactions may begin in the magnet coil, ECAL or HCAL.
While electrons and muons can be efficiently removed from the \tpiK ~sample,
the   semi-leptonic tau decays containing neutral pions
are problematic because
at E$_{cm}$=91~$GeV$ the boost of the tau causes significant overlap
in the ECAL of the charged pion hadronic shower and the electromagnetic
showers from the $\pi^0$ decay to photons.  We exploit
the fact that electromagnetic interactions begin in the coil with a
higher probability than hadronic interactions. The presampler thereby
provides an effective
veto against decays containing neutral pions.
 
\subsection{ \mbox{ \tel}  ~identification}
Typical candidate for the $\tel$ decays is characterized by
a single track which deposits its energy in a narrow cone before the end
of the ECAL.
A tau jet is identified as a $\tel$ candidate if each of the following
requirements are met.
\begin{description}
\item[E1.] The jet has one or two good charged tracks.
If the jet has two tracks, the higher momentum
track is taken as the electron candidate.
\item[E2.] The ECAL energy associated to the candidate track divided
by the track momentum, E/p, is between 0.7 and 2.
\item[E3.] The electrons shower is expected not to be too wide.
Hence, it is required that at least $90\%$ of the energy of the ECAL
cluster associated
with the candidate track is deposited within three lead-glass blocks.
\item[E4.] The ECAL energy not associated to the candidate
track, $\Excess$, is less than 4\% of the beam energy. This removes
tau decays containing neutral pions and highly radiative tau pair events.
\item[E5.] The difference between the track and the
presampler signal furthest away in $\phi$ but still assigned
to the jet, $\dphimax$, is less than $5^\circ$. This effectively removes
tau decays containing neutral pions.
\item[E6.] To reduce background coming from muons or from $\tn$
hadronic decays,
the number of HCAL layers containing a signal which are
assigned to the jet is less than two.
\item[E7.] If the opposite tau jet
contains a single charged track with a momentum greater than $75\%$ of the
beam energy  then
the acoplanarity between the two jets must be greater than $0.1^\circ$.
The acoplanarity is defined in the r-$\phi$ plane.
This suppresses residual background from Bhabha events
where one of the two electron energies is mismeasured.
\item[E8.] The variable used for the polarization study of $\tel$ decays
is the total electromagnetic energy associated with the jet
divided by the beam energy, $x_e=E_{cls}/\eb$. $E_{cls}$ is corrected
for the energy lost in the  material in front of ECAL where the correction
uses presampler information.
The $x$-parameter used in the fit is expected to be between 0 and 1.
It can exceed 1 due to finite detector resolution. Values far
above 1 might happen due to bad track reconstruction which is not fully
taken into account in our parametrization of the resolution (see
Sect.~\ref{sect-DR}). Tau candidates with very high $x$ are few, but still
may affect the result of the ML fit because of their extremely low
probability.
In order to restrict ourselves to a kinematic region with high and well
understood electron identification efficiency, we require measured
$x_e$ to be above 0.05 and below 1.1.
\item[E9.]
Fiducial requirements are imposed to remove
candidates entering ECAL regions
where the four subassemblies of the calorimeter come in contact.
Good EB performance is important not only for the measurement of the electron
energy, but also for electron pion identification using E/p cuts, and for
Bhabha events rejection.
The small gaps in the calorimeter coverage in these contact regions
allow some Bhabha events to be accepted as \tel ~candidates.
Hence, a candidate is also removed if the track in the opposite
jet points to one of these regions.
 
The PB subdetector is extremely important for the pion identification,
but it is used also to correct the electron energy (see below in
Sect.~\ref{sect-DR}).
If the track  enters a
region of the presampler which is not fully operational, the tau jet
is also removed.
\item[E10.] The jet is not identified as a \tpiK ~candidate
according to the criteria listed below.
\end{description}

\subsection{ \mbox{ \tmu}  ~identification}
The signature of this decay is an isolated charged track characterized
by a small energy deposition in the ECAL and HCAL, and with associated
signals in
the MUON chambers, consistent with a passage of a minimum-ionizing particle.
The \tmu ~candidate jet must satisfy each of the following requirements.
\begin{description}
\item[M1.] There is exactly  one good charged track in the tau jet.
\item[M2.]
At least two of the following three criteria which characterize the passage of
minimum ionizing particle must be met:
\begin{description}
\item($i$) two or more hits in the MUON chambers are associated to the track,
\item($ii$) the number of HCAL layers containing associated hits, $\Nhc$, is
greater than three and $\Nhl$ is less than two,
\item($iii$) the ECAL energy in the jet is less than 2~$GeV$.
\end{description}
\item[M3.] In order to further ensure the minimum ionizing particle
characteristic, the jet is discarded if $\Nhc$ is greater than two and
$\Nhl$ is greater than or equal to three.
\item[M4.] The combined invariant mass of the track (assuming a
charged pion) and all the ECAL clusters in the jet (assuming $\gamma$)
is less than 0.3~$GeV$. This is calculated after subtracting 0.5~$GeV$
from the cluster nearest to the track in order to account for
energy deposition from a minimum ionizing particle. This suppresses
contamination from semi-leptonic tau decays accompanied by neutral pions.
\item[M5.] Residual background from $\eemm$ events is
suppressed by rejecting jets in which the opposite jet is consistent with
a  muon, and the sum of track momenta and ECAL energies, excluding the
expected energy deposition of 0.5~$GeV$, is greater than $80\%$ of the beam
energy. The opposite jet is considered to be a muon if:
\begin{description}
\item($i$) one of the first two conditions of cut M2 is satisfied,
\end{description}
or if,
\begin{description}
\item($ii$) the track extrapolates to a region not fully
covered by any of the HCAL or MUON chambers.
\end{description}
\item[M6.] A cut on the variable $x_\mu$
 used for the polarization study of $\tmu$ events.
It is defined as $x_\mu=(p_{trk}+E_{cls}-0.5)/\eb$ where
$p_{trk}+E_{cls}$ is the sum of track momentum and ECAL energies associated
to the jet, and 0.5~$GeV$ is the expected energy deposition of the muon.
As in the electron case (see cut E8)
the muon candidate is required to have $x_\mu$ above 0.05 and below 1.1
in order to ensure a reliable and well understood identification efficiency.
\item[M7.]
An additional fiducial requirement is imposed to suppress background
from $\mu$-pair events having tracks close to an anode or cathode wire
plane of the jet chamber. This source of background is not well
simulated.
A cut rejecting muon tracks near the anode wire planes has been
included in the 1990 analysis~\cite{OPALPL}. Later this cut was removed
from the TP standard
cuts because it was believed that the momentum resolution near
the anode wires has been improved and the remaining deficiency is
properly described in GOPAL. However, a recent study~\cite{Clayton}
has demonstrated that there is still a problem,
near the anode wires and even near the cathode wires. Due to the
deteriorated resolution in those regions, $\mu$-pair events which
should have been rejected by the cuts T6 and M5,
still infiltrate to our sample more than expected
from the MC.
Consequently, a
candidate track that is within 0.5$^{\circ}$ of an anode plane
or within 0.3$^{\circ}$ of a cathode plane is
removed if both tau jets are identified as muons or
if one jet is identified as a muon
and the opposite tau jet is not identified at all.
\item[M8.] The jet is not accepted as a $\tmu$ candidate if it also
satisfies the  \tpiK
 ~criteria listed below.
\end{description}

\subsection{ \mbox{ \tpiK}  ~identification}
The difficulty in the identification of this channel arises from the fact
that its signature is not highly distinct from those of other channels,
and in particular hadronic channels with neutral pions.
 
A \tpiK~ candidate must satisfy the following requirements.
\begin{description}
\item[P1.] There is exactly one good charged track in the jet.
\item[P2.] The total cluster energy associated to
the track is less than $80\%$ of the reconstructed track momentum.
This removes electrons and other contaminants having large
deposited energy in ECAL.
\item[P3.]
 $\Excess$ is less than $2\%$ of the beam energy, and
$\dphimax$ is below $0.5^\circ$.
This rejects semi-leptonic tau decays containing a neutral pion
using the same variables described in the \tel ~selection
criteria E4 and E5.
 
\item[P4.] In order to suppress the contamination from $\tmu$ decays,
only tracks extrapolating to
a geometrical region in which there is active MUON chamber coverage are
selected. Furthermore, if a track extrapolates to a region where HCAL is
active, the jet is rejected if
\begin{description}
\item($i$) the number of layers hit out of the
four MUON chamber layers and the last three HCAL layers,
$\Nhm$, is more than two,
\end{description}
and or,
\begin{description}
\item($ii$) $\Nhm  = 2$ and $\Nhl$ is below 3.
If the track extrapolates to a region which is not fully covered
by HCAL then the jet is
rejected if the number of hit MUON layers is greater than 2.
\end{description}
\item[P5.] The variable used in the polarization study of \tpiK
~is $x_h=p_{trk}/\eb$.
For the  same arguments given for  the electron and muon (cuts E8 and M5)
the pion candidate is required to have  $0.05<x_h<1.1$.
This ensures a reliable and well understood identification efficiency.
\item[P6.]
Fiducial requirements are imposed to remove
candidates entering ECAL regions
where the eight subassemblies of the calorimeter come in contact (See
E9). Tau jets are also removed if the track enters a
region of the presampler which is not fully operational.
\end{description}

\subsection{ \mbox{ \tro} ~identification}
The \tro ~channel is not included in the present analysis. We utilize however,
the results obtained by OPAL in other analyses (see Ref.~\cite{TN142}
and ~\cite{TN203}) to combine with our findings.
Hence, here  we describe only briefly its selection procedure.
Due to the fact the  \rhon ~decays into a charged pion
and a neutral pion, the signature of the \tro ~decay is
a single  charged  track accompanied by
ECAL energy deposition consistent with the interaction
of the two photons from the neutral pion decay.
 
Two analysis methods to identify rho decays were used in OPAL.
The two \tro
~identification algorithms (for Sample I and II) are similar to those
used to measure the \tn \ra h \piz ~branching ratio~\cite{bib-OPALhpizbr}.
The selection of sample I~\cite{TN203} is based on an attempt to
reconstruct from the ECAL clusters the
neutral pion which is then combined with
the charged track to form a \rhon -meson candidate.
 
The second \tro ~selection procedure~\cite{TN142} uses a different \piz ~identification
algorithm. It finds clusters of activity in the ECAL which are restricted to a
maximum size of 2$\times$2 blocks in $\theta$ and $\phi$ plane. Such clusters
contain 99\% of the energy of an electromagnetic shower and 95\% of the
energy of a hadronic shower. These clusters which are limited in size
differ from those used
in the \tautau selection or the \tel , \tmu ~and \tpiK ~identifications.
Unlike the sample I selection,
this algorithm accepts  as \piz ~candidates {\it single} clusters which are
not associated to the track.

\subsection{Summary of $\t$ Decay Identification}
\begin{figure}[p]
\epsfysize=18cm.
\epsffile[30 140 710 730]{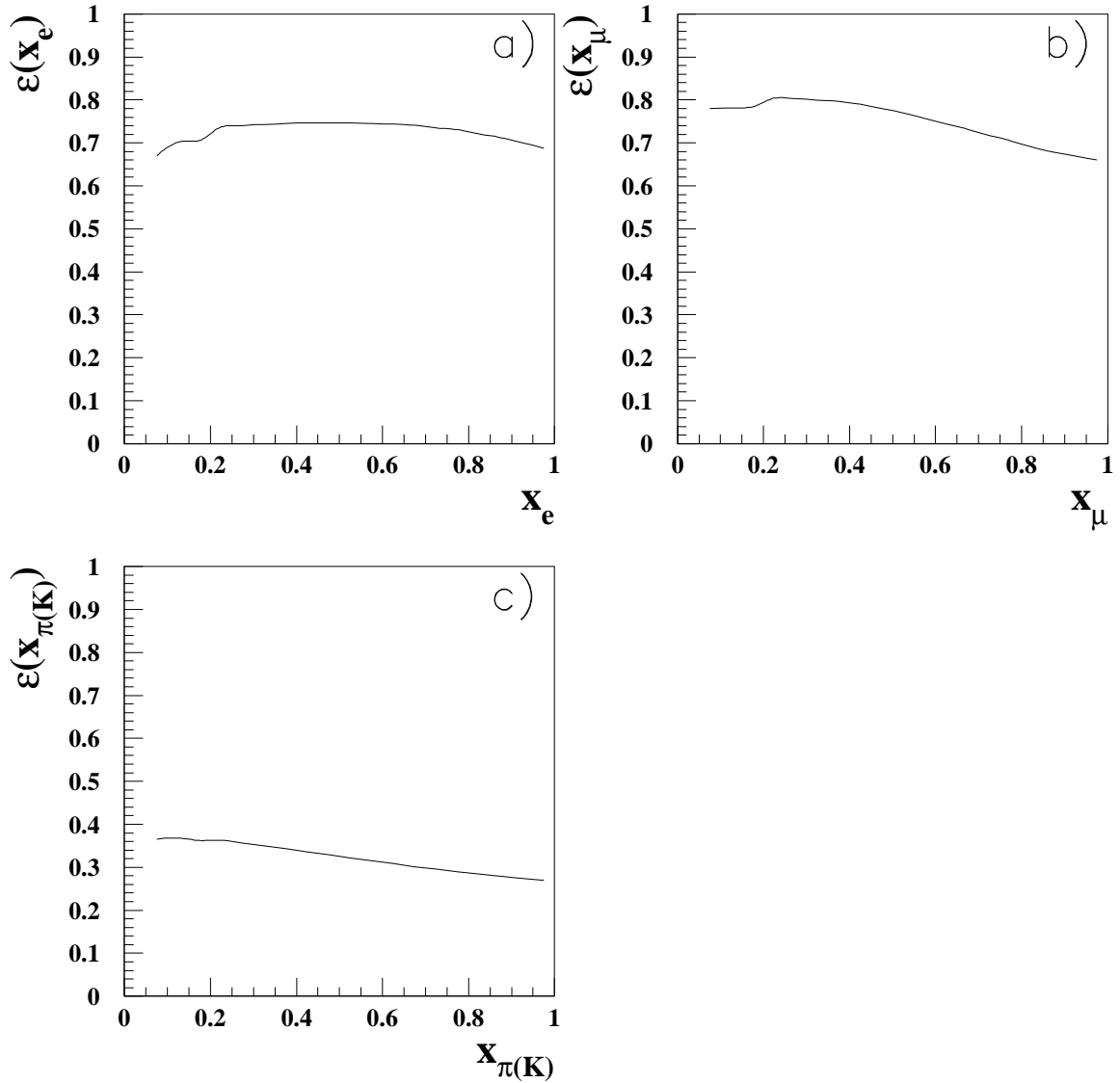}
\caption[Global efficiencies as a function of $x$.]
{Global efficiencies as a function of $x$ for
a) \tel  b) \tmu  and c) \tpiK ~decays.
The efficiencies  include the 93\% tau-pair selection efficiency
and are after the fiducial cut of $|\cos\theta|<0.68$.
The complex behavior in the low and high $x$ regions of a) and
b) are caused by requirements designed to remove two-photon,
Bhabha and mu-pair events.}
\label{effemp}
\end{figure}

 The typical global efficiencies obtained by the selections
described above are 65.8\%, 72.1\% and 30.4\% for
  the \tel, $\tmu$ and $\tpiK$ , respectively.
These numbers include the approximately 93\%
preselection efficiency within the fiducial region of the analysis
 and ignore
decay or detection correlations. The corresponding background levels
are 5.0\%, 2.5\% and 7.2\%, respectively.
Plots of the global efficiencies  as a function of $x$
are shown in Fig.~\ref{effemp}.
 
The results from the \tro ~decay are based on two samples of
\tro ~candidate
events. Each sample is independently selected and values for
\pta ~and \aplfb ~are determined separately.
Both selections have an efficiency of
approximately 50\% and about one third of the combined number
of candidates is common to both samples.
Sample II candidates are selected only from the 1991 and 1992 data which
contain 27353 tau-pair events.
The efficiency for selecting sample I is
47.5\% where the background is 21.2\%. For sample II the efficiency
is  57.5\% and the background 22.5\%.
 
In Table~\ref{evtab} we list the number of $\tn$-pair events
in the data, according to their $\tn$ decay mode identifications.
The percentage efficiencies and background are also presented
for each event classification within $|\cos\theta|<0.68$.
The efficiencies include the contributions from the tau-pair
selection.
 The label `0' refers to the case where the $\tn$ decay was not
 identified at all. The events in which none of the two jets was
 identified (23\% of the whole sample) are not used in the analysis.
As is evident from this table, the identification efficiencies of
 tau jets are not independent. The requirements that remove two-photon,
Bhabha and mu-pair events
introduce these correlations. Such correlations are, however,
 taken into account in the analysis.

Fig.~\ref{scan} 
shows a "typical" $\tt$ selected event
in the detector\footnote{ The plot is produced with 
the event viewing facility, GROPE, which is the graphic processor
of the OPAL reconstruction program, ROPE.}.
In the figure we see an
$e^+e^- \to \t^+\t^-$ event selected by our
criteria.
One of the tau decays into a one charged particle 
while on its back side we find a collimated 3 charged particles jet. 

\begin {table} [hb]
\begin{center}
 \begin{tabular}{|l r||c|c|c|c|c|}  \cline{1-2}
$\tau_1$ decay  &$\downarrow$    & \multicolumn{5}{c}{ } \\ \cline{1-3}
No. & 0        & 6833 &\multicolumn{4}{c}{ } \\ \cline{1-4} \cline{1-4}
No. & e        & 3585 &  574  & \multicolumn{3}{c}{ } \\
$\epsilon$ & & 54. (19.)  &  46.  & \multicolumn{3}{c}{ } \\
$\beta$        & &  5. &  12. &\multicolumn{3}{c}{ } \\ \cline{1-5} \cline{1-5}
No. & $\mu$  & 3897 & 1288 & 487 & \multicolumn{2}{c}{ } \\
$\epsilon$ & & 58. (23.)  &  58.  & 46. & \multicolumn{2}{c}{ } \\
$\beta$    & &  3. &  6. & 5. & \multicolumn{2}{c}{ } \\ \cline{1-6} \cline{1-6}
No. & $\pi(K)$ & 1145 &  346 & 385 &  55 & \multicolumn{1}{c}{ }\\
$\epsilon$ &   & 26. (9.)  &  21. & 25. & 11. & \multicolumn{1}{c}{ } \\
$\beta$    &   &   7.&  11. & 8. & 14. & \multicolumn{1}{c}{ } \\ \hline
No. & $\rho$ & 5879 & 1960 & 1985& 623 & 1621   \\
$\epsilon$ & & 51. (19.) &  44. & 51. & 20. & 41. \\
$\beta$    & & 27.  & 31.  & 29. & 32. & 48.   \\
 \hline \hline
$\tau_2$ decay & $\rightarrow$  & 0    &   e  & $\mu$ & $\pi(K)$ & $\rho$
 \\ \hline
\multicolumn{6}{c}{ }
\end{tabular}
\caption[$\tn$-pair events classified according to
their identification]
{Classification of the $\t$-pair events with $\act<0.68$
where the first figure in each box represent their number.
 The  efficiency ($\epsilon$) and  background ($\beta$) given in percentages are presented
underneath. The efficiencies include the
contributions from the tau-pair selection  $\sim93\%$.
The efficiencies in the first column (i.e.
when only one tau is identified) are subdivided into two classes: the first
is for the case where the opposite tau does not decay into one of  the four
channels; the second (in brackets) is where the opposite tau decays
into one of the four
channels but is not identified.
The label `0' refers to the case where the $\tn$ decay was not
identified.}
\label{evtab}
\end{center}
\end{table}
%
\newpage
\pagestyle{headings}
\setcounter{page}{89}
\begin{figure}[p]
\epsfysize=17cm.
\epsffile[0 0  700 840]{fig5_7.eps}
\caption[A display of a typical $\tt$ event in the OPAL detector.]
{A display of typical $\tt$ event in the OPAL detector. 
Tracks in the central tracking
system are shown in red. Small magenta boxes show hits in the TOF system. 
Clusters of energy in the lead glass ECAL are shown as blue boxes, 
of size proportional to their energy.
Similarly, clusters of energy in the HCAL are drawn in green.
Penetrating charged particle tracks, which are candidates for muons, are
shown as light-blue arrows. 
In this event ,one of the taus (on the right side) has decayed to a 
penetrating muon, and the other to an electron.
}
\label{scan}
\end{figure}

\chapter{Physics Analysis}
\label{chap-anal}
\section{The Fit Procedure}
\label{secfit}
For the analysis of the  $\tel$, $\tmu$ and $\tpiK$ decays, \pta ~ and
\aplfb ~are determined
using an event-by-event ML fit to the data of
the theoretical distribution (Eq.~\ref{dsig3}).
In this method, the following  expression is minimized,
\beq
\label{ML}
W=-2\ln{\cal L}=-2\sum_{n=1}^N \ln \left\{ \frac{1}{\sijp}
\frac{d^3\sijp}{d\cst\:d\xip\:d\xjp} \right\}_n,
\eeq
where ${\cal L}$ is the likelihood function. The sum
in Eq.~\ref{ML} runs over all $N$ selected $\tn$-pair events, where at 
least
one of the two $\tn$'s has been identified as
a $\tel$, $\tmu$ or $\tpiK$ candidate.
The term in the logarithm is the differential cross-section
for a production $\t$-pairs where one $\t$ decays into channel $i$
and the second one decays into channel $j$ ( Eq.~\ref{dsig3}).
It is normalized to one,
and corrected for {\em radiation}, {\em $x$ resolution}, {\em efficiency}
and {\em background} effects, hence it will be denoted as primed $\sigma$.
This term corresponds to the case where both $\tn$ decays are
identified. When only one decay is identified, it must be replaced
by $(1/\sip)d^2\sip/d\cst\;d\xip$.
 
When both $\tn$ decays are identified, the corrected cross-section
can be written in the following form,
\bea
\label{corcst}
\frac{16}{3}\frac{1}{\sijp}\frac{d^3\sijp}{d\cst\:d\xip\:d\xjp} & = &
 \sum_{hel=\pm}\{ [(1\pm\pta)(1+\cstsq)+
 \eitovth(\afb\pm\aplfb)\cst] \nonumber \\
 & \times & \epstij [\hi\epsdi+\beti] \nonumber \\
 & \times & [\hj\epsdj+\betj] \} \nonumber \\
 & + & \bntij.
\eea
Here $\epstij$ is the efficiency for selecting a $\tn$-pair event
in which one tau decays via channel $i$ and the other via $j$,
$\epsdi$ is the efficiency for identifying channel $i$
which for some of the decay combinations  depends
on the decay mode in the opposite jet and on the helicity.
The $\beti$ is the distribution of background events from other $\tn$ decay
 channels  when applying the channel $i$ selection procedure and
$\bntij$ corresponds to background contributions from events which are
not tau-pairs. The function $\hi$ is the $x$ distribution
for channel $i$, after applying corrections for
detector resolution and radiative effects to  $f_i(x_i)\pm g_i(x_i)$.
It is given by the following
convolution function
\beq
\label{hxd}
\hi=C_i\int_0^1 [\fix(\xxi^0)\pm\gix(\xxi^0)]\ri\rxx d\xxi^0.
\eeq
where the normalization factor $C_i$ was introduced to satisfy,
\beq
\label{normh}
\int_0^{\infty}\hi d\xip=1.
\eeq
The $\xxi^0$ is the $\xxi$ before resolution effects are taken into account.
The functions $\fix$ and $\gix$ are the Born distributions listed
in Chapter ~\ref{chap-theory},
Eqs.~\ref{wl},~\ref{wxh},~\ref{wx},~\ref{fgnor}. The correction term
$\ri$ is due to radiative effects and $\rxx$ is the momentum resolution
function.
 
If only one tau decay is identified, the corrected differential cross-section
takes a simpler form,
\bea
\label{corcso}
\frac{16}{3}\frac{1}{\sip}\dsigtwp & = & \sum_{hel=\pm}\{ [(1\pm\pta)(1+\cstsq)+
 \eitovth(\afb\pm\aplfb)\cst] \nonumber \\
 & \times & \epstpi [\hi\epsdi+\beti] \} \nonumber \\
 & + & \bntij,
\eea
where $\epstpi$ is given by the following expression,
\bea
\label{epsp}
\epstpi  &=&\epsti \\ \nonumber
&+&\frac{1}{1-\sum_j B_j} \sum_{j\neq i}B_j\int d\xjp
\epstij\hj [1-\epsdj].
\eea
Here, the first term is the efficiency for selecting a $\t$-pair event in
which one tau decays via channel $i$ and the other via a channel not
included in our analysis. The second term is the contribution of events
where the other channel is included in our analysis but is not identified
due to inefficiency. The coefficients $B_j$ are the $\t$ branching
fractions. They were taken from the OPAL measurements~\cite{OPALPL}.
The summation in Eq.~\ref{epsp} is over the three decay channels
considered in this analysis.
 
The normalization factor in Eq.~\ref{ML}, $\sijp$ (or $\sip$ in the
case of only one identified tau) are the total corrected cross sections
evaluated in every step of the fitting procedure.
 
Our fitted parameters are $\pta$ and $\aplfb$.
In a former work of OPAL~\cite{bib-z0par}
the values of $\afb$($s$) have been determined using the distribution
of $\cst$  alone. In the present analysis we have used these reported
values as an input.
Since $\afb$ has a strong $s$ dependence
(see Fig.~\ref{fig_assecm}), we divide
the available c.m. energy range as shown in Fig.~\ref{figecm}
into 7 intervals. For each interval
we calculate the average of the OPAL $\afb$ measurements taken during
1989-1992. In our global fit we use for each event the corresponding
$\afb$ value according to its c.m. energy. The energy intervals, the
average energies, the sample sizes and the corresponding
$\afb$ values are listed in
Table~\ref{eint}. The uncertainties in the $\afb$ values were taken
into account in the calculation of the overall systematic error.
 
\begin {table} [htb]
\begin{center}
\begin{tabular}{||c|c|r|c||}  \hline  \hline
$\ecm$  & $\ecma$  & \# of   & $\afb$    \\
($GeV$)  &  ($GeV$)   &  events &  (\%) \\
 \hline
87.7 - 88.7 & 88.398 &   149 & $-29.0 \pm 5.1$ \\
88.7 - 89.7 & 89.372 &   337 & $ -8.4 \pm 4.0$ \\
89.7 - 90.7 & 90.228 &   555 & $ -7.8 \pm 3.2$ \\
90.7 - 91.7 & 91.276 & 28034 & $  1.2 \pm 0.5$ \\
91.7 - 92.7 & 92.046 &   790 & $  5.1 \pm 2.8$ \\
92.7 - 93.7 & 93.072 &   478 & $ 10.8 \pm 3.4$ \\
93.7 - 94.7 & 93.914 &   320 & $ 14.6 \pm 4.0$ \\
\hline \hline
\end{tabular}
\caption[Forward-backward asymmetries.]{Forward-backward asymmetries used in the fit and numbers of
events for 7 c.m. energy intervals. }
\label{eint}
\end{center}
\end{table}
 
In performing the fit, the functions $h^{\pm}_{i(j)}$, $\epsilon_{i(j)}$,
$\beta^{\pm}_{i(j)}$, ${\cal E}^{\pm}_{i(ij)}$ and
$\beta^{non-\tn}_{i(ij)}$ as well as the normalization
factors are calculated for each event.
The $h^{\pm}_{i(j)}$ functions are obtained from
the theoretical decay spectra corrected for radiative effects using large
MC \tautau ~event samples generated without detector
simulation and corrected for resolution effects using
resolution functions measured in data samples of \ee\ra\ee events,
low energy ``single electron'' events from highly radiative Bhabha scattering,
and ~\ee\ra\mumu events.
The  $\epsilon_{i(j)}$, ${\cal E}^{\pm}_{i(ij)}$,
$\beta^{\pm}_{i(j)}$ and $\beta^{non-\tn}_{i(ij)}$ functions
are parametrized in terms of simple
functions, in most cases low order polynomials, using few parameters.
The parameters are determined from fits to the pertinent
MC samples and subsequently corrected using appropriate
 control samples from the data.
The uncertainties in the parameters, their correlations and the
corrections are taken into
account in the treatment of systematic errors which are discussed
in detail below.
For this we are
using the standard error propagation rule,
\beq
\label{ERP}
\Delta A=\sqrt{\sum_{i,j}\frac{\partial A}{\partial p_i}
   \frac{\partial A}{\partial p_j}\Delta p_i \Delta p_j \rho_{ij}}\;\;,
\eeq
where $A$ is the final result ($\pta$ or $\aplfb$), $p_i$, $p_j$ are the
parameters used to describe the correction function and $\rho_{ij}$ is the
correlation matrix between them. The partial derivatives
$\frac{\partial A}{\partial p_i}$ are calculated numerically, by
modifying the parameter by its uncertainty, by repeating our fit and
calculating the change in the final result $A$.
 
As suggested in Sect.~\ref{sec-wformalism} in order to check the
parameters characterizing the structure of the charged weak decay
of the $\tn$ one has also to treat them as free parameters.
In that case  we use Eqs.~\ref{w2l} and ~\ref{dsig3p} to modify
Eq.~\ref{corcst}  in the following way:
\bea
\label{corcstmod}
\frac{d^3\sijp}{d\cst\:d\xip\:d\xjp} & = \frac{3\sigma}{64}\times\{
&   \;\;h^+_ih^+_j [{\cal C}(1+\xi_i\xi_j)-{\cal P}(\xi_i+\xi_j)] \nonumber \\
&&+ h^-_ih^-_j [{\cal C}(1+\xi_i\xi_j)+{\cal P}(\xi_i+\xi_j)] \nonumber \\
&&+ h^+_ih^-_j [{\cal C}(1-\xi_i\xi_j)-{\cal P}(\xi_i-\xi_j)] \nonumber \\
&&+ h^+_ih^-_j [{\cal C}(1-\xi_i\xi_j)+{\cal P}(\xi_i-\xi_j)]  \\
&+ 2(\rho-0.75)&\;\; [f^i_2(\phi^j_+ ({\cal C}-{\cal P}) +
\phi^j_- ({\cal C}+{\cal P}) \nonumber \\
        &&         +f^j_2(\phi^i_+ ({\cal C}-{\cal P})
        + \phi^i_- ({\cal C}+{\cal P})] \nonumber \\
&+ 2(\delta-0.75)&\;\; [\xi_i g^i_2(\phi^j_{+} ({\cal C}-{\cal P})
- \phi^j_- ({\cal C}+{\cal P}) \nonumber \\
         &&        +  \xi_j g^j_2(\phi^i_{+} ({\cal C}-{\cal P})
         - \phi^i_- ({\cal C}+{\cal P})]\} \nonumber
\eea
where  $\cal C,\cal P$, $h^{\pm}_{i(j)}$ and $\phi_{i(j)}^{\pm}$
are defined as following:
\bea
\label{corcstmod2}
{\cal C} &=& 1+ \cstsq + \eitovth \afb \cst \nonumber \\
{\cal P} &=& \pta(1+ \cstsq) + \eitovth \aplfb \cst \nonumber \\
h^{\pm}_{i} &=& f^i_1(\xxi)+\frac{3}{4}f^i_2(\xxi)\pm
[g^i_1(\xxi)+\frac{3}{4}g^i_2(\xxi)] \\
\phi^i_{\pm}&=& f^i_1(\xxi)+\rho f^i_2(\xxi)
\pm \xi_i [g^i_1(\xxi) +\delta g^i_2(\xxi)]\nonumber
\eea
 
In the first two terms of Eq.~\ref{corcstmod} (first two lines) there is
an exact repetition on Eq.~\ref{dsig3} where a $V-A$ structure is assumed
for the $\tn$ decay. These terms are corrected for radiation and
detector (resolution, efficiency and purity) effects as  in
Eq.~\ref{corcst}.
For the other six terms,
which vanish in the case when $V-A$ is exact,
we cannot use the standard correction procedure as our
calculated correction are helicity dependent.
Therefore in order to correct these terms we used corrections
averaged over the two helicity states. For example,
\mbox{$r_i(\xxi)=0.5\cdot [r_i^+(\xxi)+r_i^-(\xxi)]$}.
 
Eqs.~\ref{corcstmod} and~\ref{corcstmod2} as it is written is applicable
to the $\tle$ events through which $\xi,\rho$ and $\delta$.
In the case of  $\tpiK$
decay only $\xih$ can be measured. In order to generalize
Eq.~\ref{corcstmod} and~\ref{corcstmod2} we equate for the hadronic decays
$f(g)^{had}=f(g)^{had}_1+f(g)^{had}_2$ where their values, not to change
Eq.~\ref{wxh} relations, are:
\bea
\label{fghad}
f^{had}_2&=&0 \nonumber \\
g^{had}_2&=&0 \nonumber \\
f^{had}_1&=&\frac{1}{1-\mhs/\mts}\\
g^{had}_1&=&\frac{2x-1-\mhs/\mts}{(1-\mhs/\mts)^2} \nonumber
\eea
\section{Radiative and Threshold Effects}
\label{sect-RCC}
For a given helicity state, initial and final state
radiation as well as radiation in the decay of the tau
affect the spectra of the tau decay products.
The radiative distortions to the spectrum for the decay mode $i$ are
taken into account using a radiative correction function, $\ri$.
 
Radiative effects are usually separated into photonic (QED) and
weak interaction effects.
The QED effects are due to all the diagrams involving real or virtual
photons. They do not depend on unknown parameters
but they usually depend on the experimental cuts.
The other effects are due to the weak corrections and depend on
unknown theoretical parameters but not on the experimental cuts.
 
Since the $\t$-polarization results are part
of the $\Z$ OPAL lineshape parameters and are used with them
in all sorts of
overall electroweak fits, they are handled in the same way.
Therefore, following the procedure used in all LEP experiments
for the  line shape analysis,
we do not apply corrections due to QED and weak radiative effects on
the measured $\pta$ or $\aplfb$. However, even for a
given helicity state, initial and final state
radiation modify the spectrum of outgoing leptons and pions.
 
The functions $\ri$ are determined from the ratio of the spectrum
containing radiative effects to the spectrum where only the
Born level cross-section is considered.
A high statistics run of the KORALZ~4.0 MC
 generator\footnote{The KORALZ~4.0 is the latest version
of this MC program in which  the radiative corrections were improved
compared to previous versions. It was used to calculate our correction
to the data. However, as all our other MC tests were done with
KORALZ~3.8, we have also used radiative corrections calculated with the 3.8
version when we have cross checked the analysis with MC events.
More details about KORALZ are given in Sect.~\ref{MC}
and Ref.~\cite{bib-koralz}},
is used  to create the radiative corrected spectrum.
 
The QED effects, and in particular the initial state radiation, depend on
the c.m. energy. We use in this analysis data taken on the $\Z$
peak as well as at 1, 2 and 3~ $GeV$ below and above the peak
(see Table~\ref{eint}). Therefore  the KORALZ~4.0 is used to
generate events distributed at the same seven c.m. energies as in
the data. We actually generated three samples of 420000 events each,
corresponding to the three decay channels investigated here. In
each sample, all the $\t$'s were forced to decay into a single  channel.
 
For the $\t$-polarization measurement on the $\Z$, the most important QED
effect is the final state radiation from the $\t$
or its decay products~\cite{CALASY} .
This would modify the decayed particle momentum and distort its
distribution from the Born term expectation.
In order to minimize this effect in our analysis, the electromagnetic
calorimeter is
used to detect the radiated photons in a $35^\circ$ cone around the track.
The energy of these photons is added to the track energy and in this way,
the effect of the final state radiation is minimized. This procedure
is done only for lepton final states, since in the case of pions,
events with extra photons are rejected anyhow, because they resemble
$\tro$ decays. The same treatment is done also for the generated
events in order to investigate the residual effect of radiative
corrections.
 
\begin{figure}[p]
\epsfysize=18cm.
\epsffile[80 137 650 677]{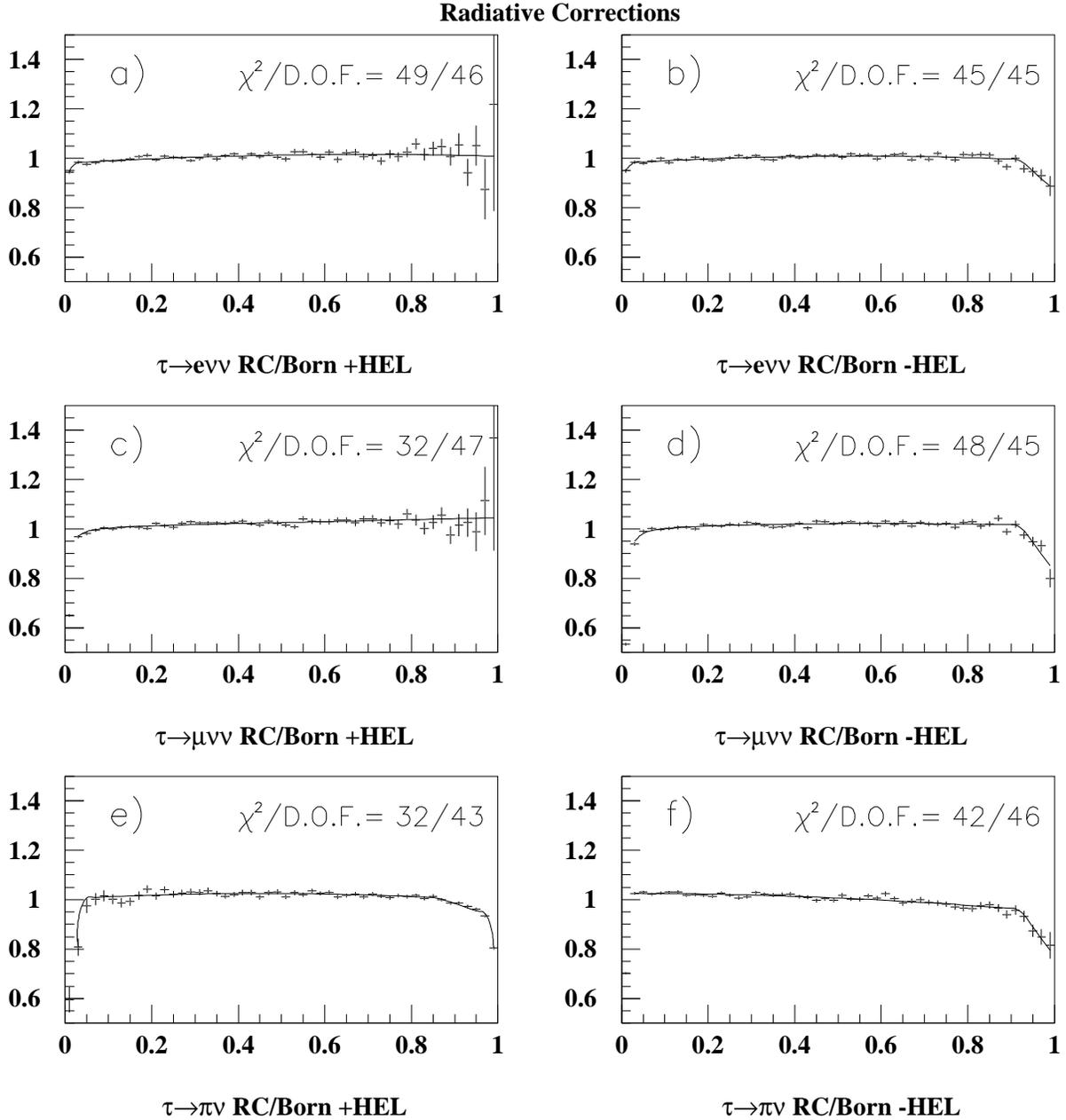}
\caption[Radiative corrections and the Born term expectation.]
{$x$ distribution of the ratio between generated MC events with
radiative corrections and the Born term expectation, plotted
with our parametrization for the radiative effects (solid lines).
Figs. a) and b)  are for the $\tel$ channel for positive
and negative helicity states respectively. Figs. c) and d) correspond to the
$\tmu$ channel and Figs. e) and f) are for the $\tpiK$ channel.}
\label{fig_rad}
\end{figure}
 
We calculated separately the radiative corrections, $r^{\pm}_i(\xxi)$,
separately for positive and negative helicity events.
In this way the corrections are independent on the polarization
measurement.
The MC distributions are divided by the corresponding lowest order diagram
expectations. The obtained ratios are the radiative corrections $\ri$
which are plotted in Fig.~\ref{fig_rad}.
These are
slightly dependent on the helicity state.
The reason of this dependence
originates partly from the different slopes of the Born spectra.
For instance, in the channel $\tpiK$ the spectrum is increasing (decreasing)
for positive (negative) helicity. The initial and final state radiation
tend to decrease the energy of the outgoing pion. Therefore, one should
expect the low $x$ range be enhanced and distorts the distribution.
 
The solid lines in Fig.~\ref{fig_rad} represent our parametrization.
We needed 3 to 4 parameters to describe each distributions.
The $\chs$ values between the distributions
and their parametrization are indicated in the figures. The first bin
in Figs.~\ref{fig_rad}c and \ref{fig_rad}d is low due to the
finite $\mu$-mass effect, which will be  discussed in the following,
and was not incorporated in the Born expression. Therefore, this
bin was not included in the $\chs$ calculation. For the $\tpiK$ channel
the first bin starts at the kinematical threshold and therefore no
depletion is presented.
 
The major uncertainty associated with these corrections arises from the
fact that the ${\cal O}(\alpha)$ in the  QED  corrections to tau decay
are included only in the leading logarithmic approximation. The effect of
this correction on the polarization asymmetries
is quantified by analyzing the MC event sample after
full simulation as  if it was measured data. The event sample is
analyzed twice, once for the full event sample and also
for a sample which excludes events with
radiated photons in the tau decay process. The difference in the results
is used as a measure of the magnitude of the correction. Following
Ref.~\cite{bib-PHOTOS}, the correction is assumed to have a systematic
error of the order of $1/\ln(\mt/m_i)$ of the correction. An additional but
small contribution to 
\newpage
\noindent
the error on $\ri$ arises from the
parameters used to describe the radiative
corrections due to the MC statistics which is large
but nevertheless limited. These parameters are obtained, together 
with their correlations,
from the fits of  the parametrization functions.
Their effect on the final $\pta$ and $\aplfb$ results was calculated
with the method described in the last section.
 
Table~\ref{radsys} summarizes the systematic errors associated with the
radiative corrections for the various decay channels and
for the whole   sample.
\begin {table} [htb]
\begin {center}
\begin {tabular}{||l||cc||cc|cc|cc||} \hline \hline
  & \multicolumn{2}{c||}{all} & \multicolumn{2}{c|}{$\tel$} &
\multicolumn{2}{c|}{$\tmu$} & \multicolumn{2}{c||}{$\tpiK$} \\
source & $\dpta$ & $\daplfb$ & $\dpta$ & $\daplfb$ & $\dpta$ & $\daplfb$ &
$\dpta$ & $\daplfb$  \\ \hline
MC  & 0.15 & 0.01 & 0.44 & 0.06 & 0.56 & 0.04 & 0.14 & 0.01 \\
LL & 0.25 & 0.03 & 0.16 & 0.03 & 0.15 & 0.04 & 0.33 & 0.02 \\
\hline
Total & 0.29 & 0.03 & 0.47 & 0.07 & 0.58 & 0.06 & 0.36 & 0.02 \\
\hline \hline
\end {tabular}
\caption[Systematic errors associated with  radiative effects.]
{Systematic errors (in \%) associated with radiative effects.
MC is the systematic error introduced by the MC finite statistics
and LL is the main contribution coming from the leading logarithm
approximation. Their quadratic sum is given in the last line.}
\label{radsys}
\end{center}
\end{table}

As can be seen  the
treatment of initial and final state radiation in KORALZ~4.0
introduces a smaller  contribution to the error than the leading logarithm
approximation. The total error assigned
to the radiative correction uncertainties of \pta~and \aplfb~are
0.29\% and 0.03\%.
 
We also investigated the possible influence of the radiative corrections
on the $\cst$ distributions.
Since we found this to be negligible we have further taken
$\ri$ to be independent of $\cst$.
 
To account for the finite muon mass in the
 $\tmu$ channel we have multiplied
the expression under the integral of Eq.~\ref{hxd} by the term $K^\pm(x)$
defined as
\beq
\label{Kpm}
K^\pm(x)= \left\{\begin{array}{ll}
 1-\left(\frac{x_m^\pm}{x}\right)^{q^\pm} & \mbox{for $x>x_m^\pm$} \\
 0                      & \mbox{for $x\leq x_m^\pm$}
\end{array} \right.
\eeq
 
\begin{figure}[htbp]
\epsfysize=5.7cm.
\epsffile[100 350 1430 530]{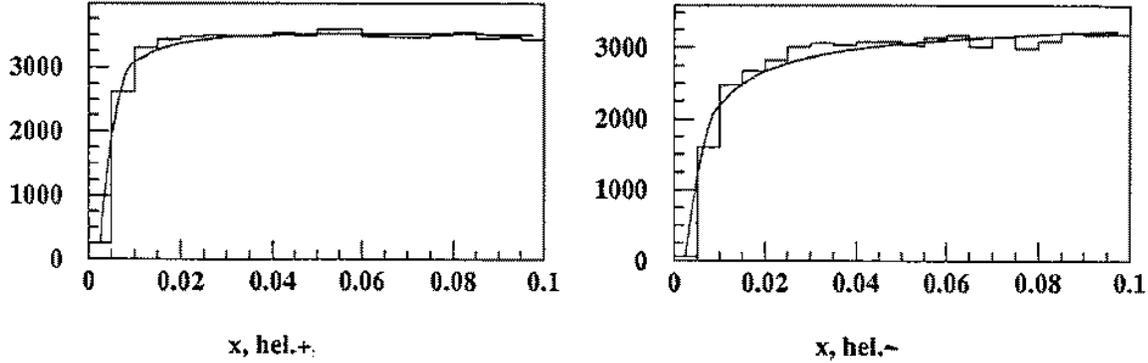}
\caption[ Parametrization of the finite $\mu$ mass
effect.] {Low $x$ distributions of positive and negative helicity
$\tmu$ events obtained from high statistics MC
sample (histograms), compared with our parametrization (solid line) of
the finite $\mu$ mass  effect.}
\label{fig_fmass}
\end{figure}

The $x_m$ and $q$ parameters are given in Table~\ref{fmass}. They were
derived from a large MC event sample. Fig.~\ref{fig_fmass} shows the
$x$ distribution of these MC events at the low $x$ region together with
our parametrization.
 
\begin {table} [htb]
\begin{center}
\begin{tabular}{||l|c|c||}  \hline  \hline
 helicity         & $x_m^{\pm}$  & $q^{\pm}$    \\  \hline
 $(+)$         & 0.0024 & 1.27   \\
 $(-)$         & 0.0047 & 1.56   \\  \hline \hline
\end{tabular}
\caption[Parametrization of the finite $\mu$-mass effect.]
{Parametrization of the finite $\mu$-mass effect.}
\label{fmass}
\end{center}
\end{table}
 
The effect of the threshold for tau decays to kaons
(Eq.~\ref{wxpik}) introduces an error
on \pta~and
\aplfb~associated with the uncertainties in the tau branching ratios to pions
and kaons and leads to errors below 0.1\%.
There is also a very small error (less than 0.1\%)
introduced from the measurement
uncertainty of \afb~\cite{bib-z0par}
~which enters the analysis (see Eq.~\ref{ptt}).
 
\section{Detector Resolution}
\label{sect-DR}
As in every experimental observation the measured quantities have
a natural spread which depends on the detector resolution.
The main influence of this in our analysis comes from the resolution in $x$
which is essential for the polarization measurement.
We are less effected by 
the inaccuracy in the measurement of $\cst$ since
our fit to $\aplfb$
is sensitive mainly to the numbers of
forward and backward scattered particles and not to the specific shape
of the $\cst$-distribution.
Our results are also not effected by the  charge misassignment effect,
which occurs only in a very small number of events.
 
A non-negligible effect on $\pta$ and $\aplfb$ can rise from the
$x$-smearing, since the fit is sensitive to very small deviations
from the true
shape of the $x$-distribution. In particular, due to this smearing, some
particles may have an unphysical $x$ value  above one,
which requires a special care.
 
To describe the resolution effect we
introduce the $x$ resolution function, $\rxx$, which is
for a given  $x^0$  the   measured $\xp$.
This function depends on the identification of the  particle.
 
For pions, $x$ is the particle momentum normalized to the beam energy.
For muons, we add to the particle momentum also the energy deposited
in the EM calorimeter associated with the $\t$-jet. From this EM
energy we subtract 0.5~ $GeV$, which is the expected energy deposited
by the muon itself. This is done
in order to include the energy of a photon which might be radiated from
the tau or  the muon.
This correction is very small in
most events,  therefore, also for muons the resolution in $x$ is
essentially  the momentum resolution.
 
For electrons however, we are using only the EM calorimeter to
 define $x$, summing up all the clusters in the $\t$-jet.
Therefore, the resolution function for electrons is completely
different than that of the $\mu$ or $\pi$.
Both resolution
\newpage
\noindent
functions, for e,  $\mu$ or $\pi$,
are investigated in two steps.
In the first step we determine a characteristic variable $y$, which
takes into account the resolution dependence on $x$ and possibly
also $\cst$.
In the second step we are using a control sample to investigate
the exact shape of the $y$ distribution.
In the following we shall discuss each resolution function separately.
\subsection{The Resolution Function for Muons and Pions}
 
\begin{figure}[htbp]
\epsfysize=13cm.
\epsffile[0 0  650 543]{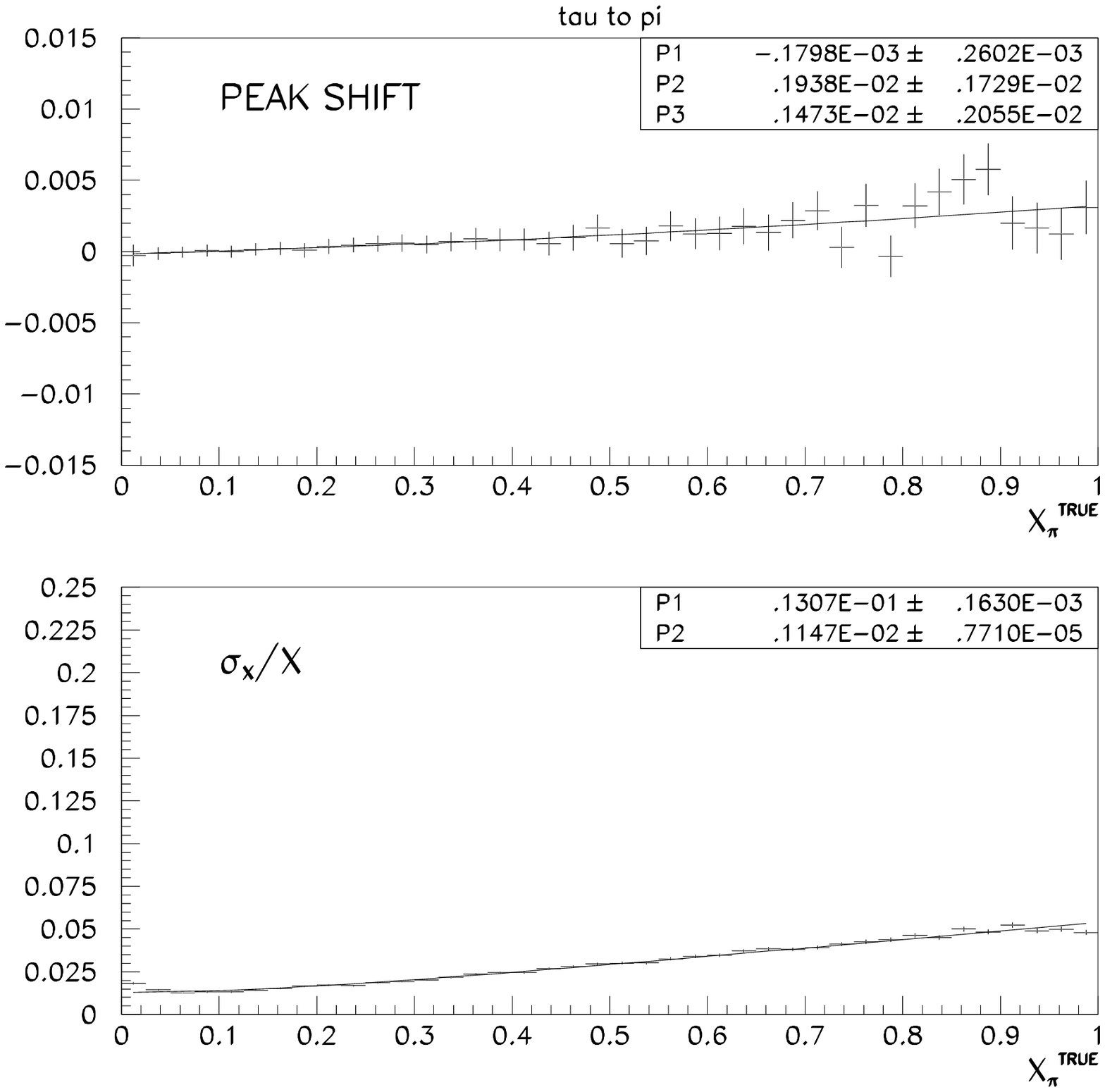}
\caption[Momentum resolution in $\tpiK$ MC]
{MC $\tpiK$ resolution parametrization with the peak $(\bar{x}-x)$
shift (top)
and the resolution $\sigma_x/x$ (bottom). The solid lines present
the fit curves.
}
\label{saspires}
\end{figure}

\begin{figure}[htbp]
\epsfysize=13cm.
\epsffile[0 0  650 543]{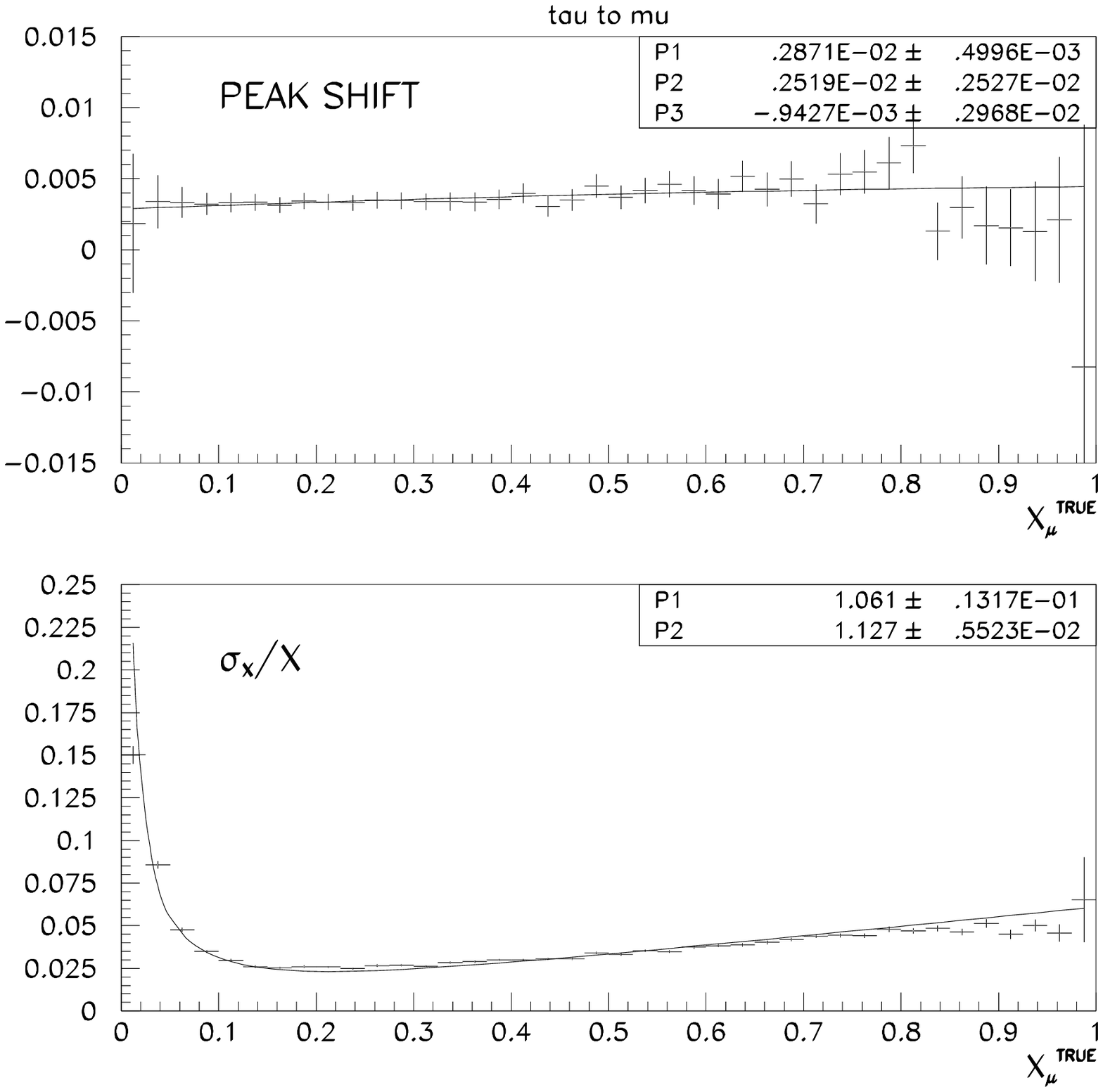}
\caption[Momentum resolution in $\tmu$ MC]
{MC $\tmu$ resolution parametrization with the peak $(\bar{x}-x)$
shift (top)
and the resolution $\sigma_x/x$ (bottom). The solid lines present
the fit curves.
}
\label{sasmures}
\end{figure}

First we examine only  the {\em shape} of $\Delta x/x$
where as before $x=p_{track}/E_{beam}$ and $\Delta x$ is the difference
between the true and measured normalized momentum, $\xp-x^0$, for each 
decay
particle respectively.
This was investigated by M.\ Sasaki~\cite{TN192} using $\tmu$ and $\tpiK$
MC events which were initially generated by KORALZ~3.8, and then
processed by GOPAL129 and ROPE403. For this study a  Gaussian distribution
of the kind
\beq
P(\xp,\bar{x})=\frac{1}{\sqrt{2\pi}\sigma(x)}
e^{-\frac{(x-\bar{x})}{2\sigma^2(x)}}
\eeq
was taken allowing also a possible shift in the average  $\bar{x}$
parametrized as
\beq
\bar{x}=a_1+a_2x^0+a_3{(x^0)}^2
\eeq
 The distribution of $\xp(x^0)$ was investigated separately
in 40 intervals between 0 to 1.
Neglecting the non Gaussian tails the $\bar{x}$ shift and the
width $\sigma(x)$ were determined for each $x$ interval.
The relation between $\frac{\sigma(x)}{x}$ and the track momentum
was taken to be
\beq
\label{sigxpi}
\frac{\sigma(x_{\pi})}{x_{\pi}}=\sqrt{b_1^2+b_2^2p_{track}^2}
\eeq
and from this fitted values for $b_1$ and $b_2$ could be obtained
(see Table~\ref{tabsasres}).
Fig.~\ref{saspires} illustrates the peak $(\bar{x}-x)$ shift and the
Gaussian width as a function of $x_{\pi}^0$, together with  the fitted
curves  estimated from the MC.
For the $\tmu$ decays, $x_{\mu}$ has a more complicated definition, namely
\beq
\xp_{\mu}=\frac{p_{track}-E_{cls}-0.5~GeV}{E_{beam}}
\eeq
where $\xp_{\mu}$ depends on both the $E_{cls}$ and the track momentum.
Therefore the resolution function for the $\tmu$ is  a
combination of the ECAL energy and the charged track
momentum resolution. This combination introduces  tuning parameters to weight
the components assuming that  this decay deposits the minimum ionizing
energy of $0.5~GeV$ in the ECAL. Hence, the width term has the form
\beq
\sigma(\xp_{\mu})=\frac{\sqrt{b_1^2\cdot\sigma_{E_{cls}}^2
+b_2^2\cdot\sigma^2_{ptrk}}}{E_{beam}}|_{E_{cls}=0.5\;GeV}
\eeq
Fig.~\ref{sasmures} illustrates $(\bar{x}-x)$  shift and the
Gaussian width as a function  of $x_{\mu}^0$, together with  the fitted
curves  estimated from the MC.
Table~\ref{tabsasres} presents the results of the fitted parameters.
As can be seen, 
a shift in the peak
appears only in the $\tmu$ decays.
This means that the energy  of 0.5~$GeV$ assumed to be deposit by the muon
in the ECAL was underestimated.
 
\begin{table}
  \begin{center}
    \begin{tabular}{|c|r|r|r|r|r|}
       \hline
  &  $a_1$ & $a_2$ & $a_3$ & $b_1$ & $b_2$ \\
       \hline
{ \tmu}     &  0.0029 & 1.0000 &  0.0000 &  1.0610 &  1.1270     \\
{ \tpiK }   &  0.0000 & 1.0000 &  0.0000 &  0.0131 &  0.0012     \\
       \hline
    \end{tabular}
  \end{center}
  \label{tabsasres}
  \caption[MC detector response parametrization]
  {The parameters for the detector response function determined from
  $\tmu$ and $\tpiK$ MC investigation.}
\end{table}
 
To obtain the final resolution function for muons and
pions we use real $\mu$-pair data  events to investigate the exact
shape of the $y_{\mu,\pi}$ distribution [$y_{\mu,\pi}=(\xp-x^0)/\sigma(x)$].
These events were selected
according to the standard criteria of the TP103 package~\cite{TP103}
with two additional requirements:
\begin{itemize}
\item The event acollinearity must be below
$0.2^\circ$. This cut was imposed in order to remove events with initial
state radiation where the outgoing muons have less than the beam
energy.
\item The track must be away from an anode (cathode) wire plane by at least
\newpage
$0.3^\circ$ ($0.5^\circ$).
\end{itemize}
 
\begin{figure}[p]
\epsfysize=18.5cm.
\epsffile[80 100 650 720]{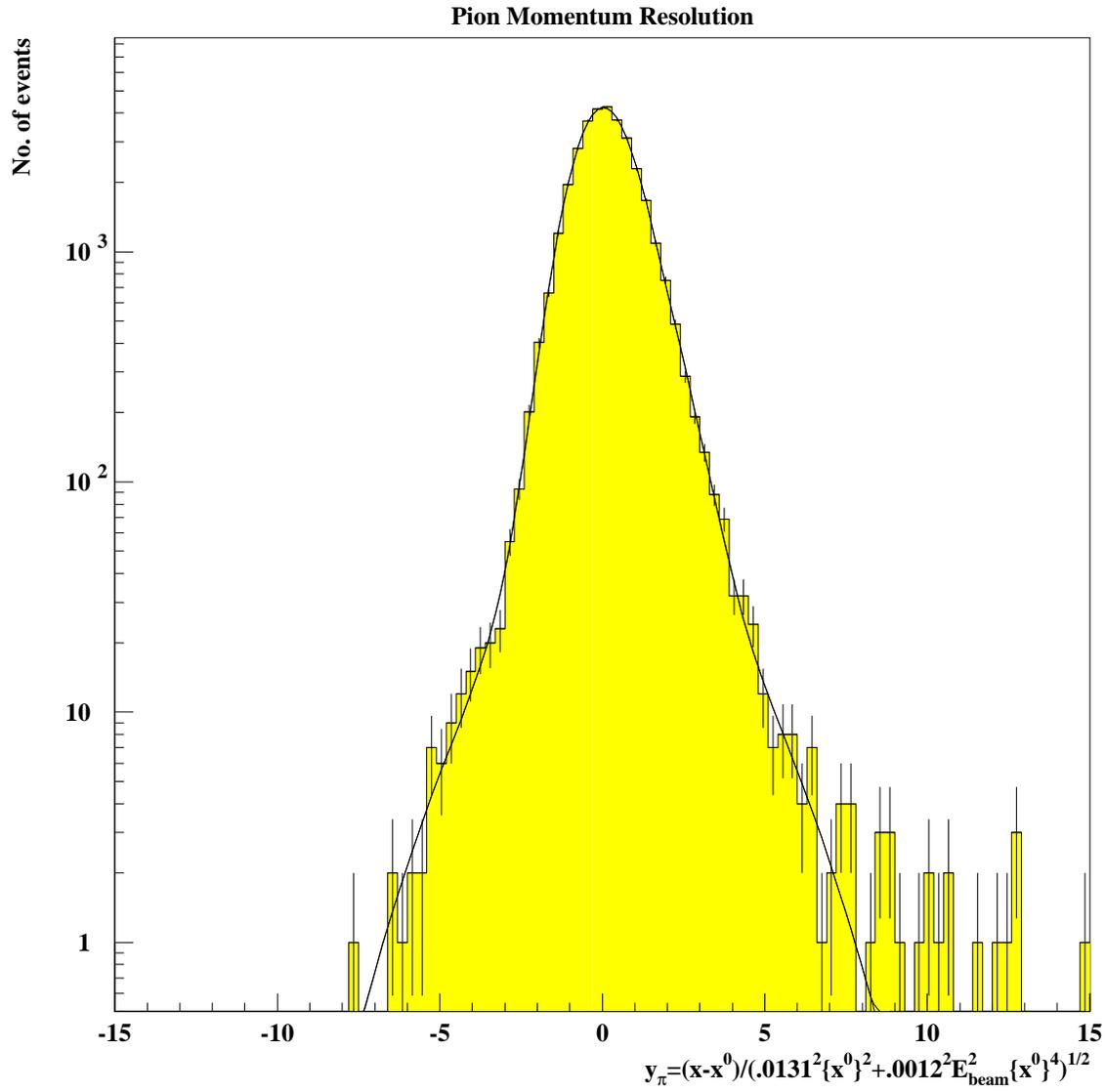}
\caption[The resolution ${\cal R}(x,\xp)$ as a function of
$y_{\pi}$.]
 {The number of events as a function of $y_{\mu}$
 given in a logarithmic scale.
 The solid line is our parametrization which represent
 the energy resolution ${\cal R}$.}
\label{pires}
\end{figure}
 
\begin{figure}[p]
\epsfysize=18.5cm.
\epsffile[80 100 650 720]{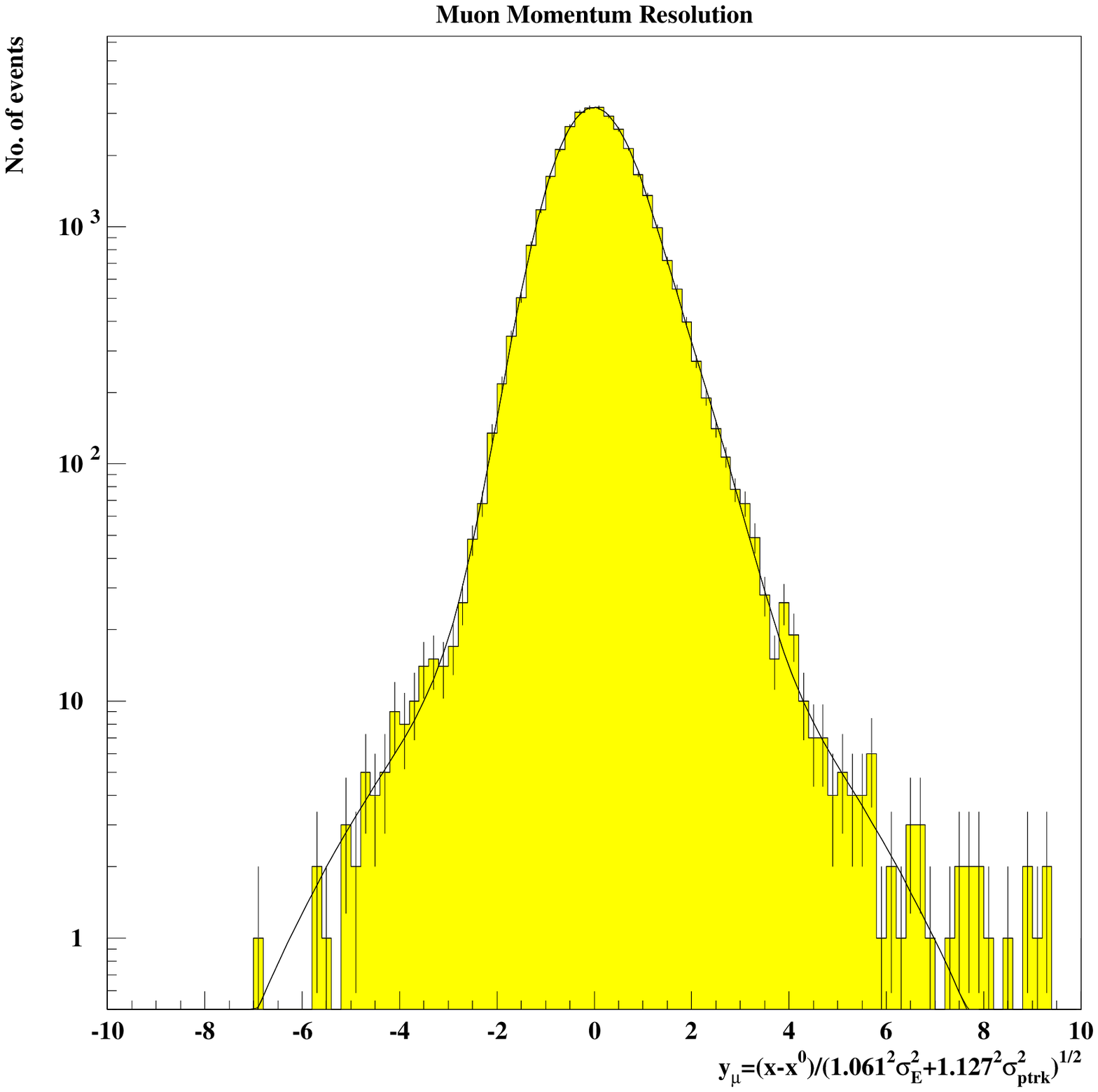}
\caption[
The resolution
${\cal R}(x,\xp)$
as a function of
$y_{\mu}$.]
 {The number of events as a function of $y_{\mu}$
 given in a logarithmic scale.
 The solid line is our parametrization which represent
 the energy resolution ${\cal R}$.}
\label{mures}
\end{figure}

Fig.~\ref{pires} and~\ref{mures}  show the resulting $y_{\pi}$
and  $y_{\mu}$
distributions respectively. These distributions contain
non-Gaussian tails. Therefore, they are  parametrized as
sums of three Gaussians with  different width, mean and normalization
parameters
\beq
\label{rhodef}
{\cal R}(x^0,\xp,\cst)= \sum_{i=1}^3
\rho_i\exp\left[-\frac{1}{2}\left(\frac{y-\bar{y}_i}{\sigma_i}\right)^2\right].
\eeq
The values of the fit parameters and their errors are
listed in Tables~\ref{resfitpi} and~\ref{resfitmu} .
The $\cof$ values of the fit are 61/53 and 54/64 for the
$\tpiK$ and $\tmu$ respectively.
 
\begin{table}[htb]
\begin {center}
\begin {tabular}{|l|c|c|c|} \hline
i & 1 & 2 & 3 \\ \hline
norm., $\rho_i$ & 1. & 0.292$\pm$0.031 &0.016 $\pm$ 0.012 \\
mean, $\bar{y}_i$ & -0.023$\pm$0.020 & 0.528$\pm$0.049 & 0.51$\pm$0.11 \\
width, $\sigma_i$ & 0.818$\pm$0.010 & 1.22$\pm$0.072 &  2.57$\pm$0.44\\
\hline
\end {tabular}
\caption[Parameterization of the  $x$  resolution of pions.]
{Parameters of the three  Gaussians describing the exact shape of the
pion $x$  resolution function.}
\label{resfitpi}
\end{center}
\end{table}
 
\begin{table}[htb]
\begin {center}
\begin {tabular}{|l|c|c|c|} \hline
i & 1 & 2 & 3 \\ \hline
norm., $\rho_i$   & 1.               & 0.2732$\pm$0.0048 & 0.0098$\pm$0.0037  \\
mean, $\bar{y}_i$ & -0.057$\pm$0.010 & 0.427$\pm$0.030 & 0.36$\pm$0.14\\
width, $\sigma_i$ & 0.7266$\pm$0.0071  & 1.127$\pm$0.029 &  2.60$\pm$0.\\
\hline
\end {tabular}
\caption[Parameterization of the  $x$  resolution of muons.]
{Parameters of the three  Gaussians describing the exact shape of
the muon  $x$  resolution function.}
\label{resfitmu}
\end{center}
\end{table}

\begin{figure}[p]
\epsfysize=18.5cm.
\epsffile[80 130 650 750]{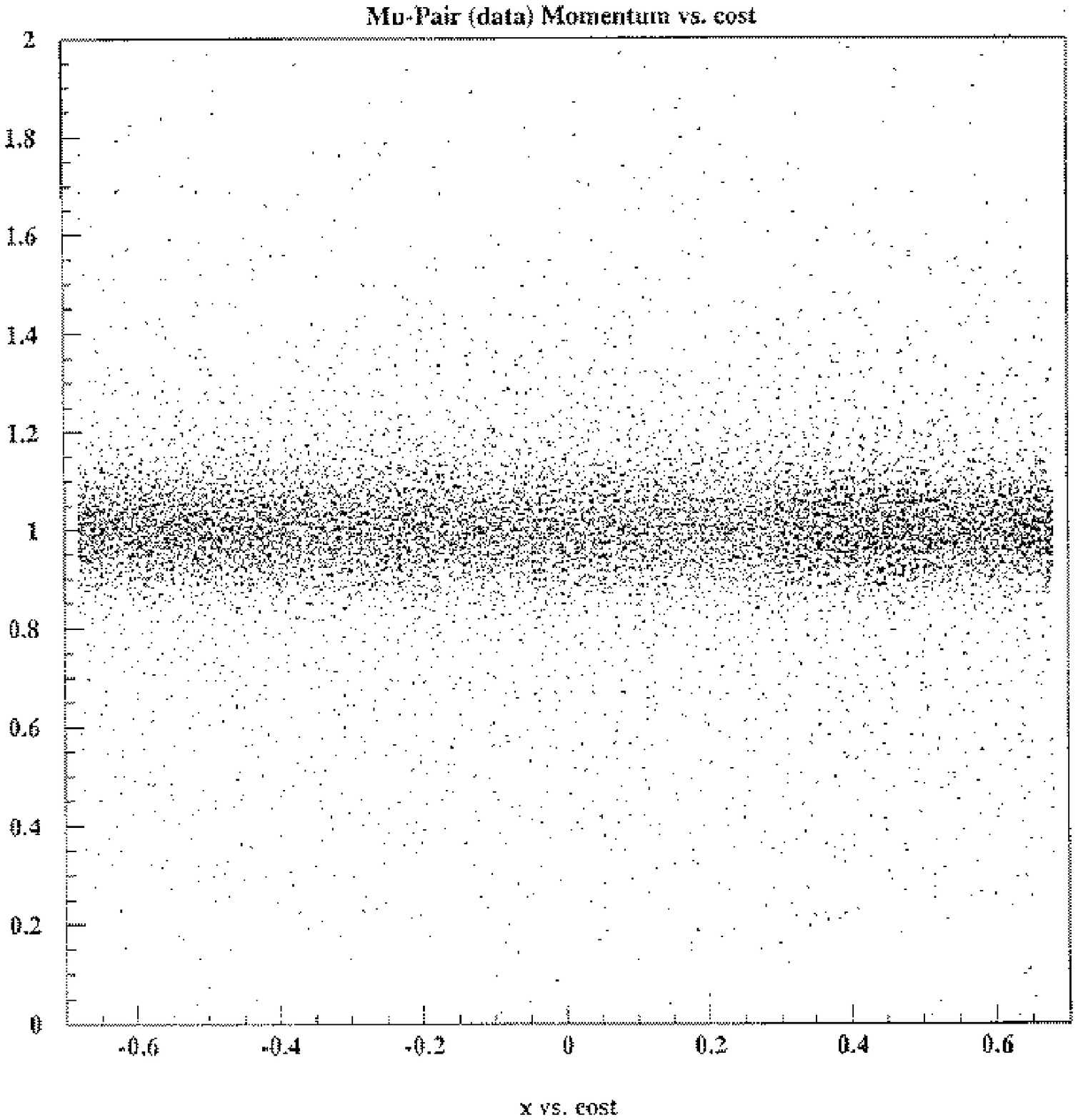}
\caption[
 $\xp$ vs. $\cst$ for data $\mu$-pairs.]
{ $\xp$ vs. $\cst$ for data $\mu$-pairs.}
\label{xvscost}
\end{figure}
 
We considered the following contributions to the systematic error
associated with the muon and pion resolution function.
\begin{itemize}
\item The uncertainties in the parameters of the resolution functions
coming from the limited statistics of the $\mu$-pair sample. We used
the errors listed in Tables~\ref{resfitpi} and~\ref{resfitmu}
including
their correlations in order to propagate them to the final results.
\item We investigated a possible dependence of the momentum
measurement on $\cst$. Such dependence could result from a possible
twist of the two end-plates of the jet chamber,
that will result in a
tracks distortion so that   the measured momentum would be larger
or smaller than the true value, depending on the charge and $\cst$
values.
In this case, there would be a net bias in the $\aplfb$ measurement.
Fig.~\ref{xvscost} shows a scatter plot of $\xp$
vs. $\cst$ for $\mu$-pairs data which looks uniform. dividing
 the $\cst$
range into 14 intervals, average over $\xp$ for each interval,
and fitting
$\bar{x}$ to $a(1+b\cst)$ we obtain $b=(-1.99\pm0.94)\cdot10^{-3}$
with $\cof$=11.8/12. We introduced this small $b$ value in our ML fit
 the effect on the final results was found to be very small (see
Table~\ref{muresys}).
\item We investigated whether there is any dependence of the
$y_{\mu,\pi}$ distribution on $\act$, by considering separate
distributions from MC $\tmu$ decays for $\act$ below and above 0.35.
The obtained RMS values were 0.0716 and 0.0759. Assuming a linear
dependence of the RMS on $\act$, we multiplied $y_{\mu,\pi}$ by
$0.996+0.011\act$ and the obtained deviations in $\pta$ and $\aplfb$
are listed in Table~\ref{muresys}.
\end{itemize}
 
Table~\ref{muresys} summarizes all the systematic errors associated
with the muon and pion momentum resolutions. This amounts a total of
$0.38\%$ and $0.21\%$ for $\pta$ and $\aplfb$, respectively.
 
\begin {table} [htb]
\begin {center}
\begin {tabular}{||l|cc||cc|cc||} \hline \hline
   &  \multicolumn{2}{c||}{whole sample} &
 \multicolumn{2}{c|}{$\tmu$} & \multicolumn{2}{c||}{$\tpi$} \\
source & $\dpta$ & $\daplfb$ & $\dpta$ & $\daplfb$ & $\dpta$ & $\daplfb$ \\
\hline
$\mu$-pair statistics       & 0.10 & 0.01 & 0.47 & 0.03 & 0.10 & 0.01 \\
$\xp-\cst$ dep.          & 0.21 & 0.21 & 1.03 & 1.17 & 0.32 & 0.72 \\
resolution $\act$ dep.      & 0.30 & 0.02 & 0.98 & 0.11 & 0.11 & 0.00 \\
\hline
Total                            & 0.38 & 0.21 & 1.50 & 1.18 & 0.35 & 0.72 \\
\hline \hline
\end {tabular}
\caption[Systematic errors of  $\mu$ and $\pi$ resolution function]
{Systematic errors (in \%) associated with the muon and pion
resolution function.}
\label{muresys}
\end{center}
\end{table}
 
Before concluding this sub-section, one should mention two other sources
of systematic errors associated with the track measurement,
which are related to the $\aplfb$ determination.
\begin{itemize}
\item Uncertainty in $\cst$ measurement may affect the $\aplfb$
determination, since events near $\cst=0$ can migrate from positive
to negative $\cst$ value and vice versa. It also introduces an
inaccuracy in the cut on $\act$ at 0.68. The $\act$ value is
obtained in our analysis by averaging the corresponding values of
the two $\t$-jets, taking into account tracks and EB clusters.
This is a measure of the $\act$ of the tau and not just the track.
Therefore, part of the inaccuracy in
$\cst$ results from the invisible $\nu$'s in the event. We
investigated the $\cst$ resolution using MC events and obtained an
uncertainty of 0.0225.
We introduced in our ML fit a resolution function in $\cst$ using an
error in $\cst$ twice as large, in order to account for possible data-MC
differences. The resulting variation in $\aplfb$ is
smaller than 0.05\%.
\item Wrong charge assignment of the $\t$-jets, resulting in a wrong
sign of $\cst$. The charge assignment is done
by the following algorithm: The charge of each $\t$-jet is defined as
the sum of its track charges. If both $\t$-jets have opposite charges,
there is no ambiguity. If one jet charge is zero, only the
second jet charge is used. If both jets have the same charge
and this charge is not zero, then the correct charge is taken from the
jet with a smaller number of tracks. In the remaining cases
where the two jets have the same number of tracks, we consider
only the most energetic track from each $\tau$-jet and take the charge
of the track with the smaller momentum.
As a result of this procedure, only 0.35\% of the events have a wrong
charge assignment, as obtained from the MC sample. This would reduce the
value of $\aplfb$ by a negligible amount of $0.007\times\aplfb$.
If we assume that in the data the situation is twice as bad, we end up
with a systematic error in $\aplfb$ of 0.16\%.
\end{itemize}
In total, the systematic error in $\aplfb$ associated with these two
sources is 0.17\%. This uncertainty is common to all decay channels.
\subsection{The Resolution Function for Electrons}
The electron energy is taken from the EM calorimeter. We are
using the corrected EB energy and we include a further correction
depending on the total presampler multiplicity in the cone. In order to
investigate the possibility of using the PB data, we used Bhabha events
and tested the EB energy normalized by the beam energy ($x_{EB}$)
as a function
of the PB multiplicity (number of hits). A decrease in the EB
energy for increasing PB multiplicity is found as expected due
to electrons which started showering in the material in front of
The result of our study is following correction to $x_e$,
\beq
\label{elxcor}
x_e=x_{\scriptscriptstyle EB}+0.02 \times
Max(N_{\scriptscriptstyle PB}-10,0)/\eb,
\eeq
where $N_{\scriptscriptstyle PB}$ is the PB multiplicity.
 
 
In order to investigate the energy resolution dependence on its
energy, we are using real Bhabha electrons data, single
electron events and $\tel$ decays. For $\tel$ electrons we are
using our analysis sample and the other
events were selected using the standard cuts of the
TP103 package~\cite{TP103}. For Bhabha events we required also
an  acollinearity below $0.2^\circ$, to reduce radiative effects
similarly to our procedure with $\mu$-pairs.
 
The Bhabha electrons  events are assumed to have the beam
energy. We do not know the real energy of the electrons from the
other sources but we can utilize their
$E/p$ and $\Delta p/p$ distributions at different energies in order to
obtain the energy resolution,
\beq
\label{dee}
\frac{\Delta x}{x}=\frac{\Delta E}{E}=\sqrt{[\Delta(\frac{E}{p})
/(\frac{E}{p})]^2-(\frac{\Delta p}{p})^2}.
\eeq
 
The width $\Delta(E/p)$ is obtained by a Gaussian fit to
the distribution of $E/p$ at the different energy points.
For the evaluation of $\Delta p/p$ we use
$\Delta p/p=C\sqrt{x^4+(0.3x)^2+0.02x}/x$ where the
factor $C=0.073\pm0.006$ is obtained from Bhabha and $\mu$-pair events.
 
\begin{figure}[p]
\epsfysize=18.5cm.
\epsffile[80 100 650 720]{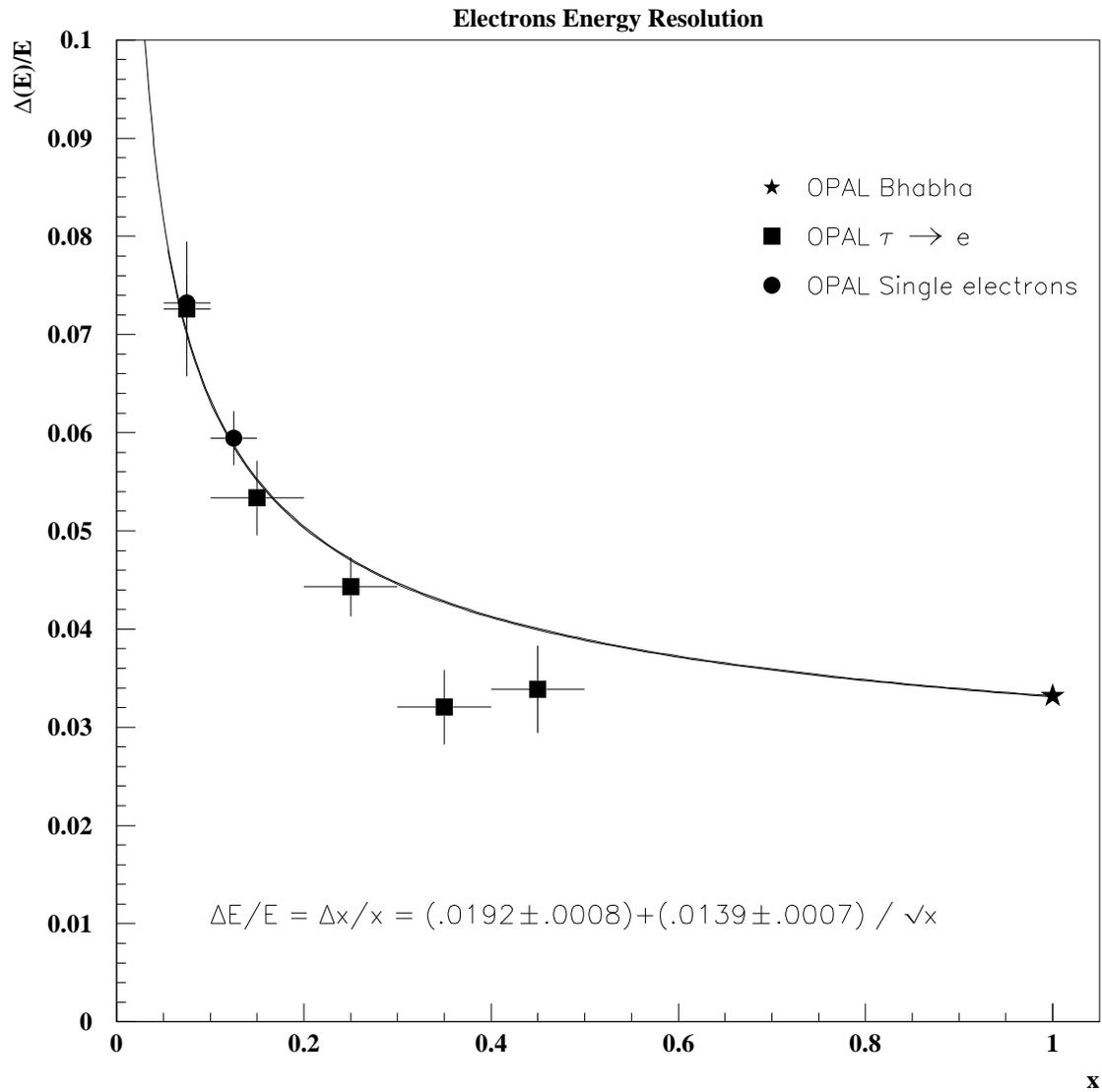}
\caption[ $\Delta E/E$ dependence on $x$.]{  $\Delta E/E$ dependence on $x$.
The last point is calculated with Bhabha
events and in the first three points it is done with
single electrons. For consistency  we plot also the $E/p$
of the $\tel$ events from our data sample. The line represents our parametrization
to this dependence.}
\label{exres}
\end{figure}
 
Fig.~\ref{exres} presents the $\Delta E/E$ dependence on $x$.
The resolution in the last point is estimated from the Bhabha
events, whereas the first three points, were calculated with
the single electrons.
As a consistency check  we also plot
the results of the electrons from $\tel$ decays,
using their $E/p$ distribution. The error bars of all these results
derived from
the statistical uncertainties of the $E/p$ widths and the
error factor $C$.
 
The $\frac{\Delta E}{E}$  values were fitted by the function
$a+b/\sqrt{x}$,
\beq
\label{dee2}
\frac{\Delta x}{x}=\frac{\Delta E}{E}=
(0.0192\pm0.0008)+(0.0139\pm0.0008)/\sqrt{x}.
\eeq
This is shown in Fig.~\ref{exres} having a $\cof$ of
 4.7 for 6 D.O.F.
 
As for the muons and pions, we define
$y_e$ for electrons as,
\beq
y_e=\frac{\xp-x^0}{0.0192x^0+0.0139\sqrt{x^0}}.
\label{ydefe}
\eeq
 
As before also here the second step is the study
of the exact resolution shape, including the tails of the distribution.
For the $\tel$ it is investigated by using Bhabha events
describing them by a sum of three
Gaussians as in Eq.~\ref{rhodef}.
 
\begin{figure}[p]
\epsfysize=18.5cm.
\epsffile[80 100 650 720]{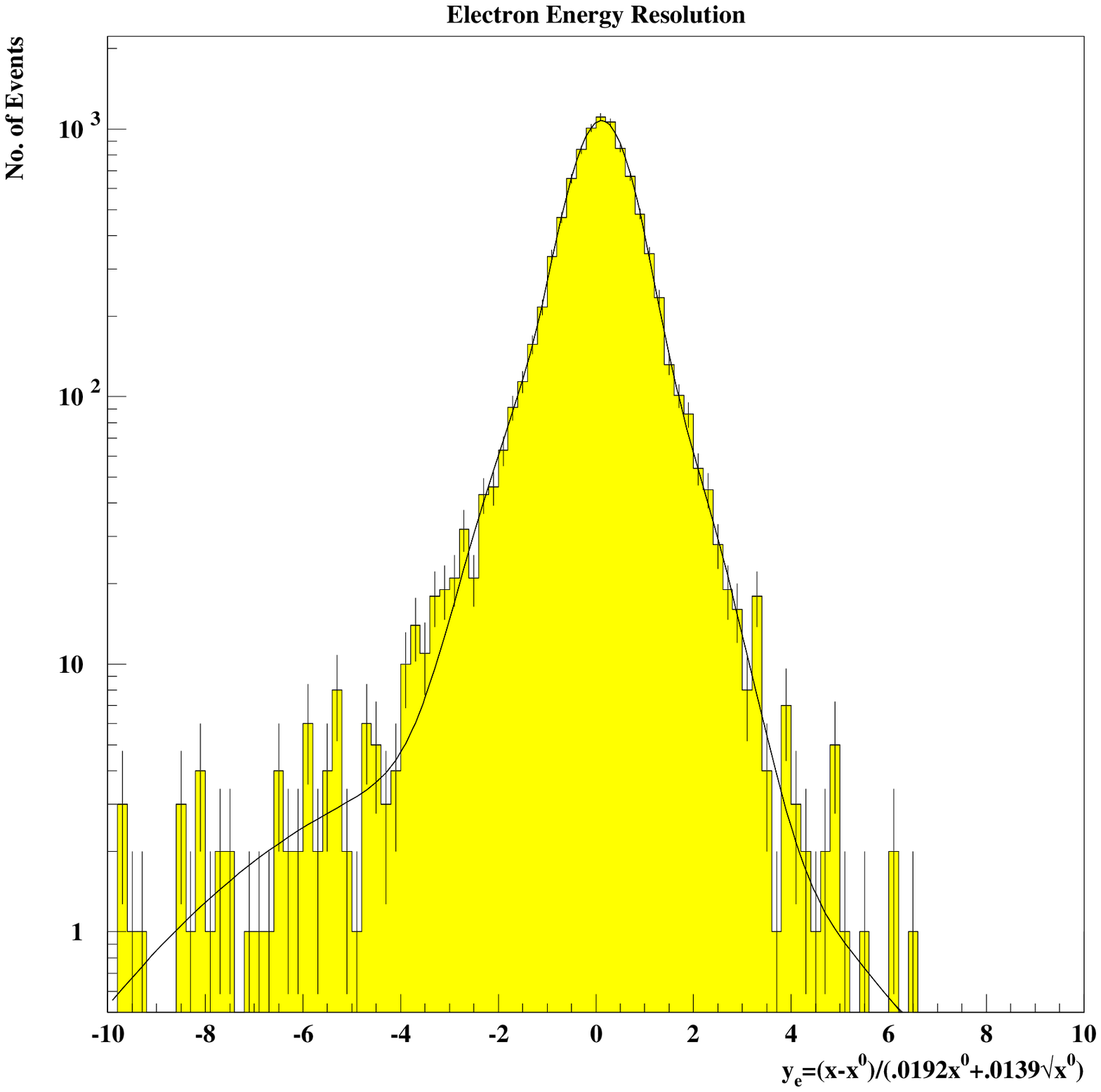}
\caption[The energy resolution  as a function of
 $y_e$.]
 {The number of events as a function of $y_e$
 given in a logarithmic scale.
 The solid line is our parametrization which represent
 the energy resolution ${\cal R}$.}
\label{eres}
\end{figure}

Fig.~\ref{eres} presents the $y_e$ distribution and our parametrization.
The values of the fit parameters and their errors are
listed in Table~\ref{resfite}. The $\cof$ of the fit is 74/58.
 
\begin{table}[htb]
\begin {center}
\begin {tabular}{|l|c|c|c|} \hline
i & 1 & 2 & 3  \\ \hline
normalization, $\rho_i$ & 1 & 0.192 $\pm$ 0.032 & 0.0067 $\pm$ 0.0018 \\
mean, $\bar{y}_i$ & 0.120 $\pm$ 0.010 & 0.00 $\pm$ 0.039 & -1.72 $\pm$ 0.52  \\
width, $\sigma_i$ & 0.571 $\pm$ 0.016 & 1.317 $\pm$ 0.067 &  4.65 $\pm$ 0.67  \\
\hline
\end {tabular}
\caption[Parameters of the electron energy resolution functions.]
{Parameters of the three Gaussians describing the electron energy
resolution functions.}
\label{resfite}
\end{center}
\end{table}
 
Two sources of uncertainties contribute to the systematic error associated
with the electron energy resolution.
\begin{itemize}
\item The errors on the energy dependence of the resolution
(see Eq.~\ref{dee2}).
\item The errors on the parameters describing the exact shape of the
resolution function, shown in Table~\ref{resfite}.
\end{itemize}
These errors were propagated properly to the final results, taking also
their correlations into account.
 
A summary of the systematic errors associated
with the electron energy resolution is presented in Table~\ref{eressys}.
The uncertainty in the calorimeter response introduces total errors of
$0.48\%$ for $\pta$ and $0.12\%$ for $\aplfb$.
 
\begin {table} [htb]
\begin {center}
\begin {tabular}{||l|cc|cc||} \hline \hline
   &  \multicolumn{2}{c|}{whole sample} &
 \multicolumn{2}{c||}{$\tel$}  \\
source & $\dpta$ & $\daplfb$ & $\dpta$ & $\daplfb$  \\
\hline
energy dependence    & 0.01 & 0.00 & 0.04 & 0.01  \\
resolution function  & 0.32 & 0.10 & 1.46 & 0.45  \\
energy scale         & 0.37 & 0.07 & 2.23 & 0.17  \\
\hline
Total                  & 0.48 & 0.12 & 2.67 & 0.48  \\
\hline \hline
\end {tabular}
\caption[Systematic errors associated with electron
resolution function.]{Systematic errors (in \%) associated with electron
resolution function.}
\label{eressys}
\end{center}
\end{table}

\section{Tau-Pair Selection Efficiency}
\label{sect-TSE}
 
The efficiency of $\tn$-pair selection, \epsti, is investigated using the
$\tn$-pair MC events and
comparing the kinematic variable distributions
before and after the selection requirements. This is done separately for
events with positive and negative $\tm$ helicities and for the various
combinations of the decays of the two taus.
 
When both $\tn$'s decay via one of the modes considered here
(e, $\mu$ or $\pi$(K)),
the efficiency is parametrized as a function of the kinematic
variables $\xip$, $\xjp$ and $\cst$.
When the selection requirements involve two or even all three
kinematic variables in a
correlated manner then these are explicitly taken into account in the
parametrization of the efficiency.
 
The efficiency used in our fit (Eqs.~\ref{corcst} and~\ref{corcso})
is expressed as a function of the {\em measured} variables,
$\xip$ and $\xjp$, on which all the cuts are applied. However, we cannot
obtain the efficiency as a function of the {\em measured} variables
since for events which were not identified as $\t$-pairs, very
often those variables are not defined. Therefore, we have to use
the corresponding {\em generated} variables $\xxi^0$ and $\xxj^0$.
As long as the variation of efficiency with the variables is weak,
it should not make any difference. However, for those $\t$-pair
selection
cuts which depend on $\xip$ and $\xjp$ variable one has to be
more careful and include this fact in the efficiency function.
The three cuts which depend on $\xip$ and $\xjp$ are T4, T5 and T6
and these are treated in the following way:
 
\begin{itemize}
\item The efficiency $\epstij$ to select $\t$-pair events where both
taus decay into the same channel is set to be 0 if $\xip+\xjp<0.4$
and $(\xip-\xjp)\sin\theta<0.04$. This corresponds to the cut against
two photon events (T4). This cut
rejects events with low visible energy namely, if both missing
transverse momentum calculated from the tracks and the from
the EM clusters are below 2~ $GeV$. It therefore affect
mainly $\t$-pair
events where both taus decay  to the same channel.
\item  The efficiency  $\epstij$ to select events where both
taus decay into electron is required to vanish when $\xip+\xjp>1.6$.
This is due to the rejection of Bhabha events with cut T5 which
influence $\t$-pair events of this type.
\item The rejection of mu-pair events by cut T6
affects mainly events where both taus decay
into muons. For such events, we require
that 
\newline
\noindent
$\epstij=0$ when ${\mbox \xip+\xjp>1.2}$. 
A possible residual effect
of this cut on events with one $\tmu$ and one $\tpiK$ decays was
taken into account by allowing a step function change in
$\epstij$ at $\xip+\xjp=1.2$.
\end{itemize}
 
To determine the $\t$-pair selection efficiency we investigate
$\t$-pair MC events  which satisfy the
three cuts mentioned above. We consider the distributions
of the kinematic parameters before and
after the $\t$-pair selection. The ratio between the distributions
after and before the selection gives the efficiency of the
selection excluding the cuts T4, T5 and T6. This procedure is done
separately for positive and negative $\tm$ helicity events.
The $\act$ dependence of the efficiency turns out to be constant over
the whole region from 0 to 0.68 and that for all decay channel combinations.
 
In the case where both taus decay into electron, muon or pion, we
have to consider in addition to $\act$ the two kinematic variables,
$\xxi$ and $\xxj$. Therefore, we consider two-dimensional efficiency
distributions in a  6$\times$6 regions in $\xxi$ and $\xxj$
and fit them  using expression of the following form
\bea
\label{epsij}
{\cal E}_{ij}(\xxi,\xxj)&=&a_1(1+a_2\xxi+a_4\xxi^2)(1+a_3\xxj+a_5\xxj^2)
\nonumber \\
& &+a_6\xxi\xxj+a_7\xxi^2\xxj+a_8\xxi\xxj^2+a_9\xxi^2\xxj^2.
\eea
For $\xxi<.05$ or $\xxj<.05$ the efficiency is reduced by a factor,
while if both $\xxi$ and $\xxj$ are $<.05$ then the efficiency is set to 0.
The parameter $a_1$ is expected to be close to 1 and all other
parameters are expected to be close to 0.
If \nolinebreak after
\newpage
\noindent
doing the fit any of
the parameters turns out to be consistent within error with its
expected value it is set to that value as a constant in the fit,
and the fit is repeated. In this way we are using just the
number of parameters necessary to obtain a reasonable fit.
 
When only one $\tn$ decays to
e, $\mu$ or $\pi$(K) and the other \tn~decay is not
identified,  the selection efficiency has two components. The first
accounts for the case where the other \tn~{\em does not
decay} via e, $\mu$ or $\pi$(K) while the second accounts for
the case where it {\em does} decay via these channels but could not
be identified.
The first component is parametrized
as a function of $\xip$ and $\cst$, checking that these two
kinematic variables are not correlated. The second component, which
arises from identification inefficiencies, can have
correlations which were taken into account.
 
\begin{figure}[p]
\epsfysize=18.5cm.
\epsffile[40 100 680 650]{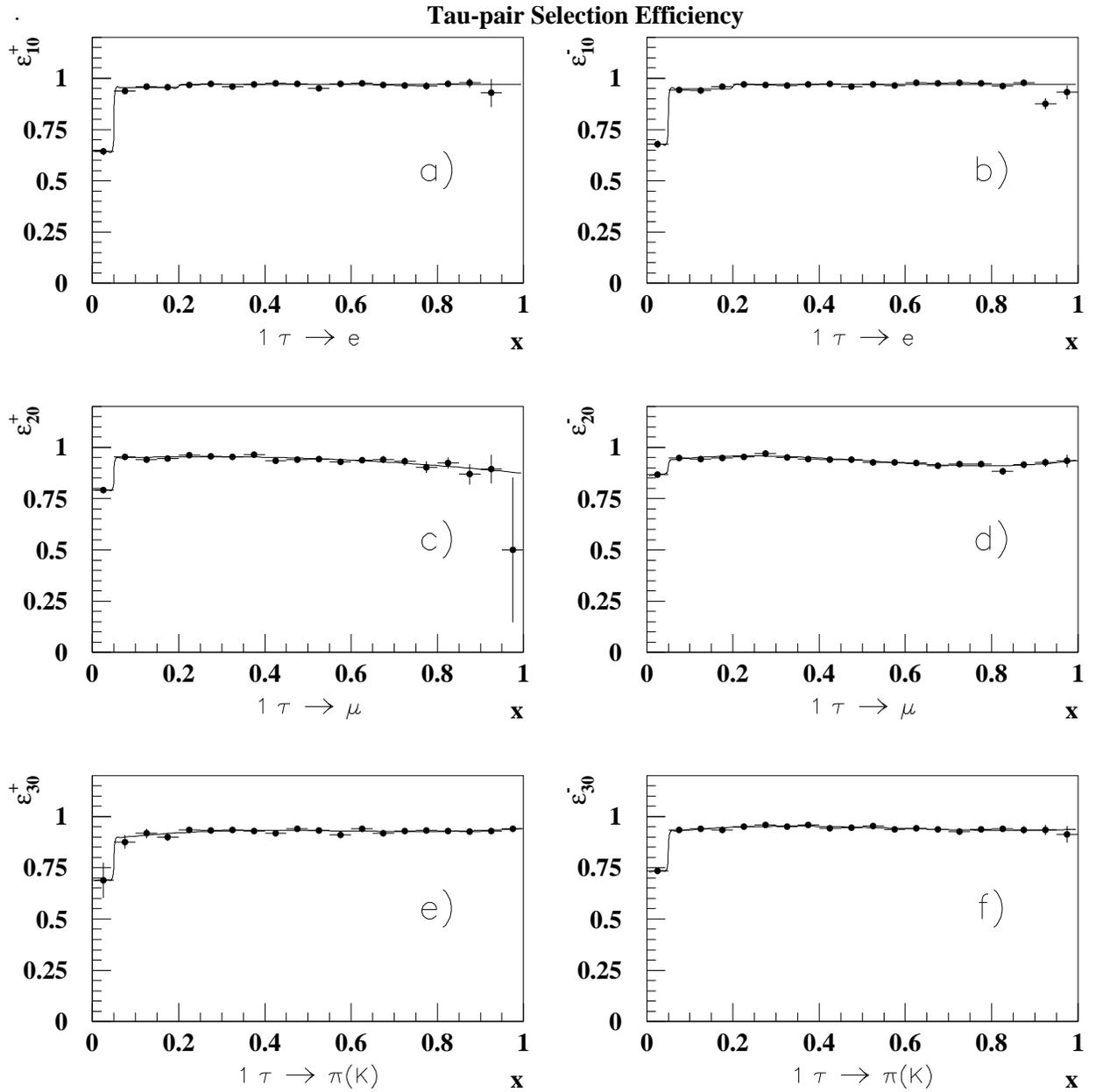}
\caption[Tau-pair selection efficiencies as functions of $x$.]
{Tau-pair selection efficiencies as functions of $x$
for events where only one tau decays into electron (Figs. a,b),
muon (Figs. c,d) or pion (Figs. e,f). Figs. a,c,e (b,d,f)
correspond to positive (negative) helicity events.}
\label{fig_eftt}
\end{figure}
 
In the cases where only one $\t$ decay is identified,
the selection efficiency ${\cal E}_i(\xxi)$ is parametrized
as a quadratic function of just one variable, $\xxi$.
For $\xxi<.05$ the efficiency, ${\cal E}_i(\xxi)$,
is further reduced by a factor.
Also here, when any parameter is obtained from the fit as
consistent with its expected value (1 for the normalization and 0 for the
others) it is set to the expected value and the fit is redone.
 
In Fig.~\ref{fig_eftt} we plot the $\t$-pair selection efficiencies
as function of $x$
for these cases where only one tau decays to $e$, $\mu$ or $\pi(K)$
together with our parametrization. In Table~\ref{tab_eftt} we list
the $\chs$ values obtained in our parametrization fits.
\begin {table} [htb]
\begin {center}
\begin {tabular}{||l|c|c||} \hline \hline
Decay channel combination & positive helicity & negative helicity \\
\hline
a) $\tel$, none   & 13.6/16 & 15.2/16 \\
b) $\tmu$, none   & 12.6/15 & 12.6/14 \\
c) $\tpiK$, none   & 11.4/14 & 9.0/14 \\
d) $\tel$, $\tel$ & 15.9/29 & 24.9/32 \\
e) $\tel$, $\tmu$ & 35.9/24 & 41.6/24 \\
f) $\tel$, $\tpiK$ & 18.9/27 & 41.7/24 \\
g) $\tmu$, $\tmu$ & 19.4/29 & 29.4/32 \\
h) $\tmu$, $\tpiK$ & 32.4/25 & 44.9/24 \\
i) $\tpiK$, $\tpiK$ & 12.1/33 & 20.0/28 \\  \hline \hline
\end {tabular}
\caption[$\cof$ of the  $\t$-pair selection efficiency fits.]
{$\cof$ coming from the parametrization fits for $\t$-pair selection
efficiency.}
\label{tab_eftt}
\end{center}
\end{table}
 
The error coming from the $\tn$-pair selection efficiency is composed of
the  MC statistics and the error
related to the resolution and scale uncertainties in the ECAL
and tracking detector.
 
Systematic uncertainties in the efficiency
resulting from the MC statistics
were investigated
by propagating the statistical errors in the parameters obtained
from the parametrization fits to the final results as explained in
Sect.~\ref{secfit}.
 
As for possible effects in the data not fully modeled by the MC,
we were concerned with the momentum and energy resolutions
being broader in the real data as compared with the MC.
In order to investigate this effect on the final
results we used our MC event sample and smeared the generated
track momenta and electron/photon energies according to our resolution
functions obtained from real data control samples (see Sect.~\ref{sect-DR}).
After replacing those smeared values with the corresponding
MC simulated ones, we recalculated the $\t$-pair selection efficiencies
and parametrized them in the same way as before.
The different tau selection efficiency parametrization,
calculated with and without additional smear to the MC, has a negligible
effect to the overall uncertainty in $\pta$ and in $\aplfb$
of the order of 0.03\% and 0.05\% respectively.
Nevertheless, we have used in the analysis the MC sample  with
an additional smear.
 
Table~\ref{efttsys} summarizes the systematic errors associated
with the $\t$-pair selection efficiency for
each decay channel separately.
\begin {table} [htb]
\begin {center}
\begin {tabular}{||l||cc|cc|cc||} \hline \hline
   & \multicolumn{2}{c|}{$\tel$} &
\multicolumn{2}{c|}{$\tmu$} & \multicolumn{2}{c||}{$\tpiK$} \\
  & $\dpta$ & $\daplfb$ & $\dpta$ & $\daplfb$ &
$\dpta$ & $\daplfb$  \\ \hline
error    & 1.10 & 0.80 & 0.59 & 0.06 & 0.36 & 0.08 \\
 \hline \hline
\end {tabular}
\caption[Systematic errors associated with $\t$-pair selection
efficiency.]{Systematic errors (in \%) associated with $\t$-pair selection
efficiency. This is essentially based on the MC statistics.}
\label{efttsys}
\end{center}
\end{table}
 
Thus, the contributions from the tau selection
efficiency to the overall uncertainty on $\pta$ and $\aplfb$
are 0.39\% and 0.19\%, respectively.
\section{Efficiencies of the $\tn$ Decay Identification}
\label{sect-TDI}
The efficiencies to identify the various $\tn$ decays are obtained in
two steps. In the first step, the MC events are used in order
to obtain the efficiency distributions for each decay channel,
separately for events with positive or negative $\tm$
helicity.
These distributions depend on the $\xp$ variables of the
identified $\tn$ and on $\cst$. However, some of the identification
requirements, such as those designed to remove Bhabha and mu-pair events,
also introduce a dependence on the
$\xp$ variable of the $\tn$ on the opposite side. This dependence
is taken into account as described below.
 
For $\tel$ decays we refer to the cut
against residual Bhabha events (cut E7)
which rejects electron candidates
when the other side has a single track with $x>.75$,
and its acoplanarity is below $0.1^\circ$. Therefore,
in evaluation of the  $\tel$ identification efficiency,
the events where the other tau decay into $e, \mu$ or $\pi$(K)
were divided to those with the $x$ of the opposite side below and above
0.75. We did not find any statistically significant
difference between the efficiency distributions of those event
subsamples. In the case where the other $\t$ do not decay into these
three channels our criteria are independent of its $x$ value.
 
For the $\tmu$ decay, the
against $\mu$-pair events (cut M5) introduces a
dependence on the opposite $\t$ properties.
 This cut rejects muon candidates
if the opposite jet is consistent with being a muon and has
$x>0.8$. Hence, for events where both taus are identified as muons
we require both $x$-values to be below 0.8. This condition is
applied already in the parametrization of $\epstij$ and therefore does not
have to be repeated.
However there might be still some residual effects of this cut on
events where the opposite jet contains a charged pion, which resemble
a muon. Therefore, we investigated the
$\tmu$ identification efficiency for the following cases.
\begin{itemize}
\item events where the other tau does not decay into $e, \mu$ or $\pi$(K);
\item events where the other tau decays into a lepton ($e,\mu$);
\item events where the other side is a $\tpiK$ decay with $x<0.8$;
\item events where the other side is a $\tpiK$ decay with $x>0.8$;
\end{itemize}
It turned out that for each case the  identification
efficiency was different. Moreover, for the third case, there
was even a difference between positive and negative $\tm$ helicity
events. This is due to the fact that the pions in positive $\tmu$
helicity events are
more energetic and thus are more sensitive to the cut.
We had, therefore, to parametrize each case separately.
 
\begin{figure}[p]
\epsfysize=18.5cm.
\epsffile[40 100 720 650]{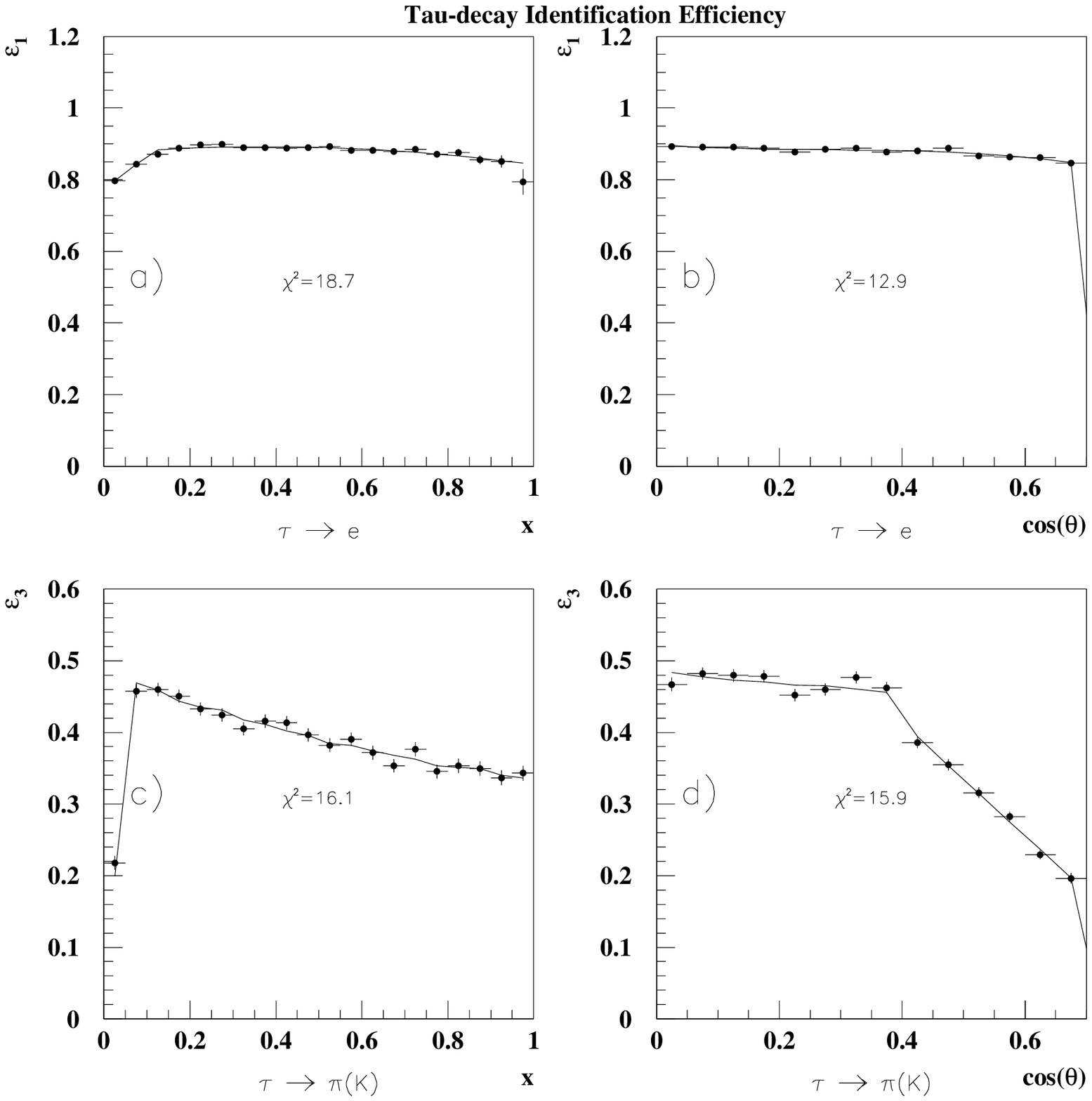}
\caption[Efficiencies of $\tel$ and $\tpiK$ identification.]
{Efficiencies of $\tel$ (a,b) and $\tpiK$ (c,d)
identification as functions of $x$ and $\act$.}
\label{efftd1}
\end{figure}

\begin{figure}[p]
\epsfysize=18.5cm.
\epsffile[40 100 720 650]{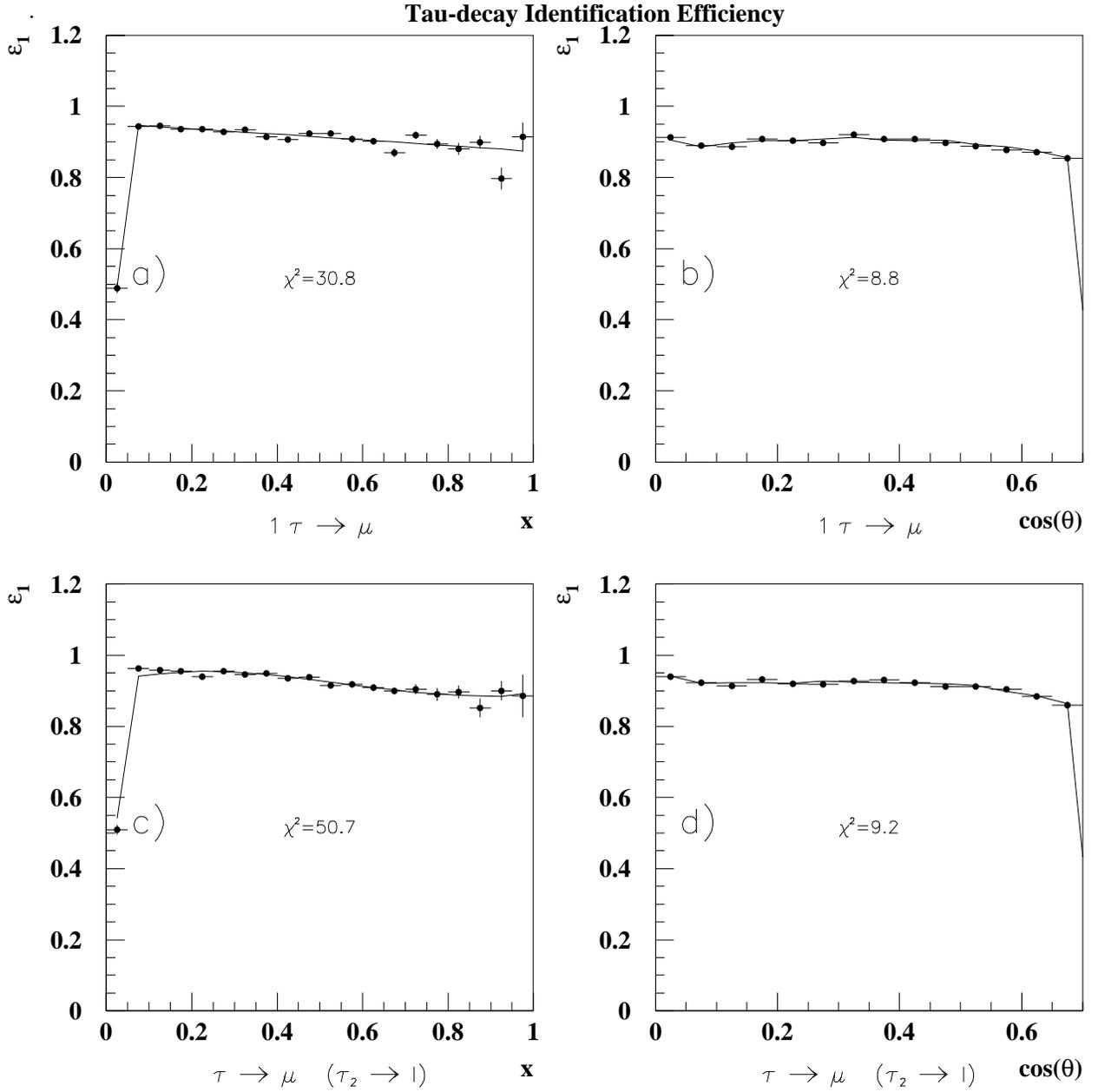}
\caption[Efficiencies of $\tmu$ identification (I).]
{Efficiencies of $\tmu$ identification as function of
$x$ and $\act$. In (a,b) the decay mode of  the second hemisphere is not
identified, where in (c,d) the second $\tn$ decays into a lepton.}
\label{efftd2}
\end{figure}

\begin{figure}[p]
\epsfysize=18.5cm.
\epsffile[40 100 720 650]{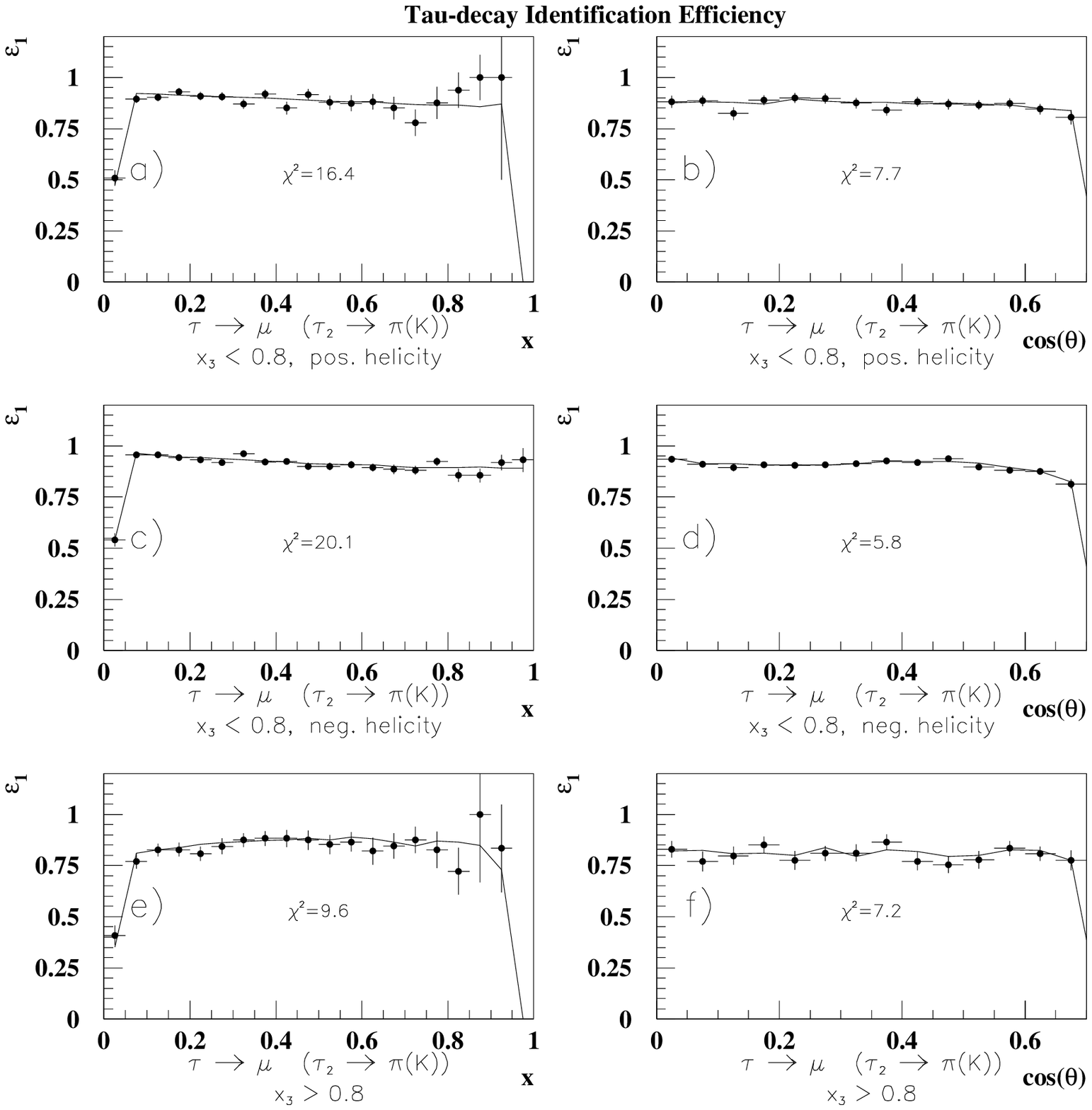}
\caption[Efficiencies of $\tmu$ identification (II).]
{Efficiencies of $\tmu$ identification as function of
$x$ and $\act$ for different cases (see also text).}
\label{efftd3}
\end{figure}

The efficiency plots for $\tel$, $\tmu$ and $\tpiK$ are given
in Figs.~\ref{efftd1}-\ref{efftd3} along with their parametrization
curves.
To allow  a possible correlation between the efficiency
dependencies on $x$ and $\cst$ we use in the analysis a two-dimensional
distribution and fit it to an expression
which depends on both. The list of $\cof$ values of all the efficiency fits
are given in Table~\ref{tab-efftd}.
 
\begin {table} [htb]
\begin {center}
\begin {tabular}{||l|c||} \hline \hline
Decay channel combination &$ $\cof$$ \\
\hline
$\tel$         & 17.7/27 \\
$\tpiK$        & 26.5/28 \\
$\tmu$, none   & 41.0/292 \\
$\tmu$, $\tle$ & 14.5/24 \\
$\tmu$, $\tpiK$, $x<0.8$, pos. helicity & 19.9/30 \\
$\tmu$, $\tpiK$, $x<0.8$, neg. helicity & 28.6/28 \\
$\tmu$, $\tpiK$, $x>0.8$               & 25.5/28 \\
\hline \hline
\end {tabular}
\caption[$\cof$ of the  $\t$ decay  identification fits.]
{$\cof$ of the parametrization fits for the $\t$ decay identification.}
\label{tab-efftd}
\end{center}
\end{table}

{\large{\bf Control samples}}
 
A shortcoming of efficiency evaluation based on MC distributions
alone comes mainly from possible effects not modelled correctly
in the detector simulation.
Therefore, events from control
samples of the data are used  to correct the
MC derived $\t$ decay identification efficiency.
 The
selection procedures of these control samples are largely
independent of the $\tn$ decay identification requirements. The efficiencies
obtained from those real data events are compared with those from MC
simulating the same process and the same selection criteria.
In this way, any bias due to the selection of the control sample
itself is minimized.
 
\begin{figure}[p]
\epsfysize=12cm.
\epsffile[60 150 630 610]{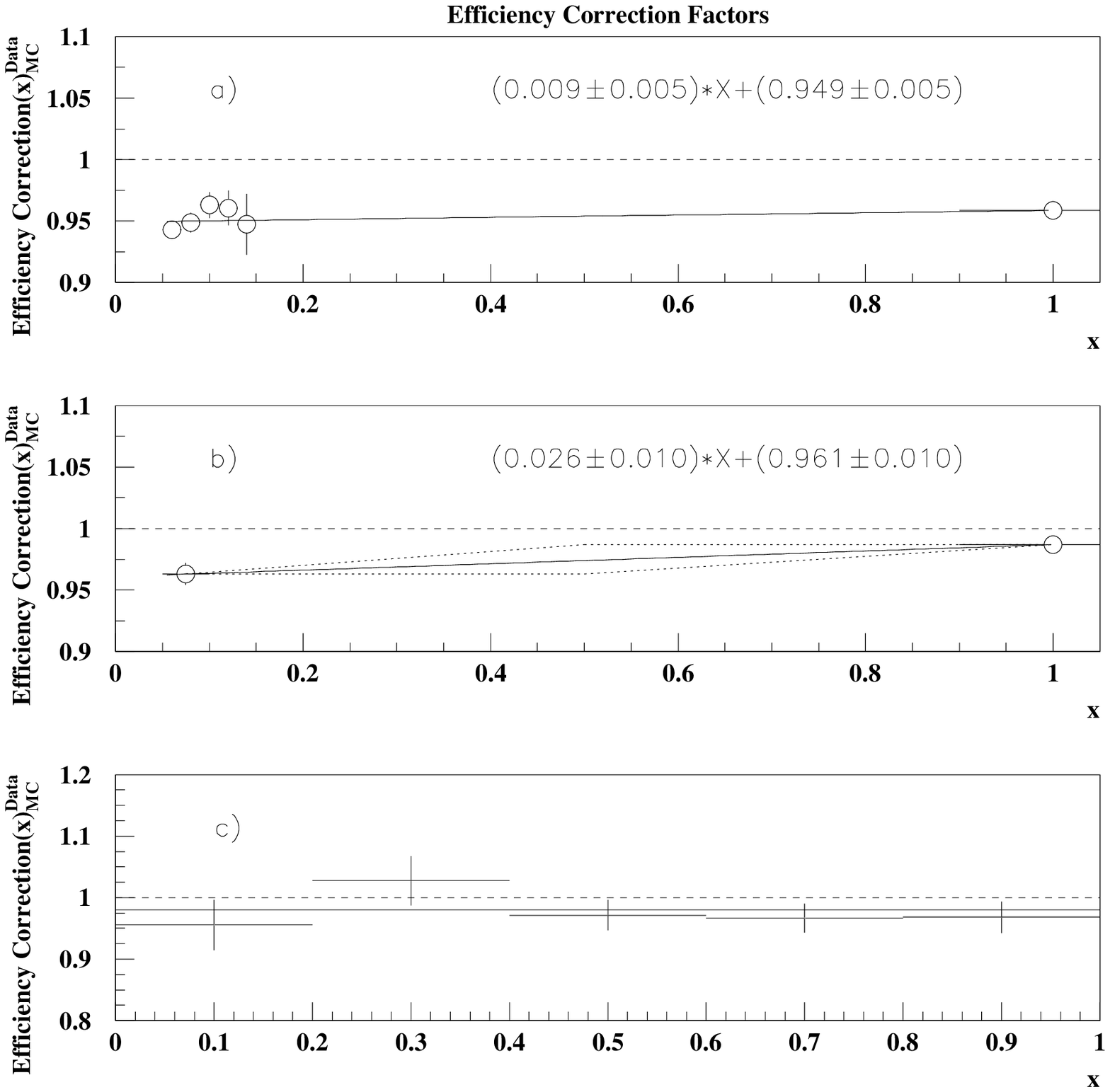}
\caption[Data corrections to the efficiencies for the tau identification]
{Corrections to the MC derived efficiencies for the tau identification
using data control samples: a) for $\tel$ decay, using single electron
and Bhabha events; b) for $\tmu$ decay, using $\ee\ra\ee\mumu$
and $\mu$-pairs; c) for $\tpiK$ decay, using $\tro$ events (see text).
The solid lines are our parametrization. The dotted lines
in b) are alternative parametrization used to estimate the
systematic uncertainties.}
\label{conteff}
\end{figure}

For $\tel$ decay,  Bhabha and single electron events are used.
Fig.~\ref{conteff}a shows the ratio between the efficiencies of
the data and MC control samples, where the low energy points are
from the single electron events and the high energy point is
from the Bhabha events. The straight line fitted to these points
parametrizes the correction which must be multiplied with the
$\tel$ MC derived efficiency.

For Bhabha events, namely for high energy electrons, the efficiency
calculated from the data is lower by 4.1\% than the efficiency calculated
with the MC Bhabha events.
For single
electron events, which are low energy electrons, the data
efficiency is lower by 5\% than the MC derived efficiency,
leading to an energy dependent
correction of $(0.949\pm 0.005)+(0.009\pm 0.005)\xp$.
 
In order to investigate the muon detection efficiency we used $\mu$-pair
events for high energy and \ggmm ~events for low energy muons.
The resulting  correction for effects not modelled in the MC,
shown in Fig.~\ref{conteff}b, is
$(0.961\pm 0.010)+(0.026\pm 0.010)\xp$.
The uncertainty of these corrections, for both electrons and muons,
is increased beyond the quoted errors of the fit parameters in order to
allow for possible non-linear interpolations between the extreme
$x$ regimes of the control samples.
 
For single charged pions or kaons there are no clean non-tau
control samples available over the momentum range of interest. Instead,
decays of the tau  into hadronic states containing neutral pions and
a single charged hadron
are used. For example, the $\pi$(K) identification requirements
that remove muons (see P4)
can be controlled using the $\tro$ sample
since these requirements are not used in the ~\tro ~identification.
The resulting correction for these
requirements is approximately uniformed equal to 0.98, but can also be
parametrized as a second order polynomial.
This ratio between data and MC is shown in Fig.~\ref{conteff}c.
We use the flat
correction function, but investigate the effect of using the
other alternative for the systematic study.
We investigated this correction also separately
for positive and negative $\cst$ values and did not find any
statistically significant difference.
 
In order to study the requirements designed to remove background
containing neutral pions (see P3) an alternative pion
selection  based on a low jet mass requirement is used. This sample
has larger background (12\%), but is clean enough to estimate the
systematic dependence of the results on the requirements on the
presampler signal and unassociated ECAL clusters. The conclusion
of this study is that the MC modelling of the
$x$ dependence of the efficiency is in excellent
agreement with the data.  This is verified using a  subsample
of the \tro ~events in which the neutral pion is well separated
from the track and with a sample of very high momenta \tpiK ~events
which had  little background from the other hadronic decays.
The statistical error obtained from this control sample investigation
is taken to be a systematic error.
The \tro ~events are also used to demonstrate that the
requirement that removes electrons (see P2),
introduces a negligible contribution to the overall systematic error.
 
To conclude, three different sources of systematic errors associated with the
decay identification efficiencies were considered.
\begin{itemize}
\item The MC statistics, i.e. the statistical errors of the parameters
obtained by the  fits to the MC efficiency
distribution.
\item Control sample statistics, i.e. the statistical errors is
the parameters describing the corrections to the MC efficiencies
obtained from the control samples. For the $\tpiK$ channel we made
an alternative fit to a non-flat line, $a+bx$, and we used the
deviation of the $b$ parameter from 0 and its statistical error.
\item Uncertainty in the interpolation of the corrections
obtained from the control samples.
This uncertainty refers only to the $\tel$ and $\tmu$ channels
where the control samples cover only very low and very high lepton
energies. In order to estimate the effect of this uncertainty,
we applied two alternative
interpolations, in addition to the straight line between the two
regions. These interpolations are described by the dotted
lines in Figs.~\ref{conteff}a,b.
\end{itemize}
 
The effect of each uncertainty was propagated to the final $\pta$
and $\aplfb$ results using the same method as used in the previous
sections, and the results are summarized
in Table~\ref{eftdsys}. For each decay channel
we list separately the effect on the $\pta$ and on the $\aplfb$
results of the whole sample, and on the results of a subsample
containing the individual decay modes.
 
\begin {table} [htb]
\begin {center}
\begin {tabular}{||l|c||cc||cc||} \hline \hline
  & &  \multicolumn{2}{c||}{whole sample} &
 \multicolumn{2}{c||}{corr. subsample} \\
source & \tn ~decay  & $\dpta$ & $\daplfb$ & $\dpta$ & $\daplfb$ \\
\hline
MC statistics       & $\tel$  & 0.52 & 0.14 & 2.36 & 0.64 \\
                    & $\tmu$  & 0.20 & 0.08 & 0.85 & 0.34 \\
                    & $\tpiK$ & 0.77 & 0.07 & 1.42 & 0.13 \\
                    & all     & 0.95 & 0.18 &      &      \\
\hline
Control sample      & $\tel$  & 0.15 & 0.00 & 0.68 & 0.01 \\
                    & $\tmu$  & 0.24 & 0.01 & 1.02 & 0.04 \\
                    & $\tpiK$ & 1.33 & 0.06 & 2.47 & 0.10 \\
                    & all     & 1.51 & 0.06 &      &      \\
\hline
Interpolation       & $\tel$  & 0.04 & 0.00 & 0.18 & 0.01 \\
                    & $\tmu$  & 0.11 & 0.00 & 0.61 & 0.04 \\
                    & all     & 0.11 & 0.01 &      &      \\
\hline \hline
All sources         & $\tel$  & 0.52 & 0.14 & 2.46 & 0.64 \\
                    & $\tmu$  & 0.38 & 0.08 & 1.46 & 0.34 \\
                    & $\tpiK$ & 1.54 & 0.08 & 2.85 & 0.16 \\
\hline
                    & Total     & 1.79 & 0.19 &      &      \\
\hline \hline
\end {tabular}
\caption[Systematic errors associated with $\tn$ decay
identification]
{Systematic errors (in \%) associated with tau decay
identification efficiencies.}
\label{eftdsys}
 
\end{center}
\end{table}
 
These various efficiency corrections contribute systematic
errors on $\pta$ and $\aplfb$ of 1.79\% and 0.19\%, respectively.

\section{Background}
\label{sect-bckg}
As seen from Eqs.~\ref{corcst} and~\ref{corcso}, the correction for
background
contamination in channel $i$
is performed separately for background from other
$\tn$ decay sources ($\beta^\pm_i$) and background from non-$\tn$ sources
($\beta^{non-\tn}_{i(j)}$).
There are two reasons for this separation of background sources.
\begin{itemize}
\item The background from other $\t$ decay channels depends on the
$\t$ helicity of the event and therefore it must be calculated
for each helicity separately.
\item The background from non-$\t$ events depends on the whole
event, whereas the background from other $\t$ decays depends only
on the $\t$-jet considered.
\end{itemize}
For these reasons we had to handle those two background types
differently (see Eqs.~\ref{corcst},~\ref{corcso}). In Section~
\ref{BOD}
we describe only the first background  and in the next
Section we shall present the second one.
 
\subsection{Background from Other Tau Decay Channels}
\label{BOD}
 
The background from other tau decays was initially investigated using
our $\t$-pair MC sample and separating between the two $\tm$ helicity
states and between five decay channels, $\tel$, $\tmu$, $\tpi$,
$\thp$ and combining all other channels together. The separation between
the sources was done in order to be able to correct for the MC
branching fractions of the different decay channels and to investigate
systematics (see below). We considered separately the dependence
of the background on the kinematic variables,
$\beti$, and its overall normalization, $\bbeti=\int d\xp_i d\cst \beti$.
 
To obtain the overall normalization, we calculate
the background fraction to channel $i$ from source $j$ by,
\beq
\label{bij}
\bbetij=\frac{n^{MC}_{ij}}{n^{MC}_{ii}}\frac{c_j}{c_i},
\eeq
where $n^{MC}_{ij}$ is the number of MC events from source $j$,
identified as coming from channel $i$. The correction factor
$c_i$ is the ratio
between the branching fractions to channel $i$ in data and MC.
For the branching fractions in the data of the $\tel$, $\tmu$
$\tpiK$ and $\thp$ channels we are using the published
OPAL results~\cite{OPALPL},~\cite{bib-OPALhpizbr}.
No correction was applied to the background
from the fifth source (all other channels). Table~\ref{brs} lists
the MC and data branching fractions.
 
\begin {table} [htb]
\begin{center}
\begin{tabular}{||l|c|c|c|c||}  \hline  \hline
                      & $\tel$ & $\tmu$ & $\tpiK$ & $\thp$  \\ \hline
 $B^{MC}$             & 18.25  & 17.76  & 12.59  & 24.77   \\
 $B^{data}$           & 17.62$\pm$0.47 & 16.80$\pm$0.41 &
                        12.22$\pm$0.59 & 26.27$\pm$0.61   \\ \hline
 $c=B^{data}/B^{MC}$  & 0.965$\pm$0.026 & 0.946$\pm$0.023 &
                        0.971$\pm$0.047 & 1.061$\pm$0.025 \\ \hline \hline
\end{tabular}
\caption[Tau branching fractions.]
{Tau branching fractions (in \%) used in our analysis.}
\label{brs}
\end{center}
\end{table}
 
The overall background fraction for channel $i$ is obtained by
summing up the contributions from the five sources,
$\bbeti=\sum_{j=1}^5 \bbetij$. Table~\ref{brfr} lists the
various contributions to the background fractions and their sums.
 
\begin {table} [htb]
\begin{center}
\begin{tabular}{||l||c|c|c||c|c|c||}  \hline  \hline
$\tm$-hel.     & \multicolumn{3}{c||}{positive}
                   & \multicolumn{3}{c||}{negative} \\ \hline
$\t$ decay & $\tel$ & $\tmu$ & $\tpi$
                   & $\tel$ & $\tmu$ & $\tpi$ \\ \hline
$\tel$ &      & 0.00 & 0.92 &      & 0.01 & 0.92 \\
$\tmu$ & 0.03 &      & 0.26 & 0.02 &      & 0.27 \\
$\tpiK$ & 2.34 & 1.01 &      & 2.68 & 1.76 &      \\
$\thp$ & 1.52 & 0.03 & 5.55 & 1.83 & 0.03 & 3.68 \\
other  & 0.19 & 0.02 & 1.57 & 0.17 & 0.01 & 1.96 \\ \hline
overall& 4.08 & 1.06 & 8.30 & 4.70 & 1.81 & 6.83 \\ \hline\hline
\end{tabular}
\caption[Background fractions from the various sources.]
{Background fractions (in \%) to the three decay channels
$\tel$, $\tmu$, $\tpiK$ (columns) from the various sources (rows).}
\label{brfr}
\end{center}
\end{table}
 
\begin{figure}[p]
\epsfysize=15cm.
\epsffile[80 100 720 650 ]{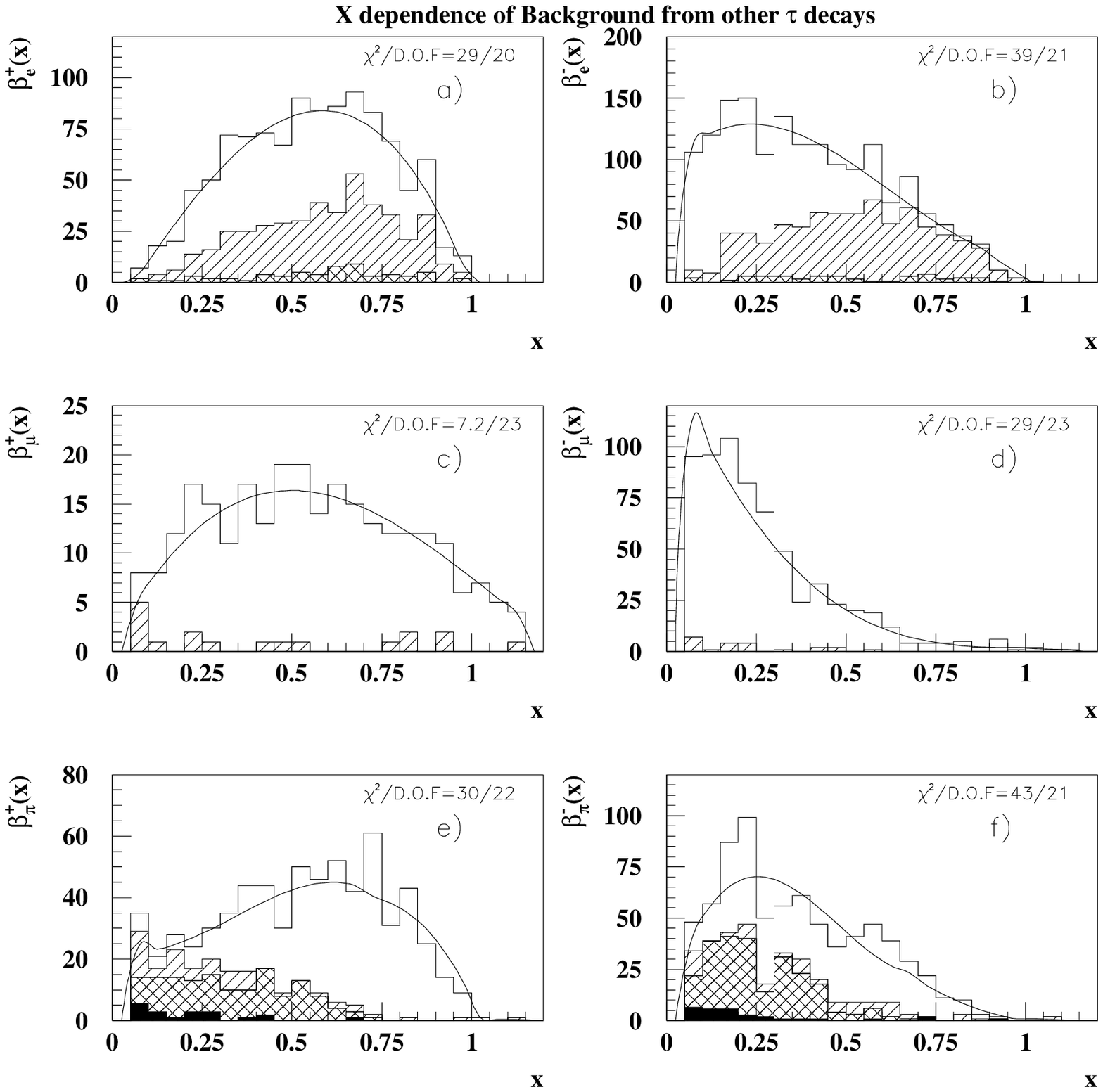}
\caption[Background from other tau decays as functions of $\xp$.]
{Background from other tau decays as functions of $\xp$
to the channel $\tel$ (Figs. a,b), $\tmu$
(Figs. c,d) and $\tpiK$ (Figs. e,f). Figs. a,c,e (b,d,f)
correspond to positive (negative) helicity events.
In Figs. a,b contamination contributions from $\tpiK$, $\tro$ and other $\t$
decay channels (including $\tmu$) are presented by open, hatched and
cross-hatched histograms correspondingly. In Figs. c,d contamination
contributions from $\tpiK$ and all other $\t$-decay
channels are shown by open and hatched histograms correspondingly.
In Figs. e,f contamination contributions from $\tro$, $\tel$, $\tmu$ and
other $\t$ decays are drawn with open, hatched, black and cross-hatched
histograms correspondingly. The curves are
our parametrization. The vertical scale is in arbitrary units.}
\label{betx}
\end{figure}
 
\begin{figure}[p]
\epsfysize=15cm.
\epsffile[60 100 720 650]{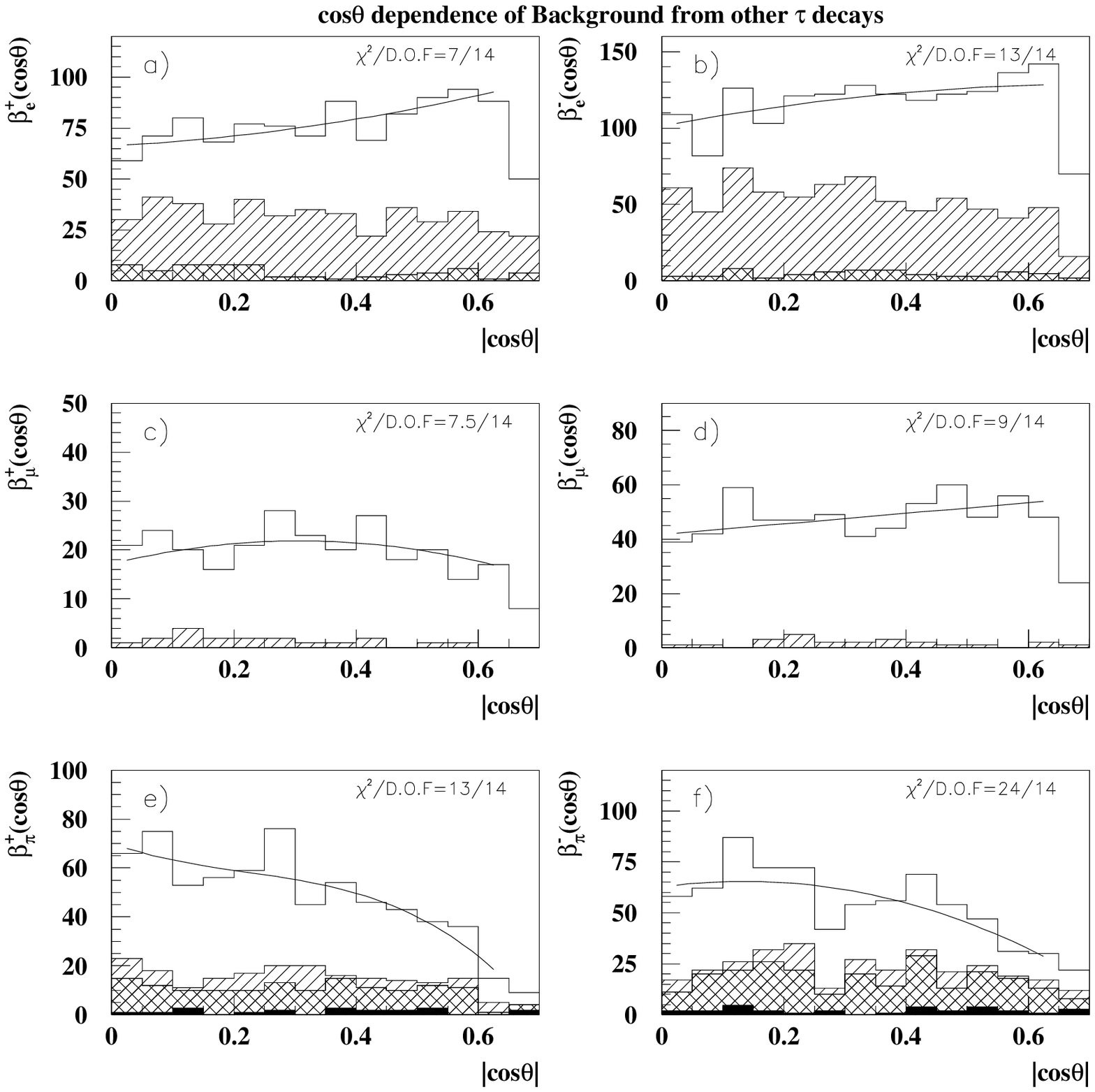}
\caption[Background from other tau decays as functions of $\act$.]
{Background from other tau decays as functions of $\act$.
to the channel $\tel$ (Figs. a,b), $\tmu$
(Figs. c,d) and $\tpiK$ (Figs. e,f). Figs. a,c,e (b,d,f)
correspond to positive (negative) helicity events.
In Figs. a,b contamination contributions from $\tpiK$, $\tro$ and other $\t$
decay channels (including $\tmu$) are presented by open, hatched and
cross-hatched histograms correspondingly. In Figs. c,d contamination
contributions from $\tpiK$ and all other $\t$-decay
channels are shown by open and hatched histograms correspondingly.
In Figs. e,f contamination contributions from $\tro$, $\tel$, $\tmu$ and
other $\t$ decays are drawn with open, hatched, black and cross-hatched
histograms correspondingly. The curves are
our parametrization. The vertical scale is in arbitrary units.}
\label{betc}
\end{figure}
 
The dependence on the kinematic variables of the background fractions
was also taken from the MC events. The $\xp$ and $\cst$ MC background
distributions for each helicity and decay channel, and from each
source were fitted to third order polynomials. The normalization
of each background source was then corrected by multiplying with
the corresponding branching fraction correction $c$.
The contributions from all sources were then summed up and normalized
to $\bbeti$. The resulting
functions, $\beti$, are plotted in Figs.~\ref{betx} and~\ref{betc},
together with the corrected MC distributions. The $\chs$ values for the
agreement between the MC distributions and their polynomial
parametrization are indicated on the plots.
 
For $\tel$, the background level from other $\tn$ decays
is $\approx4.7\%$, differing slightly between positive and negative
$\tm$ helicity events, and originating mainly from $\tpiK$ and
$\tro$ decays. For $\tmu$ this background is $\approx1.5\%$ coming
from $\tpiK$ decays, and the background to $\tpiK$ is $\approx7.8\%$,
mainly from $\tro$, $\tn\ra\pi K^0\nu$ and $\tel$ decays. The MC
expectations for these background levels are
checked against the data by studying the tau jets that are identified
in more than one decay channel and from studies of
contaminations with
Bhabha and muon-pair control samples.
 
The following contributions to the systematic error associated with
the background from tau sources were considered.
\begin{itemize}
\item The uncertainties in the parameters of the background functions,
originating from the  statistics of our MC sample.
An error on the $\tn\ra\pi K^0\nu$ background is estimated by
varying the MC expectation by $\pm 100\%$.
\item
A small ($<$0.1\%) additional contribution to the errors
arising from the experimental
uncertainties in the tau branching ratios has been
included to correct the MC numbers.
\item Possible disagreement between the data and MC simulation.
This was investigated in two ways.
\begin{enumerate}
\item We used our Bhabha and $\mu$-pair control samples and counted
the number of particles identified as pions. We compared these
numbers for data and MC, and the ratios were 0.72$\pm$0.48 for
electrons and 1.17$\pm$0.39 for pions. These ratios represent
correction factors to be multiplied with the MC estimates
of the background to $\tpiK$ from $\tel$ and $\tmu$ decays.
\item We counted tau decays which were identified by two decay
algorithms and compare it to MC. Almost all those identified as a pion
and electron or as a pion and muon are pions (from MC). Therefore we can get
a correction of the pions background  to the electrons and muons.
These corrections turn out to be 0.41$\pm$0.09 and 0.58$\pm$0.08 respectively.
50\% of those identified as a rho and pion or electron or muon are
$\rho$'s. Therefore we can get a correction to the background from rho
to electron, muon
and pion which are 1.30$\pm$0.10, 5.0$\pm$1.0 and 1.09$\pm$0.09.
\end{enumerate}
 
These corrections are applied in the systematic studies.
We use either their uncertainties or their
difference from the factor one (whichever greater) and see the corresponding
effects on the final results.
\end{itemize}
 
Table~\ref{tbacksys} summarizes the systematic errors associated
with the background from other tau decays.
The contribution of the
branching fraction uncertainties actually
affect not only the background estimates but
are used also in the expression for
$\epstpi$ (Eq.~\ref{epsp}) and in the estimate of the $\pi/K$ mass
threshold effect.
 
\begin {table} [htb]
\begin {center}
\begin {tabular}{||l||cc||cc|cc|cc||} \hline \hline
  & \multicolumn{2}{c||}{all} & \multicolumn{2}{c|}{$\tel$} &
\multicolumn{2}{c|}{$\tmu$} & \multicolumn{2}{c||}{$\tpiK$} \\
source & $\dpta$ & $\daplfb$ & $\dpta$ & $\daplfb$ & $\dpta$ & $\daplfb$ &
$\dpta$ & $\daplfb$  \\ \hline
MC  & 0.29 & 0.04 & 0.50 & 0.09 & 0.44 & 0.06 & 0.45 & 0.05 \\
Data  & 0.58 & 0.09 & 1.49 & 0.06 & 0.25 & 0.16 & 0.10 & 0.01 \\
 \hline
Diff  & 0.32 & 0.02 & 0.15 & 0.08 & 0.22 & 0.01 & 0.49 & 0.02 \\
 \hline \hline
Total        & 0.72 & 0.10 & 1.58 & 0.13 & 0.55 & 0.17 & 0.67 & 0.05 \\
 \hline
\end {tabular}
\caption[Systematic errors of background from
other tau decays]
{Systematic errors (in \%) associated with background from
other tau decays and tau branching fractions.
MC, Data, and BR stand for MC statistics, the difference between
data and MC and Branching ratio fraction respectively.}
\label{tbacksys}
\end{center}
\end{table}
 
These background sources
contribute errors of 0.72\% and 0.10\% on ~\pta  ~and ~\aplfb ,
respectively.
\subsection{Background from non-tau Events}
\label{BNT}
Concerning the residual non-$\tn$ background after the selection
criteria, the following sources are considered.
\begin{itemize}
\item $\ee\ra\mumu$. This source mainly contaminates  events where
one $\tn$ is identified as $\tmu$ decay
and the other as $\tpiK$ or when one tau
is identified as a  $\tmu$ and the other is not identified, at the
level of $0.4\%$ and $1.3\%$, respectively.
\item $\ee\ra\ee\mumu$. This source contributes only to events
with two identified muons or one identified muon and an unidentified tau,
 at the level of $1.6\%$ and $0.3\%$ respectively.
\item $\ee\ra\ee$. This source contributes background to events where only one
$\tn$ decay is identified as e or $\pi$(K) and the other tau decay is
unidentified, and events where both
decays are identified as $\tel$. The corresponding contamination
levels are $0.3\%$, $0.2\%$ and $0.3\%$, respectively.
\item $\ee\ra\ee\ee$. This source contaminates only events with one
electron where the opposite tau decay is unidentified or two identified
electrons, at the level of $0.3\%$ and $2.6\%$,
respectively.
\item $\ee\ra q\bar{q}$. The background from this source is
negligible for all channels considered here.
\end{itemize}
The $\beta^{non-\tn}_{i(j)}(\xip,\xjp,\cst)$ correction functions for
each source were initially determined using  the corresponding
MC event samples. In practice
the correlations between \xp$_{i(j)}$ ~and \cst ~are small and the
$\beta^{non-\tn}_{i(j)}$ functions factorize into products
of simple functions.  These were adjusted  by factors
obtained from comparisons between the data and MC
when the tau selection requirements designed to suppress these
backgrounds were loosened.
 
Table~\ref{ntb} lists
the five sources which were considered where for each source
the size of the MC sample is quoted.
These numbers were
estimated from the integrated luminosity collected by OPAL during
1990-1992. In addition, we list the expected background fraction
for each combination of decay channels.
 
\begin {table} [htb]
\begin {center}
\begin {tabular}{||l||c|c|c|c|c||c||} \hline \hline
source  & $ee\ra\mu\mu$ & $ee\ra q\bar{q}$ & $ee\ra ee$ &
          $ee\ra ee\mu\mu$ & $ee\ra eeee$ & total \\ \hline
MC    & 200000 & 702400 & 46187$^\ast$ & 50.23 pb$^{-1}$ &
                   42.98 pb$^{-1}$ &  \\ \hline
$e$, none     & 0.05$\pm$0.02 & 0.00          & 0.31$\pm$0.12 &
                   0.00          & 0.25$\pm$0.08 & 0.61$\pm$0.15 \\
$\mu$, none     & 1.29$\pm$0.09 & 0.00          & 0.00          &
                   0.26$\pm$0.08 & 0.00          & 1.55$\pm$0.12 \\
$\pi$, none     & 0.00          & 0.00          & 0.18$\pm$0.13 &
                   0.00          & 0.00          & 0.18$\pm$0.13 \\
$e$, $e$   & 0.00          & 0.00          & 0.28$\pm$0.28 &
                   0.00          & 2.60$\pm$0.65 & 2.88$\pm$0.71 \\
$e$, $\mu$   & 0.14$\pm$0.05 & 0.00          & 0.00          &
                   0.00          & 0.00          & 0.14$\pm$0.05 \\
$\mu$, $\mu$   & 0.05$\pm$0.02 & 0.00          & 0.00          &
                   1.60$\pm$0.40 & 0.00          & 1.65$\pm$0.40 \\
$\mu$, $\pi$   & 0.40$\pm$0.16 & 0.00          & 0.00          &
                   0.00          & 0.00          & 0.40$\pm$0.16 \\
\hline \hline
\multicolumn{7}{l} { } \\
\multicolumn{7}{l} {$^\ast$ numbers correspond to $\act<0.9$} \\
\end {tabular}
\caption[Non-tau background sources.]
{Non-tau background sources and their relative
contributions (in \%) to the various combinations of $\t$ decay channels.
The combinations not mentioned in this table have no background.}
\label{ntb}
\end{center}
\end{table}
 
\begin{figure}[p]
\epsfysize=17cm.
\epsffile[60 100 650 700]{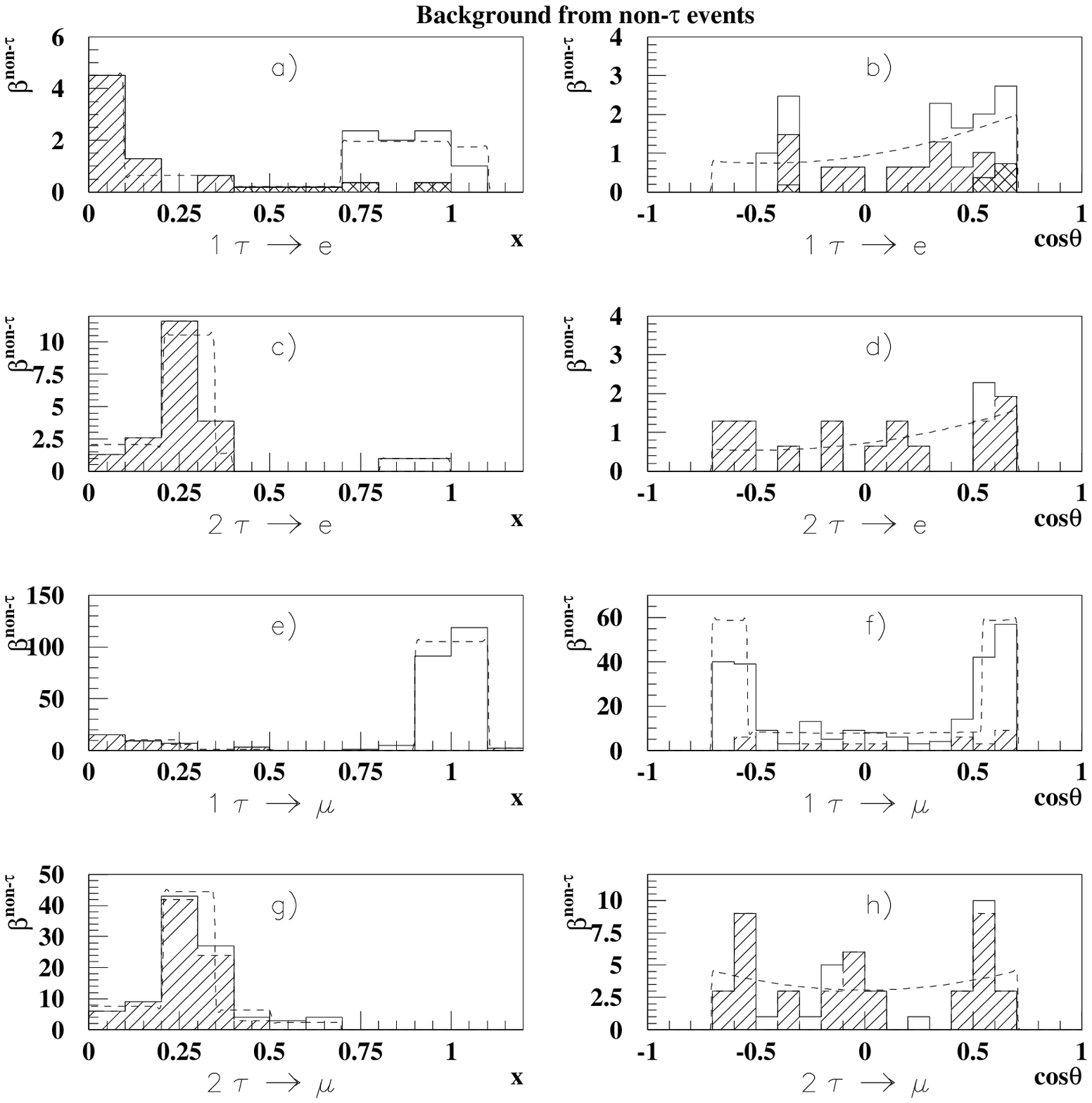}
\caption[Background from non-tau sources.]
{Background from non-tau sources to events with one
identified $\tel$ and the other side not identified (a,b);
events with two identified $\tel$ (c,d);
events with one identified $\tmu$ and the other side not identified (e,f);
events with two identified $\tmu$ (g,h).
Figs. a,c,e,g (b,d,f,h) show the $\xp$ ($\cst$) dependence.
In Figs. a-d, the open, hatched and cross-hatched histograms represent the
background from Bhabha, $ee\ra eeee$ and $\mu$-pairs respectively.
In Figs. e-h, the open and hatched histograms are the background from
$\mu$-pairs and $ee\ra ee\mu\mu$ respectively. The dashed lines are
our parametrization. The vertical scale is in arbitrary units.}
\label{fig_ntb}
\end{figure}

The $\xp$ and $\cst$ dependence was taken from the MC distributions
and parametrized by simple functions. Fig.~\ref{fig_ntb} shows these
distributions for those decay channel combinations where
the overall non-tau background is above 0.5\%.
 
Two sources of systematic errors associated with the non-tau background
were considered.
\begin{itemize}
\item Finite statistics of our MC event samples, illustrated by the
given errors in Table~\ref{ntb}. To
estimate their effect, we varied both the background levels and the
parametrization of their $\xp$ ans $\cst$ dependence within the
statistical errors and examined the effect on the final results.
\item Possible incompatibility between MC and data. This was
investigated by J. Clayton~\cite{Clayton} using background enriched
subsample where he obtained correction factors for all the
non-tau background sources. As we did for  the
tau background, we apply these correction factors only for the
systematic study. We use either their uncertainties or their
difference from the value one (whichever greater) and
observe the corresponding effects on the final results.
\end{itemize}
 
Table~\ref{ntbacksys} summarized the systematic errors associated
with the non-tau background.
 
\begin {table} [htb]
\begin {center}
\begin {tabular}{||l||cc||cc|cc|cc||} \hline \hline
  & \multicolumn{2}{c||}{all} & \multicolumn{2}{c|}{$\tel$} &
\multicolumn{2}{c|}{$\tmu$} & \multicolumn{2}{c||}{$\tpiK$} \\
source  & $\dpta$ & $\daplfb$ & $\dpta$ & $\daplfb$ & $\dpta$ & $\daplfb$ &
$\dpta$ & $\daplfb$  \\ \hline
MC      & 0.48 & 0.16 & 1.56 & 0.60 & 1.20 & 0.17 & 0.33 & 0.15 \\
Diff    & 0.48 & 0.08 & 0.72 & 0.26 & 1.97 & 0.26 & 0.07 & 0.03 \\
 \hline
Total & 0.68 & 0.18 & 1.72 & 0.65 & 2.31 & 0.31 & 0.34 & 0.15 \\
 \hline \hline
\end {tabular}
\caption[Systematic errors associated with the non-tau
background.]{Systematic errors (in \%) associated with the non-tau
background. MC and Data stand for the MC statistics and the difference between
data and MC.}
\label{ntbacksys}
\end{center}
\end{table}

The contribution to the
errors on \pta ~and \aplfb ~from the non-tau background corrections
are 0.68\% and 0.18\%, respectively.

\chapter{Results and Cross Checks}
\label{chap-RCC}
\section{Maximum Likelihood Fit Results}
The ML fit, applied to the $\tel$, $\tmu$ and
$\tpiK$ decay channels yields the following results,
\bea
\label{mlres}
\pta  & = & (\PTEMPF \pm \PTEMPST (stat))\% \nonumber \\
\aplfb& = & (\APFBEMPF \pm \APFBEMPST (stat))\%
\eea
where the correlation between the two numbers is 0.03.
 
\subsection*{The \tn~ W $\nu_{\tn}$ Coupling}
Since the $\tn$ decays via weak interaction, where parity is not conserved,
the
angular distribution of its decay products in the $\tn$ rest frame depends
strongly  both on the $\tn$ spin orientation  and the characteristics
of its charged weak decay process. The results given in the previous
paragraph were calculated using the theoretical Eq.~\ref{dsig3} which assume
a $V-A$ structure for the $\tn$ decay. However, following
Sect.~\ref{sec-wformalism} one can use the more general expression
presented in
Eq.~\ref{dsig3p}, and extract not only the $\tn$ production parameters
but also its decay parameters.
Our results of a six parameters ML fit to the two polarization asymmetries and
the Michel parameters $\rho,\delta,\xi$ and $\xih$ are:
 
\bea
\label{mlm6res}
\pta  & = & (-\MFPTAU \pm \MFDPTAU (stat))\% \nonumber \\
\aplfb& = & (-\MFAPOLFB \pm \MFDAPOLFB (stat))\% \nonumber \\
\xi   & = & \MFXI \pm \MFDXI (stat)  \\
\xih  & = & \MFXIH \pm \MFDXIH (stat) \nonumber \\
\rho  & = & \MFRHO \pm \MFDRHO (stat) \nonumber \\
\delta& = & \MFDELTA \pm \MFDDELTA (stat) \nonumber \\
\eea
which are consistent with
a  $(V-A)(V-A)$ structure (see Table~\ref{Tab-Michel}).

The existing measurements of the Michel parameters are already precise
enough
to exclude interaction different from $V-A$ in the $\ell\nulb$ vertex
($\ell$=e,$\mu$),
therefore one can use  Eqs.~\ref{helndef} to evaluate the helicity
of the $\nut$ from its relations with all last four parameters.
%
This was further checked by fixing the $l\bar{\nu_l}$ vertex structure
and allow only one free parameter to determine the characteristic of the
$\tn\bar{\nu_{\tn}}$ vertex.
The fit results are:
 
\bea
\label{mlm3res}
\pta  & = & (-\MCPTAU \pm \MCDPTAU (stat))\% \nonumber \\
\aplfb& = & (-\MCAPOLFB \pm \MCDAPOLFB (stat))\% \nonumber \\
h_{\nu_{\t}} & = & \MCXI \pm \MCDXI (stat)  \\
\eea
The measurement of the chirality parameter and
the leptonic Michel parameters are all consistent with a $h_{\nu_{\t}}=-1$
which is expected from a pure $V-A$ coupling at the $\t\bar{\nu_{\t}}$ vertex.
It is also an indication for the $\nu_{\t}$ being a left handed particle.
 
\section{Summary of Systematic Studies}
 
Table~\ref{systab} summarizes the systematic errors in the ML
analysis. Each error listed in Table~\ref{systab}
is a combined result of several related contributions.
Since are no correlations between entries on different rows of the
table, they are combined in quadrature to give the overall
error listed in the last row.
Thus, the values of the systematic errors of \pta ~and \aplfb ~using
the $\tel$, $\tmu$ and $\tpiK$ decays are measured to be
\bea
\label{mlsys}
\Delta\pta  & = &  \pm \PTEMPSY\% (sys.) \nonumber \\
\Delta\aplfb & = & \pm \APFBEMPSY\% (sys.).
\eea

\begin {table} [htb]
\begin{center}
\begin{tabular}{||l|c|c||}  \hline \hline
 Source                          & $\dpta$ & $\daplfb$ \\
                                 & (\%)    & (\%)      \\ \hline
 $\tn$ decay identification      & 1.79    & 0.19     \\
 backg. from other $\tn$ decays  & 0.72    & 0.10     \\
 radiative effects               & 0.29    & 0.03     \\
 backg. from non-$\tn$ events    & 0.68    & 0.18     \\
 $\tn$-pair selection efficiency & 0.39    & 0.20     \\
 calorimeter response            & 0.48    & 0.12    \\
 tracking response               & 0.38    & 0.21     \\
 $\afb$  and K mass threshold    & 0.10    & 0.07     \\
 $\cst$ and charge measurement   & 0.00    & 0.17     \\ \hline
 Total                           & 2.19    & 0.43      \\ \hline \hline
\end{tabular}
\caption[Summary of systematic uncertainties.]
{Summary of systematic uncertainties in the
determination of \pta ~and \aplfb ~using \tel , \tmu ~and \tpiK
~channels .}
\label{systab}
\end{center}
\end{table}
 
The largest contribution to the systematic error of
$\pta$ is derived from the uncertainties in
the efficiency of the $\tn$ decay mode identification. For the most part
these arise from
the limited statistics of the control samples, and in particular,
the $\tro$ sample used to correct the efficiency of $\tpiK$
identification.
Therefore, one can expect that the systematic errors
will decrease as more data is collected.
 
\section{Monte Carlo checks}
\subsection{MC test of polarization biases}
 
In order to check for potential biases in the fitting method,
the analysis was performed on the $\tn$-pair MC sample.
The ML fit
is performed on the MC events, removing the corrections to the
$\t$ identification efficiencies derived from the data, and using
resolution parameters extracted exclusively from MC events.
Non-tau background correction is also switched off, since our
sample contains only $\t$-pair events.
The input values for $\pta$ and $\aplfb$ are $-14.0$\% and $-10.5$\%,
respectively and the fit returns values of ($-13.9\pm$1.0)\% and
($-8.9\pm$1.3)\%.
 
\begin{figure}[htbp]
\epsfysize=16cm.
\epsffile[40 100 720 650]{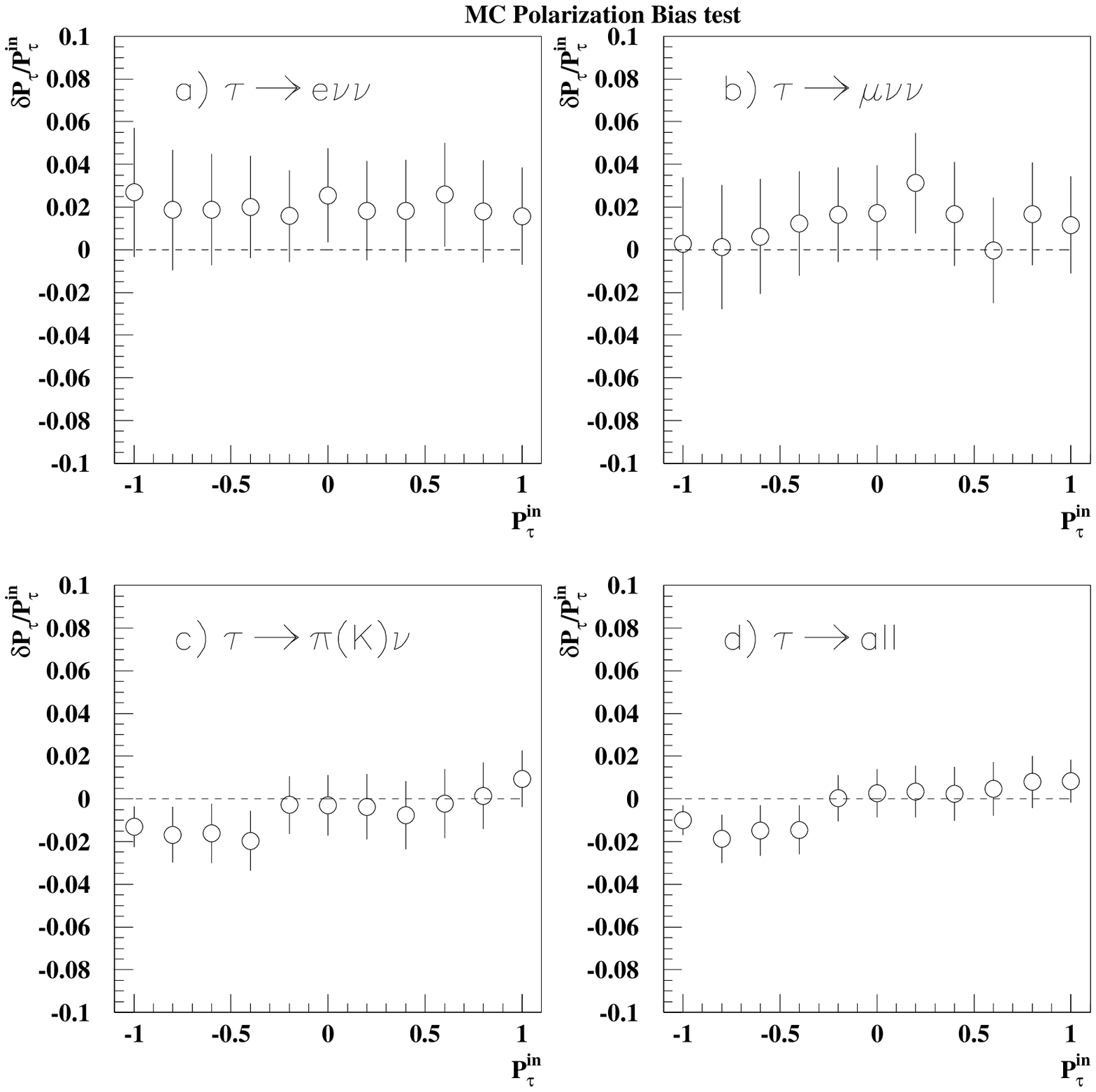}
\caption[Measurement of $\pta$ in MC with different input $\pta$.]
{Test of Bias in the $\pta$ measurement. The error in the measurement of
 $\pta$ as a function of the MC input polarization.
Fig. a) - c) are for the $\tel$, $\tmu$ and $\tpiK$ respectively and
d) present the results of all three channels together.}
\label{corrlin}
\end{figure}

The fit results are listed
in Table~\ref{mcres} which is similar to Table~\ref{table-datres} for the
data.
\begin {table} [htb]
\begin{center}
\begin{tabular}{||l|c|c|c||}  \hline \hline
               & $\pta$  & $\aplfb$ & Number of Events \\ \hline
Decay Channels & (\%)  & (\%) &  \\ \hline
All decays        & -13.9 $\pm$ 1.0 &  -8.9 $\pm$ 1.3 & 93462 \\
\hline \hline
1 $\t$ identified   & -13.9 $\pm$ 1.3 &  -8.5 $\pm$ 1.6 & 67199 \\
 \hline
{\em (a)} $\tel$, none        & -11.9 $\pm$ 3.2 &  -5.5 $\pm$ 3.6 & 27058 \\
{\em (b)} $\tmu$, none        & -15.6 $\pm$ 3.2 &  -8.1 $\pm$ 3.6 & 27120 \\
{\em (c)} $\tpiK$, none        & -13.9 $\pm$ 1.7 & -9.6 $\pm$ 2.0 & 13021 \\
\hline \hline
2 $\t$'s identified & -13.9 $\pm$ 1.7 &  -9.5 $\pm$ 2.1 & 26263 \\
\hline
{\em (d)} $\tel$, $\tel$      &  -8.9 $\pm$ 5.7 & -15.0 $\pm$ 6.6 &  4684 \\
{\em (e)} $\tel$, $\tmu$      &  -13.7 $\pm$ 3.6 &  -8.3 $\pm$ 4.2 & 10328 \\
{\em (f)} $\tel$, $\tpiK$      & -16.7 $\pm$ 3.2 &  -8.2 $\pm$ 4.0 &  3183 \\
{\em (g)} $\tmu$, $\tmu$      & -19.3 $\pm$ 9.4 &  0. $\pm$ 11. &  4289 \\
{\em (h)} $\tmu$, $\tpiK$      & -13.4 $\pm$ 3.3 &  -9.1 $\pm$ 4.0 &  3195 \\
{\em (i)} $\tpiK$, $\tpiK$      & -9.4 $\pm$ 6.3 & -19.5 $\pm$ 8.2 &   584 \\
\hline \hline
Average             & -13.9 $\pm$ 1.0 &  -8.9 $\pm$ 1.3 &       \\
$\cof$              &     3.2/8       &     4.4/8       &       \\
$\chs$ probability (\%) &     92.4        &      81.8       & \\ \hline \hline
\end{tabular}
\caption[Polarization results in MC sample.]
{$\t$ polarization results (in \%) from our MC sample. The average values
the $\cof$ and the $\chs$ probabilities correspond to the combined
results of the
9 different channels {\em (a)} to {\em (i)}. The input values
are  $\pta=-14.0$\% and $\aplfb=-10.5$\%.}
\label{mcres}
\end{center}
\end{table}
Although these values are consistent with the
input values, we have changed the input parameters
from purely negative  to purely positive $\tm$
helicity and studied the dependence of the measured asymmetries
on the input ones. Fig.~\ref{corrlin}
illustrates this investigation for the three individual channels
as well as for the whole sample.
We find no evidence for a  bias in our measurement method.
 
\subsection{Check of the various parametrization using MC}
 
\begin{figure}[p]
\epsfysize=16cm.
\epsffile[40 150 720 700]{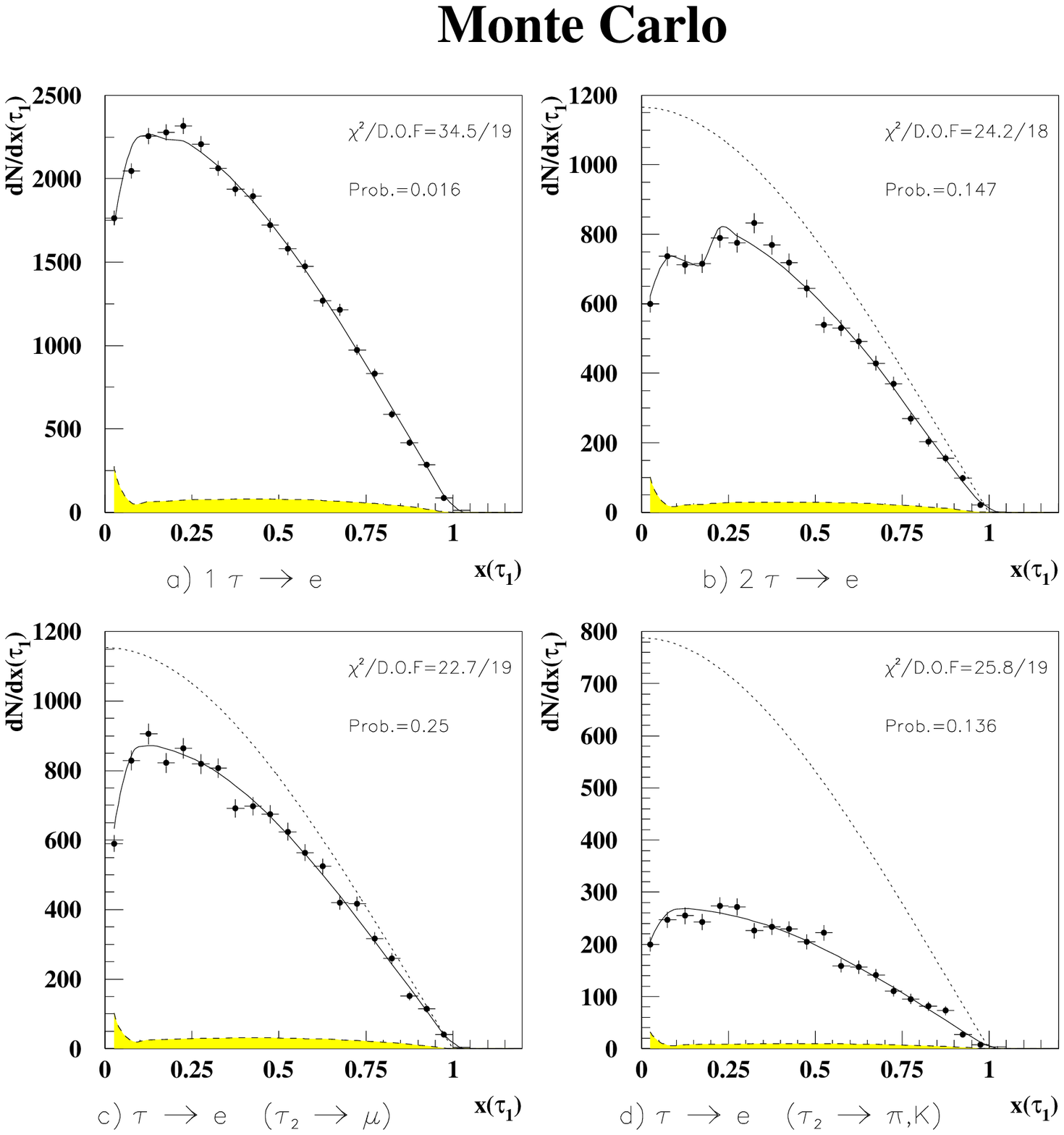}
\caption[MC $x$ distributions of the $\tel$ decay channel.]
{MC $x$ distributions for the $\tel$ decay channels and for
various decays on the opposite side,
compared with the theoretical predictions (solid lines) with all
corrections included, taking the $\pta$ value obtained from
the global ML fit. The filled area represent the background part of the
spectrum. The dotted lines are the uncorrected theoretical curves.}
\label{eresulmc}
\end{figure}

\begin{figure}[p]
\epsfysize=16cm.
\epsffile[40 150 720 700]{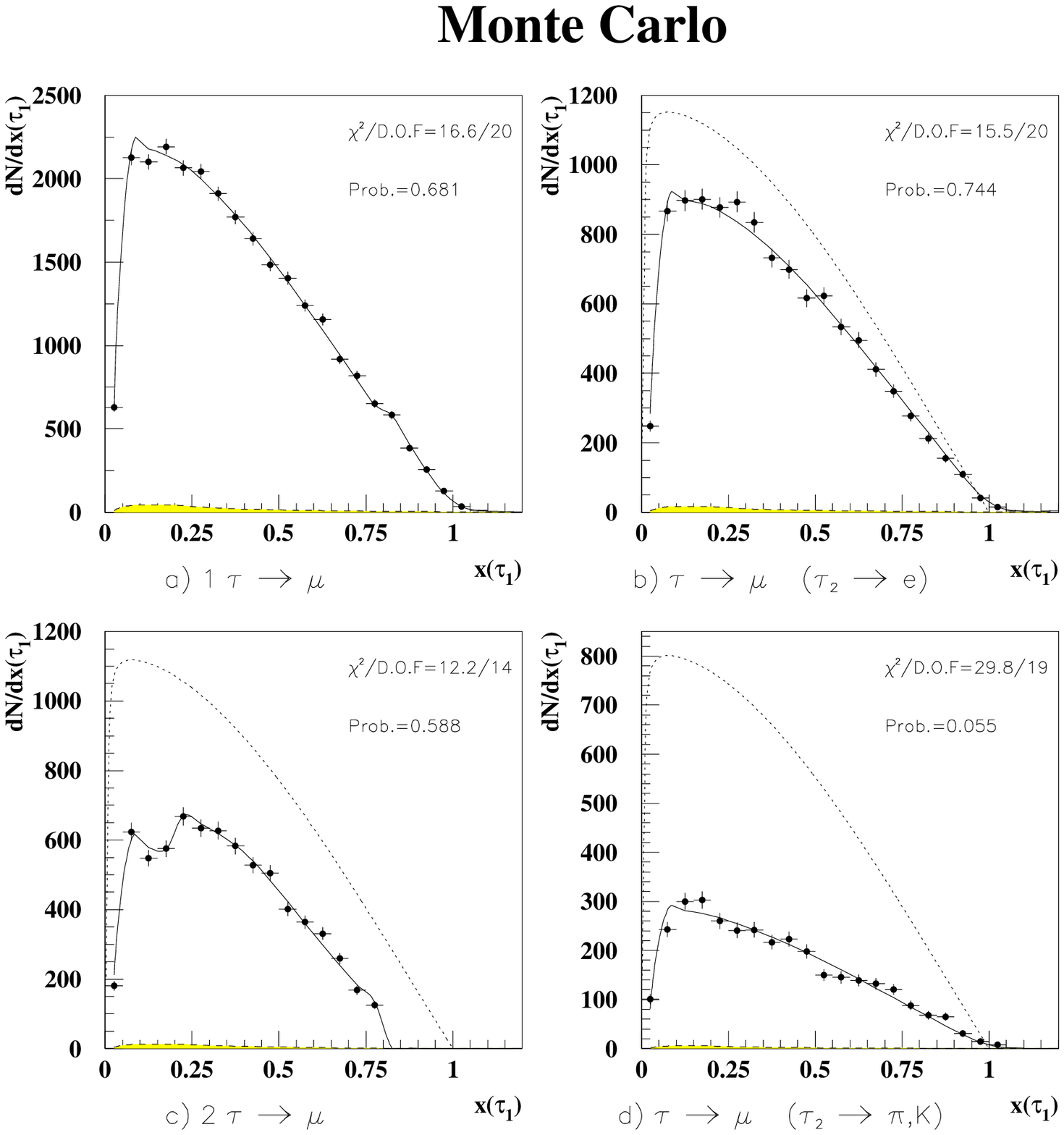}
\caption[MC $x$ distributions for the $\tmu$ decay channel.]
{MC $x$ distributions of the $\tmu$ decay channels and for
various decays on the opposite side,
compared with the theoretical predictions (solid lines) with all
corrections included, taking the $\pta$ value obtained from
the global ML fit. The filled area represent the background part of the
spectrum. The dotted lines are the uncorrected theoretical curves.}
\label{mresulmc}
\end{figure}

\begin{figure}[p]
\epsfysize=16cm.
\epsffile[40 150 720 700]{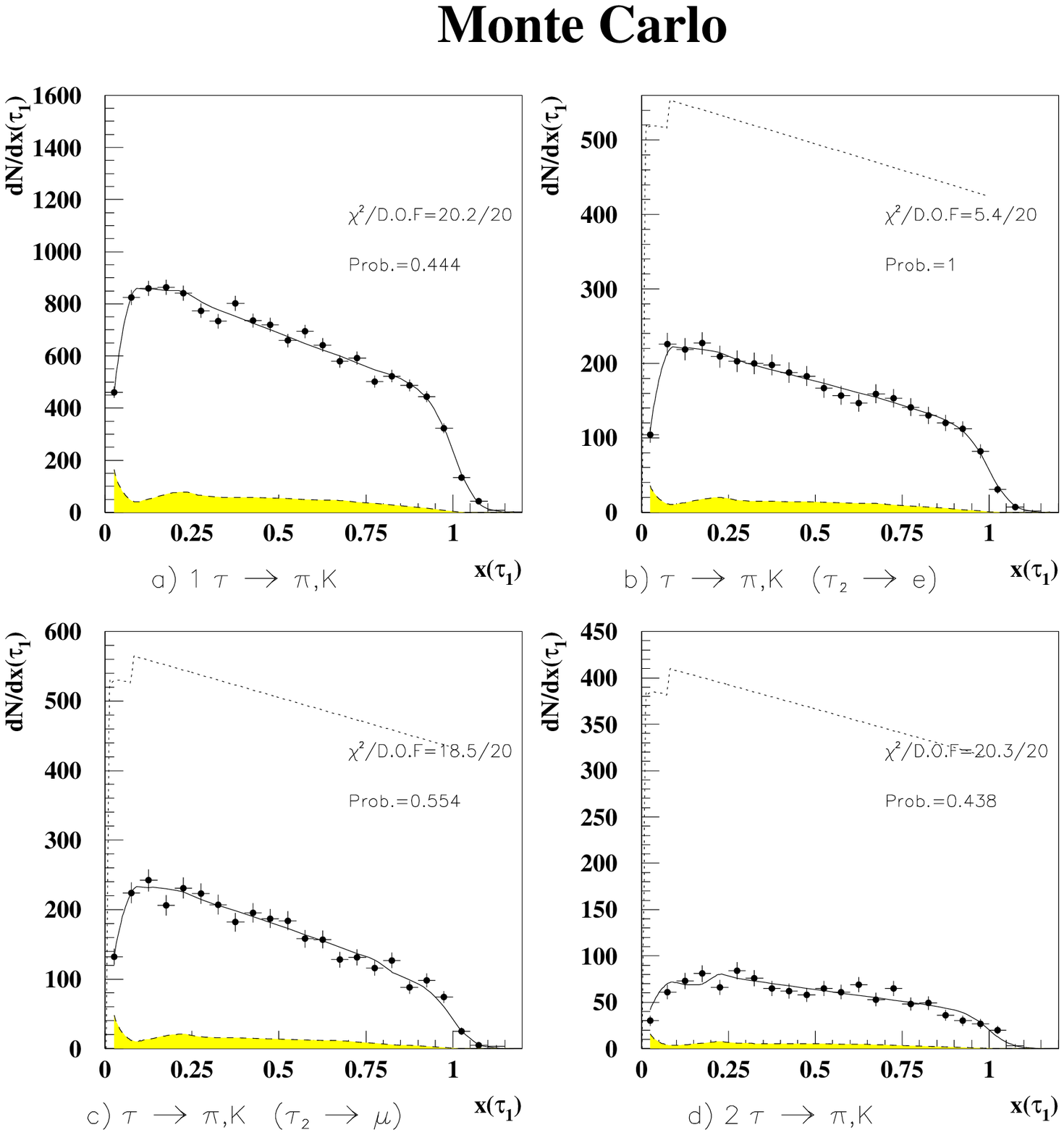}
\caption[MC $x$ distributions of the $\tpiK$ decay channel.]
{MC $x$ distributions for the $\tpiK$ decay channels and for
various decays on the opposite side,
compared with the theoretical predictions (solid lines) with all
corrections included, taking the $\pta$ value obtained from
the global ML fit. The filled area represent the background part of the
spectrum. The dotted lines are the uncorrected theoretical curves.}
\label{presulmc}
\end{figure}

\begin{figure}[p]
\epsfysize=16cm.
\epsffile[40 150 720 700]{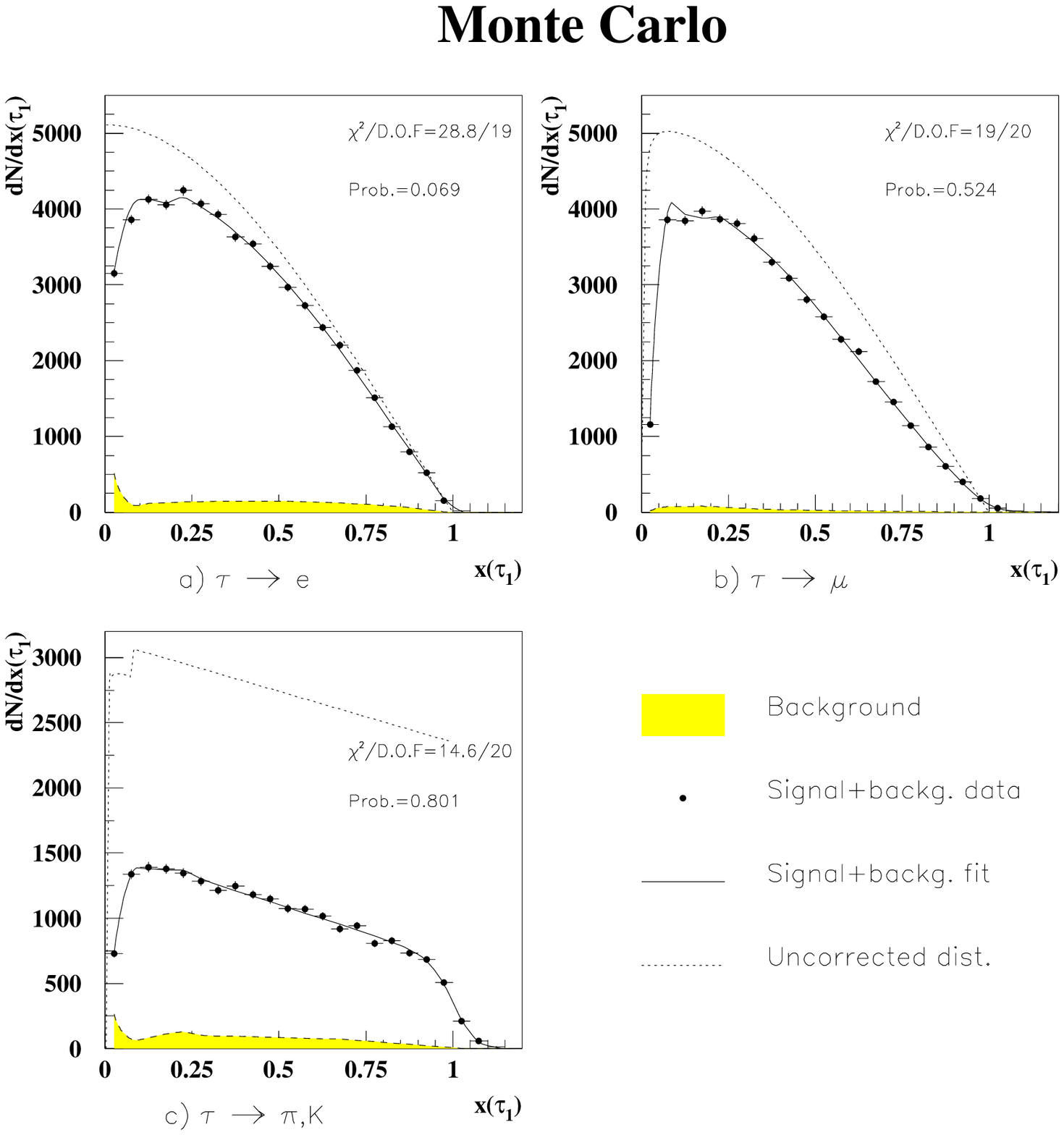}
\caption[MC $x$ distributions of the $\tel$, $\tmu$ and $\tpiK$.]
{MC $x$ distributions for the $\tel$, $\tmu$ and $\tpiK$ decays
channels, summing over all decays on the other side and
compared with the theoretical predictions (solid lines) with all
corrections included, taking the $\pta$ value obtained from
the global ML fit. The filled area  represent the background part of
the spectrum. The dotted lines are the uncorrected theoretical curves.}
\label{empresmc}
\end{figure}

For an overall check of our parametrization of the various corrections
obtained from MC, we plot the MC $x$ distributions separately for each
decay combination and compare it
with the corresponding corrected theoretical curves
(Figs.~\ref{eresulmc},~\ref{mresulmc},~\ref{presulmc},~\ref{empresmc}).
The $\chs$ values indicated on the figures show that
the agreement between the MC distributions and the theoretical
curves is good.
 
\section{Cross Checks}
\subsection{Statistical error examination}
 
In order to convince ourselves that the statistical errors obtained from the
fit are correct, we used a large sample of generated events without
detector simulation and perform a simple ML
fit to each decay,
ignoring the other side and ommiting the corrections. We only
applied the $x>0.05$ and $\act<0.68$ requirements. The errors obtained,
after scaling them for the different numbers of events, were
consistent with the statistical error obtained in our global fit
within 3\%.
 
\subsection{Separate checks with the different decay topologies}
 
The data has also been classified
into nine independent subsamples, corresponding to all
possible combinations of both $\tn$ decays excluding an identified
$\tro$ decay. Doing this,  nine independent results for $\pta$ and
$\aplfb$ are obtained.
These are listed in Table~\ref{table-datres}.
 The weighted mean of these values are within 0.3\%
of the global fit results and the
$\chs$ probabilities for the consistency between the nine results for
$\pta$ and $\aplfb$ are 17.2\% and 81.6\%, respectively.
 
\begin {table} [htb]
\begin{center}
\begin{tabular}{||l|c|c|c||}  \hline \hline
Decay Channel combination & $\pta$  & $\aplfb$ & Number of Events \\
                          & (\%)    & (\%)     &                  \\ \hline
Global fit value    &$-13.5\pm$ 2.9 &$-11.0 \pm$ 3.5 & 12397 \\
\hline \hline
1 \tn ~identified   &$-17.5\pm$ 3.5 &$-10.2 \pm$ 4.3 & 9250  \\
\hline
{\em (a)} $\tel$, none        &$-24.7\pm$ 8.9 &$-17.  \pm$10.  & 3585  \\
{\em (b)} $\tmu$, none        &$-8.7 \pm$ 7.9 &$-10.8 \pm$ 9.0 & 3897  \\
{\em (c)} $\tpiK$, none        &$-18.4\pm$ 4.4 &$ -8.1\pm$ 5.5 & 1768  \\
\hline \hline
2 \tn's identified &$-5.5  \pm$ 5.0 &$-12.6 \pm$ 5.9 & 3123  \\ \hline
{\em (d)} $\tel$, $\tel$      &   9.  $\pm$16.  &$ -28.\pm$18.  &  565  \\
{\em (e)} $\tel$, $\tmu$      & $-16. \pm$10.  &   3.  $\pm$12.  & 1288  \\
{\em (f)} $\tel$, $\tpiK$      & $-9.5 \pm$ 9.7 &$ -23.\pm$12.  &  346  \\
{\em (g)} $\tmu$, $\tmu$      &  7.  $\pm$24.  &   2. $\pm$27.  &  484  \\
{\em (h)} $\tmu$, $\tpiK$      &  6.1  $\pm$ 8.9 &$ -13. \pm$11.  &  385  \\
{\em (i)} \tpiK, \tpiK      & $-29.  \pm$20.  & $-21.  \pm$26.  &   55  \\
\hline \hline
Average             & $-13.2\pm$ 2.9 & $-11.1 \pm$ 3.4 &       \\  \hline
$\cof$              &    11.6/8       &     4.4/8       &       \\
$\chs$ probability (\%) &     17.2        &      81.6       & \\ \hline \hline
\end{tabular}
\caption[Results for the
nine independent subsamples.]
{$\tn$ polarization results for the
nine ($a$ to $i$) independent subsamples, corresponding to all
possible combinations of both \tn ~decays to e, $\mu$ or $\pi$(K).
The average values,
the $\cof$ and the $\chs$ probability correspond to the combined results of
these 9 different $\t$ decay channels. }
\label{table-datres}
\end{center}
\end{table}
 
\subsection{Graphical evaluation of the fit quality}
 
\begin{figure}[p]
\epsfysize=16cm.
\epsffile[40 150 720 700]{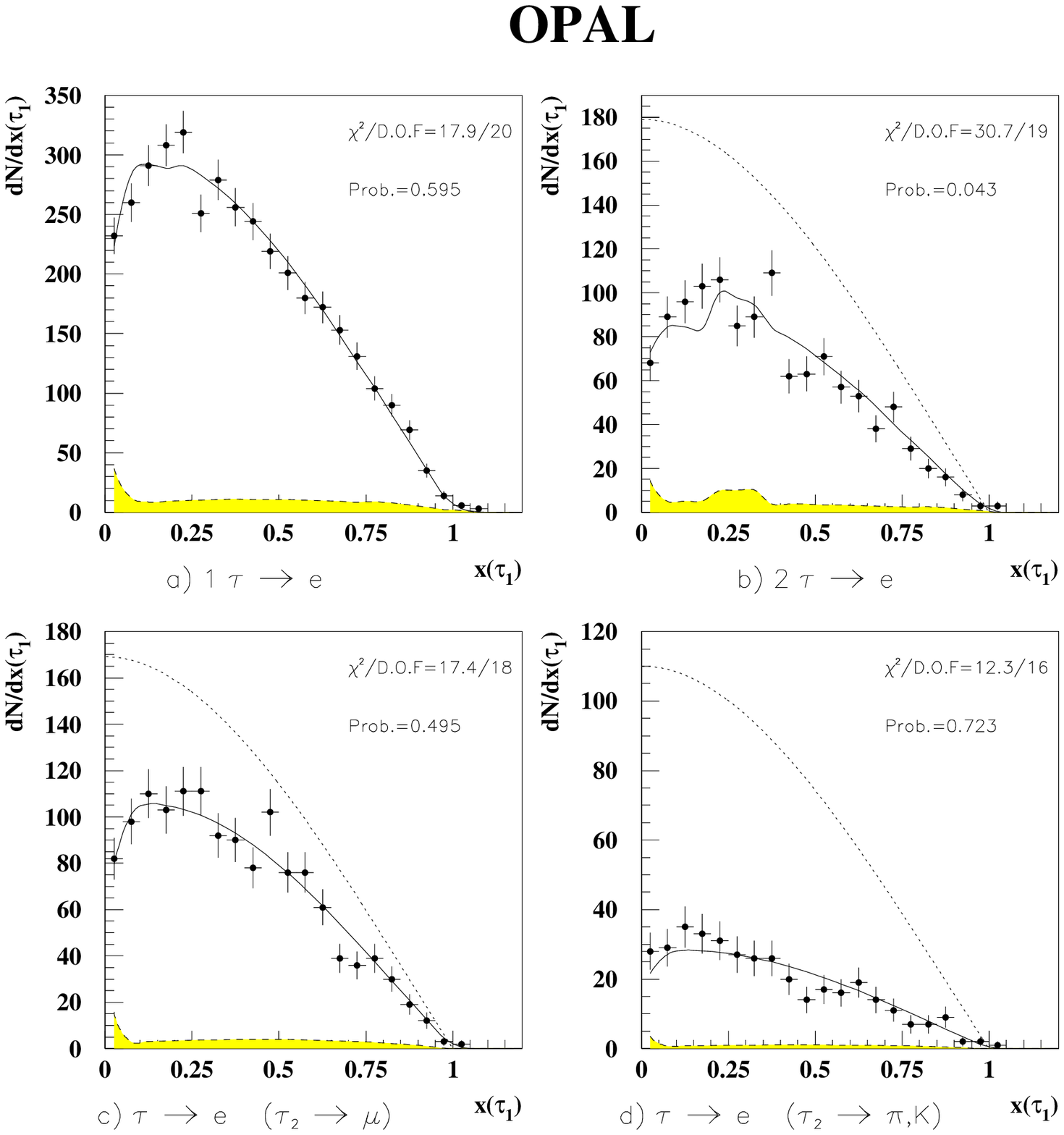}
\caption[$x$ distributions of the $\tel$ decay channel]
{$x$ distributions for the $\tel$ decay channels and for
various decays on the opposite side,
compared with the theoretical predictions (solid lines) with all
corrections included, taking the $\pta$ value obtained from
the global ML fit. The filled area represent the background part of the
spectrum. The dotted lines are the uncorrected theoretical curves.}
\label{eresults}
\end{figure}

\begin{figure}[p]
\epsfysize=16cm.
\epsffile[40 150 720 700]{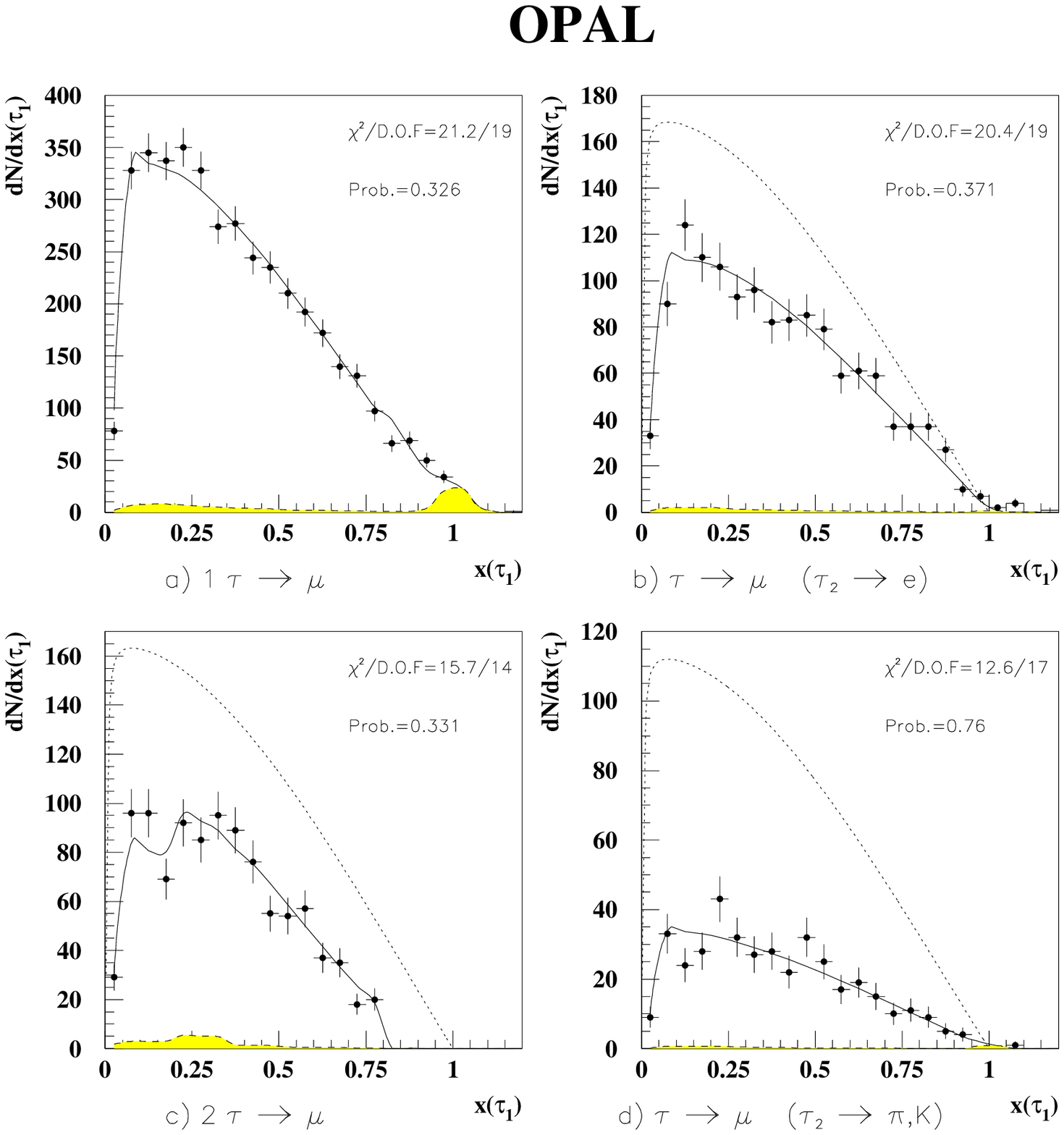}
\caption[$x$ distributions of the $\tmu$ decay channel]
{$x$ distributions for the $\tmu$ decay channels and for
various decays on the opposite side,
compared with the theoretical predictions (solid lines) with all
corrections included, taking the $\pta$ value obtained from
the global ML fit. The filled area represent the background part of the
spectrum. The dotted lines are the uncorrected theoretical curves.}
\label{mresults}
\end{figure}

\begin{figure}[p]
\epsfysize=16cm.
\epsffile[40 150 720 700]{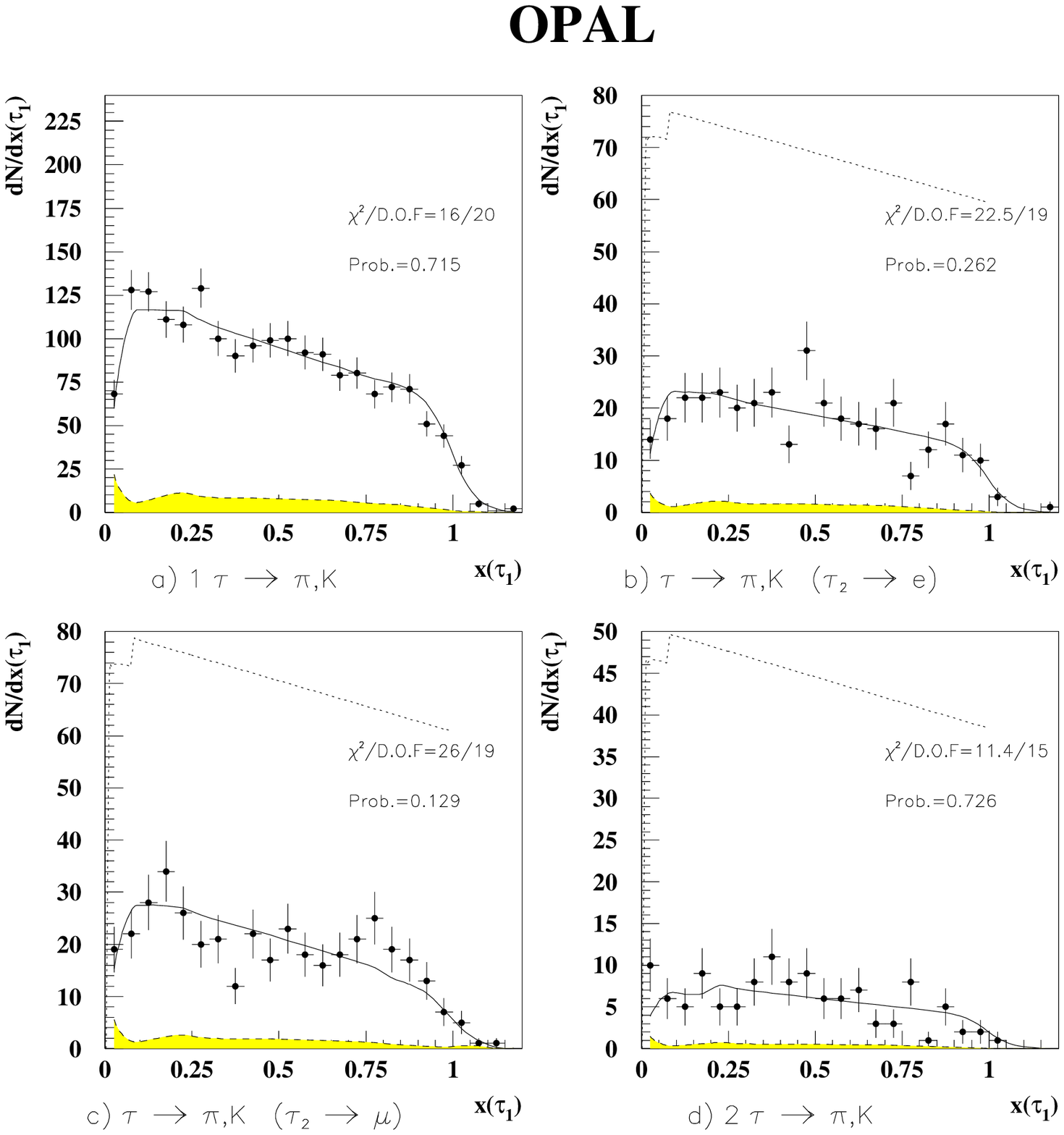}
\caption[$x$ distributions of the $\tpiK$ decay channel]
{$x$ distributions for the $\tpiK$ decay channels and for
various decays on the opposite side,
compared with the theoretical predictions (solid lines) with all
corrections included, taking the $\pta$ value obtained from
the global ML fit. The filled area represent the background part of the
spectrum. The dotted lines are the uncorrected theoretical curves.}
\label{presults}
\end{figure}

\begin{figure}[p]
\epsfysize=16cm.
\epsffile[40 150 720 700]{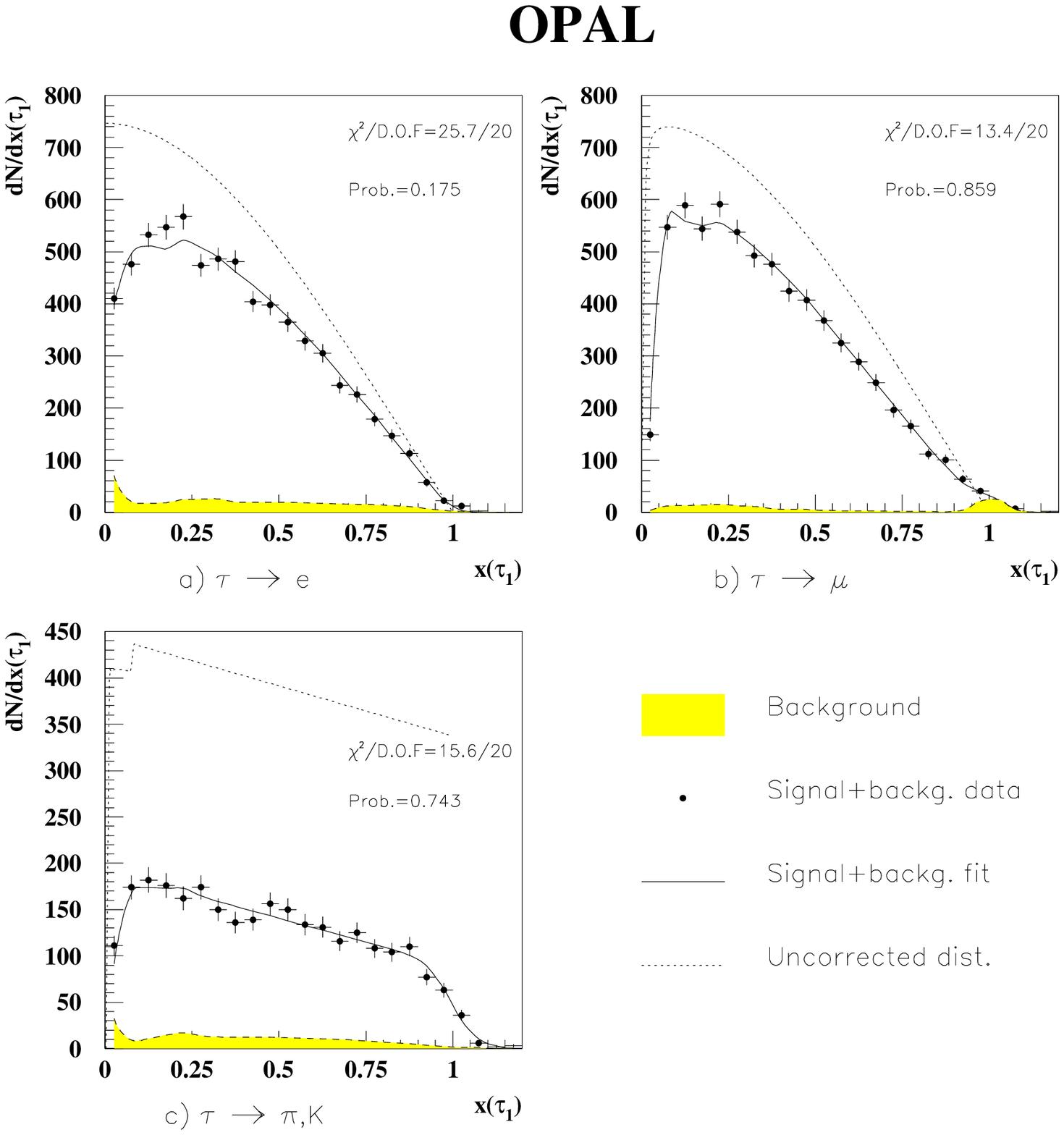}
\caption[$x$ distributions of the $\tel$, $\tmu$ and $\tpiK$ decays.]
{$x$ distributions for the $\tel$, $\tmu$ and $\tpiK$ decay
channels, summing over all decays on the other side and
compared with the theoretical predictions (solid lines) with all
corrections included, taking the $\pta$ value obtained from
the global ML fit. The filled area represent the background part of
the spectrum. The dotted lines are the uncorrected theoretical curves.}
\label{empreslt}
\end{figure}

In order to evaluate the fit quality, we plot the $x$ distributions
separately for each decay combination and compare it
with the corresponding corrected theoretical curves
(Figs.~\ref{eresults},~\ref{mresults},~\ref{presults}).
The bins corresponding to $x > 1$ are not used in the fit, since they have low
statistics and are dominated by the non-tau background which has large
uncertainties.
We note that the complex behavior of the theoretical
distributions in the low $x$ region
where both taus decay to electrons
or both decay to muons is the result of the need to
remove two-photon events.
The effect of the kaon threshold in the \tpiK ~decays is
also evident on these plots.
The $x$ distributions for \tel , \tmu ~and \tpiK, ~summing over all
decays of the opposite tau are presented in Fig.~\ref{empreslt}.
The normalization of the theoretical curves is obtained by a fit to
the experimental distributions.
The low  $\chs$ values verify
the good agreement between the experimental distributions and the theoretical
curves.
The filled regions
in the figures represent the background and the
dotted lines are the uncorrected theoretical distributions. One can
see that the background contributions are small,
and the overall corrections to
the theoretical curves, excluding the two lowest $x$ values, are
smooth, and do not depend strongly on $x$.

\subsection{A $\chs$ fit consistency check}
 
Another check of the ML fitting method is a $\chs$ fit of the $x$
distributions to a linear combination of the corrected theoretical
curves for positive and negative helicity. The $\chs$ fit is
meaningful only for the case where there is enough entries
(at least 10) in each of the bins. Therefore, it
was only applied to the case where just one tau was identified.
The results which are listed in Table~\ref{postfit} are consistent
with the ML fits listed in Table~\ref{table-datres}.
 
\begin {table} [htb]
\begin{center}
\begin{tabular}{||l|c||}  \hline \hline
Decay Channel & $\pta$  \\ \hline
$\tel$        & -21.1 $\pm$ 9.1  \\
$\tmu$        &  -4.2 $\pm$ 8.1  \\
$\tpiK$        & -20.1 $\pm$ 4.6  \\ \hline \hline
\end{tabular}
\caption[Tau polarization results calculated by $\chs$ fit.]
{Tau polarization results calculated by $\chs$ fits to
events where just one tau was identified.}
\label{postfit}
\end{center}
\end{table}
 
\subsection{Measurements of polarization as a function of $\cst$}
 
\begin{figure}[ht]
\epsfxsize=16.8cm.
\epsfysize=7cm.
\epsffile[40 400 680 650]{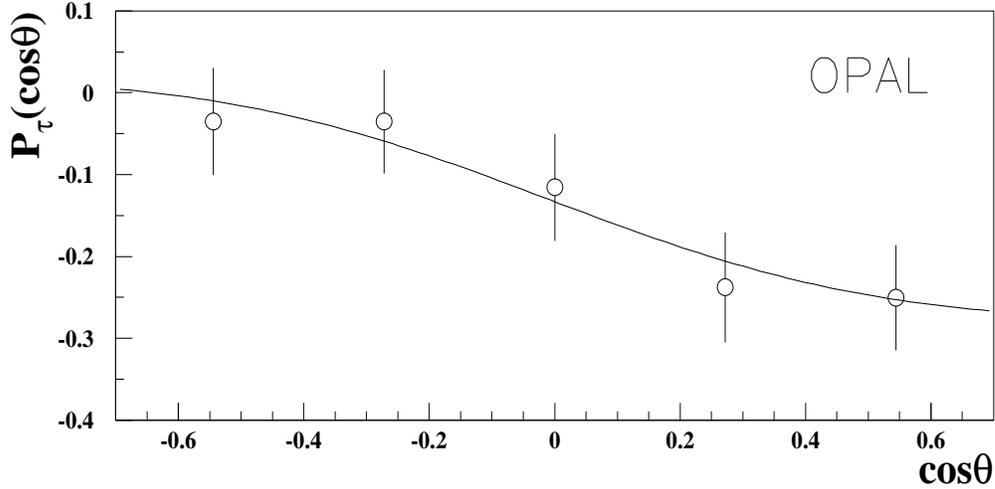}
\caption[Tau polarization  as function of $\cst$.]
{Measured tau polarization results as function of $\cst$. The solid
line is  our fit result (see text).}
\label{polcst}
\end{figure}

The results were also checked by dividing the $\cst$ range
into five intervals to determine the tau polarization in each of them
separately using the ML  fit.
The results are plotted in Fig.~\ref{polcst}. The solid line in the
figure describes result of a $\chs$ fit of the tau polarization
expected dependence on $\cst$ (Eq.~\ref{ptt}). The OPAL result for the
$\Z$ peak
$\afb=1.2\%$ ~\cite{bib-z0par} is used
(see Table~\ref{eint}).
We obtained $(\pta=-13.9\pm 2.9(stat.))\%$ and
$\aplfb=(-10.8\pm 3.5(stat.))\%$, which is in excellent agreement with our nominal results.
The fitted $\chs$ is 0.7 for 3 degrees of freedom and the correlation
between \pta ~and \aplfb ~is 0.027.

\subsection{A test of the $\afb$ influence}
 
We also modified our fitting procedure allowing the seven $\afb$ values
to be free parameters in the fit.
They are consistent with the values which have
been used in our standard fit, where the largest deviation is
1.8$\sigma$. Our statistical errors for $\afb$ are twice as large
since we use only events in the barrel region and only those
events where at least one tau was identified. 
\newpage
 
The corresponding
results for $\pta$ and $\aplfb$ were $(-13.5\pm2.9(stat.))\%$ and
$(-10.9\pm3.5(stat.))\%$ respectively, in an excellent agreement with our
standard global fit values.
 
\section{  \mbox{ \pta} ~and  \mbox{ \aplfb}
 ~from  \mbox{ \tro} ~decays}
 
In order to combine our results with those measured from
the \mbox{\boldmath \tro} ~decays,
we  quote here the OPAL  \pta~ and \aplfb~ measured
from this decay channel.
The two \tro ~analyses done by OPAL
has used a least squares fit to determine \ptat~ in five $\cst$ intervals,
and than
extracted the \pta ~and \aplfb ~parameters
in a least squares fit of the \ptat ~dependence on \cst.

The results of these two \tro ~analyses were combined, taking
care of the fact that these two data sets have an overlap of
40\%.
The  \tro ~OPAL's result is then:
\newpage
\bea
\label{rhoresfin}
\pta  & = & (\PTRHO \pm \PTRHOST \pm \PTRHOSY)\% \nonumber \\
\aplfb & = & (\APFBRHO \pm \APFBRHOST \pm \APFBRHOSY)\%
\eea
 
It is worthwhile to know that there is no overlap between
the events used in our analysis for $\t$ decay to e, $\mu$ or $\pi$(K)
and the events used by OPAL for the study of \tro.
Specifically, in our analysis the events of the type
$\ell$-$\rho$ (i.e. one tau decays to a
lepton and
the other tau decays to a $\rho$-meson) are discarded.
On the other hand in the \tro ~analyses events of the type
$\pi$-$\rho$ were discarded.
For events of the type
$\rho$-$\rho$  each identified
$\tro$ decay was weighted in the \tro ~analyses by 0.5.
Table~\ref{ptsumres} summarize the results of the $\t \to$ e, $\mu$ or $\pi$(K)
channels (ML fit),
the results of the $\tro$ channel ($\chs$ fit) and the combined results.

\section{Comparison With Previous Measurements}
 
\subsection{Previous determination of the Michel parameters}
Muon decay has been used for determination of the structure of the
charged weak interaction since the 1950's~\cite{Michel,Fetscher}.
The Particle Data Group~\cite{PDG}
values for the Michel parameters measured in
muon decays are
$\rho=0.7518\pm0.0026$~\cite{DERENZO},
$\xi=1.0027\pm0.0079\pm0.0030$~\cite{BELTRAMI}, and
$\delta=0.755\pm0.009$~\cite{BULKE}. The ARGUS collaboration has recently
published a measurement of two of the Michel parameters evaluated in
leptonic tau decays~\cite{ARGUS}.
For $\rho_{\tn \to \mu}$ and $\rho_{\tn \to e}$ they obtain
from the $e$ and $\mu$ momentum spectra the values of
$0.76\pm0.07\pm0.06$ and
$0.79\pm0.08\pm0.06$ respectively. Assuming
$\xi_{\tn \to \mu}=\xi_{\tn \to e}$ they have measured
$\xi=0.90\pm0.15\pm0.10$.
The ALEPH collaboration have used the correlation in $\t$ pairs where both
$\tn$'s decay into hadrons ($\pi$ or $\rho$) to measure $\xih$.
Their values are~\cite{ALEPHXI}:
$\xi_{\pi}=0.95\pm0.11\pm0.05$, $\xi_{\rho}=1.03\pm0.11\pm0.05$
or assuming $\xi_{\pi}=\xi_{\rho}$ they obtained $0.99\pm0.07\pm0.04$.
\begin {table} [htbp]
\begin{center}
\begin{tabular}{||l|l||c||c||}  \hline \hline
 $\t$ decay & Exp. & $\pta$ (\%)         & $\aplfb$ (\%)       \\ \hline
 \tel       & L3       & -12.7$\pm$9.7$\pm$6.2   & -- \\
            & DELPHI   & -12$\pm$27$\pm$8        & --\\
            & ALEPH    & -22.5$\pm$8.5$\pm$4.5   & --\\
            & OPAL 1   & 20$\pm$13$\pm$8
                       & -16$\pm$19                \\
            & OPAL 2   &  -8.5 $\pm$ 5.8$\pm$4.5
                       &  -10.4 $\pm$ 6.6$\pm$1.3  \\ \hline
 \tmu       & L3       &  -2.0$\pm$ 10.1$\pm$5.5 & --\\
            & DELPHI   &  -5$\pm$18$\pm$7        & --\\
            & ALEPH    &  -15.4$\pm$6.5$\pm$2.9  & --\\
            & OPAL 1   &  -17$\pm$16$\pm$10
                       &  -88$\pm$22               \\
            & OPAL 2   &  -8.0 $\pm$ 5.4$\pm$3.3
                       &  -10.1 $\pm$ 6.2$\pm$1.3  \\ \hline
 \tpiK      & L3       &  -14.8$\pm$ 9.7$\pm$6.2 & --\\
            & DELPHI   &  -35$\pm$11$\pm$7       & --\\
            & ALEPH    &  -13.3$\pm$3.1$\pm$1.8  & --\\
            & OPAL 1   &  -8$\pm$10$\pm$7
                       &  -34$\pm$16               \\
            & OPAL 2   &  -14.3 $\pm$ 3.7$\pm$3.0
               &  -10.9 $\pm$ 4.5$\pm$0.8           \\ \hline
All         & L3       &  -13.2$\pm$ 2.6$\pm$2.1 & --\\
            & DELPHI   &  -24$\pm$7              & --\\
            & ALEPH    &  -14.3$\pm$2.3          & -9$\pm$2 \\
            & OPAL 1   &  -9.7$\pm$8.4
                       &  -22$\pm$10               \\
            & OPAL 2   &  -11.3 $\pm$ 3.3
                       &  -10.6 $\pm$ 3.3            \\ \hline 
\end{tabular}
\caption[$\tn$ polarization results.]
{The LEP experiments $\tn$ polarization published
results for the \tel ,\tmu ~and \tpiK ~channels.
When two errors are quoted the first is the statistical
and second is the systematic error, when one error is quoted it is
the combined statistical and systematic error.
The L3~\cite{L3PL} and ALEPH~\cite{ALE2PL} published the study of 1990-1991
data using five decay channels of $\t$ (e, $\mu$, $\pi(K)$, $\rho$
and $a_1$). DELPHI's publication~\cite{DELPHI} is based on the analysis of
four decay channels (e, $\mu$, $\pi(K)$ and $\rho$) from the years
1990-1991. The OPAL 1 is the OPAL first publication~\cite{OPALPL} based on the
analysis of three decay channels (e, $\mu$ and $\pi(K)$) during 1990.
OPAL 2, is the present analysis using 1990-1992  e, $\mu$ and $\pi(K)$
decay channels.
Note that  for the comparison with the other numbers the individual
results quoted  assume, {\em incorrectly},
that there are no correlations between the taus. These
correlations are fully taken into account in the global ML
fit result. }
\label{ptres}
\end{center}
\end{table}

\subsection{Results to compare with other measurements}
 
In the analyses published by OPAL in the past~\cite{OPALPL} and by other
LEP collaborations~\cite{DELPHI}-\cite{ALE2PL},
one did not look at the whole $\t$-pair event. Instead,
one did consider each decay separately, ignoring the opposite side.
In order to be able to compare our results with those of other
analyses, we repeated our ML fit adopting their policy.
For that, we had to recalculate those corrections depending on the
whole event, such as $\t$-pair selection efficiencies and non-tau
background. Each event, where both $\t$'s were identified, was handled
as two independent events.
In Table~\ref{ptres} we list the results for each decay channel
separately. One should note, however, that the
results for the $\tel$, $\tmu$ and $\tpiK$ decays
in this table cannot be interpreted as independent polarization
measurements because the same tau-pair event can contribute to two
channels.

\subsection{LEP tau polarization results}
\begin{figure}[htb]
\epsfysize=9cm.
\epsffile[52 421 577 735]{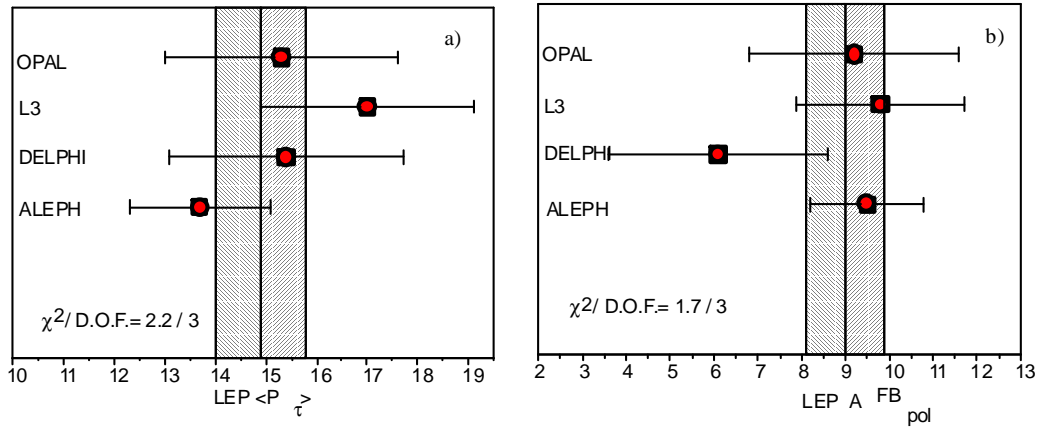}
\caption[LEP 1990-1992 ~$\pta$ and $\aplfb$  (preliminary) results]
{LEP results for~ $\pta$ and $\aplfb$. The  ALEPH,
DELPHI and L3 are preliminary results given at the {\small MORIOND} 94
conference, whereas the OPAL figures are the updated results sent for
publication. The shaded areas are the averages of the 4 experiments.}
\label{figlepres}
\end{figure}

The following Table~\ref{Tab-moriond},
presents the up to date but preliminary  $\pta$ and $\aplfb$  results of 
all
other 3 LEP detectors presented at the La Thuile and Moriond Conferences
in March 1994
~\cite{Moriond94}. The OPAL figures are those  of the
present analysis combined with the $\tro$ figures~\cite{paper}.
In averaging the results we repeat the procedure discussed by the LEP
tau polarization group in Ref.~\cite{Moriond94}.
Three common effects, which affect all the tau decay mode equally,
were taken into account:
\begin{itemize}
\item A dependence of the results on $\ecm$ due to a direct $\gamma$ and
$\gamma$ - $\Z$ interference contributions to the $\Z$ intermediate state.
These were corrected so that the results quoted correspond to $\ecm=M_Z$.
\item Electromagnetic initial and final state radiative corrections.
\item Born level mass terms leading to helicity flip configurations.
\end{itemize}
As shown by the theoretical calculations in Refs.
~\cite{was,bib-koralz,bib-PHOTOS,bib-ZFITTER}
the combined effect of these three
is to reduce the strength of the measured \pta ~(\aplfb) by about
0.3\% (0.2\%).
Therefore the experiments results (including ours) were corrected
by adding this amount  to the measured $\pta$ and $\aplfb$.
The four experiments results were combined
without extracting a common
systematic error.
 The $\cof$ for the average is
2.2/3  and 1.7/3  for $\pta$ and $\aplfb$ respectively. The results of the
four  experiments and their average are summarized in Fig.~\ref{figlepres}.
 
\begin {table} [htb]
\begin{center}
\begin{tabular}{|c|c|c|}  \hline \hline
 Experiment & $\pta$          & $\aplfb$        \\ \hline
               & (\%)            & (\%)            \\ \hline
OPAL           &$-15.3 \pm 2.3$     & $-9.2 \pm 2.4$  \\  
L3             &$-17.4 \pm 2.1$     & $-9.8 \pm 1.9$  \\
DELPHI         &$-15.4 \pm 2.3$     & $-6.1 \pm 2.5$  \\
ALEPH         &$-13.7 \pm 1.4$     & $-9.5 \pm 1.3$ \\ \hline
LEP Average    &$-15.0 \pm 0.9$     & $-9.1 \pm 0.9$  \\   \hline \hline
\end{tabular}
\caption[LEP 1990-1992 ~$\pta$ and $\aplfb$  (preliminary) results]
{LEP results for~ $\pta$ and $\aplfb$. The OPAL figures are the updated
results where the figures of all other three experiments are preliminary
results presented at {\small MORIOND} 94.}
\label{Tab-moriond}
\end{center}
\end{table}

\chapter{Summary}
\label{chap-SUM}
 
We have measured the average tau polarization and its forward backward
asymmetry using the tau decay channels $\tel$, $\tmu$ and $\tpiK$.
The results of these measurements (taking into account all corrections) are
\bea
\label{mlresfin}
\pta  & = & (\PTEMPF \pm \PTEMPST \pm \PTEMPSY)\% \nonumber \\
\aplfb & = & (\APFBEMPF \pm \APFBEMPST \pm \APFBEMPSY)\%
\eea
where the first error is statistical and the second systematic.
Our Maximum Likelihood fit method takes into account correlations between the
decay distributions of the $\tm$ and the $\tp$ in the same event
due to their opposite
helicity. The theoretical expectation had to be corrected for
radiative effects, detector resolution, efficiencies of $\t$-pair
selection and tau decay identification, and background originating from
other $\t$ channels and non-$\t$ events. These correction were
calculated from Monte Carlo events and whenever possible,
checked with control measured data.
 
A summary of the measurements from all channels is
presented in Table~\ref{ptsumres}.
\begin {table} [htb]
\begin{center}
\begin{tabular}{||l|c|c||}  \hline \hline
 Decay Channel & $\pta$          & $\aplfb$        \\ \hline
               & (\%)            & (\%)            \\ \hline
 $\tel,\mu\bar{\nu}_{\mu}\nu_{\t},\pi(K)\nu_{\t}$ present analysis
               &  \PTEMPF$\pm$\PTEMPST$\pm$\PTEMPSY
               &   \APFBEMPF$\pm$\APFBEMPST$\pm$\APFBEMPSY \\ \hline \hline
 OPAL \tro     &   \PTRHO$\pm$\PTRHOST$\pm$\PTRHOSY
               &   \APFBRHO$\pm$\APFBRHOST$\pm$\APFBRHOSY    \\ \hline \hline
  OPAL combined results
               & \PTALL$\pm$\PTALLST$\pm$\PTALLSY
               & \APFBALL$\pm$\APFBALLST$\pm$\APFBALLSY    \\ \hline
\end{tabular}
\caption[OPAL combined \t ~polarization  results.]
{OPAL combined $\tn$ polarization results.}
\label{ptsumres}
\end{center}
\end{table}
The results are quoted for
$\sqrt s =m_Z$. A very small correction ($1\%$ of the \pta),
which takes into account the
fact that some data was collected off the peak of the Z$^0$
resonance, is made. Our final result,
combined with the OPAL \tro ~results are:\\
\bea
\label{resfin}
\pta  = & -\lamt = & (\PTALL \pm \PTALLST \pm \PTALLSY)\% \nonumber \\
\aplfb =  &-\frac{3}{4} \lame = & (\APFBALL \pm \APFBALLST \pm \APFBALLSY)\%.
\eea

Following  
these results we obtain:
\bea
\lamt = & \frac{2\vt\at}{\vt^2+\at^2} = & 0.153 \pm 0.023\\ \nonumber
\lame = & \frac{2\ve\ae}{\ve^2+\ae^2} = & 0.122 \pm 0.032 \\ \nonumber
\eea
where the errors include both statistical and systematic uncertainties.
These are already corrected,
(using ZFITTER ~\cite{bib-ZFITTER}),
for the contribution of the intermediate
$\gamma$, $\gamma$-Z$^0$ interference and other photonic radiative
corrections. Therefore they can be interpreted in terms
of the vector and axial-vector couplings within the improved Born
approximation.
In  the context of the SM $\lamt$ and $\lame$ yield a measured values for
the coupling ratios of
\bea
\vt/\at &  = & \VOVAT \pm \VOVATSI \\ \nonumber
\ve/\ae & = & \VOVAE \pm \VOVAESI
\eea
where as before the error includes  both statistical and
systematic contributions.
 
These measured values of
\pta ~and \aplfb ~correspond to the following values of \efswsq
~(see Eq. ~\ref{vovad}):
\bea
\efswsq (\pta)   & = & 0.2308 \pm 0.0030 \\ \nonumber
\efswsq (\aplfb) & = & 0.2345 \pm 0.0040 .
\eea
These values are in a good agreement with the hypothesis
of lepton universality.
Therefore they  can be averaged
(with $\chs=0.56$)
to give ,
\bea
v/a & = &  0.072 \pm  0.010  \\  \nonumber
\efswsq & = & \SINWALL \pm \SINWALLSI
\eea
 
\begin{figure}[p]
\epsfysize=19cm.
\epsffile[19 20 576 822]{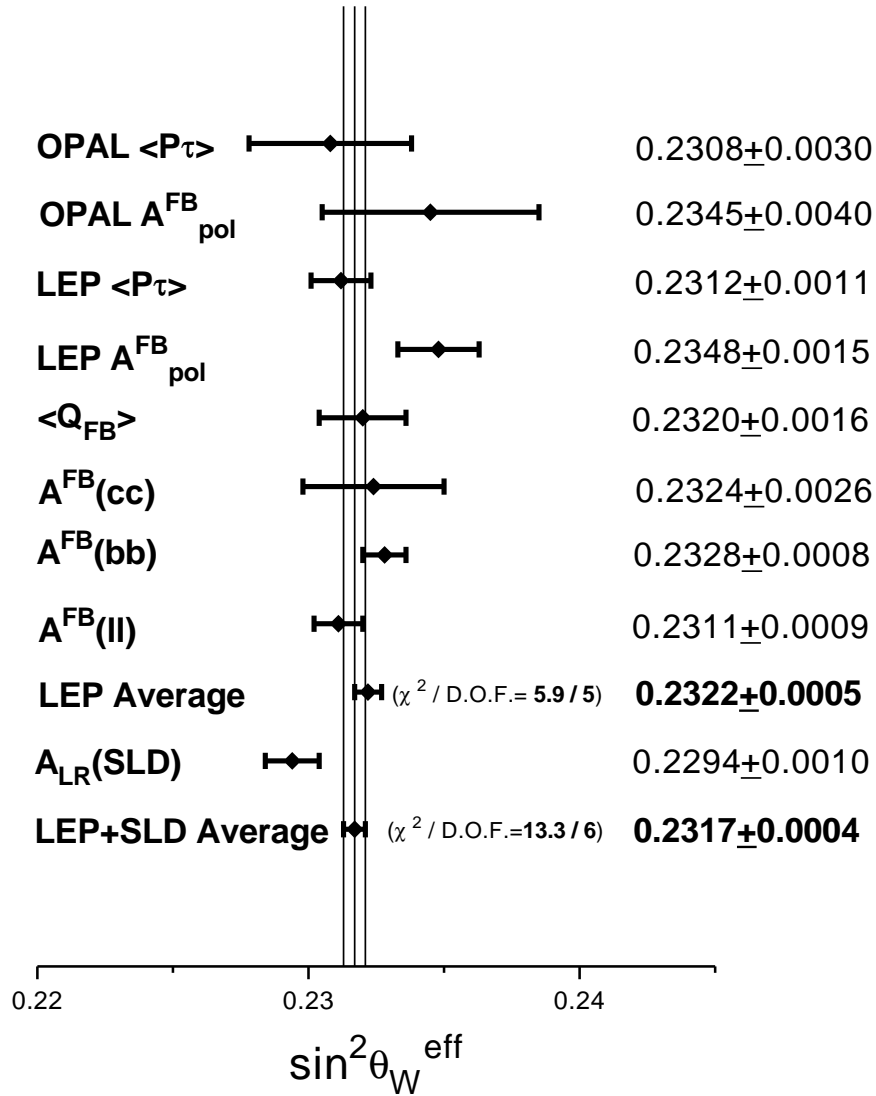}
\caption[Comparison of several determinations of 
the weak mixing angle]
{Comparison of several determinations of the weak mixing angle
from asymmetries as was presented 
in the MORIOND 94 conference. The values of LEP 
were modified using the results of the present analysis. 
Combined results of the LEP experiments and a combined LEP-SLD 
results are obtained as weighted averages assuming no correlations.}
\label{fig-sinw}
\end{figure}

Fig.~\ref{fig-sinw} presents the result of various determinations of
\efswsq measured at LEP and SLD. These were calculated from
the observed forward
backward asymmetries of leptons ($\afb(\ell \ell))$ and quark pairs
($\afb(b\bar{b}),\;\; \afb(c\bar{c})$ and $\langle Q_{FB}\rangle$-
the inclusive forward backward asymmetry of negative quarks),
the tau polarization and its forward backward asymmetry.
 
The figures of the
charge asymmetries are those which were presented as LEP results in the
Moriond 94 conference~\cite{sinw}.
The results given  for the $\pta$ and $\aplfb$  are the modified
OPAL numbers using the results of the present work. The obtained LEP
average (with $\cof=5.9/5$) is
\beq
\efswsq(LEP)   =  0.2322 \pm 0.0004 \nonumber
\eeq
For a comparison, also  the SLD value of
$\efswsq$ derived  from their measurement
of the left-right asymmetry, $A_{LR}$ is given.
This asymmetry is defined in a fully polarized electrons beam as
\beq
A_{LR}=\frac{\sigma_L-\sigma_R}{\sigma_L+\sigma_R}=\lame
\eeq
which means that it probes the same couplings as the $\aplfb$.
One can see from Fig.~\ref{fig-sinw} that while there is a good
agreement betwen all LEP
results, the present SLD value for $\efswsq$~\cite{SLD} is
shifted from the LEP numbers.
Therefore as expected,  the $\chs$ obtained when combining
the measured weak mixing angle
of LEP and SLD there is no good agreement
($\cof=13.3/6$).
The average value being
\beq
\efswsq(LEP+SLD)   =  0.2317 \pm 0.0003 \nonumber
\eeq

\newpage
\newpage
\bibliographystyle{plain}

\begin{thebibliography}{60}
 
\bibitem {GWS1} S.~L. Glashow, Nucl.\ Phys. {\bf 22} (1961) 579.
 
\bibitem {GWS2} S.~Weinberg, Phys.\ Rev.\ Lett. {\bf 19} (1967) 1264.
 
\bibitem {GWS3} A.~Salam, in proceeding of the Eight Nobel Symposium, on
Elementary Particle Theory,  ed.\ N.\ Svartholm (1968) 367.
 
\bibitem {CALASY} S.~Jadach and Z.~W\c{a}s in
  "The $\tau$ Polarization Measurement",
  {\it Z physics at LEP1}, CERN Yellow Report 89-08,
  eds.\ G.\ Altarelli \etal
  Vol.\ 1 (1989) p.\ 235, and an update of august 1990.
 
\bibitem{bib-IBA}    M. Consoli, W. Hollik and F. Jegerlehner in
                  {\it Z Physics at
                 LEP1}, CERN 89-08, ed.
                 ~G. Altarelli \etal Vol. 1 (1989) 7.
 
\bibitem{bib-z0par}  OPAL Collaboration:\\
                 M.Z. Akrawy \etal Phys. Lett. {\bf B240} (1990) 497. \\
                 G. Alexander \etal Z.\ Phys.\ {\bf C52} (1991) 175.  \\
                 R. Akers \etal
                  Z. Phys. {\bf C61} (1994) 19.
 
 
\bibitem{OPALPL} OPAL Collaboration,
                 G. Alexander \etal Phys.\ Lett.\ {\bf B266} (1991) 201.
 
 
\bibitem{ALE1PL} ALEPH Collaboration,
                 D. Decamp \etal Phys. Lett. {\bf B265} (1991) 430.
\bibitem{DELPHI} DELPHI Collaboration,
                 P. Abreu \etal Z.\ Phys.\ {\bf C55} (1992) 555.
\bibitem{L3PL}   L3 Collaboration,
                 O. Adriani \etal Phys.\ Lett.\ {\bf B294} (1992) 466.
 
\bibitem{ALE2PL} ALEPH Collaboration,
                 D. Buskulic \etal
                  Z.\ Phys.\ {\bf C59} (1993) 369.
\bibitem{DELPH2} DELPHI Collaboration,
                 contribution to XVI International Symposium on Lepton-Photon
                 Interactions, Cornell University, August 1993.
\bibitem{L32} L3 Collaboration,
                 contribution to the Int. Europhysics Conf. on High Energy
                 Physics, Marseille, July 1993.
 
\bibitem{TN203} A.~Beck,
{"$\tau$-Polarization and its
   Forward-Backward Asymmetry From The $\tro$ channel
   Using Dedicated Clustering"},
       OPAL technical note {\bf TN203} (1994).
 
\bibitem{TN142} M.~Rosvick,
 {"Results of the Victoria $\tro$ analysis"},
       OPAL technical note {\bf TN142} (1993).
 
\bibitem{TN202} G.~Bella and E.~Etzion,
       "Analysis of the \tn ~Polarization and its Forward-Backward
       Asymmetry Using a Global Fit Method",
       OPAL technical note {\bf TN202} (1994).
 
\bibitem{PN126} A.~Beck,G.~Bella,E.~Etzion,M.~Roney,M.~Rosvick
               and M.~Sasaki,
               "Measurement of the tau lepton polarization and its
               forward-backwad asymmetry at LEP",
                  OPAL Physics Note, {\bf PN126} (1994).
 

\bibitem{paper}  OPAL Collaboration,
    R.\ Akers \etal
   "Measurement of the Tau Lepton Polarization and its
        Forward-Backward Asymmetry from Z0 Decays",
    CERN-PPE/94-120 (1994),
   submitted to Z.\ Phys.\ C .

 
 
 
 
 
\bibitem {LEP1} P.\ Schmuser,
"Basic course on accelerator optics",
{\it CERN acceleration school}, General accelerator physics
Aarhus Denmark, Sep. 86, ed. S. Turner
CERN 87-10,  (1987) 1.
 
\bibitem {LEP2} A.\ Hofmann,
Nucl.\ Phys.\ {\bf B3} (1988) 511.
 
\bibitem {LEP3} J.\ J.\ Tresher,
"LEP: machine and experiment",
{\it XXIV International conference on High energy Physics},
eds. R.\ Kotthaus and J.\ H.\ Kuhn, Munich, Germany,
(1988) 387.
 
\bibitem {LEP4}
"Large Electron Positron storage ring - Technical notebook"
CERN (1988) 1.
 
 
\bibitem{ROPAL1}
   OPAL Technical Proposal, CERN/LEPC/{\bf 83-4}.
 
\bibitem{ROPAL2} OPAL Collaboration, K.~Ahmet \etal
Nucl.\ Inst.\ and
Meth.\ {\bf A305} (1991) 275.
 
\bibitem{RFSTPHYS}
          OPAL Collaboration, M.Z.~Akrawy \etal
          Phys. Lett. {\bf B231} (1989) 530.
 
\bibitem{FILTER} D.G.~Charlton \etal
Nucl.\ Inst.\ and Meth.\ {\bf A325} (1993) 129 .
 
 
\bibitem{ZEBRA}
 R.~Brun, M.~Goossens and J.~Zoll, ZEBRA Dynamic Data Structure and Memory
 Manager, Program Library, CERN.
 
\bibitem {unity} A.\ Zee,
{\it "Unity of forces in the universe"}, World Scientific (1982).
 
\bibitem {generations}
 
 Mark II at SLAC and all four LEP detectors at CERN
 have measured this number of families
(with $m_{\nu}<\frac{1}{2}M_Z$).
See for example the OPAL publication ref.~\cite{RFSTPHYS}.
 
\bibitem {Report} B.C.\ Barish and R.\ Stroynowski,
Phys.\ Rep.\  {\bf 157} (1988) 1.
 
\bibitem{Tsai} Y.\ S.\ Tsai,
Phys.\ Rev.\ {\bf D4} (1971) 2821.
 
\bibitem{dicus} D.\ A.\ Dicus, Phys.\ Rev.\ {\bf D8} (1973) 890.
 
\bibitem{koniuk} R.\ Koniuk, R.\ Leroux, N.\ Isgur,
Phys.\ Rev.\ {\bf D17} (1978) 2915.
 
\bibitem {Augustin} J.E.\ Augustin,
"Use of $\tau$ Hadronic Decay to Study Polarization Near $Z^0$ Pole
at LEP", ECFA/LEP 29, {\it Proceeding of the LEP summer study,
 CERN 79-01}, vol.\ 2 (1979) 499.
 
\bibitem{goggi} G.\ Goggy,
"Study of Heavy Lepton Polarization in $e^+e^-$ Annihilation at the
$Z^0$ Pole", ECFA/LEP 2, {\it Proceeding of the LEP summer study,
 CERN 79-01}, vol.\ 2  (1979) 483.
 
\bibitem {gogg} G.\ Goggy,
Letters al Nuovo Cimento , Vol {\bf 24} (1979) 2.
 
\bibitem{was} S.~Jadach and Z.~W\c{a}s,
Acta Physica Polonica, Vol {\bf B15} (1984) 1151.
 
\bibitem{Dorfan} D.\ Dorfan,
"Polarization Experiments without Polarized Beams:
$e^+e^- \rightarrow \tau^+\tau^-$", SLAC PUB. {\bf 3407} (1984).
 
\bibitem{chauveau} J.\ Chauveau,
 "Testing the Standard Model by Measuring the $\tau$ Helicity at $Z^0$",
{\it Physics at LEP}, eds.\ J.\ Elis R.\ Pecci, CERN report 86-02 (1986).
 
\bibitem{Burger} G.~ Burgers and W.~Hollik,
"The shape and size of Z resonance",
{\it Polarization at LEP, CERN report 88-06}, eds.\ G.\ Alexander \etal
CERN, Geneva (1988) 136.
 
\bibitem{Hagiwara} K.\ Hagiwara, A.D.\ Martin, D.\ Zeppenfeld,
Phys.\ Lett.\ {\bf B235} (1990).
 
 
 
\bibitem{TN031} E.\ Etzion,
"$\swsq$ Measurements in the $e^+e^- \to \t^+\t^-$
Channel Using $\t \to \mu$ Decay", OPAL technical note {\bf TN031}
 (1991), and M.Sc. thesis, Tel Aviv University (June 1991).
 
 
\bibitem{Higgs} G.\ Alexander, C.\ Milst\`{e}ne, W.\ Hollik,
Z.\ Phys.\ {\bf C52} (1991) 283.
 
\bibitem{Ellis} J.\ Ellis and M.K.\ Gaillard,
CERN 76-18 (1976) p21.
 
\bibitem {halzen} see e.g. F.\ Halzen, A.\ D.\ Martin,
{\it Quarks \& leptons}, Wiley-Interscience (1984).
 
 
\bibitem{WMAT} W.\ Fetscher, H.J.\ Gerber, K.F.\ Johnson,
               Phys.\ Lett.\ {\bf B173} (1986) 102.
 
\bibitem{Privitera} P. Privitera, Phys. Lett. {\bf B288} (1992) 227.
 
\bibitem{Michel} L. Michel, Proc. Phys. Soc. {\bf A63} (1950) 514.
 
\bibitem{OPALPK} OPAL Collaboration, contribution to the Int. Europhys.
                 conf. on High Energy Physics, Marseille, July 1993.
 
\bibitem{Gulotv} A. Gulotvin,
                 "New Determination of Michel Parameters in Leptonic
                 $\tau$ Decays",
                 Contribution of the Second Workshop on Tau Lepton
                 Physics, Ohio (September 1992).
 
 
 
\bibitem{ABE}    G. Alexander, G. Bella and E. Etzion,
                 "The statistical Errors of $\tau$-Asymmetries and their
                 Correlations",
                 OPAL Technical Note {\bf TN062} (1992) .
 
\bibitem{BELLA} G. Bella,
    "The Weight Ratio Method and its Application to Identify
      $\tau$-Decay Modes",
 OPAL Technical Note {\bf TN058} (1992).
\bibitem{TP103}  The Tau Platform (TP) Manual, Version 1.02.
 
\bibitem{bib-gopal} J. Allison \etal Nucl.\ Inst.\ and Meth. {\bf A317}
                 (1992) 47.
 
\bibitem{bib-geant}
J.~Allison \etal Comp.\ Phys.\ Comm. {\bf 47} (1987) 55;\\
 R.~Brun, F.~Bruyant, M.~Maire, A.~C.~McPherson, and P.~Zanarini,
 {\it GEANT3}, CERN  DD/EE/{\bf84-1} (1987).
 
 
 
 
\bibitem{bib-koralz}  S. Jadach, B.F.L Ward and Z. W\c{a}s, Comp.\ Phys.\ Comm.\
                 {\bf 66} (1991) 276; \\
                 S. Jadach, J.H. K\"{u}hn and Z. W\c{a}s, Comp.\ Phys.\ Comm.\
                 {\bf 64} (1991) 275.
 
\bibitem{bib-Jetset}  T.~Sj\"{o}strand, Comp. Phys. Comm. {\bf 39} (1986) 347;\\
 M.~Bengtsson and T.~Sj\"{o}strand,
 Comp.\ Phys.\ Comm.\ {\bf 43} (1987) 367; \\
 M.~Bengtsson and T.~Sj\"{o}strand,
 Nucl.~Phys. {\bf B289} (1987) 810 (JETSET).
 
\bibitem{bib-OPALQCD}
 OPAL Collaboration, M.~Z.~Akrawy \etal
  Z.\ Phys.\ {\bf C47} (1990) 505.
 
\bibitem{bib-babamc}
 M.~Bohm, A.~Denner and W.~Hollik, Nucl. Phys. {\bf B304}
 (1988) 687;\\
 F.~A.~Berends, R.~Kleiss and W.~Hollik, Nucl.\ Phys.\ {\bf B304}
 (1988) 712.
 
\bibitem{bib-vermas}
 R.~Battacharya, J.~Smith and G.~Grammer,  Phys.\ Rev.\ {\bf D15} (1977) 3267;\\
 J.~Smith, J.~A.~M.~Vermaseren and G.~Grammer, Phys.\ Rev.\ {\bf D15} (1977) 3280.
 
\bibitem{bib-LINESHAPE0} OPAL Collaboration, G. Alexander \etal
   ~Z. Phys. {\bf C52} (1991) 175.
 
 
\bibitem{Clayton} J. Clayton, private communication.
 
\bibitem{bib-OPALhpizbr} OPAL Collaboration,
Phys.\ Lett. {\bf B328} (1994) 207.
 
 
 
 
 
 
 
\bibitem{select} G.\ Alexander and E.\ Etzion,
"Some Aspects Concerning the Selection of
$Z^0 \rightarrow \tau^+\tau^-$ Events", CERN August 1989.
 
\bibitem{bib-PHOTOS} E.~Barberio, B.~van~Eijk and Z.~W\c{a}s,
                     Comp.\ Phys.\ Comm. {\bf 66} (1991) 115.
 
 
\bibitem{TN192} M. Sasaki,
    "Improved Extraction Method Used for
Tau Polarization Measurement",
 OPAL Technical Note {\bf TN192} (1993).
 
 
 
 
 
 
 
 
 
 
\bibitem{Moriond94} J. Harton \etal Tau Polarization Group of LEP,
"Combining the LEP tau polarisation results",
LEPTAU {\bf 94-01} and OPAL technical note {\bf TN226} (1994).
 
\bibitem{Fetscher} W.\ Fetscher, H.J.\ Gerber and K.F.\ Johnson,
Phys.\ Lett.\ {\bf B173} (1986) 102.

\bibitem{PDG} Review of Particles Properties, K.\ Hikasa \etal
(Partcle Data Group), Phys.\ Rev. {\bf D45} (1992) Part II. 
 
\bibitem{DERENZO} S.\ E. Derenzo, Phys. Rev. {\bf 181} (1969) 1854.
 
\bibitem{BELTRAMI} 
I.\ Beltrami \etal  Phys. Lett. {\bf B194} (1987) 326.
 
\bibitem{BULKE} B.\ Bulke \etal Phys. Rev.  {\bf D37} (1988) 587.
 
\bibitem{ARGUS} ARGUS Collaboration, Albrecht \etal
Phys.\ Lett. {\bf B324} (121), 1994.
 
\bibitem{ALEPHXI} ALEPH Collaboration, D. Buskulic, \etal
Phys.\ Lett. {\bf B321} (1994) 168.
 
\bibitem{bib-ZFITTER} D.\ Bardin \etal CERN-TH. {\bf 6443/92} (1992) (ZFITTER).
 
 
\bibitem{sinw} The LEP Electroweak Working Group,
"Updated Parameters of the \Z Resonance from Combined Preliminary Data
of the LEP Experiments",  LEPEWWG/{\bf 94-01} and OPAL Technical Note
{\bf TN235} (1994).
 
\bibitem{SLD} SLD Collaboration, K.\ Abe \etal
"Precise Measurement of the Left-Right Cross Section Asymmetry in Z
Boson Production by $e^+e^-$ Collisions"  SLCA-PUB {\bf 6456} (1994),
submitted to Phys.\ Rev.\ Lett.
\end{thebibliography}

\include{ack}
\end{document}